\documentclass[11pt,a4paper]{article}
\pdfoutput=1

\usepackage{jheppub-nosort,amsthm}

\usepackage{amsmath,amssymb,amsthm,amscd}
\usepackage{physics}
\usepackage{tikz}
\tikzset{node distance=2cm, auto}
\usepackage{epsfig}
\usepackage{psfrag,comment}
\usepackage{graphicx}
\usepackage{caption}
\usepackage{subcaption}
\usepackage{mathrsfs}
\usepackage{xcolor}
\usepackage[normalem]{ulem}
\usepackage{upgreek}
\usepackage{rotating}

\usepackage{url}
\newcommand\doilink[1]{\href{http://dx.doi.org/#1}{#1}}
\newcommand\arxivlink[1]{\href{http://arxiv.org/abs/#1}{#1}}

\newcommand{\CE}{{\mathcal E}}
\newcommand{\CF}{{\mathcal F}}

\newcommand{\CN}{{\mathcal N}}

\newcommand{\CP}{{\mathcal P}}

\newcommand{\CS}{{\mathcal S}}

\newcommand{\CZ}{{\mathcal Z}}

\newcommand{\NCC}{{\mathscr C}}

\newcommand{\NCF}{{\mathscr F}}

\newcommand{\NCP}{{\mathscr P}}

\newcommand{\NCU}{{\mathscr U}}

\def\BN{{\mathbb N}}
\def\BZ{{\mathbb Z}}
\def\BR{{\mathbb R}}
\def\BC{{\mathbb C}}
\def\BP{{\mathbb P}}

\newcommand{\be}{\begin{equation}}
\newcommand{\ee}{\end{equation}}
\newcommand{\ba}{\begin{aligned}}
\newcommand{\ea}{\end{aligned}}
\newcommand{\bea}{\begin{eqnarray}}
\newcommand{\eea}{\end{eqnarray}}
\newcommand{\bean}{\begin{eqnarray*}}
\newcommand{\eean}{\end{eqnarray*}}

\def\r{\right\rangle}

\def\1{\boldsymbol{\mathsf{1}}}
\def\0{|\1\r}
\def\im{{\mathbb{I}}{\mathrm{m}}\,}
\def\re{{\mathbb{R}}{\mathrm{e}}\,}
\def\Tr{{\mathbb{T}}{\mathrm{r}}\,}

\newcommand{\rme}{{\mathrm{e}}}
\newcommand{\rmi}{{\mathrm{i}}}
\newcommand{\rmd}{{\mathrm{d}}}

\renewcommand{\mod}{\mbox{~mod~}}
\DeclareMathOperator{\arccosh}{arccosh}

\def\XXint#1#2#3{{\setbox0=\hbox{$#1{#2#3}{\int}$}
     \vcenter{\hbox{$#2#3$}}\kern-.5\wd0}}

\newenvironment{rcases}
  {\left.\begin{aligned}}
  {\end{aligned}\right\rbrace}

\makeatletter
\newsavebox\myboxA
\newsavebox\myboxB
\newlength\mylenA

\newcommand*\widebar[2][0.75]{%
    \sbox{\myboxA}{$\m@th#2$}%
    \setbox\myboxB\null% Phantom box
    \ht\myboxB=\ht\myboxA%
    \dp\myboxB=\dp\myboxA%
    \wd\myboxB=#1\wd\myboxA% Scale phantom
    \sbox\myboxB{$\m@th\overline{\copy\myboxB}$}%  Overlined phantom
    \setlength\mylenA{\the\wd\myboxA}%   calc width diff
    \addtolength\mylenA{-\the\wd\myboxB}%
    \ifdim\wd\myboxB<\wd\myboxA%
       \rlap{\hskip 0.8\mylenA\usebox\myboxB}{\usebox\myboxA}%
    \else
        \hskip -0.5\mylenA\rlap{\usebox\myboxA}{\hskip 0.5\mylenA\usebox\myboxB}%
    \fi}
\makeatother

\newdimen\tableauside\tableauside=1.0ex
\newdimen\tableaurule\tableaurule=0.4pt
\newdimen\tableaustep
\def\phantomhrule#1{\hbox{\vbox to0pt{\hrule height\tableaurule width#1\vss}}}
\def\phantomvrule#1{\vbox{\hbox to0pt{\vrule width\tableaurule height#1\hss}}}
\def\sqr{\vbox{%
  \phantomhrule\tableaustep
  \hbox{\phantomvrule\tableaustep\kern\tableaustep\phantomvrule\tableaustep}%
  \hbox{\vbox{\phantomhrule\tableauside}\kern-\tableaurule}}}
\def\squares#1{\hbox{\count0=#1\noindent\loop\sqr
  \advance\count0 by-1 \ifnum\count0>0\repeat}}
\def\tableau#1{\vcenter{\offinterlineskip
  \tableaustep=\tableauside\advance\tableaustep by-\tableaurule
  \kern\normallineskip\hbox
    {\kern\normallineskip\vbox
      {\gettableau#1 0 }%
     \kern\normallineskip\kern\tableaurule}%
  \kern\normallineskip\kern\tableaurule}}
\def\gettableau#1{\ifnum#1=0\let\next=\null\else
\squares{#1}\let\next=\gettableau\fi\next}
\subheader{\hfill
{\small{\textsf{CERN-TH-2021-096}}}
}

\title{From Minimal Strings towards Jackiw--Teitelboim Gravity: On their Resurgence, Resonance, and Black Holes}

\author[a]{Paolo~Gregori,}
\affiliation[a]{CAMGSD, Departamento de Matem\'atica, Instituto Superior T\'ecnico,\\ Universidade de Lisboa, 1049-001 Lisboa, Portugal\\}
\emailAdd{gregori@tecnico.ulisboa.pt}

\author[b,a]{Ricardo~Schiappa\,}
\affiliation[b]{Theoretical Physics Department, CERN,\\
CH-1211 Gen\`eve 23, Switzerland}
\emailAdd{ricardo.schiappa@tecnico.ulisboa.pt}

\abstract{
Two remarkable facts about Jackiw--Teitelboim two-dimensional dilaton-gravity have been recently uncovered: this theory is dual to an ensemble of quantum mechanical theories; and such ensembles are described by a random matrix model which itself may be regarded as a special (large matter-central-charge) limit of minimal string theory. This work addresses this limit, putting it in its broader matrix-model context; comparing results between multicritical models and minimal strings (\textit{i.e.}, changing in-between multicritical and conformal backgrounds); and in both cases making the limit of large matter-central-charge precise (as such limit can also be defined for the multicritical series). These analyses are first done via spectral geometry, at both perturbative and nonperturbative levels, addressing the resurgent large-order growth of perturbation theory, alongside a calculation of nonperturbative instanton-actions and corresponding Stokes data. This calculation requires an algorithm to reach large-order, which is valid for arbitrary two-dimensional topological gravity. String equations---as derived from the Gel'fand--Dikii construction of the resolvent---are analyzed in both multicritical and minimal string theoretic contexts, and studied both perturbatively and nonperturbatively (always matching against the earlier spectral-geometry computations). The resulting solutions, as described by resurgent transseries, are shown to be resonant. The large matter-central-charge limit is addressed---in the string-equation context---and, in particular, the string equation for Jackiw--Teitelboim gravity is obtained to next derivative-orders, beyond the known genus-zero case (its possible exact-form is also discussed). Finally, a discussion of gravitational perturbations to Schwarzschild-like black hole solutions in these minimal-string models, regarded as deformations of Jackiw--Teitelboim gravity, is included---alongside a brief discussion of quasinormal modes.
}


\arxivnumber{2108.11409}

\begin{document}

%%%%%%%%%%%%%%%%%%%%%%%%%%%%%%%%%%%%%%%%%%%%%%%%%%%%%%%%%%%%%%%%%
%%%%%%%%%%%%%%%%%%%%%%%%%%%%%%%%%%%%%%%%%%%%%%%%%%%%%%%%%%%%%%%%%
\maketitle
%%%%%%%%%%%%%%%%%%%%%%%%%%%%%%%%%%%%%%%%%%%%%%%%%%%%%%%%%%%%%%%%%
%%%%%%%%%%%%%%%%%%%%%%%%%%%%%%%%%%%%%%%%%%%%%%%%%%%%%%%%%%%%%%%%%

\vfill

\eject

\allowdisplaybreaks

%%%%%%%%%%%%%%%%%%%%%%%%%%%%%%%%%%%%%%%%%%%%%%%%%%%%%%%%%%%%%%%%%
%%%%%%%%%%%%%%%%%%%%%%%%%%%%%%%%%%%%%%%%%%%%%%%%%%%%%%%%%%%%%%%%%
\section{Introduction and Summary}\label{sec:intro}
%%%%%%%%%%%%%%%%%%%%%%%%%%%%%%%%%%%%%%%%%%%%%%%%%%%%%%%%%%%%%%%%%
%%%%%%%%%%%%%%%%%%%%%%%%%%%%%%%%%%%%%%%%%%%%%%%%%%%%%%%%%%%%%%%%%

Recent years have witnessed great interest and developments on an ensemble-averaged version of the renowned AdS/CFT correspondence \cite{m97}, which relates the Sachdev--Ye--Kitaev (SYK) quantum-mechanical model \cite{sy92, k15, ks17} at low temperatures to Jackiw--Teitelboim (JT) two-dimensional (2d) dilaton gravity \cite{t83, j84, h85, lgk93}---itself localizing the 2d bulk geometry to AdS$_{2}$, whilst placing its non-trivial dynamics on the one-dimensional fluctuating boundary \cite{ap14, ms16, j16, msy16, emv16, sw17} (see, \textit{e.g.}, \cite{r18, t20, f20}, for reviews on SYK, 2d dilaton gravities, and further references). This endeavor finally brought AdS/CFT to the strict quantum-mechanical realm. But, more importantly perhaps, it has created a large body of evidence supporting the possibility that the SYK model is in fact describing a \textit{quantum} black hole. Indeed, the SYK model is maximally chaotic as its Lyapunov exponents saturate the black-hole bound $\lambda_{\text{L}} = 2\pi k_{\text{B}} T_{\text{BH}}/\hbar$ \cite{ss13, mss15, p15}. Further, its spectral form factor displays random-matrix universality across time, in the by-now famous ``slope-ramp-plateau'' behavior \cite{cghps4t16, sss18}. Progress has also been substantial and fast-paced on the JT-gravity side of this (approximate) correspondence and we refer to, \textit{e.g.}, \cite{mq18, hj18, bmv19a, mt19, lmz19, ipvw19, nsv19, mstv19b, ammz19, iktv20, sy20, w20a, giky20, ms20, jm21, cs21, ssy21, hw21} for a very limited list of references. 

It is precisely in JT (quantum) gravity \textit{per se} that we are interested in, and which we turn to in the following. Fairly recently, in seminal work, Saad, Shenker, and Stanford \cite{sss19} have shown that JT gravity is actually dual to a \textit{random ensemble} of quantum mechanical systems---rather than to, as could have been expected at first, a particular, fixed quantum mechanical system. In other words, at least in dimension $d=2$, \cite{sss19} implies that there are clean instances where holography should be properly thought-of as \textit{ensemble}-holography\footnote{The idea of ``ensemble holography'' has also been uplifted from bulk dimension $d=2$ to $d=3$, in recent work \cite{acht20, mw20, pt20, cj20, h:mt20, mms21, ddkmm21, bkoz21, adkly21, cm21}, albeit it is unclear if it may hold within higher-dimensional settings \cite{mv20}.}. Making use of this duality to a random matrix model, \cite{sss19} further gave very explicit matrix-integral representations for certain sets of correlation-functions (of JT partition-function insertions). In particular, their approach allows for in-principle all-genus computations of said correlators. Finally, it was also shown in \cite{sss19} that these\footnote{Other perturbative, asymptotic series within the JT-gravity context were addressed in, \textit{e.g.}, \cite{gps21, gpps21}.} genus expansions are in fact \textit{asymptotic} (as expected within string theory on general grounds), which, via \textit{resurgence}, opens the door to the explicit analysis of \textit{all} nonperturbative\footnote{Other distinct approaches addressing the nonperturbative content of SYK/JT include \cite{akv19, b19, nsv19, l19}.} content of JT gravity. Let us also note that this matrix model formulation has been further generalized along several  directions; \textit{e.g.}, \cite{av19, sw19, i19, s19, bmv19b, h:mt20, w20b, aajnt20, sssy21, gjk21, bk21, syy21, bu21}.

The fact that (closed) string perturbation theory is asymptotic has been long known \cite{gp88} and, in fact, the precise nature of this asymptotic growth was a precursor to the discovery of D-branes \cite{s90}. Historically, detailed analyses of the large-order growth of perturbation theory---and its relation to the nonperturbative/instanton content of a given theory---began a long time ago with the seminal quantum-mechanical work in \cite{bw69, bw73, bpv78a, bpv78b}. Constructing upon these analyses, early work addressing the large-order growth of (closed) string perturbation theory itself was later carried-through within the context of double-scaled/minimal string-theory; see \cite{d91, gz91, d92, ez92, ez93}. More recently, now already under the resurgence umbrella \cite{e81}, the large-order growth of string perturbation theory, alongside that of its instanton/D-brane sectors, was studied in detail in both double-scaled/minimal and topological string theories \cite{m06, msw07, m08, msw08, ps09, gikm10, kmr10, dmp11, asv11, sv13, as13, cesv13, gmz14, cesv14, c15, csv16, cms17}. In particular, it was shown that the \textit{complete} string-theoretic asymptotics is in fact resurgent. In a nut-shell, resurgence means that chosen \textit{any} sector of the theory we wish to focus upon, perturbative or nonperturbative alike, one can find, possibly buried deep in its asymptotics, the ``resurgence'' of \textit{any other} sector---and hence, iteratively, the ``resurgence'' of the \textit{full} nonperturbative content of said theory. We refer to \cite{abs18} for an introduction to resurgence and transseries, and a complete list of references. 

In this work, we wish to initiate a nonperturbative resurgent\footnote{Note that, throughout this work, by ``resurgence'' we always imply the resurgence associated to the asymptotic growth of string perturbation theory; \textit{i.e.}, the $g_{\text{s}}$-string-coupling perturbative series. Of course there may always be asymptotic growths associated to other types of perturbations; \textit{e.g.}, in a cutoff parameter as in \cite{gpps21}---albeit this latter perturbation ends up leading to ``linear'' resurgence (associated to classical special functions and their connection formulae), rather than our present $g_{\text{s}}$ ``nonlinear'' resurgence (associated to nonlinear special functions \cite{c03} and with non-trivial Stokes data \cite{bssv22}); see \cite{abs18} for more on the distinctions.} analysis of JT gravity, alongside the would-be multicritical and/or minimal strings leading up to it---which, in fact, implies we must first address this multicritical/minimal string connection. Besides the aforementioned duality of JT gravity to random matrices, it was shown in \cite{sss19} (also following unpublished work \cite{ss19}) that one may further think of JT gravity as the multicritical-order $k \to +\infty$ limit of minimal\footnote{Specifically, we always refer to $c<1$ minimal string theory. The case of $c=1$ was later addressed in \cite{bp20}.} string theory. This was therein checked at the level of the planar matrix-model eigenvalue density, and this line of research received further support in \cite{t:mt20} now matching correlation functions between minimal string theory and matrix model sides of the duality (at both fixed and large $k$). A further growing body of evidence along these lines may be found in \cite{h:mt20, m20, km20, tuw20, gv21, msy21, fjkk21}. Furthermore, on top of these, there are two other research lines contributing to the many synergies between JT gravity and minimal strings. On the one hand, a substantial amount of work \cite{os19, os20a, os20c, os21a, os21b, os21c} has given yet another viewpoint towards JT gravity. Herein, one may now further think of JT gravity as a particular 2d topological gravity \cite{dw90, w91, k92, iz92, dw18}, where the (usually ``free'') Korteweg--de~Vries (KdV) times are set to very specific values \cite{os19}. In this topological-gravity context, the difference to minimal strings is that now there is an infinite set of KdV times turned-on (in contrast to finitely many); and this is also perfectly consistent with the above $k \to +\infty$ perspective (we shall make this very precise in the main text). On the other hand, another substantial amount of work \cite{j19, j20b, j20c, jr20, j21a, j21b} aims at placing JT gravity within the proper framework of type 0A minimal strings \cite{kms03b}, \textit{i.e.}, aims at a very specific nonperturbative completion of the JT perturbative expansion (also by building upon complex matrix models \cite{djm92a, djm92b, djm91}). In particular, this line of research has produced remarkable numerical results for nonperturbative spectral density, spectral form factor, and quantum mechanical eigenfunctions \cite{j20b, j21b}. A common theme in all aforementioned lines of research is to find proper \textit{nonperturbative} grounds for JT gravity (and perhaps multicritical and/or minimal strings as one moves along). Our goal in this work is precisely to approach the \textit{complete} nonperturbative content of JT gravity---alongside multicritical and minimal strings---\textit{via resurgence}. Building on previous resurgent analyses of multicritical gravities \cite{msw07, m08, msw08, gikm10, asv11, sv13, as13} we shall extend that work into generic multicritical and minimal string theories, as well as the latter's JT gravity limit \cite{eggls23}.

The precise contents of this paper are as described in the following. We begin in section~\ref{sec:spec_curv} by addressing the perturbative and nonperturbative contents of both double-scaled multicritical models \cite{dgz93} (in subsection~\ref{subsec:multicritical}) and minimal string models \cite{ss03} (in subsection~\ref{subsec:minimal}); and in both instances strictly from the point-of-view of the matrix-model spectral geometry. What this entails is that one may first explicitly write down the \textit{generic} spectral curve for either multicritical or minimal strings. Then, using the topological recursion \cite{eo07a}, one may compute higher-genera perturbative expansions\footnote{The complete perturbative and nonperturbative data we have computed, via the topological recursion, is included in appendix~\ref{app:top_rec}; subappendices~\ref{subapp:pert} and~\ref{subapp:nonpert}, respectively.}; and further, using matrix-model instanton calculus \cite{msw07, msw08}, one may compute instanton actions and Stokes data---always out of spectral-curve input alone. With this set-up in hand we then turn in subsection~\ref{subsec:deformations} to making JT gravity explicit as the large\footnote{By large minimal-matter central-charge we shall always mean \textit{negatively} large.} minimal-matter central-charge, or large multicritical order, of minimal string theory. Building upon the preceding discussions, this can be achieved at \textit{both} perturbative \textit{and} nonperturbative levels. An analogous limit may also be taken at the level of multicritical double-scaled models (rather than minimal strings) and we compute its spectral curve and spectral density---leaving open a possibly new ensemble-holographic correspondence. One could ask what types of limits could be possible in the case of the two-matrix model, corresponding to arbitrary $(p,q)$ minimal matter, and this is also very briefly discussed. Moving-on to section~\ref{sec:string_eqs}, we readdress the perturbative and nonperturbative contents of both double-scaled multicritical models \cite{gm90a, ds90, bk90, d90, gm90b} (in subsection~\ref{subsec:multi_string_eq}) and minimal string models \cite{ss03, ss04a} (in subsection~\ref{subsec:minimal_string_eq}); this time around from the point-of-view of the string equations. This requires first recalling the Gel'fand--Dikii construction \cite{gd75} of KdV potentials, and the asymptotic nature of the resolvent, which we first address in subsection~\ref{subsec:GDKdV_pot}. Using standard perturbative and nonperturbative \textit{ans\"atze} (\textit{e.g.}, \cite{m08, asv11}) one may then efficiently generate higher-genera perturbative expansions, compute instanton actions, and perturbative expansions around their respective instanton sectors. These results are in complete agreement with those in section~\ref{sec:spec_curv}. One then proceeds to set-up transseries solutions (in both multicritical and minimal-string models) and realizes that their respective transseries must be \textit{resonant} \cite{gikm10, asv11, abs18, bssv22}. This is shown by constructing the \textit{generic} string equation beyond genus zero---including explicit computation of generic instanton actions---, and it implies that the complete nonperturbative content of all these models goes beyond what was determined by spectral geometry. Constructing the minimal-string string-equations out of the multicritical string-equations entails a change to the so-called conformal background \cite{mss91, ss03} and a specific choice of KdV times---this is carefully discussed in subsection~\ref{subsec:general-KdV} (alongside a discussion of nonperturbative KdV flows in ``massive'' models). One may then tune all times to the JT-gravity KdV-times and, because we have previously constructed arbitrary KdV potentials beyond genus zero in subsection~\ref{subsec:multi_string_eq}, construct the part of the JT-gravity string-equation yielding nonperturbative data in subsection~\ref{subsec:JT_string_eq} (and where the form of the \textit{exact} JT-gravity string-equation is also discussed; going beyond the known genus-zero result \cite{os19, j19, t:mt20, h:mt20, tuw20}). Explicit computation of the JT instanton-actions further shows that this theory must \textit{also} be described by a resonant transseries. In section~\ref{sec:large_order} we test the plethora of nonperturbative results obtained in the previous sections, addressing the resurgent large-order behavior of perturbation theory (very much in line with tests previously done for the Painlev\'e~I and~II equations in \cite{msw07, m08, msw08, gikm10, asv11, sv13, bssv22}; see \cite{abs18} for a review). One first needs to generate large-order perturbative data, which is done by adapting a (computationally efficient) algorithm from \cite{z07, z08} (see as well \cite{os19}) and implementing it in our case. The precise way in which we do this, allows our version of the algorithm to be applicable to \textit{arbitrary} 2d topological gravity \cite{dw18}. The resurgent asymptotics of multicritical models is then addressed in subsection~\ref{subsec:multi_large_order}, and that of minimal strings in subsection~\ref{subsec:minimal_large_order}. In both cases we find excellent agreement between instanton actions and Stokes data, as predicted analytically, versus large-order asymptotics. We conclude our work in section~\ref{sec:black_holes}, addressing gravitational (scalar-field) perturbations to specific minimal-string theoretic spacetime-geometries (which are set-up in subsection~\ref{subsec:JT_perturb}), namely pure Anti-de~Sitter (AdS) spacetimes in subsection~\ref{subsec:AdS_perturb} and Schwarzschild-AdS spacetimes in subsection~\ref{subsec:black-hole_perturb}, always regarded as deformations of JT gravity. Given the interest of JT gravity within higher-dimensional black-hole dynamics \cite{ao93, fnn00, nsstv18, mtv18, mstv19a}, this analysis provides further support to a renewed interest of minimal strings towards JT-deformed higher-dimensional black-hole dynamics \cite{w20a}. In fact, in this context, the bulk gravity theory of minimal string theory is 2d dilaton gravity with a hyperbolic-sine dilaton potential \cite{t:mt20} (see as well \cite{koy17}). In both aforementioned cases this is studied perturbatively in the large matter-central-charge expansion, and quasinormal modes are briefly discussed (see as well \cite{kkm04, bsb20}). For the particular case of Schwarzschild-AdS spacetimes, there is a tantalizing solution at matter central-charge $c=0$. One final appendix~\ref{app:harm_resolv} discusses the resolvent of the quantum mechanical harmonic oscillator, mostly for pedagogical completeness of presentation.

Having set-up the resurgent, nonperturbative structure of multicritical models and minimal strings, as they lead up to JT gravity, one may ask what could be the next steps. One future research line could address extending the present bosonic/hermitian matrix-model story to its supersymmetric/unitary matrix-model counterpart (where the hierarchy of string equations gets built over the Painlev\'e~II equation rather than the Painlev\'e~I equation as herein); see, \textit{e.g.}, \cite{dss90, ps90a, ps90b, dkkmm03, kms03b, ss03, ss04b}. A resurgent analysis of the Painlev\'e~II equation has been carried through in \cite{m08, sv13}. Furthermore, much has been recently written on JT supergravity and minimal superstrings, \textit{e.g.}, \cite{sw19, j20a, j20b, m20, os20b, j20c, jrs21, j21a}. Hence, an extension of our analyses along these lines seems a rather viable next step. Another (on-going) future research line could build closer to the present (bosonic) work. One obvious future step would be to do the very same in the \textit{strict} large minimal-matter central-charge limits of both backgrounds. In closely-related upcoming work \cite{eggls23} we address the resurgent structure of JT gravity (also exploring its relation to the Kontsevich matrix model \cite{k92}, which we do not discuss in the present work). Further, in unpublished work \cite{gos21}, we address the strict large-multicritical-order $k \to +\infty$ limit of multicritical models (\textit{i.e.}, as opposed to the minimal-string models, which lead to the JT spectral curve). This ensemble of works sets-up the resurgent structure of the many incarnations of JT gravity. Having all this information in place, the final step would be to analyze the nonperturbative structure of, \textit{e.g.}, the eigenvalue spectral densities or the spectral form-factors in all these models. This may be \textit{directly} accomplished in the resurgent-transseries setting by considering the different resummations one may tackle: Borel resummations of the perturbative series at play---or Borel--\'Ecalle resummations of the transseries at play---, for analytical albeit computationally-implemented results; or transasymptotic resummations, for \textit{fully analytical} nonperturbative results. Hopefully we will report on these in the near future.

%%%%%%%%%%%%%%%%%%%%%%%%%%%%%%%%%%%%%%%%%%%%%%%%%%%%%%%%%%%%%%%%%
%%%%%%%%%%%%%%%%%%%%%%%%%%%%%%%%%%%%%%%%%%%%%%%%%%%%%%%%%%%%%%%%%
\section{Spectral Curves: Perturbative and Nonperturbative}\label{sec:spec_curv}
%%%%%%%%%%%%%%%%%%%%%%%%%%%%%%%%%%%%%%%%%%%%%%%%%%%%%%%%%%%%%%%%%
%%%%%%%%%%%%%%%%%%%%%%%%%%%%%%%%%%%%%%%%%%%%%%%%%%%%%%%%%%%%%%%%%

Let us begin by setting up some notation on hermitian one-matrix models (see, \textit{e.g.}, \cite{g91, gm93, dgz93, m04, am20a} for reviews relevant within our context). Let $M$ be an $N \times N$ matrix and $V(z)$ an (for the moment) arbitrary potential function. The matrix-model partition function is
\be
\label{eq:ZN}
\CZ_N = \frac{1}{\text{vol} \left( \text{U}(N) \right)} \int \rmd M\, \rme^{- \frac{1}{g_{\text{s}}} \Tr V(M)},
\ee
\noindent
where $\text{vol} \left( \text{U}(N) \right)$ is the volume factor of the gauge group and $g_{\text{s}}$ is the string coupling. In the large-$N$ 't~Hooft limit \cite{th74, bipz78} the matrix-model free energy $F = \log \CZ$ has a perturbative (asymptotic) genus expansion
\be
\label{eq:F}
F \simeq \sum_{g=0}^{+\infty} F_g (t)\, g_{\text{s}}^{2g-2},
\ee
\noindent
with $t = g_{\text{s}} N$ the 't~Hooft coupling. A key set of correlation functions in the matrix model, which will later play a main role in the topological recursion construction \cite{eo07a, eo08, eo09}, are the (connected) ``$h$-point multi-resolvent'' correlation functions
\be
\label{eq:Wh-multi}
W_{h} \left( x_1, \ldots, x_h \right) = \ev{\Tr \frac{1}{x_1-M} \cdots \Tr \frac{1}{x_h-M}}_{(\text{conn})}.
\ee
\noindent
These are the generating functions of multi-trace (``standard'') correlation functions,
\be
\label{eq:Wh-multi-trace}
W_{h} \left( x_1, \ldots, x_h \right) = \sum_{\left\{ n_i \ge 1 \right\}} \frac{1}{x_1^{n_1+1} \cdots x_{h}^{n_h+1}} \ev{\Tr M^{n_1} \cdots \Tr M^{n_h}}_{(\text{conn})},
\ee
\noindent
which themselves also have perturbative (asymptotic) expansion
\be
\label{eq:Wh-genus}
W_{h} \left( x_1, \ldots, x_h \right) \simeq \sum_{g=0}^{+\infty} W_{g,h} \left( x_1, \ldots, x_h; t \right) g_{\text{s}}^{2g+h-2}.
\ee

We shall be interested in (the double-scaling limit of) one-cut matrix models in the following. In this case the $\lambda$-eigenvalue spectral density $\rho (\lambda)$ has support $\NCC = (a,b)$, and saddle points of \eqref{eq:ZN} will be described by (genus zero) spectral curves which are double-sheeted coverings of $\BC$ with single-cut $\NCC$. Such saddle geometries may be swiftly constructed starting with the one-point correlator $W_{1} (x)$ (up to a factor of $\frac{1}{N}$ this is the matrix-model \textit{resolvent}), which in the planar limit relates to the eigenvalue spectral density via\footnote{It is also sometimes useful to use the inverse relation, where the eigenvalue spectral density is computed via
\be
\rho (\lambda) = - \frac{1}{2\pi\rmi t} \left\{ W_{0,1} \left(\lambda+\rmi\epsilon\right)  - W_{0,1} \left(\lambda-\rmi\epsilon\right) \right\}.
\ee
}
\be
\label{eq:W01}
W_{0,1} (x) = t \int \rmd\lambda\, \frac{\rho(\lambda)}{x-\lambda}.
\ee
\noindent
The genus-zero, one-cut, matrix-model spectral curve $y(x)$ is then written as
\be
\label{eq:spect_curv}
y(x) = V'(x) - 2\, W_{0,1} (x) \equiv M(x) \sqrt{\left( x-a \right)\left( x-b \right)},
\ee
\noindent
where, for \textit{polynomial} potentials $V(z)$, the moment function is given by \cite{ackm93}
\be
M (x) = \oint_{(0)} \frac{\rmd z}{2\pi\rmi}\, \frac{V'(1/z)}{1-z x}\, \frac{1}{\sqrt{\left( 1-a z \right) \left( 1-b z \right)}},
\ee
\noindent
and where one still needs to specify the endpoints of the cut, $\left\{ a,b \right\}$. For these, the large-$x$ asymptotics of the resolvent \eqref{eq:W01}, $W_{0,1} (x) \sim \frac{1}{x}$ (where we assumed the eigenvalue density is normalized to one), immediately yields two conditions for two unknowns:
\be
\oint_{\NCC} \frac{\rmd z}{2\pi\rmi}\, \frac{z^{n}\, V'(z)}{\sqrt{\left( z-a \right)\left( z-b \right)}} = 2t\, \delta_{n1}, \qquad n=0,1.
\ee
\noindent
We shall later see how these expressions may be very compactly written for large classes of potentials, when in the double-scaling limit for multicritical and minimal string models.

One more quantity we need to set up is the holomorphic effective potential, $V_{\text{h;eff}}'(x) = y(x)$. It is the leading contribution to the large $N$ expansion of the matrix integral,
\be
\CZ_N = \frac{1}{N!} \int \prod_{i=1}^{N} \frac{\rmd \lambda_i}{2\pi}\, \exp \left( - \frac{1}{g_{\text{s}}} \sum_{i=1}^{N} V_{\text{h;eff}}(\lambda_i) + \cdots \right),
\ee
\noindent
such that its real part yields the effective potential $V_{\text{eff}}(\lambda) = \re \int_{a}^{\lambda} \rmd x\, y(x)$ (\textit{i.e.}, it relates to the force felt by a given eigenvalue). The imaginary part of the spectral curve, on the other hand, takes us back to the eigenvalue spectral density as\footnote{\label{ftnt:rho-t}As displayed this equation holds off-criticality, with an explicit factor of $t$. Later, in the double-scaling limit, we will absorb this factor into the normalization of the spectral density, hence no longer appearing explicitly.}
\be
\label{eq:spec_dens}
\rho (\lambda) = \frac{1}{2\pi t}\, \im y(\lambda).
\ee
\noindent
The effective potential will be relevant later-on when computing instanton actions out of spectral geometry, as they follow via \cite{msw07, msw08} (see as well \cite{d91, d92, ss03, hhikkmt04, st04, iy05, iky05})
\be
\label{eq:MM-inst-action}
A = V_{\text{h;eff}}(x_0) - V_{\text{h;eff}}(b) = \int_{b}^{x_0} \rmd x\, y(x).
\ee
\noindent
Herein $x_0$ is the non-trivial saddle describing the one-instanton sector of the matrix model (the pinching of the spectral curve, or the ZZ-brane contribution in the minimal string \cite{ss03}). It is explicitly given by
\be
\label{eq:MM-saddle-pt}
V_{\text{h;eff}}'(x_0) = 0 \qquad \Rightarrow \qquad y(x_0) = 0 \qquad \Rightarrow \qquad M(x_0) = 0
\ee
\noindent
(with $x_0$ outside\footnote{Hence sometimes these are called $B$-cycle instantons. There are also $A$-cycle instantons; see, \textit{e.g.}, \cite{gz90a, ps09}.} the cut, $x_0 \not\in (a,b)$, and using the explicit form of the spectral curve \eqref{eq:spect_curv}).

Finally, let us introduce the macroscopic-loop operator of length $\beta$ \cite{bdss90},
\be
\label{eq:macro-loop-MM}
Z (\beta) = \ev{\Tr \rme^{-\beta M}}.
\ee
\noindent
The reason for this particular notation is, of course, because the matrix-model macroscopic-loop corresponds to the thermal partition-function of JT gravity \cite{sss19, os19}. Its spectral form factor then follows as $\sim Z \left(\beta+\rmi T\right) Z \left(\beta-\rmi T\right)$. The macroscopic loop easily relates to the one-point correlator $W_{1} (x)$,
\be
W_{1} (x) = - \int_{0}^{+\infty} \rmd\beta\, \rme^{\beta x}\, Z(\beta);
\ee
\noindent
hence, at leading order, to the eigenvalue spectral density\footnote{The comment in footnote~\ref{ftnt:rho-t} is also relevant herein, when we later apply this formula in the double-scaling limit.},
\be
\label{eq:rho-to-Zbeta}
Z_{0,1} (\beta) = t \int_{0}^{+\infty} \rmd\lambda\, \rme^{-\beta\lambda}\, \rho (\lambda).
\ee

%%%%%%%%%%%%%%%%%%%%%%%%%%%%%%%%%%%%%%%%%%%%%%%%%%%%%%%%%%%%%%%%%
\subsection{Multicritical One-Matrix Models}\label{subsec:multicritical}
%%%%%%%%%%%%%%%%%%%%%%%%%%%%%%%%%%%%%%%%%%%%%%%%%%%%%%%%%%%%%%%%%

Consider a one-matrix model whose large $N$ 't~Hooft limit is in a one-cut phase described by the genus-zero spectral curve \eqref{eq:spect_curv}, and zoom-in on one of the endpoints of the cut, say, on $a$. The resulting behavior of the spectral curve upon this zoom-in will very much depend on the behavior of the moment function at this very same point. For generic potential $V(z)$, hence generic moment function $M(x)$, the spectral curve just scales as $y \sim \sqrt{x-a}$. But there are special choices of potential functions $V(z)$, hence special choices of the moment function $M(x)$, that lead to so-called order-$k$ multicritical points in the matrix model \cite{dgz93}. Namely, if $M(x)$ has a zero of order $k-1$ at $x=a$, \textit{i.e.},
\be
M(x) = \left( x-a \right)^{k-1} m(x)
\ee
\noindent
with $m(x)$ regular non-vanishing at $x=a$, then the matrix model spectral curve \eqref{eq:spect_curv} becomes
\be
\label{eq:spect_curv_multicritical}
y(x) = m(x) \left( x-a \right)^{k-\frac{1}{2}} \left( x-b \right)^{\frac{1}{2}}.
\ee
\noindent
This curve defines a multicritical point of order $k$ in the one-matrix model (see, \textit{e.g.}, \cite{dgz93} for a thorough discussion and a list of references).

A particularly rich way of approaching a multicritical point is via the double-scaling limit. In the vicinity of an order-$k$ multicritical point, the matrix-model genus-$g$ free energies diverge as \cite{dgz93}
\be
F_g (t) \sim f_g \left( t-t_{\text{c}} \right)^{- \frac{2k+1}{2k} \left( 2g-2 \right)} + \cdots
\ee
\noindent
(we are just displaying the singular part near the critical point $t \to t_{\text{c}}$). The full perturbative expansion of the free energy \eqref{eq:F} may still be kept regular at criticality as long as one implements the \textit{double}-scaling limit: take $g_{\text{s}} \to 0$ at the same time as approaching $t_{\text{c}}$, but in such a way as to keep the double-scaled string-coupling $\kappa_{\text{s}}$ fixed; where
\be
\label{eq:double-scaled-string-coupling}
\kappa_{\text{s}} := g_{\text{s}} \left( t-t_{\text{c}} \right)^{- \frac{2k+1}{2k}}.
\ee
\noindent
The double-scaled free energy now has the perturbative (asymptotic) genus expansion\footnote{But to keep in line with standard notation we shall revert back to denote $f_g \to F_g$ in the following.}
\be
\label{eq:Fds}
F_{\text{ds}} \simeq \sum_{g=0}^{+\infty} f_g\, \kappa_{\text{s}}^{2g-2}.
\ee
\noindent
This double-scaled free energy describes a minimal model CFT coupled to 2d gravity \cite{dgz93} (\textit{i.e.}, a minimal string theory---and we shall say more about the world-sheet theory and corresponding closed-string background(s) in the next subsection). Recall that the minimal series \cite{bpz84} is labelled by two coprime integers $\left( p,q \right)$ leading to central charge
\be
\label{eq:(p,q)-c}
c_{p,q} = 1 - 6\, \frac{\left( p-q \right)^2}{pq} < 1.
\ee
\noindent
This is not quite what one gets starting with the \textit{one}-matrix model \eqref{eq:ZN}. In this one-matrix model case, an order $k$ multicritical point leads to a double-scaled model of $\left( 2,2k-1 \right)$ minimal matter coupled to 2d gravity (the full minimal series may be found starting off with a \textit{two}-matrix model, and we shall say slightly more about this in subsection~\ref{subsec:deformations}) \cite{dgz93}.

The study of double-scaled multicritical models, historically denoted by 2d quantum gravity or 2d string theory, was first carried out in \cite{gm90a, ds90, bk90, d90, gm90b}. In the $\left( 2,2k-1 \right)$ one-matrix model case, the \textit{matter} central charge is (increasingly negative)
\be
\label{eq:(2,2k-1)-c}
c_{k} = 1 - 3\, \frac{\left( 2k-3 \right)^2}{2k-1},
\ee
\noindent
the string critical-exponent is
\be
\gamma = - \frac{1}{k},
\ee
\noindent
and the double-scaled string coupling $\kappa_{\text{s}}$ relates to the ``double-scaled variable'' $z$ (more on this variable in section~\ref{sec:string_eqs}; for the moment just a convenient parametrization) as
\be
\label{eq:kappa-to-z}
\kappa_{\text{s}}^2 = z^{\gamma-2}.
\ee

%%%%%%%%%%%%%%%%%%%%%%%%%%%%%%%%%%%%%%%%%%%%%%%%%%%%%%%%%%%%%%%%%
\subsubsection*{Spectral Geometries and Perturbative Expansions}
%%%%%%%%%%%%%%%%%%%%%%%%%%%%%%%%%%%%%%%%%%%%%%%%%%%%%%%%%%%%%%%%%

An alternative approach to the spectral-geometry framework for solving (large $N$) matrix models is that of orthogonal polynomials \cite{biz80} (see, \textit{e.g.}, \cite{dgz93, m04} for reviews). Herein, one introduces orthogonal polynomials $p_n (x)$ (orthogonal with respect to the natural positive-definite measure constructed from the potential function), such that $p_n (x) = x^n + \cdots$. They satisfy recursion relations of the type (herein we are considering a symmetric potential function on $\BR$)
\be
\label{eq:OP-recursion}
p_{n+1} (x) = x\, p_{n} (x) - r_{n}\, p_{n-1} (x),
\ee
\noindent
where the $\left\{ r_{n} \right\}$ are recursion coefficients dictated by the potential. In the 't~Hooft limit these recursion coefficients become functions of $t$ and $g_{\text{s}}$, satisfying a finite-difference equation dubbed the string-equation. In the double-scaling limit the (now continuous) recursion parameters double-scale to the \textit{specific heat} $u (\kappa_{\text{s}})$ of the multicritical model, whereas the matrix-model string equation becomes the multicritical \textit{string equation} (a nonlinear ordinary differential equation (ODE) for $u (\kappa_{\text{s}})$, to which we shall return in section~\ref{sec:string_eqs}) \cite{gm90a, ds90, bk90, d90, gm90b}. In this way, and using the $z$-variable we have introduced above, $z = \kappa_{\text{s}}^{-\frac{2k}{2k+1}}$, the free energy is obtained (to all orders) as
\be
\label{eq:u-to-F_ds}
F_{\text{ds}}'' (z) = - \frac{1}{2} u (z).
\ee

At leading (spherical) order one finds 
\be
\label{eq:planar_u_&_F}
u (\kappa_{\text{s}}) = \kappa_{\text{s}}^{-\frac{2}{2k+1}} + \cdots \qquad \text{ and } \qquad F_{\text{ds}} (\kappa_{\text{s}}) = - \frac{1}{2}\frac{1}{\left( 2+\frac{1}{k} \right) \left( 1+\frac{1}{k} \right)}\, \frac{1}{\kappa_{\text{s}}^{2}} + \cdots,
\ee
\noindent
whereas the full (asymptotic) perturbative expansion of the specific heat is \cite{gm90b}
\be
\label{eq:u_pert_exp}
u(z) \simeq z^{-\gamma} \left( 1 + \sum_{g=1}^{+\infty} \frac{u_g}{z^{\left(2-\gamma\right)g}} \right) = z^{\frac{1}{k}} \left( 1 + \sum_{g=1}^{+\infty} \frac{u_g}{z^{\left(2+\frac{1}{k}\right)g}} \right),
\ee
\noindent
such that the asymptotic behavior in $z$, at infinity, corresponds to the planar (spherical) limit. For the free energy, the standard string-theoretic genus expansion \eqref{eq:Fds} holds.

The next step is to actually \textit{compute} the coefficients of the different genus expansions at play---the free energy \eqref{eq:Fds}, the connected correlation functions \eqref{eq:Wh-genus}, the specific heat \eqref{eq:u_pert_exp}---to \textit{all orders} beyond the aforementioned planar limit. All these coefficients may, in principle, be iteratively computed through the topological recursion construction \cite{eo07a, eo08, eo09}, whose key ingredient is the spectral curve $y(x)$ in \eqref{eq:spect_curv}, and to which we now turn. Note that the double-scaled spectral geometry is not simply described by the spectral curve \textit{at the} multicritical point \eqref{eq:spect_curv_multicritical}; rather one has to perturb it away from the exact critical point and consider its scaling-form \cite{dgz93}. In the case of the one-matrix model $(2,2k-1)$, this resulting double-scaled spectral curve has its single-cut running from $-\infty$ to $-1$, and its explicit expression given by\footnote{Note that we are using a different overall normalization as compared to \cite{dgz93}; convenient for what follows.} \cite{dgz93}
\bea
y_k (x) &=& - \left(-1\right)^{k} k \int_{-1}^x \rmd s\, \frac{s^{k-1}}{\sqrt{x-s}} = \nonumber \\
&=& 2k\, {}_2F_1 \left( 1-k, 1; \left. \frac{3}{2}\, \right| x+1 \right) \sqrt{x+1}
\label{eq:DiFra_spect_curve}
\eea
\noindent
Herein ${}_2 F_{1} \left( a,b;c \left.\right| z \right)$ is the Gauss hypergeometric function (see, \textit{e.g.}, \cite{olbc10}). In the second line we have conveniently rewritten this double-scaled spectral curve by splitting it into a square-root part corresponding to the infinite cut, and a polynomial part (essentially given by the hypergeometric function, whose power-series truncates) corresponding to the moment function.

Let us make a small parenthesis to consider the general $(p,q)$ case---which will yield helpful formulae for what follows. In this case, it is convenient to introduce a uniformization variable $\upzeta \in \BC$ and write the spectral curve in terms of two polynomials $P_p(\upzeta)$ and $Q_q(\upzeta)$ of degrees $p$ and $q$, respectively, as
\bea
x &=& P_p (\upzeta), \\
y &=& Q_q (\upzeta).
\eea
\noindent
These two polynomials satisfy the well-known Poisson relation \cite{dgz93, eo07a}
\be
\label{eq:poissonPQ}
p\, P_p (\upzeta)\, Q_q' (\upzeta) - q\, Q_q (\upzeta)\, P_p' (\upzeta) = \text{constant}.
\ee
\noindent
Now, if $P_p(\upzeta)$ is set to be the $p$th Chebyshev polynomial of the first kind $T_{p} (\upzeta)$, defined by $T_{p} \left( \cos \theta \right) = \cos p \theta$, $P_p(\upzeta) \equiv T_p(\upzeta)$, then there is a polynomial solution of \eqref{eq:poissonPQ} whenever $q$ is of the form
\be
\label{eq:q_cond}
q = \left( 2m+1 \right) p \pm 1, \qquad m \in\mathbb{Z}.
\ee
\noindent
Such solution is explicitly given by \cite{dgz93}
\be
Q_q (\upzeta) = \sum_{n=0}^{m} \binom{q/p}{n}\, T_{q-2pn} (\upzeta).
\ee
\noindent
The class of models specified by \eqref{eq:q_cond} clearly includes the $(2,2k-1)$ multicritical one-matrix models, for which we therefore obtain an alternative expression for the spectral curve
\bea
\label{eq:multicrit_spect_curve_T-x}
x &=& 2 \upzeta^2 - 1, \\
\label{eq:multicrit_spect_curve_T-y}
y &=& - 2^{\frac{3}{2}-k} \left(-1\right)^k k\, B (k,\frac{1}{2})\, \sum_{n=0}^{\left[ \frac{k-1}{2} \right]} \binom{k-1/2}{n}\, T_{2k-1-4n}(\upzeta).
\eea
\noindent
The choice of the prefactor in \eqref{eq:multicrit_spect_curve_T-y} (where $B(a,b)$ is the Euler beta function \cite{olbc10}), as compared to the equivalent expression right above, ensures that we match the conventions used in \cite{gm90b}, which will be relevant in section~\ref{sec:string_eqs}. It is then easy to check that the two spectral curves \eqref{eq:DiFra_spect_curve} and \eqref{eq:multicrit_spect_curve_T-x}-\eqref{eq:multicrit_spect_curve_T-y} are now equivalent, since the following relation holds
\be
\label{eq:spec_curv_relation}
{}_2F_1 \left( 1-k, 1; \left. \frac{3}{2}\, \right| 2\upzeta^2 \right) \upzeta = - 2^{-k}\, (-1)^{k}\, B (k,\frac{1}{2})\, \sum_{n=0}^{\left[\frac{k-1}{2}\right]} \binom{k-1/2}{n}\, T_{2k-1-4n} (\upzeta).
\ee
\noindent
In particular, this allows us to use a more compact ``hypergeometric notation'' for the $(2,2k-1)$ multicritical spectral curve which, combining \eqref{eq:multicrit_spect_curve_T-x}-\eqref{eq:multicrit_spect_curve_T-y} and \eqref{eq:spec_curv_relation}, yields
\bea
\label{eq:multicrit_spect_curve_hyp-x}
x &=& 2 \upzeta^2 - 1, \\
\label{eq:multicrit_spect_curve_hyp-y}
y &=& 2\sqrt{2}\, k\, {}_2F_1 \left( 1-k, 1; \left. \frac{3}{2}\, \right| 2\upzeta^2 \right) \upzeta,
\eea
\noindent
or, equivalently, \eqref{eq:DiFra_spect_curve} above:
\be
\label{eq:multicrit_spect_curve_hyp2}
y_{k} (x) = 2k\, {}_2F_1 \left( 1-k, 1; \left. \frac{3}{2}\, \right| x+1 \right) \sqrt{x+1}.
\ee

We are now ready to apply the topological recursion \cite{eo07a} to our spectral curve \eqref{eq:multicrit_spect_curve_hyp-x}-\eqref{eq:multicrit_spect_curve_hyp-y} (see appendix~\ref{app:top_rec} for an overview and more details). This procedure first requires an ``upgrade'' of the multi-resolvents \eqref{eq:Wh-multi}, or, to be more precise, of their genus-expansion coefficients \eqref{eq:Wh-genus}. This corresponds to trading the multi-resolvent correlators $W_{g,h}$ by the multi-differentials $\widehat{W}_{g,h}$; for the moment  simply defined in terms of the coefficients in \eqref{eq:Wh-genus} as
\be
\label{eq:init_cond}
\widehat{W}_{0,1} \left(\upzeta\right) := - 2\upzeta\, y(\upzeta), \qquad \widehat{W}_{0,2} \left(\upzeta_1,\upzeta_2\right) := \frac{1}{\left(\upzeta_1-\upzeta_2\right)^2},
\ee
\noindent
and
\be
\label{eq:hatted-def}
\widehat{W}_{g,h} \left(\upzeta_1,\ldots,\upzeta_h\right) := 2^{2h}\, \upzeta_1 \cdots \upzeta_h\, W_{g,h} \left( 2\upzeta_1^2-1, \ldots, 2\upzeta_h^2-1 \right)
\ee
\noindent
for all other values of $g$ and $h$. The multi-differential coefficient $\widehat{W}_{g,h}(\upzeta_1,J)$, with $J=\left\{ \upzeta_2, \ldots ,\upzeta_h \right\}$, is then obtained through the following (topological recursion) recursive relation \cite{eo07a, eo08, eo09}:
\bea
\widehat{W}_{g,h} \left(\upzeta_1,J\right) &=& \underset{\upzeta \to 0}{\text{Res}}\, \Bigg\{ - \frac{1}{4 y(\upzeta)}\, \frac{1}{\upzeta_1^2-\upzeta^{2}}\, \left( \widehat{W}_{g-1,h+1} \left(\upzeta,-\upzeta,J\right) + \right. \nonumber \\
&&
\hspace{75pt}
+ \sum_{\substack{m+m'=g \\ I\sqcup I'=J}}' \left. \widehat{W}_{m,|I|+1} \left(\upzeta,I\right)\, \widehat{W}_{m',|I'|+1} \left(-\upzeta,I'\right) \right) \Bigg\}.
\label{eq:top-rec-hat}
\eea
\noindent
The initial conditions for this recursion are \eqref{eq:init_cond} alongside $\widehat{W}_{g,h}=0$ for $g<0$. The notation in the sum is explained in the topological-recursion overview of appendix~\ref{app:top_rec}. Furthermore, besides recursively yielding multi-resolvents, this recursion may also be used to compute the perturbative coefficient of the free energy in \eqref{eq:F}, $F_g$ (for $g>1$). They are obtained via
\be
\label{eq:free-en-hat}
\left(2-2g \right) F_g := \underset{\upzeta \to 0}{\text{Res}}\, \Phi \left(\upzeta\right)\, \widehat{W}_{g,1} \left(\upzeta\right).
\ee
\noindent
Herein $\Phi \left(\upzeta\right)$ is a primitive of $y(\upzeta)\, x'(\upzeta)$. Using this procedure one may finally compute free-energy coefficients for the order-$k$ multicritical model, going beyond the genus-zero result in \eqref{eq:planar_u_&_F}. The first few such coefficients are:
\bea
\label{eq:multicrit_free_en-2}
F_2 (k) &=& \frac{1}{1440 k} \left(2k+3\right) \left(k-1\right), \\
\label{eq:multicrit_free_en-3}
F_3 (k) &=& \frac{1}{362880 k^3} \left(2k+3\right) \left(k-1\right) \left(4 k^3 + 196 k^2 + 189 k + 31 \right), \\
\label{eq:multicrit_free_en-4}
F_4 (k) &=& \frac{1}{43545600 k^5} \left(2k+3\right) \left(k-1\right) \times \\
&&
\times \left(- 3376 k^6 - 2340 k^5 + 75564 k^4 + 151529 k^3 + 102342 k^2 + 26769 k + 2312\right). \nonumber
\eea
\noindent
More perturbative data for the $(2,2k-1)$ one-matrix models can be found in appendix~\ref{app:top_rec}. Note that all these coefficients precisely match the string-equation results obtained in \cite{gm90b} (up to the considered perturbative orders). We shall get back to such string equation methods in section~\ref{sec:string_eqs}. For the moment, let us next address what type of \textit{nonperturbative} information may be also obtained from the spectral curve.

%%%%%%%%%%%%%%%%%%%%%%%%%%%%%%%%%%%%%%%%%%%%%%%%%%%%%%%%%%%%%%%%%
\subsubsection*{Spectral Geometries and Instanton Actions}
%%%%%%%%%%%%%%%%%%%%%%%%%%%%%%%%%%%%%%%%%%%%%%%%%%%%%%%%%%%%%%%%%

Given that (in the large $N$ 't~Hooft limit) the matrix-model free energy $F$ is computed as an asymptotic expansion \eqref{eq:F}, whose coefficients grow factorially fast $F_g \sim \left( 2g \right)!$ \cite{gp88, s90}, resurgence dictates \cite{abs18} that there will be (multi) instanton corrections in the matrix model \cite{m06, msw07, m08, msw08, ps09, gikm10, asv11, sv13}. Of course these instanton contributions precisely correspond to the old picture of eigenvalue tunneling \cite{d91, d92}. The free energy one-instanton contribution may be computed directly from spectral geometry. Its structure is \cite{msw07}
\be
\label{eq:MM-F(1)}
F^{(1)} \simeq g_{\text{s}}^{\frac{1}{2}}\, S_1\, \rme^{- \frac{A}{g_{\text{s}}}} \left.\Big\{ F^{(1)}_{1} + g_{\text{s}}\, F^{(1)}_{2} + \cdots \right.\Big\},
\ee
\noindent
where the several quantities at play are computed in terms of spectral-curve data alone \cite{msw07} (we are still assuming the one-cut setting \eqref{eq:spect_curv}). The instanton action (which we have discussed before, including the definition of the non-trivial/pinched saddle $x_0$) is \eqref{eq:MM-inst-action},
\be
\label{eq:MM-inst-action-2}
A = V_{\text{h;eff}}(x_0) - V_{\text{h;eff}}(b) = \int_{b}^{x_0} \rmd x\, y(x),
\ee
\noindent
whereas the one-loop contribution around the one-instanton sector $F^{(1)}_{1}$---multiplied by the Stokes coefficient\footnote{To be fully precise one should actually consider the appropriate Borel residue or Stokes vector \cite{abs18}, but for the purposes of this work we shall stick with the simplified terminology of ``Stokes \textit{coefficients}''.} $S_1$ (see \cite{asv11})---is
\be
\label{eq:MM-stokes-coeff}
S_1 \cdot F^{(1)}_{1} = \frac{b-a}{4} \sqrt{\frac{1}{2\pi M'(x_0) \left.\Big[ \left( x_0-a \right) \left( x_0-b \right) \right.\Big]^{\frac{5}{2}}}}.
\ee
\noindent
For the purpose of this paper we have to consider a specification of this formula to the infinite-cut case (\textit{e.g.}, see \eqref{eq:multicrit_spect_curve_hyp2}),
\be
\label{eq:MM-stokes-coeff-inf}
S_1 \cdot F^{(1)}_{1} = \frac{1}{4} \sqrt{\frac{1}{2\pi M'(x_0)  \left( x_0-b \right) ^{\frac{5}{2}}}},
\ee
where $b$ is the position of the remaining endpoint of the cut. The two-loop contribution $F^{(1)}_{2}$ is also computed in \cite{msw07} (corresponding to what is denoted by $\mu_1 \cdot \mu_2$ and given by formulae (3.61) and (3.63) in that reference), resulting in a lengthy expression which we shall herein use in examples, but refer the reader back to \cite{msw07} for the explicit result. Higher-loop contributions are worked out in examples in \cite{m08, asv11}, and multi-instantons in \cite{m08, msw08, asv11}. These formulae may also be obtained through a nonperturbative generalization of the topological recursion, which is briefly mentioned in appendix~\ref{app:top_rec} and will be presented in detail in a closely-related paper \cite{eggls23}.

%%%%%%%%%%%%%%%%%%%%%%%%%%%%%%%%%%%%%%%%%%%%%%%%%%%%%%%%%%%%%%%%%
\begin{figure}[t!]
\centering
     \begin{subfigure}[h]{0.48\textwidth}
         \centering
         \includegraphics[width=\textwidth]{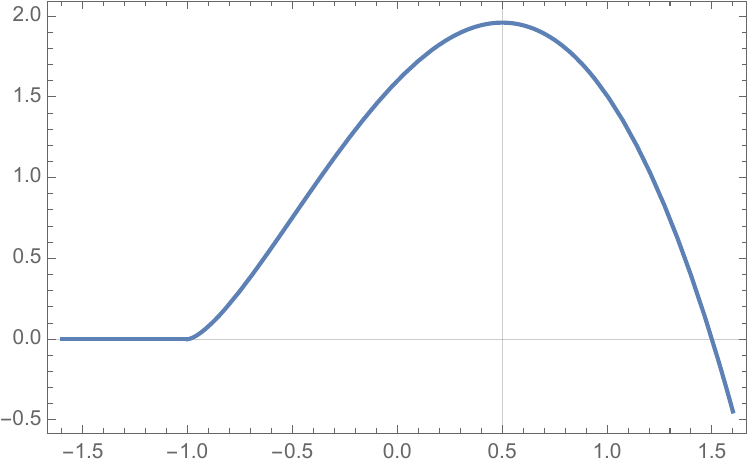}
     \end{subfigure}
\hspace{4mm}
     \begin{subfigure}[h]{0.47\textwidth}
         \centering
         \includegraphics[width=\textwidth]{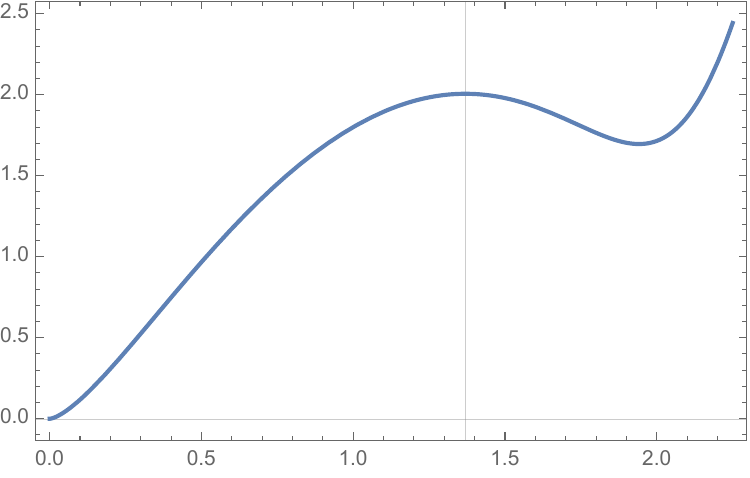}
     \end{subfigure}
\caption{On the left image, we plot the real part of the holomorphic potential for the $(2,3)$ multicritical theory. It displays a cut from $-\infty$ to $-1$ and a maximum at the non-trivial saddle $x_0=\frac{1}{2}$ (the vertical axis). The height at the saddle yields the instanton action $A_{(2,3)} = \frac{4}{5}\sqrt{6}$. On the right image, we plot the absolute value of the holomorphic effective potential for the $(2,5)$ multicritical theory, evaluated on the complex plane along the ray starting at $x=-1$ and passing through the non-trivial saddle $x_{0,+}=\frac{1}{4}\left(1+\rmi \sqrt{5}\right)$ (the vertical axis). It displays a maximum at the saddle, with height given by the absolute value of the instanton action, $\abs{A_{(2,5)}}=\frac{6}{7} \cdot 30^{\frac{1}{4}}$.}
\label{fig:multicrit_eff_pot}
\end{figure}
%%%%%%%%%%%%%%%%%%%%%%%%%%%%%%%%%%%%%%%%%%%%%%%%%%%%%%%%%%%%%%%%%

The generic $(2,2k-1)$ multicritical spectral curve \eqref{eq:multicrit_spect_curve_hyp2} has $k-1$ non-trivial saddles \eqref{eq:MM-saddle-pt}, which correspond to the solutions of (recall this is a degree-$(k-1)$ hypergeometric \textit{polynomial})
\be
\label{eq:sad-point-equat}
{}_2F_1 \left( 1-k, 1; \left. \frac{3}{2}\, \right| x+1 \right) = 0.
\ee
\noindent
They are associated to an equal number of distinct instanton actions \eqref{eq:MM-inst-action-2} and corresponding Stokes data \eqref{eq:MM-stokes-coeff-inf}. There is, however, no closed-form general solution to the hypergeometric zeroes \eqref{eq:sad-point-equat}, in which case one must proceed via step-by-step enumeration of the non-trivial saddles in each example. Let us show this explicitly for the first two members in the multicritical series, namely the $k=2$ or $(2,3)$ and the $k=3$ or $(2,5)$ multicritical theories. The first has spectral curve \eqref{eq:multicrit_spect_curve_hyp2} given by
\be
\label{eq:y-(2,3)}
y_{(2,3)} (x) = \frac{4}{3} \left(1-2x\right) \sqrt{x+1},
\ee
\noindent
which has a single non-trivial saddle located at $x_0=\frac{1}{2}$. The corresponding holomorphic effective potential is
\be
V_{\text{h;eff}}(x) = \frac{8}{15} \left(3-2x\right) \left(x+1\right)^{\frac{3}{2}}.
\ee
\noindent
This exhibits the expected maximum at $x_0$, as depicted in figure~\ref{fig:multicrit_eff_pot}. It also yields the instanton action \eqref{eq:MM-inst-action-2}
\be
\label{eq:A-(2,3)SG}
A_{(2,3)} = \int_{-1}^{1/2} \rmd x\, y(x) = \frac{4}{5}\sqrt{6}.
\ee
\noindent
Using formula \eqref{eq:MM-stokes-coeff-inf} and its extension to higher loops \cite{msw07}, it is also straightforward to obtain Stokes data alongside the contributions of the first few loops\footnote{See appendix~\ref{app:top_rec} for more data on this example.} around the one-instanton sector. These are:
\bea
\label{eq:S1F1SG(2,3)}
S_1 \cdot F^{(1)}_{1} &=& \frac{\rmi}{4 \cdot 6^{\frac{3}{4}} \sqrt{\pi}}, \\
\label{eq:F12F11SG(2,3)}
\frac{F^{(1)}_{2}}{F^{(1)}_{1}} &=& - \frac{37}{32 \sqrt{6}}, \\
\label{eq:F13F11SG(2,3)}
\frac{F^{(1)}_{3}}{F^{(1)}_{1}} &=& \frac{6433}{12288}.
\eea
\noindent
Next, we consider the spectral curve of the $(2,5)$ multicritical model. Via \eqref{eq:multicrit_spect_curve_hyp2} it is given by
\be
\label{eq:y-(2,5)}
y_{(2,5)}(x) = \frac{2}{5} \left(3-4x+8x^{2}\right) \sqrt{x+1}.
\ee
\noindent
In this case we have two non-trivial saddles,
\be
x_{0,+} = \frac{1}{4} \left(1+\rmi\sqrt{5}\right), \qquad x_{0,-} = \frac{1}{4} \left(1-\rmi \sqrt{5}\right).
\ee
\noindent
Given the holomorphic effective potential (with local maxima now located on the complex plane; see figure~\ref{fig:multicrit_eff_pot})
\be
V_{\text{h;eff}}(x) = \frac{4}{35} \left(15-12x+8x^2\right) \left(x+1\right)^{\frac{3}{2}}
\ee
\noindent
these then yield two distinct instanton actions,
\be
\label{eq:A-(2,5)SG}
A_{(2,5),\pm} = \frac{6}{7} \sqrt{5 \pm \rmi \sqrt{5}}.
\ee
\noindent
The Stokes coefficient and first few loops\footnote{See appendix~\ref{app:top_rec} for more data on this example.} around the one-instanton sector for the ``plus'' action are:
\bea
\label{eq:S1F1SG(2,5)}
S_1 \cdot F^{(1)}_{1} &=& - \frac{\left(25+5\rmi \sqrt{5}\right)^{\frac{1}{4}}}{2\sqrt{2\pi}\left(\sqrt{5}-5\rmi\right)^{\frac{3}{2}}}, \\
\label{eq:F12F11SG(2,5)}
\frac{F^{(1)}_{2}}{F^{(1)}_{1}} &=& \frac{63-8\rmi\sqrt{5}}{48\sqrt{5+\rmi\sqrt{5}}},\\
\label{eq:F13F11SG(2,5)}
\frac{F^{(1)}_{3}}{F^{(1)}_{1}} &=& \frac{13375-10963\rmi\sqrt{5}}{46080}.
\eea

%%%%%%%%%%%%%%%%%%%%%%%%%%%%%%%%%%%%%%%%%%%%%%%%%%%%%%%%%%%%%%%%%
\begin{figure}[t!]
\centering
     \begin{subfigure}[h]{0.45\textwidth}
         \centering
         \includegraphics[width=\textwidth]{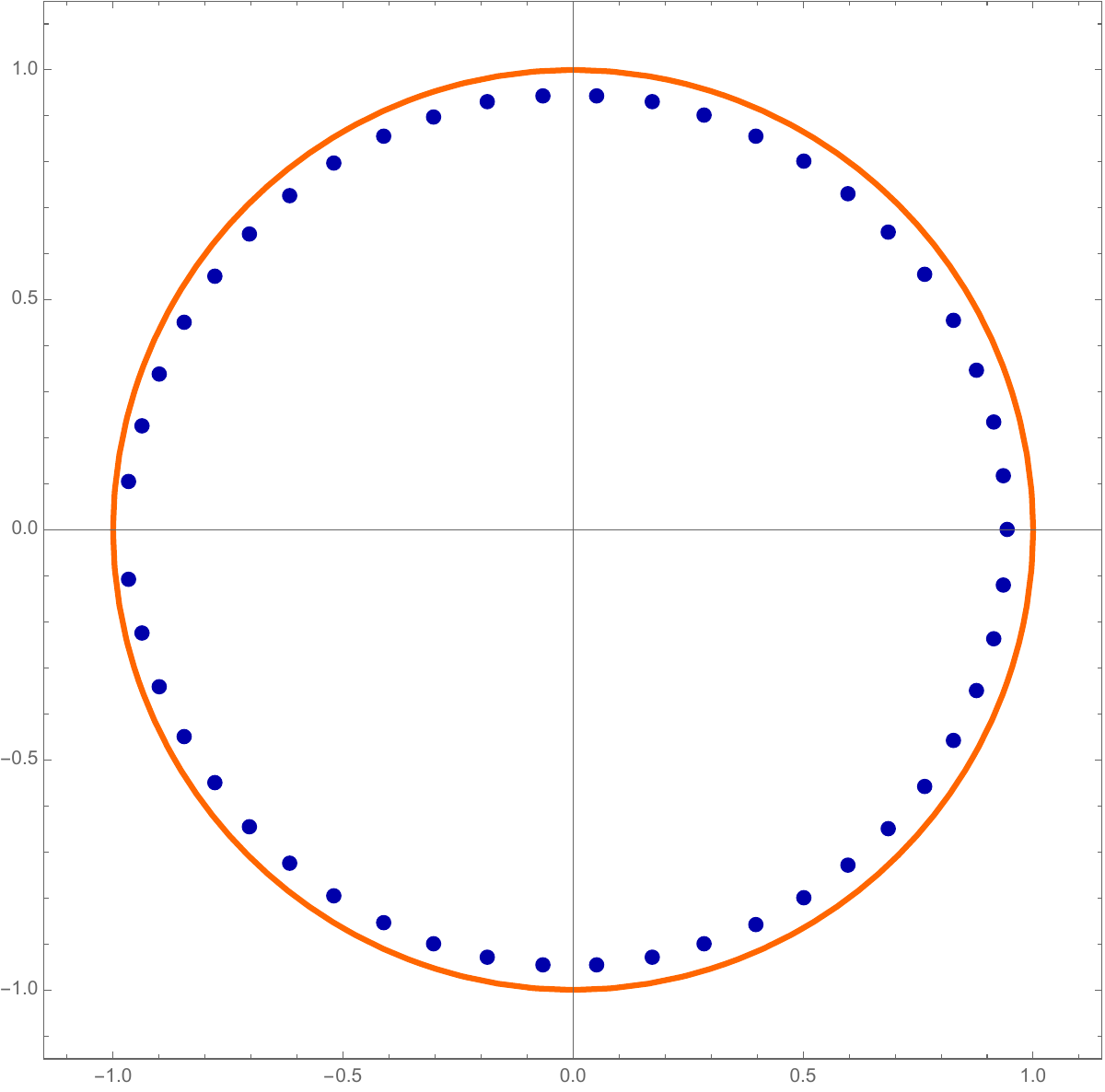}
     \end{subfigure}
\hspace{6mm}
     \begin{subfigure}[h]{0.45\textwidth}
         \centering
         \includegraphics[width=\textwidth]{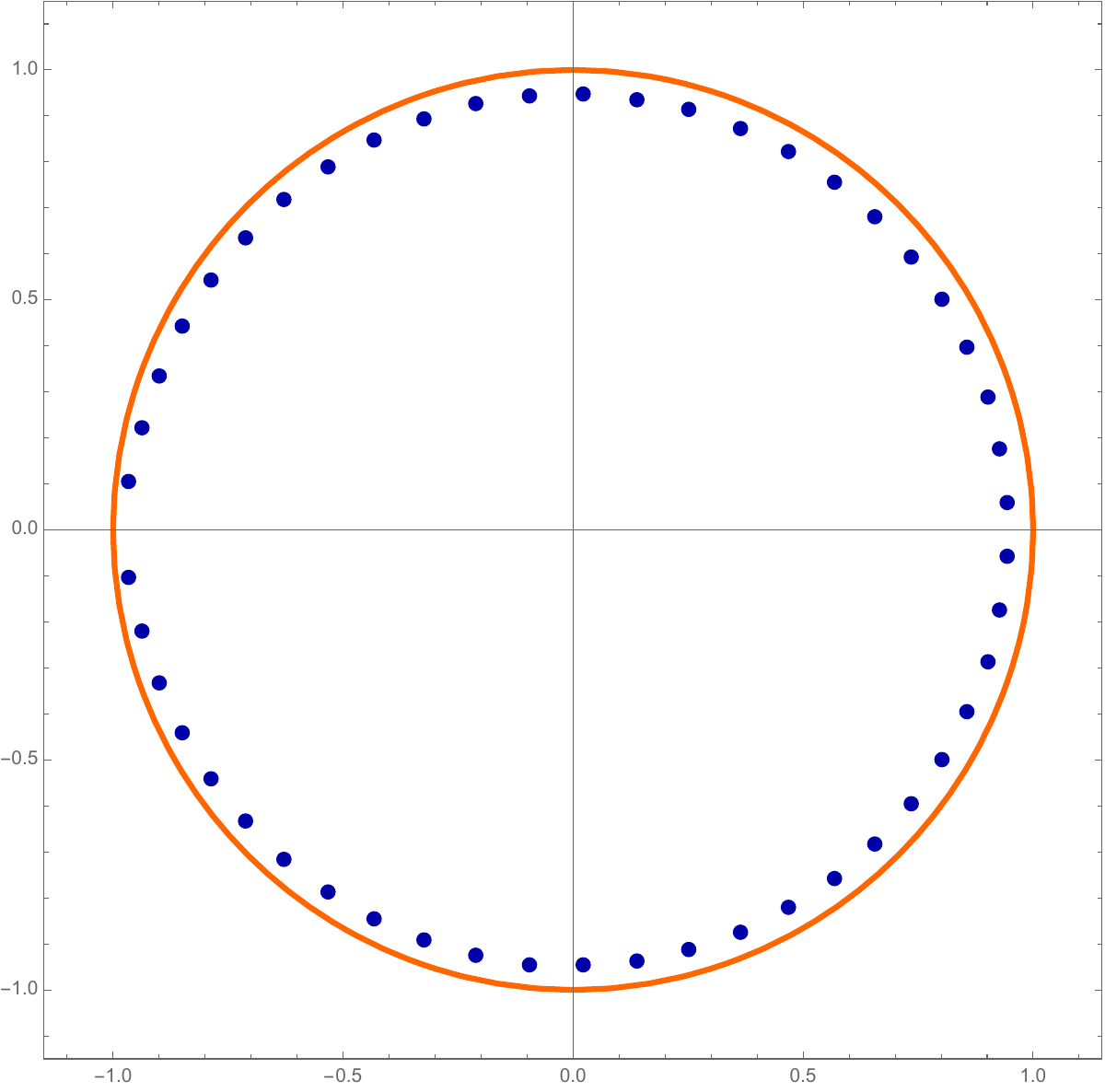}
     \end{subfigure}
\caption{Non-trivial saddles of the order-$k$ multicritical spectral curve \eqref{eq:multicrit_spect_curve_hyp2} on the complex plane. On the left image we plot the $k=50$ case, with $49$ non-trivial saddles $x_0$ given by \eqref{eq:MM-saddle-pt}; on the right image the $k=51$ case with $50$ non-trivial saddles $x_0$. Note how even $k$ ensures a real saddle. As $k$ grows these saddles asymptote to the unit circle, also plotted in both figures.}
\label{fig:multicrit_saddles}
\end{figure}
%%%%%%%%%%%%%%%%%%%%%%%%%%%%%%%%%%%%%%%%%%%%%%%%%%%%%%%%%%%%%%%%%

In the general case, the spectral geometry \eqref{eq:multicrit_spect_curve_hyp2} of the order-$k$ multicritical theory yields $k-1$ non-trivial saddles, $x_0$, corresponding to an equal number of distinct instanton actions via \eqref{eq:MM-inst-action-2}. These saddles are generically complex, appearing in complex-conjugate pairs. If $k$ is even there is always a real saddle, whereas if it is odd all saddles are complex. As $k$ grows, the saddles asymptote to the unit circle from inside its disk \cite{zsw12}. These properties are illustrated in figure~\ref{fig:multicrit_saddles}. Now, explicit determination of the corresponding $k-1$ instanton actions further requires the holomorphic effective potential, which, for the $(2,2k-1)$ multicritical model, is given by:
\be
V_{\mathrm{h;eff}} (x) = \frac{4k}{3}\, {}_2F_1 \left( 1, 1-k; \left. \frac{5}{2}\, \right| x+1 \right) \left(x+1\right)^{\frac{3}{2}}.
\ee
\noindent
Stokes data (one-loop around the one-instanton sector) associated to each non-trivial saddle $x_0$ also follows,
\be
\label{eq:multicritical-stokes}
S_1 \cdot F^{(1)}_{1} = \frac{1}{8} \sqrt{\frac{3}{2\pi}} \sqrt{\frac{1}{k \left(1-k\right) {}_2F_1 \left( 2, 2-k; \left. \frac{5}{2}\, \right| x_0+1 \right) \left(x_0+1\right)^{\frac{5}{2}}}}.
\ee
\noindent
We will return to this discussion of non-trivial saddles and their associated instanton actions in section~\ref{sec:string_eqs}---in particular realizing that spectral geometry as discussed above is only yielding \textit{half} the nonperturbative structure. The the other half, associated to \textit{resonance}, will be best understood within the setting of string equations. 

%%%%%%%%%%%%%%%%%%%%%%%%%%%%%%%%%%%%%%%%%%%%%%%%%%%%%%%%%%%%%%%%%
\subsection{Minimal String Theories}\label{subsec:minimal}
%%%%%%%%%%%%%%%%%%%%%%%%%%%%%%%%%%%%%%%%%%%%%%%%%%%%%%%%%%%%%%%%%

Up until now we have focused on the matrix-integral construction of multicritical models. One other, rather illuminating, approach is to start from the string-theoretic world-sheet theory instead \cite{m03, ss03, kopss04, mmss04} (see, \textit{e.g.}, \cite{n04, ss04a} for thorough discussions and a list of references). This world-sheet theory couples ensemble: a $(p,q)$ minimal model\footnote{We focus solely on bosonic string theory.} CFT with central charge $c<1$, specifying the ``spacetime geometry'' dictated by ``minimal matter''; a Liouville theory specifying the gauge-fixed conformal gauge, with action
\be
\label{eq:Liouville-action}
\CS_{\text{L}} [\varphi] = \frac{1}{4\pi} \int \rmd^2 \sigma \left( \left( \partial_{a} \varphi \right)^2 - 4\pi \mu\, \rme^{2b\varphi} \right)
\ee
\noindent
(herein $b>0$ is the Liouville coupling constant and $\mu$ is the bulk cosmological constant); and gauge-fixing reparametrization ghosts with usual $\mathfrak{bc}$-CFT action
\be
\CS_{\text{gh}} [\mathfrak{b},\mathfrak{c}] = \frac{1}{\pi} \int \rmd^2 \sigma\, \mathfrak{c} \bar{\partial} \mathfrak{b}.
\ee
\noindent
The ``matter'' central charge is \eqref{eq:(p,q)-c}; the Liouville central charge is $c_{\text{L}} = 1 + 6 Q^2 > 25$ (with $Q$ the Liouville background charge $Q = b + \frac{1}{b}$ required by quantum conformal invariance); and the $\mathfrak{bc}$-ghosts central charge is $c_{\text{gh}} = -26$. The standard string-theoretic requirement of vanishing total central charge translates to:
\be
\label{eq:b2=p/q}
b^2 = \frac{p}{q}.
\ee
\noindent
Note that the closed-string background constructed in this way is \textit{not} exactly the same as the natural matrix-integral closed-string background \cite{mss91, ss03}: in the former case, the spectral geometry is obtained when the exact multicritical point is deformed by the lowest dimension operator in the theory; whereas in the latter case, usually known as the \textit{conformal background} \cite{mss91}, only the bulk cosmological constant is turned on.

The nonperturbative content of minimal string theory is semiclassically described by D-branes \cite{s90, p94}, which also factorize into minimal CFT D-branes and Liouville D-branes. On the other hand, Liouville D-branes split into FZZT-branes \cite{fzz00, t00} and ZZ-branes \cite{zz01}. The FZZT-branes arise in a \textit{continuous} family parametrized by the boundary cosmological constant $\mu_{\text{B}}$ appearing in the boundary Liouville interaction
\be
\CS_{\text{B}} [\varphi] = \mu_{\text{B}} \oint \rme^{b\varphi},
\ee
\noindent
or, equivalently, and as we shall see more convenient in the following, parametrized by the (uniformization) variable $\upzeta = \cosh \frac{1}{p} \arccosh \mu_{\text{B}}$, as in $\ket{\upzeta}_{\text{FZZT}}$ with $\upzeta \in \BC$. The ZZ-branes\footnote{ZZ-branes may actually be written as the ``difference'' between two FZZT-branes \cite{zz01, m03}, in the sense that
\be
\label{eq:ZZ=FZZT-diff}
\ket{m,n}_{\text{ZZ}} = \ket{\upzeta = \cos \pi \left( \frac{m}{p} - \frac{n}{q} \right)}_{\text{FZZT}} - \ket{\upzeta = \cos \pi \left( \frac{m}{p} + \frac{n}{q} \right)}_{\text{FZZT}}.
\ee
\noindent
Herein both FZZT boundary states have the same value of $\mu_{\text{B}} = (-1)^{m} \cos \pi b^2 n$ \cite{m03}.}, on the other hand, arise in a \textit{discrete} (finite) family parametrized by two integers, $m$ and $n$, as in\footnote{The last condition is just implementing the (minimal model) reflexivity $\ket{m,n}_{\text{ZZ}} = \ket{p-m,q-n}_{\text{ZZ}}$.}
\be
\label{eq:ZZ-labeled}
\ket{m,n}_{\text{ZZ}} \qquad \text{ with } \qquad 1 \le m \le p-1, \quad 1 \le n \le q-1, \quad q m - p n > 0.
\ee
\noindent
As we shall see next, the main point for us---from the above swift overview---is not so much the world-sheet description in itself, but rather the fact that FZZT disk amplitudes result in a (closed form) spectral-geometry description of the minimal-string target space \cite{ss03}.

%%%%%%%%%%%%%%%%%%%%%%%%%%%%%%%%%%%%%%%%%%%%%%%%%%%%%%%%%%%%%%%%%
\subsubsection*{Spectral Geometries and Perturbative Expansions}
%%%%%%%%%%%%%%%%%%%%%%%%%%%%%%%%%%%%%%%%%%%%%%%%%%%%%%%%%%%%%%%%%

Finding the minimal-string target space goes as follows \cite{ss03}. Denote the FZZT disk amplitude by $\Phi (\mu_{\text{B}})$, and define (appropriate for spectral geometry) variables $x$ and $y$ by
\bea
\label{eq:x-upzeta}
x &:=& \mu_{\text{B}} = \cosh p \arccosh \upzeta, \\
\label{eq:y-upzeta}
y &:=& \frac{\partial \Phi}{\partial \mu_{\text{B}}} = \cosh q \arccosh \upzeta.
\eea
\noindent
Then they satisfy (essentially by definition) the algebraic relation
\be
\label{eq:TpTq-spect-curv}
\CE_{p,q} (x,y) \equiv T_{p} (y) - T_{q} (x) = 0,
\ee
\noindent
where $T_{p} (x)$ are again Chebyshev polynomials of first kind, $T_{p} \left( \cos \theta \right) = \cos p \theta$. This algebraic equation in $x,y$ describes a $p$-sheeted covering\footnote{In other words, at large $x$ and $y$ the equation \eqref{eq:TpTq-spect-curv} is approximately of the form $y^p \approx x^q$.} of the complex $x$-plane, which is a genus-zero Riemann surface $\Sigma_{p,q} \subset \BC^2$ with $\frac{1}{2} \left( p-1 \right) \left( q-1 \right)$ singularities corresponding to pinched $A$-cycles. These singularities occur when $x,y$ satisfy both \eqref{eq:TpTq-spect-curv} and $\partial_x \CE = 0 = \partial_y \CE$, \textit{i.e.},
\be
\label{eq:xy-singularities}
\begin{rcases}
x_{mn} &= (-1)^m \cos \frac{n \pi p}{q} \quad \\
y_{mn} &= (-1)^n \cos \frac{m \pi q}{p} \quad
\end{rcases} \quad \text{with} \quad 1 \le m \le p-1, \,\,\,\, 1 \le n \le q-1, \,\,\,\, q m - p n > 0.
\ee
\noindent
Note how this is the exact same $m,n$ labeling as for the ZZ-branes \eqref{eq:ZZ-labeled}. Equation \eqref{eq:TpTq-spect-curv} is hence the spectral curve describing the matrix-model eigenvalue distribution and target space of the minimal string \cite{ss03} (we will very soon compare with the multicritical spectral curve equation \eqref{eq:multicrit_spect_curve_hyp2}---or with its uniformization \eqref{eq:multicrit_spect_curve_hyp-x}-\eqref{eq:multicrit_spect_curve_hyp-y}). Actually, in order to be fully rigorous, herein ``target space'' does not quite imply the explicit eigenvalues but rather it refers to the (related) ``pure'' Liouville-direction closed-string background\footnote{Other backgrounds may be obtained by adding adequate operators to the world-sheet action, with couplings $\left\{ t_i \right\}$, resulting in (\textit{singularity preserving}---if keeping all $A$-cycles pinched, \textit{i.e.}, changing $B$-cycles only---or \textit{singularity destroying}---if opening up $A$-cycles, \textit{i.e.}, opening up cuts in the dual matrix model) deformations of the Riemann surface $\Sigma_{p,q}$ \cite{kopss04}. For these general classes of backgrounds, uniformization would schematically still be of the type \eqref{eq:uniformization-p} and \eqref{eq:uniformization-q} (\textit{e.g.}, degree-wise) but we would not necessarily be dealing with Chebyshev polynomials (nor would we necessarily have a single uniformization variable).}, better known as the \textit{conformal background} \cite{mss91}---where all closed-string operators have been turned off except for the cosmological constant. In particular, as hinted above, $\upzeta$ is in fact the uniformization parameter of $\Sigma_{p,q}$, \textit{i.e.},
\bea
\label{eq:uniformization-p}
x &=& T_{p} (\upzeta), \\
\label{eq:uniformization-q}
y &=& T_{q} (\upzeta),
\eea
\noindent
solves the above algebraic equation and maps (uniformizes) $\BC \to \Sigma_{p,q}$. This map holds and is one-to-one except at the aforementioned singularities which now, written in terms of the uniformizing variable, are
\be
\label{eq:upzeta-singularities}
\upzeta_{mn}^{\pm} = \cos \pi \left( \frac{m}{p} \pm \frac{n}{q} \right).
\ee
\noindent
Note how these are the exact same points as in the ``FZZT-difference'' yielding ZZ-branes \eqref{eq:ZZ=FZZT-diff}. 

As already mentioned, in this work we are mainly interested in the one-matrix model case corresponding to $\left( 2,2k-1 \right)$ minimal matter. In this case the spectral curve \eqref{eq:TpTq-spect-curv} simplifies as $T_{2} (y) = 2y^2-1$ and it follows that\footnote{This also yields the eigenvalue spectral density \eqref{eq:spec_dens} as $\rho (\lambda) = \frac{1}{2\sqrt{2}\pi}\, \im \sqrt{1 +  T_{2k-1} (\lambda)}$ (recall footnote~\ref{ftnt:rho-t}).}
\be
\label{eq:T2Tk-spect-curv}
y^2 = \frac{1}{2} \left( 1 + T_{2k-1} (x) \right).
\ee
\noindent
For example, the $k=2$ or $(2,3)$ minimal-string theory has spectral curve (this is basically the multicritical \eqref{eq:y-(2,3)} once one rescales $y \mapsto - \frac{3}{4\sqrt{2}}\, y$)
\be
\label{eq:k=2_minimal_spec_curv}
y^2 = \frac{1}{2} \left( 1 + T_{3} (x) \right) = \frac{1}{2} \left( 2x-1 \right)^2 \left( x+1 \right).
\ee
\noindent
For the $k=3$ or $(2,5)$ minimal-string theory one finds instead (compare with \eqref{eq:y-(2,5)}; now clearly distinct)
\be
\label{eq:k=3_minimal_spec_curv}
y^2 = \frac{1}{2} \left( 1 + T_{5} (x) \right) = \frac{1}{2} \left( 4x^2-2x-1 \right)^2 \left( x + 1 \right).
\ee
\noindent
These spectral curves are double-covers of the complex $x$-plane, of the general form \eqref{eq:spect_curv}, but where the (double-scaled) cut now runs from $-\infty$ to $-1$. The moment function $M(x)$ in \eqref{eq:k=2_minimal_spec_curv} and \eqref{eq:k=3_minimal_spec_curv} is clearly polynomial and immediately yields the non-trivial/pinched saddle(s) $x_0$ in \eqref{eq:MM-saddle-pt}, which are of course in one-to-one correspondence with the aforementioned singularities/pinched cycles \eqref{eq:xy-singularities} or \eqref{eq:upzeta-singularities}. Note that this is a rather dramatic improvement with respect to the multicritical case \eqref{eq:sad-point-equat} where we had no general solution for the hypergeometric zeroes. Herein, these general expressions \eqref{eq:xy-singularities} and \eqref{eq:upzeta-singularities} also simplify in the one-matrix model setting. The $\Sigma_{k}$ Riemann surface now has $\left(k-1\right)$ pinched $A$-cycles, located at $y=0$ and
\be
\label{eq:T2Tk-saddles}
x_{n} = - \cos \frac{2\pi n}{2k-1}, \qquad 1 \le n \le k-1.
\ee
\noindent
The one-matrix model minimal-string spectral curve \eqref{eq:T2Tk-spect-curv} is then, explicitly,
\be
\label{eq:T2Tk-spect-curv-explicit}
y^2 = 2^{2k-3} \left( x+1 \right) \prod_{n=1}^{k-1} \left( x-x_{n} \right)^2.
\ee

It is important to stress that the spectral curve of the $(2,2k-1)$ minimal-string--- written as an algebraic curve in \eqref{eq:T2Tk-spect-curv} or \eqref{eq:T2Tk-spect-curv-explicit}, or via uniformization in \eqref{eq:uniformization-p}-\eqref{eq:uniformization-q}---is \textit{not} equivalent to the spectral curve of the $(2,2k-1)$ multicritical theory---written as an algebraic curve in \eqref{eq:multicrit_spect_curve_hyp2}, or via ``hypergeometric'' uniformization in \eqref{eq:multicrit_spect_curve_hyp-x}-\eqref{eq:multicrit_spect_curve_hyp-y}. Their distinction is particularly clear when the uniformization is implemented via Chebyshev polynomials: whereas in both cases \eqref{eq:uniformization-p} and \eqref{eq:multicrit_spect_curve_T-x} are indeed the same, $T_2 (\upzeta) = 2\upzeta^2-1$, this is \textit{not} true for the $y$ variable. In the minimal string we find $y$ given by a \textit{single} Chebyshev polynomial \eqref{eq:uniformization-q}, as $y = T_{2k-1} (\upzeta)$; but the multicritical theory instead displays $y$ as a \textit{sum} over specific odd-index Chebyshev polynomials all the way up to $T_{2k-1} (\upzeta)$, as in \eqref{eq:multicrit_spect_curve_T-y}. The two spectral curves are however related via a non-trivial change of the closed-string background \cite{mss91, ss03}; these are the multicritical and conformal backgrounds we have already alluded to. One way to implement this change of background is to start off with the order-$k$ multicritical theory and consider its most general perturbation\footnote{One may ask why to start off from the multicritical theory---with a seemingly more intricate spectral curve---and then try to reach the minimal string theory---with a seemingly simpler spectral curve. The reason for this has to do with the string equations to be discussed in section~\ref{sec:string_eqs}, where the precise opposite occurs: string equations are simpler (and organized according to the KdV hierarchy) for the multicritical theory than for minimal strings.}, trying to reach the minimal string in the conformal background. The generally-perturbed spectral curve is given by
\be
\label{eq:spect_curv_pertMM}
y(x) = \sum_{j=1}^k t_j\, y_j(x),
\ee
\noindent
where the $y_j(x)$ are the spectral curves of the $(2,2j-1)$ multicritical models, written as in, \textit{e.g.}, \eqref{eq:multicrit_spect_curve_hyp2}, and the $t_j$ couplings are the so-called KdV times (more on these in section~\ref{sec:string_eqs}). The original multicritical theories of subsection~\ref{subsec:multicritical} are obtained by simply setting all KdV times to zero, except for $t_k=1$ (\textit{i.e.}, nothing changed). The spectral curve of the $(2,2k-1)$ minimal string in the conformal background is instead recovered from a very specific choice of KdV times \cite{mss91, ss03}
\be
\label{eq:multi-to-minimal}
y_{\text{ms}}(x) = \sum_{p=0}^{\left[\frac{k-1}{2}\right]} t_{k-2p} (k)\, y_{k-2p}(x),
\ee
\noindent
where\footnote{One property which will be useful later-on is that $t_p (k) = 0$ when $p>k$.}
\be
\label{eq:min-str-kdv-times}
t_p (k) = \frac{2^{p-\frac{5}{2}} \left(2k-1\right) \sqrt{\pi}}{\Gamma \left(p+1\right) \Gamma \left(\frac{3-k-p}{2}\right) \Gamma \left(\frac{2+k-p}{2}\right)}.
\ee
\noindent
This choice has the effect of cancelling all lower Chebyshev polynomials produced by the sum \eqref{eq:multi-to-minimal} via the many \eqref{eq:multicrit_spect_curve_T-y}, leaving us at the end solely with the $T_{2k-1} (\upzeta)$ polynomial in \eqref{eq:uniformization-q}.

We may now go back to our minimal string theory and compute its genus-$g$ free energies. In fact, having an explicit expression for the spectral curve, it is a straightforward exercise to run the topological recursion and obtain several coefficients in the genus expansion of the free energy for the $(2,2k-1)$ minimal string theory, 
\bea
\label{eq:minimal_free_en-2}
F_2 (k) &=& \frac{1}{8640}\, \frac{k \left(k-1\right)}{\left(2k-1\right)^2} \left(30 - 67 k + 110 k^2 - 86 k^3 + 43 k^4\right), \\
\label{eq:minimal_free_en-3}
F_3 (k) &=& \frac{1}{58060800}\, \frac{k \left(k-1\right)}{\left(2k-1\right)^4} \times \\
&&
\times \left(336000 - 1432000 k + 4251576 k^2 - 8784420 k^3+ 14206510 k^4 - 17769995 k^5 + \right. \nonumber \\
&&
\left.
+ 17500403 k^6 - 13101230 k^7 + 7247840 k^8 - 2648355 k^9 + 529671 k^{10}\right), \nonumber \\
\label{eq:minimal_free_en-4}
F_4 (k) &=& \frac{1}{31603654656000} \frac{k \left(k-1\right)}{\left(2k-1\right)^6} \times \\
&&
\times \left(1280240640000 - 7604222976000 k + 29932974316800 k^2 - 86544914688000 k^3 + \right. \nonumber \\
&&
+ 202262973821712 k^4 - 391913713370816 k^5+643422480514408 k^6 - \nonumber \\
&&
- 900500072437616 k^7 + 1077995918402833 k^8 - 1099742332191848 k^9 + \nonumber \\
&&
+ 949539278079604 k^{10} - 684464640926248 k^{11} + 403772326359038 k^{12} - \nonumber \\
&&
- 188563656370936 k^{13} + 66122182536188 k^{14} - 15673806936136 k^{15} + \nonumber \\
&&
\left.
+ 1959225867017 k^{16}\right). \nonumber
\eea
\noindent
More perturbative data for the $(2,2k-1)$ minimal string theories can be found in appendix~\ref{app:top_rec}.

%%%%%%%%%%%%%%%%%%%%%%%%%%%%%%%%%%%%%%%%%%%%%%%%%%%%%%%%%%%%%%%%%
\subsubsection*{Spectral Geometries and Instanton Actions}
%%%%%%%%%%%%%%%%%%%%%%%%%%%%%%%%%%%%%%%%%%%%%%%%%%%%%%%%%%%%%%%%%

The matrix-model spectral-curve interpretation of the ZZ/FZZT D-brane results discussed in the preceding paragraph is that \cite{ss03} the FZZT disk amplitude corresponds---by definitions \eqref{eq:x-upzeta} and \eqref{eq:y-upzeta}---to the matrix-model holomorphic effective potential
\be
\Phi (x) = \int^x \rmd x'\, y(x') = V_{\text{h;eff}} (x),
\ee
\noindent
and the ZZ-branes correspond to eigenvalue tunneling---which is why they are located at the $(m,n)$ singularities/pinched $A$-cycles---, such that their disk amplitudes are given by the periods along the (corresponding) conjugate $B$-cycles
\be
\Phi_{mn} = \oint_{B_{mn}} \rmd x\, y(x).
\ee
\noindent
This expression gives a spectral-geometry interpretation of the ``FZZT-difference'' yielding ZZ-branes \eqref{eq:ZZ=FZZT-diff}, and shows how the ZZ disk amplitude essentially relates to the matrix-model eigenvalue instanton action(s) \eqref{eq:MM-inst-action} or \eqref{eq:MM-inst-action-2}. Let us make these data more precise.

\begin{figure}[t!]
\centering
     \begin{subfigure}[h]{0.48\textwidth}
         \centering
         \includegraphics[width=\textwidth]{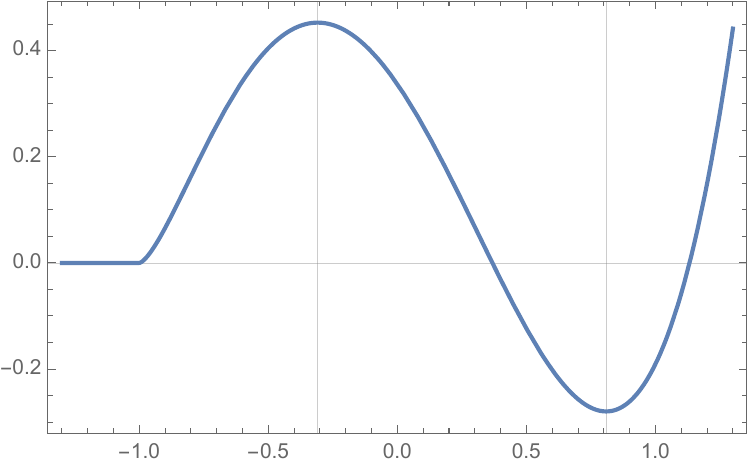}
     \end{subfigure}
\hspace{4mm}
     \begin{subfigure}[h]{0.48\textwidth}
         \centering
         \includegraphics[width=\textwidth]{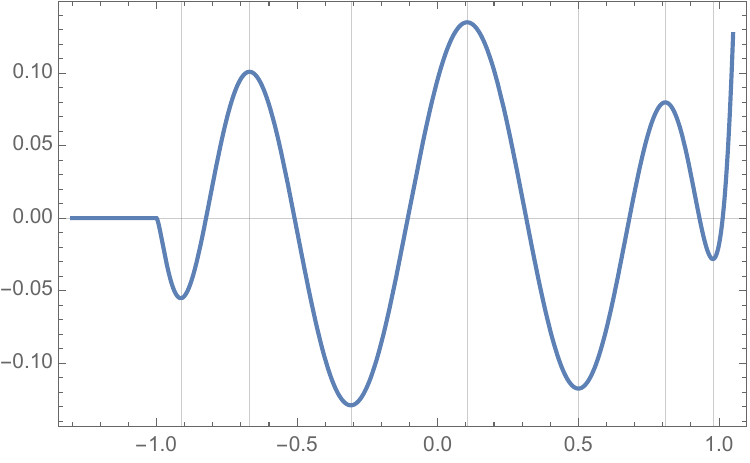}
     \end{subfigure}
\caption{These images plot the real part of the holomorphic effective potential \eqref{eq:min-str-pot} for the $(2,5)$ minimal string (left) and for the $(2,15)$ minimal string (right). Note how, in contrast with the $(2,2k-1)$ multicritical models, all the non-trivial saddles (the vertical axes) lie upon the real line, in a clear oscillating pattern.}
\label{fig:minstring_eff_pot}
\end{figure}

The minimal-string holomorphic effective potential is readily obtained by integrating the respective spectral curve \eqref{eq:T2Tk-spect-curv}, and one finds\footnote{Half-integer Chebyshev's are \textit{not} polynomials, and are best defined via their hypergeometric representation.}
\be
\label{eq:min-str-pot}
V_{\text{h;eff}} (x) = \frac{1}{2k+1}\, T_{\frac{2k+1}{2}} (x) - \frac{1}{2k-3}\, T_{\frac{2k-3}{2}} (x).
\ee
\noindent
Using standard properties of Chebyshev polynomials, \textit{e.g.}, \cite{olbc10}, it is straightforward to check that the above holomorphic effective potential has $\left(k-1\right)$ non-trivial saddles, precisely at the pinched cycles $x_n$ specified in \eqref{eq:T2Tk-saddles}. This is illustrated in two examples in figure~\ref{fig:minstring_eff_pot}, but, unlike in the previous subsection (where we were in fact a bit restricted to examples, as those in figure~\ref{fig:multicrit_eff_pot}), we may now proceed in full generality due to the many properties of Chebyshev polynomials.

Let us then follow the generic procedure we already applied to the multicritical models, and, this time around, obtain nonperturbative data for the $(2,2k-1)$ minimal string theory out of spectral geometry data alone. Evaluating the potential \eqref{eq:min-str-pot} at the saddle points \eqref{eq:T2Tk-saddles}, we immediately obtain all instanton actions for the $(2,2k-1)$ minimal string\footnote{See as well \cite{ez93, akk03} for the unitary series, and \cite{iky05} for the (double-scaled) generic two-matrix model series.} \cite{ss03, st04},
\be
\label{eq:min-string-action-n,k}
A_{(n,k)} = \left(-1\right)^{k+n} \left(\frac{1}{2k+1}+\frac{1}{2k-3}\right) \sin\frac{2\pi n}{2k-1}.
\ee
\noindent
Similarly, we can obtain the one-loop contribution (including Stokes coefficient) around each of these saddles $x_n$, by using the general formula \eqref{eq:MM-stokes-coeff-inf}. The moment function for the $(2,2k-1)$ minimal string is first readily obtained from the spectral curve \eqref{eq:T2Tk-spect-curv} (making use of recurrence and multiplication formulae for Chebyshev polynomials, \textit{e.g.}, \cite{olbc10}) as
\be
\label{eq:cheby-moment-fct}
M(x) = \frac{1}{\sqrt{2}}\, \frac{T_{k}(x)+T_{k-1}(x)}{x+1}.
\ee
\noindent
Plugging it into \eqref{eq:MM-stokes-coeff-inf} it follows
\be
\label{eq:min-string-stokes-n,k}
S_1 \cdot F^{(1)}_{1} = \frac{1}{4 \sin \frac{n \pi}{2k-1}} \sqrt{\frac{\left(-1\right)^{k+n+1}}{2 \pi \left(2k-1\right)}\, \cot\frac{n \pi}{2k-1}}.
\ee
\noindent
All these formulae are fully generic, which is a nice improvement on the results in subsection~\ref{subsec:multicritical}.

To illustrate the above formulae, let us consider the $(2,5)$ minimal string example (bear in mind that the $(2,3)$ minimal string is identical to the $(2,3)$ multicritical model, up to a rescaling of $y$). The $(2,5)$ spectral curve \eqref{eq:k=3_minimal_spec_curv} has non-trivial saddles at
\be
x_1 = -\cos \frac{2\pi}{5}, \qquad x_2=-\cos \frac{4\pi}{5},
\ee
\noindent
corresponding to two distinct and real instanton actions\footnote{One should not confuse the minimal-string actions labeling \eqref{eq:min-string-action-n,k} with the minimal-matter labeling of the actions used in the previous subsection. Hopefully which is which should be quite clear from context.}
\be
\label{eq:A-(2,5)_ms_SG}
A_{(1,3)} = \frac{5}{21}\, \sqrt{\frac{1}{2}\left(5+\sqrt{5}\right)}, \qquad A_{(2,3)} = - \frac{5}{21}\, \sqrt{\frac{1}{2}\left(5-\sqrt{5}\right)}.
\ee
\noindent
We can then write Stokes data, and the first few\footnote{More data may be found in appendix~\ref{app:top_rec}.} contributions around the first instanton action above, as
\bea
\label{eq:S1F1SG_ms_(2,5)}
S_1 \cdot F^{(1)}_{1} &=& \frac{\rmi}{2^{\frac{1}{4}} \left(5-\sqrt{5}\right)^{\frac{3}{4}} \sqrt{5 \left(\sqrt{5}-1\right) \pi}}, \\
\label{eq:F12F11SG_ms_(2,5)}
\frac{F^{(1)}_{2}}{F^{(1)}_{1}} &=&- \frac{1}{120}\, \sqrt{\frac{107791 + 49201\sqrt{5}}{2\sqrt{5}}}, \\
\label{eq:F13F11SG_ms_(2,5)}
\frac{F^{(1)}_{3}}{F^{(1)}_{1}} &=& \frac{7 \left(196475 + 69697 \sqrt{5}\right)}{288000}.
\eea
\noindent
This simple example already illustrates two important (nonperturbative) features shared by all $(2,2k-1)$ minimal string theories (and setting them apart from the corresponding multicritical models). First, all instanton actions associated to the $\left(k-1\right)$ pinched $A$-cycles are \textit{real}. Second, their overall sign \textit{oscillates}, making all these theories globally nonperturbatively unstable.

%%%%%%%%%%%%%%%%%%%%%%%%%%%%%%%%%%%%%%%%%%%%%%%%%%%%%%%%%%%%%%%%%
%%%%%%%%%%%%%%%%%%%%%%%%%%%%%%%%%%%%%%%%%%%%%%%%%%%%%%%%%%%%%%%%%

%%%%%%%%%%%%%%%%%%%%%%%%%%%%%%%%%%%%%%%%%%%%%%%%%%%%%%%%%%%%%%%%%
\subsection{Minimal Gravities as Deformations of JT Gravity}
\label{subsec:deformations}
%%%%%%%%%%%%%%%%%%%%%%%%%%%%%%%%%%%%%%%%%%%%%%%%%%%%%%%%%%%%%%%%%

As first pointed out in \cite{sss19} (also referencing unpublished work \cite{ss19}), there is a very interesting connection between the $(2,2k-1)$ minimal string and JT gravity: the latter is the $k \to +\infty$ limit of the former (see as well \cite{t:mt20, h:mt20, m20, km20, tuw20, gv21, msy21, fjkk21, os19, os20a, os20c, os21a, os21b, os21c, j19, j20b, j20c, jr20, j21a, j21b}). This was shown in \cite{sss19} solely at the level of the eigenvalue spectral density (in the dual matrix model), but which immediately translates to the present spectral-curve scenario as we will discuss in the following. Having in mind the constructions in the previous subsections, this then allows us to make the aforementioned $k \to +\infty$ limit explicit now at \textit{both} perturbative \textit{and} nonperturbative levels.

The calculation in \cite{sss19} (marginally adapted to our chain of thought) is essentially the following. Start off with the spectral curve for the $(2,2k-1)$ minimal string, \eqref{eq:T2Tk-spect-curv}. Its corresponding eigenvalue spectral-density is, via \eqref{eq:spec_dens} (compare as well with the moment function \eqref{eq:cheby-moment-fct}),
\be
\label{eq:spec-dens-cheb}
\rho_k (\lambda) = \frac{\left(-1\right)^{k+1}}{2\sqrt{2}\pi}\, \frac{T_{k}(-\lambda) - T_{k-1}(-\lambda)}{\sqrt{-\lambda-1}}.
\ee
\noindent
Using the ``explicitly trigonometric'' version of the spectral curve \eqref{eq:T2Tk-spect-curv}, \textit{i.e.}, $y = T_{\frac{2k-1}{2}} (x)$, the above spectral density is immediately equivalently rewritten as
\be
\label{eq:spec-dens-trig}
\rho_k (\lambda) = \frac{\left(-1\right)^{k+1}}{2\pi}\, \sinh \left\{ \frac{2k-1}{2}\, \arccosh \left(-\lambda\right) \right\}.
\ee
\noindent
In the JT-gravity literature, \textit{e.g.}, \cite{sss19}, it is customary to use a different normalization from \eqref{eq:spec-dens-trig}, and to perform a rescaling and a shift of the eigenvalue spectrum $\lambda$---in such a way that the spectral density is always positive with support on the positive real axis. This resulting density is
\be
\label{eq:SSS-eigen-dens}
\rho_k (\lambda) = \frac{1}{4\pi^2}\, \sinh \left\{ \frac{2k-1}{2}\, \arccosh \left( 1 + \frac{\lambda}{\kappa} \right) \right\}.
\ee
\noindent
This result may also be obtained from a world-sheet theory computation; see, \textit{e.g.}, \cite{sss19}---in which case the rescaling constant $\kappa$ is related to the Liouville bulk cosmological-constant $\mu$ in \eqref{eq:Liouville-action} via (recall \eqref{eq:b2=p/q})
\be
\kappa^2 = \frac{\mu}{\sin \left( \pi b^2 \right)}.
\ee
\noindent 
Standard JT gravity conventions set the bulk spacetime curvature to $R=-2$, which sets $\mu = \frac{1}{4\pi b^2}$ \cite{sss19}. The $k \to +\infty$ limit of the minimal string \eqref{eq:SSS-eigen-dens} at fixed $\lambda_{\text{JT}} = \frac{1}{\pi b^2}\, \lambda$ then immediately follows and yields the JT-gravity spectral density \cite{sss19}
\be
\label{eq:JT-eigen-dens}
\rho (\lambda) = \frac{1}{4\pi^2} \sinh \left\{ \frac{2k-1}{2}\, \arccosh \left( 1 + \frac{\lambda}{\kappa} \right) \right\} \qquad \underbrace{\longrightarrow}_{k \to +\infty} \qquad \rho_{\text{JT}}(\lambda_{\text{JT}})= \frac{1}{4\pi^2} \sinh \left( 2\pi \sqrt{\lambda_{\text{JT}}} \right).
\ee
\noindent
Via \eqref{eq:rho-to-Zbeta}, its Laplace transform yields the JT-gravity disk partition-function \cite{msy16, sw17, sss19},
\be
\label{eq:JT-Z_disk}
Z_{\text{disk}} (\beta) = \frac{\rme^{\frac{\pi^2}{\beta}}}{\sqrt{16 \pi \beta^3}}
\ee
\noindent
(which to a great extent was the beginning of this story). This is, of course, the leading contribution to \eqref{eq:macro-loop-MM} (as already mentioned, the matrix-model macroscopic-loop \cite{bdss90} matches the JT-gravity thermal partition-function \cite{sss19, os19}). For the record, the disk partition function of the order-$k$ minimal string is given by \cite{fzz00, sss19}
\be
\label{eq:JT-MS-Z_disk}
Z_{\text{disk}} (\beta) = \frac{1}{4\pi b^2 \beta}\, \rme^{\kappa\beta}\, K_{1/b^2} \left( \kappa\beta \right)
\ee
\noindent
(where $K_{\nu} (z)$ is the modified Bessel function of the second kind; see, \textit{e.g.}, \cite{olbc10}).

Translating this story into the spectral-curve language is absolutely straightforward due to \eqref{eq:spec_dens}. Writing the minimal-string spectral curve via uniformization \eqref{eq:uniformization-p}-\eqref{eq:uniformization-q} it immediately follows\footnote{Herein we align ourselves with the standard JT-gravity literature---which in fact we have already done starting with the eigenvalue spectral densities \eqref{eq:SSS-eigen-dens} and \eqref{eq:JT-eigen-dens}---and uniformize $x$ as $x = \upzeta^{2}$, instead of with the usual Chebyshev polynomial $x = T_2(\upzeta) = 2\upzeta^2-1$ as we did earlier. This simply amounts to a shift and a rescaling of the spectral curve, hence does not affect much any of our considerations.}
\be
\label{eq:min-string-to-JT-spec-curv}
\begin{cases}
&x = \upzeta^{2} \\
&y = \frac{(-1)^{k-1}}{2\pi}\, T_{2k-1} \left(\frac{2\pi\upzeta}{2k-1}\right)
\end{cases} 
\qquad
\underset{k \to +\infty}{\longrightarrow}
\qquad
\begin{cases}
&x = \upzeta^{2}\\
&y = \frac{1}{2\pi} \sin \left( 2\pi\upzeta \right)
\end{cases}
\ee
\noindent
This limit immediately yields the spectral curve of the JT-gravity matrix model \cite{sss19}. Running the topological recursion on the JT spectral curve is also straightforward. The first few higher-genus free energies of JT gravity are (more on appendix~\ref{app:top_rec}) 
\bea
\label{eq:JT_free_en-2}
F_2 &=& \frac{43 \pi^{6}}{2160}, \\
\label{eq:JT_free_en-3}
F_3 &=& \frac{176557 \pi^{12}}{1209600}, \\
\label{eq:JT_free_en-4}
F_4 &=& \frac{1959225867017 \pi^{18}}{493807104000}.
\eea
\noindent
In each we recognize the Weil--Petersson volumes $V_{2,0}$, $V_{3,0}$ and $V_{4,0}$ (see, \textit{e.g.}, \cite{sss19, eggls23}). More details and a complete resurgent analysis of this JT string-theoretic asymptotic expansion are addressed in detail in a closely-related paper \cite{eggls23}.

What we want to do in here is to build upon the previous minimal-string subsection and make the large (minimal matter) central charge, $k \to +\infty$ limit explicit at both perturbative (the genus-$g$ free energies) and nonperturbative (the instanton action and Stokes data) levels. Start by considering the minimal string (shifted/rescaled) spectral curve in \eqref{eq:min-string-to-JT-spec-curv}, before the large-$k$ limit, and expand it in inverse-powers of the multicritical order as
\bea
\label{eq:rescal-spect-curv}
y (x) &=& \frac{(-1)^{k-1}}{2\pi}\, T_{2k-1} \left(\frac{2\pi\sqrt{x}}{2k-1}\right) = \\
&=& \frac{1}{2\pi} \sin \left(2\pi\sqrt{x}\right) + \frac{1}{k^{2}}\, \frac{\pi^{2}}{6} x^{3/2} \cos \left(2\pi\sqrt{x}\right) + \frac{1}{k^{3}}\, \frac{\pi^{2}}{6} x^{3/2} \cos\left(2\pi\sqrt{x}\right) + \nonumber \\
&&
+ \frac{1}{k^{4}} \left\{ \frac{\pi^{2}}{40} x^{3/2} \left(5+3\pi^{2}x\right) \cos (2\pi\sqrt{x}) - \frac{\pi^{5}}{36} x^{3} \sin (2\pi\sqrt{x}) \right\} + \cdots. \nonumber
\eea
\noindent
We shall denote this version of the spectral curve as the ``rescaled'' $(2,2k-1)$ minimal string. It explicitly shows how one may regard the $(2,2k-1)$ minimal string as a deformation of JT gravity. Of course if this is true at the level of the spectral curve, then it propagates to all perturbative and nonperturbative data we have encountered so far. For example, upon the string-coupling rescaling\footnote{Recall from appendix~\ref{app:top_rec} that the minimal-string free energies behave as
\be
F_g (k) = \frac{k \left(k-1\right)}{\left(2k-1\right)^{2g-2}}\, \NCF_{g} (k),
\ee
\noindent
with polynomial $\deg \NCF_{g} (k) = 6g-8$. Further, there is an extra factor of $2$ as the JT spectral-curve \eqref{eq:min-string-to-JT-spec-curv} was uniformized with $x=\upzeta^2$ (and not $x=2\upzeta^2-1$).} $\kappa_{\text{s}} \to \frac{16\pi^3}{\left(2k-1\right)^2}\, \kappa_{\text{s}}$ in \eqref{eq:Fds}, the minimal-string free energies \eqref{eq:minimal_free_en-2}-\eqref{eq:minimal_free_en-3}-\eqref{eq:minimal_free_en-4} may now be written as deformations of the JT free energies \eqref{eq:JT_free_en-2}-\eqref{eq:JT_free_en-3}-\eqref{eq:JT_free_en-4},
\bea
\label{eq:minimal-to-JT-F2}
F_2 (k) &=& \frac{43\pi^6}{2160} + \frac{1}{k^2}\, \frac{139\pi^6}{8640} + \frac{1}{k^3}\, \frac{139\pi^6}{8640} + \frac{1}{k^4}\, \frac{49\pi^6}{3456} + \cdots, \\
\label{eq:minimal-to-JT-F3}
F_3 (k) &=& \frac{176557\pi^{12}}{1209600} + \frac{1}{k^2}\, \frac{2313247\pi^{12}}{7257600} + \frac{1}{k^3}\, \frac{2313247\pi^{12}}{7257600} + \frac{1}{k^4}\, \frac{6625247\pi^{12}}{11612160} + \cdots, \\
\label{eq:minimal-to-JT-F4}
F_4 (k) &=& \frac{1959225867017\pi^{18}}{493807104000} + \frac{1}{k^2}\, \frac{1828160015713\pi^{18}}{131681894400} + \frac{1}{k^3}\, \frac{1828160015713\pi^{18}}{131681894400} + \nonumber \\
&&
+ \frac{1}{k^4}\, \frac{13034463953869\pi^{18}}{376233984000} + \cdots.
\eea
\noindent
At nonperturbative level, and upon the very same rescaling now adequately applied to the one-instanton corrections in \eqref{eq:MM-F(1)}, the minimal-string instanton action \eqref{eq:min-string-action-n,k} and Stokes data \eqref{eq:min-string-stokes-n,k} have the large-$k$ expansions
\bea
A_{(n,k)} &=& \left(-1\right)^{n+1} \left( \frac{n}{4\pi^2} + \frac{1}{k^2} \left( \frac{n}{4\pi^2}-\frac{n^3}{24} \right) + \frac{1}{k^3} \left( \frac{n}{4\pi^2}-\frac{n^3}{24} \right) + \right. \nonumber \\
&&
\label{eq:spectral-geo-JT-grav-Ank}
\left.
+ \frac{1}{k^4}\, \frac{7}{4} \left( \frac{n}{4\pi^2}-\frac{n^3}{24}+\frac{n^5\pi^2}{840} \right) + \cdots \right), \\
S_1 \cdot F^{(1)}_{1} &=& \sqrt{\frac{\left(-1\right)^{n+1}}{2\pi}} \left( \frac{1}{n^{\frac{3}{2}}} - \frac{1}{k^4}\, \frac{\pi^4\, n^{\frac{5}{2}}}{480} - \frac{1}{k^5}\, \frac{\pi^4\, n^{\frac{5}{2}}}{240} - \right. \nonumber \\
&&
\label{eq:spectral-geo-JT-grav-S1F1}
\left.
- \frac{1}{k^6}\, \frac{\pi^4\, n^{\frac{5}{2}}}{12096} \left( 63+2\pi^2 n^2 \right) + \cdots \right).
\eea
\noindent
In the strict large-$k$ limit, this precisely matches the nonperturbative structure of JT gravity \cite{eggls23}. For convenience of the reader, a summary of the above (strict) large matter-central-charge limit of minimal string theory is included in table~\ref{table:tab1}.

%%%%%%%%%%%%%%%%%%%%%%%%%%%%%%%%%%%%%%%%%%%%%%%%%%%%%%%%%%%%%%%%%
\begin{table}[h!]
\centering
\begin{tabular}{ |c|c|c|c| } 
\hline
& $(2,2k-1)$ MS & $(2,2k-1)$ rMS & JT \\
\hline
$y(x)$ & $T_{\frac{2k-1}{2}}(x)$ & $\frac{(-1)^{k-1}}{2\pi}\, T_{2k-1} \left( \frac{2\pi\sqrt{x}}{2k-1} \right)$ & $\frac{\sin \left( 2\pi\sqrt{x} \right)}{2\pi}$ \\
$x_n$ & $- \cos \frac{2\pi n}{2k-1}$ & $\frac{\left(2k-1\right)^2}{8\pi^{2}} \left( 1 - \cos \frac{2\pi n}{2k-1} \right)$ & $\frac{n^{2}}{4}$ \\
$A_{(k,n)}$ & $(-1)^{k+n}\, \frac{4k-2}{\left( 2k+1 \right) \left( 2k-3 \right)}\, \sin\frac{2\pi n}{2k-1}$ & $(-1)^{n+1}\, \frac{\left(2k-1\right)^2}{16\pi^3}\, \frac{4k-2}{\left( 2k+1 \right) \left( 2k-3 \right)}\, \sin\frac{2\pi n}{2k-1}$ & $\frac{(-1)^{n+1}\,n}{4\pi^2}$ \\ 
$S_1 \cdot F_1^{(1)}$ & $\frac{1}{4 \sin\frac{\pi n}{2k-1}}\, \sqrt{\frac{\left(-1\right)^{k+n+1}}{2\pi \left(2k-1\right)}\, \cot \frac{\pi n}{2k-1}}$ & $\frac{1}{4 \sin\frac{\pi n}{2k-1}}\, \sqrt{\frac{8\pi^2 \left(-1\right)^{n}}{\left(2k-1\right)^3}\, \cot\frac{\pi n}{2k-1}}$ & $\frac{\sqrt{\left(-1\right)^n}}{\sqrt{2\pi}\, n^{\frac{3}{2}}}$ \\ 
\hline
\end{tabular}
\caption{This table compares nonperturbative data (non-trivial saddles, instanton actions, and one-instanton Stokes data) for the $(2,2k-1)$ minimal string (MS), the $(2,2k-1)$ ``rescaled'' minimal string (rMS), and JT gravity (JT). The JT-gravity data may be obtained from the ``rescaled'' minimal-string data by a straightforward $k \to +\infty$ limit.
}
\label{table:tab1}
\end{table}
%%%%%%%%%%%%%%%%%%%%%%%%%%%%%%%%%%%%%%%%%%%%%%%%%%%%%%%%%%%%%%%%%

Given the simple limiting match between minimal-string and JT free energies in \eqref{eq:minimal-to-JT-F2}-\eqref{eq:minimal-to-JT-F3}-\eqref{eq:minimal-to-JT-F4}, one may ask if something analogous occurs for the order-$k$ multicritical models with free energies computed in \eqref{eq:multicrit_free_en-2}-\eqref{eq:multicrit_free_en-3}-\eqref{eq:multicrit_free_en-4}. If one considers the string-coupling rescaling\footnote{Recall from appendix~\ref{app:top_rec} that the multicritical free energies behave as
\be
F_g (k) = \frac{1}{k^{2g-3}} \left(2k+3\right) \left(k-1\right) \NCF_{g} (k),
\ee
\noindent
with polynomial $\deg \NCF_{g} (k) = 3g-6$.} $\kappa_{\text{s}} \to \frac{\alpha}{\sqrt{k}}\, \kappa_{\text{s}}$ in \eqref{eq:Fds}, with $\alpha \in \BR$ unspecified by our analysis, the multicritical free energies may be written as (note how now the $k$ inverse-power expansion truncates, unlike in the minimal string case)
\bea
\label{eq:multicrit_free_en-1/k-2}
F_2 (k) &=& \frac{\alpha^{2}}{720} + \frac{\alpha^{2}}{1440 k} - \frac{\alpha^{2}}{480 k^2}, \\
\label{eq:multicrit_free_en-1/k-3}
F_3 (k) &=& \frac{\alpha^{4}}{45360} + \frac{11 \alpha^{4}}{10080 k} + \frac{281 \alpha^{4}}{181440 k^2} - \frac{337 \alpha^{4}}{362880 k^3} - \frac{67 \alpha^{4}}{45360 k^4} - \frac{31 \alpha^{4}}{120960 k^5}, \\
\label{eq:multicrit_free_en-1/k-4}
F_4 (k) &=& - \frac{211 \alpha^{6}}{1360800} - \frac{1007 \alpha^{6}}{5443200 k} + \frac{13243 \alpha^{6}}{3628800 k^2} + \frac{192821 \alpha^{6}}{21772800 k^3} + \frac{18503 \alpha^{6}}{6220800 k^4} - \frac{99569 \alpha^{6}}{14515200 k^5} - \nonumber \\
&&
- \frac{275633 \alpha^{6}}{43545600 k^6} - \frac{15599 \alpha^{6}}{8709120 k^7} - \frac{289 \alpha^{6}}{1814400 k^8}.
\eea
\noindent
These expressions seem to indicate that also the $(2,2k-1)$ multicritical spectral-curve may admit a non-trivial large-$k$ limit. This expectation may be made precise by looking directly at the spectral curve, as we just did for the minimal-string/JT-gravity limit. Due to its nature as a sum over Chebyshev polynomials, \eqref{eq:multicrit_spect_curve_T-x}-\eqref{eq:multicrit_spect_curve_T-y}, this is now slightly more intricate and we first unravel this structure and recast the spectral curve as
\bea
\label{eq:multicrit_spect_curv_re-x}
x &=& T_2(\upzeta), \\
\label{eq:multicrit_spect_curv_re-y}
y &=& - \sqrt{2\pi}\, \sum_{n=1}^{2k-1} \left(-1\right)^{n} 2^{n-1}\, \frac{\Gamma \left( n+1 \right)}{\Gamma \left( n+\frac{1}{2} \right)}\, \binom{k}{n}\, \upzeta^{2n-1}.
\eea
\noindent
The above expression together with the aforementioned rescaling of the string coupling, makes it clear that the only way to obtain a sensible $k \to +\infty$ limit is via the rescaling
\be
\label{eq:multicrit_spect_curv_scal}
y(\upzeta) \to \frac{1}{\sqrt{k}}\, y \left( \sqrt{\frac{\alpha}{k}}\, \upzeta \right).
\ee
\noindent
The limit, which needs is to be taken coefficient by coefficient, is then (see, \textit{e.g.}, \cite{h64})
\be
\lim_{k\to+\infty} \frac{1}{\sqrt{k}}\, y \left( \sqrt{\frac{\alpha}{k}}\, \upzeta \right) = 2 D\!F \left(\sqrt{2\alpha}\, \upzeta\right),
\ee
\noindent
where $D\!F(z)$ is the Dawson integral defined via the imaginary error function (see, \textit{e.g.}, \cite{olbc10})
\be
D\!F(z) = \rme^{-z^{2}}\, \int_{0}^{z} \rmd t\, \rme^{t^{2}} = \frac{1}{2} \sqrt{\pi}\, \rme^{-z^{2}}\, \text{Erfi} \left( z \right).
\ee
\noindent
The resulting spectral curve for the large (matter) central-charge multicritical theory is\footnote{Albeit herein there is now no ``standard notation'' in the literature, we chose to keep it just as in the previous JT situation for ease of comparison, \textit{i.e.}, we are still uniformizing $x = \upzeta^2$.}
\be
\label{eq:dawson_spect_curv}
y(x) = 2 D\!F \left( \sqrt{2\alpha x} \right).
\ee
\noindent
The eigenvalue spectral density corresponding to this curve is, via \eqref{eq:spec_dens}, obtained via the error function as
\be
\rho (\lambda) = \frac{1}{2\sqrt{\pi}}\, \rme^{2\alpha\lambda}\, \text{Erf} \left( \sqrt{2\alpha\lambda} \right),
\ee
\noindent
and the spectral density growth is now milder than in JT gravity. Its Laplace transform yields the very simple infinite matter-central-charge multicritical disk-partition-function (the precise bulk theory\footnote{Whereas in \cite{sss19} one starts off with the JT-gravity disk partition-function and obtains the corresponding matrix-model spectral density, herein we started off with the matrix-model spectral density and obtained a disk partition function. But we still need to identify the bulk gravity theory which yields this specific partition-function. One may wonder if this bulk gravity theory would be in any way related to one of the large central-charge multicritical models described in \cite{am20b, abm21, m21}; in particular if to the one discussed in \cite{m21}.} is addressed in \cite{gos21})
\be
Z_{\text{disk}} (\beta) = \frac{\sqrt{\alpha}}{\sqrt{2\pi\beta} \left(\beta-2\alpha\right)}.
\ee
\noindent
In this framework, the $(2,2k-1)$ multicritical spectral curve may now be regarded as a deformation of the above ``Dawson spectral curve'' \eqref{eq:dawson_spect_curv}. This is simply implemented by considering its inverse-powers expansion (akin to what we did with the minimal string spectral curve in \eqref{eq:rescal-spect-curv}),
\bea
\left(-1\right)^{k} y_k (x) &=& 2 D\!F \left(\sqrt{2 \alpha x}\right) + \frac{1}{4k} \left\{ \left( 4 \alpha x-1 \right) \sqrt{2 \alpha x} - \left( 16 \alpha^2 x^{2} - 8 \alpha x - 1 \right) D\!F \left(\sqrt{2 \alpha x}\right) \right\} + \nonumber \\
&&
+ \frac{1}{96 k^2}\, \bigg\{ \sqrt{\pi} \left( 144 \alpha^2 x^2 - 128 \alpha x + 5 \right) \rme^{-2 \alpha x}\, \text{Erfi} \left( \sqrt{2 \alpha x} \right) - \nonumber \\
&&
- 2 \sqrt{2 \alpha x} \left( 36 \alpha x - 23 + 40\, {}_3F_3 \left( 2, 2, 2; 1, 1, \left. \frac{3}{2} \,\right| - 2 \alpha x \right) - \right. \nonumber \\
&&
\left. \left.
- 12\, {}_4F_4 \left( 2, 2, 2, 2; 1, 1, 1, \left. \frac{3}{2} \,\right| - 2 \alpha x \right)
\right) \right\} + \cdots.
\eea

\begin{figure}[t!]
\centering
         \includegraphics[width=0.65\textwidth]{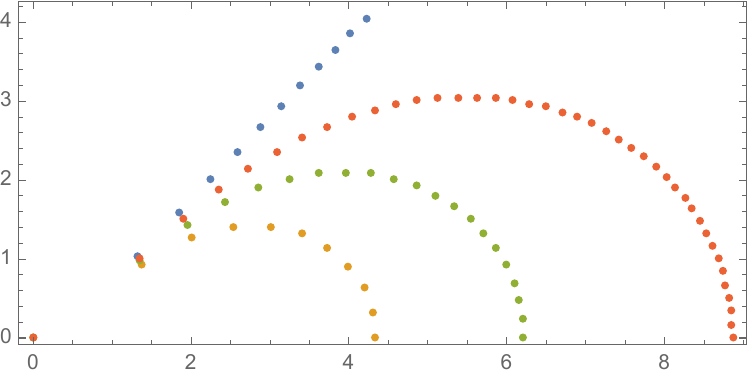}
\caption{Herein we plot the first eleven non-trivial zeroes of the ``Dawson spectral curve'' \eqref{eq:dawson_spect_curv} (blue), in the first quadrant of the complex plane. Notice how the zeroes of the $k=20$ (yellow), $k=40$ (green), and $k=80$ (red) multicritical models approach them as $k$ grows, when their corresponding spectral curve \eqref{eq:multicrit_spect_curv_re-x}-\eqref{eq:multicrit_spect_curv_re-y} is appropriately rescaled via \eqref{eq:multicrit_spect_curv_scal}.}
\label{fig:dawson_zeros}
\end{figure}

One may now work directly in the strict large-$k$ limit. At perturbative level, running the topological recursion is straightforward and the first few higher-genus free energies of ``Dawson multicritical gravity'' are (more on appendix~\ref{app:top_rec}) 
\bea
F_2 &=& \frac{\alpha^2}{720}, \\
F_3 &=& \frac{\alpha^4}{45360}, \\
F_4 &=& -\frac{\alpha^6}{1360800}.
\eea
\noindent
This indeed matches against \eqref{eq:multicrit_free_en-1/k-2}-\eqref{eq:multicrit_free_en-1/k-3}-\eqref{eq:multicrit_free_en-1/k-4}. At nonperturbative level, and just as for all multicritical models, there is now no closed-form expression for the locations of the non-trivial instanton saddles. One may still numerically analyze their location as we did earlier, as they are given by the (non-trivial) zeroes of the spectral curve \eqref{eq:dawson_spect_curv}. These are an infinite set of complex zeroes (the same as for the imaginary error function; see as well \cite{fcc73}) and they arrange themselves along two symmetry arrays, asymptotically approaching the $\pm \frac{\pi}{4}$ rays on $\BC$ (see figure~\ref{fig:dawson_zeros}). For the convenience of the reader, a summary of the (strict) large (matter) central-charge limit of multicritical models is included in table~\ref{table:tab2}.

%%%%%%%%%%%%%%%%%%%%%%%%%%%%%%%%%%%%%%%%%%%%%%%%%%%%%%%%%%%%%%%%%
%%%%%%%%%%%%%%%%%%%%%%%%%%%%%%%%%%%%%%%%%%%%%%%%%%%%%%%%%%%%%%%%%

%%%%%%%%%%%%%%%%%%%%%%%%%%%%%%%%%%%%%%%%%%%%%%%%%%%%%%%%%%%%%%%%%
\clearpage
\newpage
\begin{sidewaystable}[h!]
\centering
\begin{tabular}{ |c|c|c|c| } 
\hline
& $(2,2k-1)$ MC & $(2,2k-1)$ rMC & D \\
\hline
$y(x)$ & $2k\, {}_2F_1 \left( 1-k, 1; \left. \frac{3}{2}\, \right| x+1 \right) \sqrt{x+1}$ & $2\, {}_2F_1 \left( 1-k, 1; \left. \frac{3}{2}\, \right| \frac{x}{k} \right) \sqrt{x}$ & $2 D\!F \left(\sqrt{2\alpha x}\right)$ \\
$A_{(k,n)}$ & $\frac{4k}{3}\, {}_2F_1 \left( 1, 1-k; \left. \frac{5}{2}\, \right| x_0+1 \right) \left(1+x_0\right)^{\frac{3}{2}}$ & $\frac{4}{3}\, {}_2F_1 \left( 1, 1-k; \left. \frac{5}{2}\, \right| x_0 \right) x_0^{\frac{3}{2}}$ & $-\frac{\rmi}{2}\, \rme^{-2\alpha x_0} \left\{2 \Gamma \left(\frac{3}{2},-2\alpha x\right) - \sqrt{\pi} \right\}$ \\ 
$S_1 \cdot F_1^{(1)}$ & $\frac{1}{8} \sqrt{\frac{3}{2\pi}} \sqrt{\frac{1}{k \left(1-k\right) {}_2F_1 \left( 2, 2-k; \left. \frac{5}{2}\, \right| x_0+1 \right) \left(x_0+1\right)^{\frac{5}{2}}}}$ & $\frac{\rmi}{8} \sqrt{\frac{3}{2\pi}} \sqrt{\frac{k}{\left(k-1\right) {}_2F_1 \left( 2, 2-k; \left. \frac{5}{2}\, \right| \frac{x_0}{k} \right) x_0^{\frac{5}{2}}}}$ & $\frac{1}{4}\, \frac{1}{\sqrt{2\pi \left( \left( 1+4\alpha x_0 \right) D\!F \left(\sqrt{2\alpha x_0}\right) - \sqrt{2\alpha x_0} \right) \alpha x_0}}$ \\ 
\hline
\end{tabular}
\caption{This table compares nonperturbative data (instanton actions and one-instanton Stokes data) for the $(2,2k-1)$ multicritical model (MC), the $(2,2k-1)$ ``rescaled'' multicritical model (rMC), and the Dawson curve (D). Note that since general formulae for the position of the $n$th non-trivial saddle $x_{0,n}$ of MC, rMC, and D are not available, the expressions for the instanton action and the one-instanton Stokes data are left as functions of the $x_0$.
}
\label{table:tab2}
\end{sidewaystable}
\clearpage
\newpage
%%%%%%%%%%%%%%%%%%%%%%%%%%%%%%%%%%%%%%%%%%%%%%%%%%%%%%%%%%%%%%%%%

As already mentioned, in order to reach a multicritical point describing arbitrary minimal-model CFT $(p,q)$ with central charge
\be
c_{p,q} = 1 - 6\, \frac{\left( p-q \right)^2}{pq} < 1
\ee
\noindent
coupled to 2d quantum gravity, one has to start off from the \textit{two}-matrix model. In this case, the string critical-exponent is
\be
\gamma = - \frac{2}{p+q-1}.
\ee
\noindent
Note how for the \textit{unitary} series $(p,q)=(m,m+1)$ this critical exponent is precisely the same\footnote{Which historically led to a bit of confusion; \textit{e.g.}, see \cite{k89} versus \cite{m:s90}.} as for the \textit{one}-matrix model multicritical series $(p,q)=(2,2m-1)$ (generically non-unitary). A full analysis of these models, in either multicritical or minimal string settings, and at either perturbative or nonperturbative level, similar to what we have done up to now in this section, is beyond the scope of this paper. Nonetheless, this is obviously a very interesting route to explore in future work; even just clarifying whether there also is a sensible large matter-central-charge limit\footnote{This may very likely be in the affirmative, given the recent results in \cite{gv21}.} in this general situation. Note how in the setting of minimal string theory we still have an exact spectral curve, given by \eqref{eq:uniformization-p}-\eqref{eq:uniformization-q}, and it seems very reasonable to simply try running the topological recursion for such $(p,q)$ minimal-string spectral curve. The resulting free energies will necessarily be of the general form $F_g \equiv F_g \left(p,q\right)$. Generating even a small sequence of such functions should then clarify if also herein these may be written as some pre-factor times some possible bi-polynomial in $p$ and $q$ of fixed, $g$-dependent degrees. Just like in our previous multicritical and minimal string examples, this would tell us what types of string-coupling rescalings would be possible and what type of matter large-$c$ limits would be possible. Hopefully we will return to this question in the near future, to understand where multicritical two-matrix models sit within the broader (extended/deformed) JT gravity landscape.

%%%%%%%%%%%%%%%%%%%%%%%%%%%%%%%%%%%%%%%%%%%%%%%%%%%%%%%%%%%%%%%%%
%%%%%%%%%%%%%%%%%%%%%%%%%%%%%%%%%%%%%%%%%%%%%%%%%%%%%%%%%%%%%%%%%
\section{String Equations: Perturbative and Nonperturbative}\label{sec:string_eqs}
%%%%%%%%%%%%%%%%%%%%%%%%%%%%%%%%%%%%%%%%%%%%%%%%%%%%%%%%%%%%%%%%%
%%%%%%%%%%%%%%%%%%%%%%%%%%%%%%%%%%%%%%%%%%%%%%%%%%%%%%%%%%%%%%%%%

It has been known for a while that spectral geometry is not the complete nonperturbative story: due to \textit{resonance} \cite{gikm10, asv11, abs18, bssv22} \textit{string equations} give rise to nonperturbative corrections beyond those seen by spectral geometry \cite{gikm10, kmr10, asv11, sv13, as13, bssv22}; and these have been validated by resurgent asymptotics to be the complete set of nonperturbative corrections in simple matrix models and minimal string theories \cite{asv11, sv13, bssv22}. We first want to check that this is indeed also the case in our \textit{generic} multicritical and minimal string models. On the road towards validating that similar such structures further occur in full-fledged JT gravity, we also need a JT-gravity string equation (at least one that goes beyond the genus-zero string-equation which has been known so far in the literature \cite{os19, j19, t:mt20}). Let us set-up all these string equations.

Let us begin by setting up some notation on resolvents of Schr\"odinger operators and their main properties (see, \textit{e.g.}, \cite{l:t20} for a rather recent review). Set $\hbar=1$, $m=\frac{1}{2}$, and consider a one-dimensional potential $u(z)$. The hamiltonian operator is
\be
\label{eq:hamiltonian-u}
\mathsf{H} = - \frac{\rmd^2}{\rmd z^2} + u (z).
\ee
\noindent
The resolvent operator of $\mathsf{H}$, alongside its integral kernel, are then defined as\footnote{In the standard Schr\"odinger setting, the energy is $E=-\lambda$.}:
\be
\label{eq:resolvents}
\mathsf{R}_\lambda (\mathsf{H}) = \frac{1}{\mathsf{H} + \lambda \1}, \qquad R_\lambda (x,y) = \bra{x} \mathsf{R}_\lambda (\mathsf{H}) \ket{y}.
\ee
\noindent
As a function of $\lambda$, the resolvent is defined on a subset of $\BC$; whereas as a function of $x$ and $y$ the integral kernel is defined on a subset of $\BR^2$. The diagonal of the integral kernel $R_\lambda (z) \equiv R_\lambda (z,z)$ computes the trace of the resolvent
\be
\Tr \mathsf{R}_\lambda (\mathsf{H}) = \int_{X \subset \BR} \rmd z\, R_\lambda (z).
\ee
\noindent
For completeness of presentation, appendix~\ref{app:harm_resolv} explicitly computes the integral kernel of the resolvent for the simple harmonic oscillator with frequency $\omega=2$, \textit{i.e.}, $u(z) = z^2$.

Albeit in the following we will focus upon the resolvent, the spectral problem associated to the hamiltonian \eqref{eq:hamiltonian-u} is also of great interest. As we shall see, the potential $u(z)$ actually satisfies a nonlinear ODE (the string equation), hence has a \textit{transcendental} nature which pretty much excludes direct solutions to the associated Schr\"odinger equation. However, it turns out its eigenfunctions $\mathsf{H}\, \psi_E (z) = E\, \psi_E (z)$ are more amenable as they basically yield the FZZT partition function of the corresponding minimal string \cite{ss03, os19}, and in the present context they may be computed in a WKB approximation by applying the topological recursion to the (minimal string) spectral curve \cite{os19, os20a}. They may also be written via determinant operators in the matrix model \cite{os19, os21c}, making the FZZT connection cleaner as in \cite{mmss04}. Standard quantities of interest in the JT gravity context then follow from $\psi_E (z)$. The spectral density is \cite{bdss90, j19, j20b}
\be
\label{eq:spectral-density-QM}
\rho (E) = \int_{-\infty}^{0} \rmd z\, \abs{\psi_{E} (z)}^2,
\ee
\noindent
and its Laplace transform yields the matrix-model macroscopic loop-operator one-point function  \eqref{eq:macro-loop-MM}. Making further use of the joint spectral density
\be
\label{eq:joint-spectral-density-QM}
\rho (E,E') = \int_{-\infty}^{0} \rmd z\, \psi_{E}^* (z)\, \psi_{E'} (z),
\ee
\noindent
one may compute the spectral form factor: it is an analytic continuation \cite{sss19} of the macroscopic-loop two-point correlation function \cite{bdss90, j20b}, including both a disconnected ``two black holes'' contribution, which essentially follows from \eqref{eq:spectral-density-QM}; and a connected ``wormhole'' contribution, which follows from (the Laplace transform of) \eqref{eq:joint-spectral-density-QM}. This spectral problem was also recently addressed numerically via a Fredholm-determinant analysis in \cite{j21b}; and  considerable progress on FZZT branes in 2d topological gravity also appeared very recently \cite{os21c}. We shall report on a resurgent FZZT analysis related to the present context in the near future.

%%%%%%%%%%%%%%%%%%%%%%%%%%%%%%%%%%%%%%%%%%%%%%%%%%%%%%%%%%%%%%%%%
\subsection{Resolvent Asymptotics and Gel'fand--Dikii KdV Potentials}\label{subsec:GDKdV_pot}
%%%%%%%%%%%%%%%%%%%%%%%%%%%%%%%%%%%%%%%%%%%%%%%%%%%%%%%%%%%%%%%%%

The remarkable results in \cite{gd75} connect the asymptotic-expansion properties of $R_\lambda (z)$ as $\lambda \to + \infty$ (\textit{i.e.}, asymptotic-expansion in inverse powers of $\lambda$) to the KdV equation (and, eventually, to multicritical models and minimal string theory). Let $\mathsf{H}_0 = - \frac{\rmd^2}{\rmd z^2}$ be the free-particle hamiltonian and consider the integral kernel of its resolvent, $R_{0} = \frac{1}{2\sqrt{\lambda}}\, \rme^{- \sqrt{\lambda}\, \abs{z}}$. Expansion of the full, ``interacting'' resolvent in powers of the potential $u(z)$ and the free-particle resolvent $R_{0}$ then naturally yield an asymptotic\footnote{A simple example revealing the asymptotic properties of this expansion can be found in appendix~\ref{app:harm_resolv}.} expansion in semi-integer powers of $\lambda$ \cite{gd75}
\be
\label{eq:resolv_asymp}
R_\lambda (z) \simeq \sum_{\ell=0}^{+\infty} \frac{R_{\ell} \left[ u \right]}{\lambda^{\ell+\frac{1}{2}}}.
\ee
\noindent
The coefficients $R_{\ell} \left[ u \right]$ are polynomials in $u$ and its derivatives. To determine these coefficients, \cite{gd75} first shows that $R_\lambda (z)$ satisfies a second-order nonlinear ODE
\be
\label{eq:nonlinear_ODE_resolv}
- 2 R_\lambda (z)\, R_\lambda'' (z) + \left( R_\lambda' (z) \right)^2 + 4 \left( u(z)+\lambda \right) \left( R_\lambda (z) \right)^2 = 1,
\ee
\noindent
which in turn implies a linear albeit third-order ODE
\be
\label{eq:linear_3ODE_resolv}
R_\lambda''' (z) - 4 \left( u(z)+\lambda \right) R_\lambda' (z) - 2 u'(z)\, R_\lambda (z) = 0.
\ee
\noindent
Replacing the asymptotic-expansion \textit{ansatz} \eqref{eq:resolv_asymp} in this linear ODE yields the recursion relation
\be
\label{eq:Rell_recursion}
R_{\ell+1}' = \frac{1}{4} R_{\ell}''' - u(z)\, R_{\ell}' - \frac{1}{2} u'(z)\, R_{\ell}.
\ee
\noindent
This simply computes the Gel'fand--Dikii KdV potentials, $R_{\ell} \left[ u \right]$; the first few of which are:
\bea
\label{eq:GD-R0}
R_0 &=& \frac{1}{2}, \\
\label{eq:GD-R1}
R_1 &=& - \frac{1}{4} u, \\
\label{eq:GD-R2}
R_2 &=& \frac{1}{16} \left( 3 u^2 - u'' \right), \\
\label{eq:GD-R3}
R_3 &=& - \frac{1}{64} \left( 10 u^3 - 10 u u'' - 5 \left( u' \right)^2 + u'''' \right);
\eea
\noindent
and so on. The KdV terminology is now evident: if the potential $u(z)$ is assumed to further depend on ``time'' $\tau$, then
\be
\label{eq:KdV}
\frac{\partial u}{\partial\tau} = \frac{\partial}{\partial z} R_{2} \left[ u \right]
\ee
\noindent
is precisely the KdV equation (up to a rescaling of the ``time'' coordinate). Natural inclusion of further Gel'fand--Dikii KdV potentials leads to the generalized KdV equations \cite{gd75}
\be
\label{eq:gener_KdV}
\frac{\partial u}{\partial\tau} = \frac{\partial}{\partial z} \sum_{k=1}^{n+1} t_k R_{k} \left[ u \right],
\ee
\noindent
where the $\{ t_k \}$ are the (possibly infinite) KdV times. For $\tau$-stationary problems, the generalized KdV equations become the (order-$2n$ nonlinear) Novikov equations
\be
\label{eq:Novikov}
\sum_{k=0}^{n+1} t_k R_{k} \left[ u \right] = 0.
\ee
\noindent
As we shall also recall in the following, these equations are intimately related to the multicritical and to the minimal-string string-equations. Furthermore, the first integrals of these equations are associated to the (mutually commuting) vector fields\footnote{In this equation we denote the $i$th $z$-derivative of a function, $f(z)$, by $f^{(i)}$.} \cite{gd75}
\be
\xi_{\ell} = \sum_{i=0}^{+\infty} R_{\ell}^{(i+1)} \left[ u \right]\, \frac{\partial}{\partial u^{(i)}}, \qquad \text{ such that } \qquad \left[ \xi_{n}, \xi_{m} \right] = 0.
\ee
\noindent
In particular, the corresponding vector flows are the (higher-order) KdV flows:
\be
\label{eq:KdV_flow}
\frac{\partial u}{\partial t_\ell} = \xi_{\ell+1} \cdot u = \frac{\partial}{\partial z} R_{\ell+1} \left[ u \right].
\ee

Solutions to the Novikov equations (hence, to multicritical and minimal string-equations) satisfy a Painlev\'e-type property in which the only moveable singularities are poles, in particular \textit{double} poles \cite{c03}. This led \cite{gz90b} to consider the potential\footnote{The matrix-model eigenvalue spectral-density of this example was computed in \cite{j19}.} $u(z) = \frac{\alpha}{z^2}$ (however herein we shall assume $\alpha \in \BC$ generic\footnote{This means we consider generic potential and not any sort of solution to the Novikov equations \eqref{eq:Novikov} \cite{gz90b}.}). Evaluating a few of the Gel'fand--Dikii KdV potentials \eqref{eq:GD-R0}-\eqref{eq:GD-R3} it is straightforward to see that for this Schr\"odinger-potential their generic form is quite simple:
\be
\label{eq:GDKdV_1/x2}
R_{\ell} \left[u\right] = \frac{\rho_{\ell} (\alpha)}{z^{2\ell}},
\ee
\noindent
with $\rho_{\ell} (\alpha)$ a degree-$\ell$ polynomial\footnote{The zeroes of these polynomials are likewise simple to list, located at $\alpha = 0, 2, \ldots, j \left( j-1 \right), \ldots, \ell \left( \ell-1 \right)$.} in $\alpha$. Plugging this back into the recursion relation \eqref{eq:Rell_recursion} \cite{gz90b},
\be
\label{eq:rho_recursion}
\rho_{\ell+1} (\alpha) = \frac{2\ell+1}{2\left( \ell+1 \right)} \left( \ell\left(\ell+1\right) - \alpha \right) \rho_{\ell} (\alpha).
\ee
\noindent
We now move one step further and explicitly solve this recursion for generic $\alpha \in \BC$. The resulting expression for the $\rho_{\ell} (\alpha)$ polynomials is (where we have further fixed the overall constant):
\be
\rho_{\ell} (\alpha) = \left( -1 \right)^{\ell} \frac{\left( 2\ell-1 \right)!!}{2^{\ell+1}\, \ell!}\, \prod_{j=1}^{\ell} \left.\big( \alpha - j \left( j-1 \right) \right.\big).
\ee
\noindent
This result precisely matches our empirical expectations. Elementary large-order resurgent analysis \cite{abs18}, or else an immediate application of the recursion relation \eqref{eq:rho_recursion}, yields the expected asymptotic (factorial) growth of the coefficients \eqref{eq:GDKdV_1/x2} in the resolvent expansion \eqref{eq:resolv_asymp}; explicitly as
\be
\label{eq:rho_large-order}
\rho_{\ell}^{(0)} \simeq \frac{\cos \left( \frac{\pi}{2} \sqrt{1+4\alpha} \right)}{\pi}\, \frac{\Gamma \left( 2\ell \right)}{2^{2\ell}} \left( 1 + \frac{2 \alpha}{2k-1} + \frac{2 \alpha \left( \alpha-1 \right)}{\left( 2k-1 \right) \left( 2k-2 \right)} + \frac{2 \alpha \left( \alpha-2 \right) \left( 2\alpha-3 \right)}{3 \left( 2k-1 \right) \cdots \left( 2k-3 \right)} + \cdots \right).
\ee
\noindent
Note in particular the instanton action and Stokes coefficient\footnote{Note that, consistently, this coefficient precisely vanishes when $\alpha = j \left( j-1 \right)$ with $j \in \BZ$.},
\be
A = 2, \qquad S_1 = 2\rmi \cos \left( \frac{\pi}{2} \sqrt{1+4\alpha} \right).
\ee
\noindent
A full one-instanton sector is making its appearance in \eqref{eq:rho_large-order}, implying that not only the resolvent expansion \eqref{eq:resolv_asymp} is asymptotic, but also that it must be enlarged into a transseries \cite{abs18}. This simple example illustrates how the asymptotic expansion of the resolvent is only the beginning of a more complicated story (a further pedagogical illustration may be found in appendix~\ref{app:harm_resolv}).

%%%%%%%%%%%%%%%%%%%%%%%%%%%%%%%%%%%%%%%%%%%%%%%%%%%%%%%%%%%%%%%%%
\subsection{String Equations for Multicritical Models}\label{subsec:multi_string_eq}
%%%%%%%%%%%%%%%%%%%%%%%%%%%%%%%%%%%%%%%%%%%%%%%%%%%%%%%%%%%%%%%%%

As already mentioned in subsection~\ref{subsec:multicritical}, orthogonal polynomial techniques lead to string equations which exactly\footnote{Whereas their original introduction \cite{biz80} was purely perturbative, via resurgence and transseries \cite{abs18} orthogonal polynomial methods may actually further compute the full (nonperturbative) content of the matrix model \cite{m08, asv11, sv13} and its double-scaled limit \cite{m08, gikm10, asv11, sv13}.} determine the matrix model or its double-scaled limit. An order-$k$ double-scaled multicritical point is hence exactly described by a \textit{nonlinear} ODE of order $(2k-2)$ \cite{gm90a, ds90, bk90, d90, gm90b}. The remarkable fact unveiled in the aforementioned references is that these string equations are precisely determined by the Gel'fand--Dikii KdV potentials of the previous subsection, $R_{\ell} \left[ u \right]$, and, in fact, very much related to the Novikov equations \eqref{eq:Novikov} \cite{gd75}.

Let $u(\kappa_{\text{s}})$ be the exact double-scaled specific heat (the two-point function of the lowest-dimension operator in the theory) and $F_{\text{ds}} = \log \CZ$ the double-scaled free energy (with $F_{\text{ds}}'' = -\frac{1}{2} u$). Recall the $z$-variable, $\kappa_{\text{s}}^2 = z^{\gamma-2}$. Then the string equation for the order-$k$ multicritical model\footnote{While this is one adequate normalization for the multicritical string-equation, it is also common to find an alternative normalization in the literature (where both $u$ and $z$ then need to be rescaled); namely,
\be
\label{eq:k+1/2_string_eq_norm}
\left( k+\frac{1}{2} \right) R_k \left[ u \right] = z.
\ee
\noindent
This normalization allows for the introduction of an action for this equation, as \cite{gz90b}
\be
\label{eq:S_action_multi}
\CS = \int \rmd z \left( R_{k+1} + z u \right).
\ee
\noindent
Variational derivation $\delta_{u} \CS = 0$ immediately yields the above unconventionally-normalized string-equation if one recalls that the KdV potentials satisfy \cite{gd75} $\frac{\delta}{\delta u} \int \rmd z\, R_{\ell+1} \left[u\right] = - \left( \ell+\frac{1}{2} \right) R_\ell \left[u\right]$.} is \cite{gm90b}
\be
\label{eq:multicritical_string_eq}
\left(-1\right)^{k} \frac{2^{k+1}\, k!}{\left( 2k-1 \right)!!}\, R_k \left[ u(z) \right] = z.
\ee
\noindent
Note that the Gel'fand--Dikii KdV potentials are predetermined, in which case the string equation only really depends on the multicritical index $k$, and is in this sense universal. For example, for the $k=2$ or $(2,3)$ multicritical theory it follows\footnote{Herein the $\frac{1}{3}$ normalization corresponds to a matrix-model origin with an even potential; for an odd potential the normalization would have turned out to be $\frac{1}{6}$ instead. Because of the symmetry of the even potential, our choice of normalization leads to a doubling of the free energy, hence the factor of $\frac{1}{2}$ in \eqref{eq:u-to-F_ds}, which is required for matching the topological recursion results.}
\be
\label{eq:k=2_string_eq}
u^2 - \frac{1}{3} u'' = z.
\ee
\noindent
This is the famous Painlev\'e~I equation, which has been previously studied via the use of resurgence and transseries (in similar contexts) in \cite{gikm10, asv11}. For the $k=3$ or $(2,5)$ multicritical theory one finds instead
\be
\label{eq:k=3_string_eq}
u^3 - u u'' - \frac{1}{2} \left( u' \right)^2 + \frac{1}{10} u'''' = z.
\ee
\noindent
This is the Yang--Lee edge singularity equation, whose solutions---in particular, solutions free of singularities on the real axis---were addressed in \cite{bmp90, c17}.

%%%%%%%%%%%%%%%%%%%%%%%%%%%%%%%%%%%%%%%%%%%%%%%%%%%%%%%%%%%%%%%%%
\subsubsection*{Perturbative Content via String Equations}
%%%%%%%%%%%%%%%%%%%%%%%%%%%%%%%%%%%%%%%%%%%%%%%%%%%%%%%%%%%%%%%%%

The same way that the topological recursion \cite{eo07a} computed the perturbative (asymptotic) genus expansion out of the spectral curve alone, in subsection~\ref{subsec:multicritical}, we may now ask if the very same may be achieved starting off with the string equation. Things are a bit trickier now: whereas we had a closed-form expression for the generic order-$k$ multicritical spectral curve, \eqref{eq:multicrit_spect_curve_hyp2}, there is now no \textit{explicit} expression for the order-$k$ multicritical string equation \eqref{eq:multicritical_string_eq}---all one has is a recursive procedure to reach it, \eqref{eq:Rell_recursion}. This is a perturbative set-back with the string equation, as compared to the spectral curve approach, but one which is compensated by the fact that more nonoperturbative information may be extracted from the string equation (as we shall see below).

Evaluating a few of the Gel'fand--Dikii KdV potentials \eqref{eq:GD-R0}-\eqref{eq:GD-R3} it is straightforward to see that, at arbitrary order, they are of the form \cite{gz90b, gz91}
\be
\label{eq:Rl-alphall}
R_{\ell} \left[ u \right] = \frac{1}{\ell}\, \alpha_{\ell\ell}\, u^{\ell} + \cdots,
\ee
\noindent
where the dots represent terms with derivatives (\textit{i.e.}, this is the single \textit{no-derivatives} monomial in the Gel'fand--Dikii KdV potential). Plugging this \textit{ansatz} into the recursion relation \eqref{eq:Rell_recursion} translates to a recursion for its coefficients\footnote{This further requires fixing one constant, which is simply done given the first few Gel'fand--Dikii polynomials.} \cite{gz90b, gz91}
\be
\label{eq:alphall_coeff}
\alpha_{\ell+1,\ell+1} = - \left( 1 + \frac{1}{2\ell} \right) \alpha_{\ell\ell} \quad \Rightarrow \quad \alpha_{\ell\ell} = (-1)^{\ell}\, \frac{\left( 2\ell-1 \right)!!}{2^{\ell+1} \left( \ell-1 \right)!} \quad \Leftrightarrow \quad \frac{1}{\ell}\, \alpha_{\ell\ell} = (-1)^{\ell}\, \frac{\left( 2\ell-1 \right)!!}{2^{\ell+1}\, \ell!}.
\ee 
\noindent
Without surprise this is precisely the (inverse) normalization coefficient of the multicritical string-equation \eqref{eq:multicritical_string_eq}. Hence the \textit{leading} (classical) multicritical string equation is rather simply
\be
\label{eq:planar_multi_string-eq}
u^k = z.
\ee
\noindent
This of course matches the planar, genus-zero results of both specific-heat and free-energy we have already seen \eqref{eq:planar_u_&_F}.

It is quite straightforward to extract the string-theoretic perturbative expansion from a given string equation. Just start with the \textit{ansatz} \eqref{eq:u_pert_exp},
\be
\label{eq:u_pert_exp-2}
u(z) \simeq z^{\frac{1}{k}} \left( 1 + \sum_{g=1}^{+\infty} \frac{u_g}{z^{\left( \frac{2k+1}{k} \right) g}} \right),
\ee
\noindent
which turns the nonlinear ODE into a nonlinear recursion relation determining the coefficients $\left\{ u_g \right\}$ (and, eventually, the free energies $\left\{ F_g \right\}$). These may be swiftly generated via symbolic computation. While this is quite simple for fixed $k$, it is slightly more intricate to find the generic $k$ solution (which would reproduce the results we obtained earlier in subsection~\ref{subsec:multicritical} via the topological recursion). One way to do this is outlined in \cite{gm90b} and their string-equation results precisely match our topological-recursion results (up to the considered perturbative orders).

For example, for the $k=2$ or $(2,3)$ multicritical theory described by the Painlev\'e~I equation \eqref{eq:k=2_string_eq}, one finds for the specific heat (see as well \cite{gm90b, jk01, msw07, msw08, gikm10, asv11})
\bea
\label{eq:pert(2,3)-u}
u_{(2,3)} (z) &\simeq& z^{\frac{1}{2}}\, \sum_{g=0}^{+\infty} \frac{u_g}{z^{\frac{5}{2} g}} = \\
&\simeq& z^{\frac{1}{2}}\left( 1 - \frac{1}{24} z^{-\frac{5}{2}} - \frac{49}{1152} z^{-5} - \frac{1225}{6912} z^{-\frac{15}{2}} - \frac{4412401}{2654208} z^{-10} - \frac{73560025}{2654208} z^{-\frac{25}{2}} - \cdots \right). \nonumber
\eea
\noindent
Furthermore, for the free energy---via \eqref{eq:u-to-F_ds} 
\be
\label{eq:u-to-F_ds-2}
F_{\text{ds}}'' (z) = -\frac{1}{2} u (z)
\ee
\noindent
---one obtains:
\be
\label{eq:pert(2,3)-F}
F_{\text{ds}} (z) \simeq - \frac{2}{15} z^{\frac{5}{2}} - \frac{1}{48} \log z + \frac{7}{2880} z^{-\frac{5}{2}} + \frac{245}{82944} z^{-5} + \frac{259553}{19906560} z^{-\frac{15}{2}} + \frac{1337455}{10616832} z^{-10} + \cdots.
\ee
\noindent
For the $k=3$ or $(2,5)$ multicritical theory described by the Yang--Lee equation \eqref{eq:k=3_string_eq}, one finds instead for the specific heat (see as well \cite{gm90b, bmp90, c17})
\bea
\label{eq:pert(2,5)-u}
u_{(2,5)} (z) &\simeq& z^{\frac{1}{3}}\, \sum_{g=1}^{+\infty} \frac{u_g}{z^{\frac{7}{3} g}} = \\
&\simeq& z^{\frac{1}{3}} \left( 1 - \frac{1}{18} z^{-\frac{7}{3}} - \frac{7}{108} z^{-\frac{14}{3}} - \frac{4199}{17496} z^{-7} - \frac{409297}{262440} z^{-\frac{28}{3}} - \frac{101108329}{9447840} z^{-\frac{35}{3}} - \cdots \right), \nonumber
\eea
\noindent
while for the free energy \eqref{eq:u-to-F_ds-2} yields 
\be
\label{eq:pert(2,5)-F}
F_{\text{ds}} (z) \simeq - \frac{9}{56} z^{\frac{7}{3}} - \frac{1}{36} \log z + \frac{1}{240}  z^{-\frac{7}{3}} + \frac{247}{54432} z^{-\frac{14}{3}} + \frac{58471}{4199040} z^{-7} + \frac{465937}{8398080} z^{-\frac{28}{3}} + \cdots.
\ee
\noindent
These results are in perfect agreement with the results in subsection~\ref{subsec:multicritical}, \eqref{eq:multicrit_free_en-2}-\eqref{eq:multicrit_free_en-3}-\eqref{eq:multicrit_free_en-4}.

%%%%%%%%%%%%%%%%%%%%%%%%%%%%%%%%%%%%%%%%%%%%%%%%%%%%%%%%%%%%%%%%%
\subsubsection*{Nonperturbative Content via String Equations}
%%%%%%%%%%%%%%%%%%%%%%%%%%%%%%%%%%%%%%%%%%%%%%%%%%%%%%%%%%%%%%%%%

We have already briefly discussed in section~\ref{sec:spec_curv} how the factorial growth of the free-energy coefficients $F_{g} \sim (2g)!$ \cite{gp88} implies the existence of D-brane multi-instantons \cite{p94}. From the point-of-view of writing down the free-energy solution, what this entails is that string-theoretic nonperturbative effects go as \cite{s90}
\be
\label{eq:D-inst-exp}
\sim \exp \left( - \frac{1}{\kappa_{\text{s}}} \right) \equiv \exp \left( - z^{\frac{2k+1}{2k}} \right).
\ee
\noindent
The famous argument in \cite{s90} also follows via a combination of the topological recursion \cite{eo07a} and resurgence \cite{abs18}---but in order to be fully specific on the nature of these multi-instanton sectors, a string equation is of help. Such equations allow for full transseries constructions \cite{abs18} and this has actually been achieved in a few examples within these contexts \cite{m08, gikm10, asv11, sv13}. In the following, let us try to be a bit more general---albeit less constructive---by building upon \cite{gz90b, gz91, ez93}.

In order to access multi-instanton sectors one solves the string equation not with the usual perturbative \textit{ansatz} \eqref{eq:u_pert_exp-2}, but rather with a nonperturbative transseries \textit{ansatz} based upon the weight of instanton effects \eqref{eq:D-inst-exp} \cite{abs18}. This was studied in great detail for the example of the $k=2$ or $(2,3)$ multicritical theory described by the Painlev\'e~I equation \eqref{eq:k=2_string_eq} in \cite{gikm10, asv11, bssv22} (but see as well \cite{msw07, m08, msw08, sv13, as13}). Given \eqref{eq:planar_u_&_F}, \eqref{eq:u_pert_exp-2}, and \eqref{eq:D-inst-exp}, the natural one-parameter transseries \textit{ansatz} for the order-$k$ multicritical theory is
\be
\label{eq:trans-kappa-u}
u \left( \kappa_{\text{s}}, \sigma \right) = \kappa_{\text{s}}^{-\frac{2}{2k+1}}\, \sum_{n=0}^{+\infty} \sigma^n\, \rme^{- \frac{n A}{\kappa_{\text{s}}}}\, \kappa_{\text{s}}^{n\beta}\, \sum_{g=0}^{+\infty} u_g^{(n)}\, \kappa_{\text{s}}^g,
\ee
\noindent
in terms of the double-scaled string coupling $\kappa_{\text{s}}$; or, equivalently---more useful for input into the string equation---, in terms of the double-scaled variable $z$ one has
\be
\label{eq:trans-z-u}
u \left( z, \sigma \right) = z^{\frac{1}{k}}\, \sum_{n=0}^{+\infty} \sigma^n \exp \left( - n A\, z^{\frac{2k+1}{2k}} \right) z^{- \frac{2k+1}{2k} n \beta}\, \sum_{g=0}^{+\infty} \frac{u_g^{(n)}}{z^{\frac{2k+1}{2k} g}}.
\ee
\noindent
Herein $\sigma$ is the (so far single) transseries parameter, $A$ is the instanton action, $\beta$ is a characteristic exponent, and the $u_g^{(n)}$ are the specific-heat perturbative coefficients around the $n$-instanton sector \cite{abs18} (\textit{e.g.}, the perturbative coefficients $u_g^{(0)} \equiv u_g$ were previously computed in examples, \eqref{eq:pert(2,3)-u} and \eqref{eq:pert(2,5)-u}). For the aforementioned Painlev\'e~I example, \eqref{eq:k=2_string_eq}, this yields:
\be
\label{eq:trans(2,3)-u}
u_{(2,3)} = \kappa_{\text{s}}^{-\frac{2}{5}}\, \sum_{n=0}^{+\infty} \sigma^n\, \rme^{- \frac{n A}{\kappa_{\text{s}}}}\, \kappa_{\text{s}}^{n\beta}\, \sum_{g=0}^{+\infty} u_g^{(n)}\, \kappa_{\text{s}}^g = z^{\frac{1}{2}}\, \sum_{n=0}^{+\infty} \sigma^n\, \rme^{- n A\, z^{\frac{5}{4}}} z^{- \frac{5}{4} n \beta}\, \sum_{g=0}^{+\infty} \frac{u_g^{(n)}}{z^{\frac{5}{4} g}}.
\ee
\noindent
For the $k=3$ or $(2,5)$ multicritical theory described by the Yang--Lee equation \eqref{eq:k=3_string_eq}, this yields instead
\be
\label{eq:trans(2,5)-u}
u_{(2,5)} = \kappa_{\text{s}}^{-\frac{2}{7}}\, \sum_{n=0}^{+\infty} \sigma^n\, \rme^{- \frac{n A}{\kappa_{\text{s}}}}\, \kappa_{\text{s}}^{n\beta}\, \sum_{g=0}^{+\infty} u_g^{(n)}\, \kappa_{\text{s}}^g = z^{\frac{1}{3}}\, \sum_{n=0}^{+\infty} \sigma^n\, \rme^{- n A\, z^{\frac{7}{6}}} z^{- \frac{7}{6} n \beta}\, \sum_{g=0}^{+\infty} \frac{u_g^{(n)}}{z^{\frac{7}{6} g}}.
\ee
\noindent
Plugging the transseries \textit{ansatz} \eqref{eq:trans(2,3)-u} into the Painlev\'e~I equation \eqref{eq:k=2_string_eq} immediately determines the instanton action(s) and characteristic exponent
\be
\label{eq:A&beta-(2,3)}
A_{(2,3)} = \pm \frac{4}{5} \sqrt{6}, \qquad \beta_{(2,3)} = \frac{1}{2}.
\ee
\noindent
The specific-heat one-instanton sector, associated with the ``plus'' instanton action, is\footnote{To be precise, the string equation does \textit{not} determine the leading coefficient $u_0^{(1)}$. This must in fact be the case, as its (arbitrary) value may always be reabsorbed into the transseries parameter $\sigma$. We are hence using the standard transseries normalization where $u_0^{(1)}=1$ \cite{asv11}, and we shall later relate this to the Stokes coefficient.}:
\bea
\label{eq:u(1)(2,3)}
u_{(2,3)}^{(1)} (z) &\simeq& z^{- \frac{1}{8}}\, \sum_{g=0}^{+\infty} \frac{u_g^{(1)}}{z^{\frac{5}{4} g}} = \\
&\simeq& z^{- \frac{1}{8}} \left( 1 - \frac{5}{32\sqrt{6}} z^{-\frac{5}{4}} + \frac{75}{4096} z^{-\frac{5}{2}} - \frac{341329}{5898240 \sqrt{6}} z^{-\frac{15}{4}} + \frac{17327233}{905969664} z^{-5} - \cdots \right), \nonumber
\eea
\noindent
and the (full) free-energy one-instanton contribution is:
\be
\label{eq:F(1)-(2,3)}
F^{(1)}_{\text{ds}} (z) \simeq - \frac{1}{12} z^{- \frac{5}{8}}\, \rme^{- \frac{4\sqrt{6}}{5} z^{\frac{5}{4}}} \left( 1 - \frac{37}{32 \sqrt{6}} z^{-\frac{5}{4}} + \frac{6433}{12288} z^{-\frac{5}{2}} - \frac{12741169}{5898240 \sqrt{6}} z^{-\frac{15}{4}} + \cdots \right).
\ee
\noindent
Similarly, plugging the transseries \textit{ansatz} \eqref{eq:trans(2,5)-u} into the Yang--Lee equation \eqref{eq:k=3_string_eq} determines (the $\pm$ signs are independent, \textit{i.e.}, there are now \textit{four} solutions for $A$)
\be
\label{eq:A&beta-(2,5)}
A_{(2,5)} = \pm \frac{6}{7} \sqrt{5 \pm \rmi \sqrt{5}}, \qquad \beta_{(2,5)} = \frac{1}{2}.
\ee
\noindent
The specific-heat one-instanton sector, associated with the ``plus-plus'' instanton action, is\footnote{Again, as in the previous example, we are using the standard transseries normalization where $u_0^{(1)}=1$ \cite{asv11}.}:
\bea
\label{eq:u(1)(2,5)}
u_{(2,5)}^{(1)} (z) &\simeq& z^{- \frac{1}{4}}\, \sum_{g=0}^{+\infty} \frac{u_g^{(1)}}{z^{\frac{7}{6} g}} = \\
&\simeq& z^{- \frac{1}{4}} \left( 1 + \frac{-15 + 8 \rmi \sqrt{5}}{48 \sqrt{5 + \rmi \sqrt{5}}} z^{-\frac{7}{6}} - \frac{197 + 31 \rmi \sqrt{5}}{1024} z^{-\frac{7}{3}} + \frac{209665 - 208893 \rmi \sqrt{5}}{442368 \sqrt{5 + \rmi \sqrt{5}}} z^{-\frac{7}{2}} - \cdots \right), \nonumber
\eea
\noindent
and the (full) free-energy one-instanton contribution is:
\be
\label{eq:F(1)-(2,5)}
F^{(1)}_{\text{ds}} (z) \simeq - \frac{5-\rmi\sqrt{5}}{60} z^{- \frac{7}{12}}\, \rme^{- \frac{6}{7} \sqrt{5+\rmi\sqrt{5}}\, z^{\frac{7}{6}}} \left( 1 - \frac{63 - 8 \rmi \sqrt{5}}{48 \sqrt{5 + \rmi \sqrt{5}}} z^{-\frac{7}{6}} + \frac{13375 - 10963 \rmi \sqrt{5}}{46080} z^{-\frac{7}{3}} - \cdots \right).
\ee
\noindent
As we shall review below, the fact that the instanton actions appear in symmetric pairs is generic \cite{gz90b, gz91} and is related to resonance \cite{gikm10, asv11, abs18, bssv22}. ``Half'' the instanton actions in \eqref{eq:A&beta-(2,3)} and \eqref{eq:A&beta-(2,5)} match what was earlier obtained in subsection~\ref{subsec:multicritical}, \eqref{eq:A-(2,3)SG} and \eqref{eq:A-(2,5)SG}. The same is true for the two and three-loop coefficients around the one-instanton sector in the one-instanton free energies \eqref{eq:F(1)-(2,3)} and \eqref{eq:F(1)-(2,5)}, precisely matching\footnote{In the case of the match between string-equation data \eqref{eq:F(1)-(2,5)}, computed for the ``plus-plus'' instanton sector, and spectral-geometry data \eqref{eq:F12F11SG(2,5)}-\eqref{eq:F13F11SG(2,5)}, computed for the ``minus-plus'' instanton sector, the ``precise match'' means ``up to an oscillating $(-1)$ factor'' at each loop order. This is also associated to resonance; \textit{e.g.}, \cite{asv11}.} against \eqref{eq:F12F11SG(2,3)}-\eqref{eq:F13F11SG(2,3)} and \eqref{eq:F12F11SG(2,5)}-\eqref{eq:F13F11SG(2,5)}, respectively.  Note, however, that string equations do not determine Stokes coefficients directly, but only from large-order asymptotics (as explained in \cite{msw07}; to which we shall return).

Obtaining full nonperturbative results at generic $k$ is now more intricate than it was in the spectral-geometry realm we addressed earlier---again mainly due to the lack of an explicit closed-form expression for the order-$k$ multicritical string equation \eqref{eq:multicritical_string_eq}. However, as we have seen earlier, some parts of the Gel'fand--Dikii KdV potentials may be written in explicit closed-form---and it turns out that in fact \textit{more} and \textit{enough} about these KdV potentials may be written in explicit closed-form to determine the arbitrary order-$k$ multicritical instanton-actions \cite{gz90b, gz91}.

From \eqref{eq:Rl-alphall} and \eqref{eq:alphall_coeff}, the generic Gel'fand--Dikii KdV potentials are of the form
\be
R_{\ell} \left[ u \right] = \frac{1}{\ell}\, \alpha_{\ell\ell}\, u^{\ell} + \cdots = (-1)^{\ell}\, \frac{\left( 2\ell-1 \right)!!}{2^{\ell+1}\, \ell!}\, u^{\ell} + \cdots.
\ee
\noindent
Moving further along these lines, explicit evaluation of a few KdV potentials hints that, at arbitrary order, their first monomials including derivatives should be of the form \cite{gz90b, gz91}
\be
\label{eq:Rl-alphall-alphali-betalj}
R_{\ell} \left[ u \right] = \frac{1}{\ell} \alpha_{\ell\ell}\, u^{\ell} + \sum_{i=1}^{\ell-1} \alpha_{\ell i}\, u^{i-1}\, u^{(2\ell-2i)} + \sum_{j=2}^{\ell-1} \beta_{\ell j}\, u^{j-2}\, u'\, u^{(2\ell-2j-1)} + \cdots.
\ee
\noindent
Herein, the first (no-derivatives) monomial was already studied; the second monomial includes a \textit{single} (``generic'') derivative factor; and the third monomial includes a single ``generic'' derivative factor \textit{and} a \textit{single}\footnote{As written, this is clearly not quite so: when $j=\ell-1$ also $u^{(2\ell-2j-1)}=u'$, and there are hence \textit{two} such factors in this monomial. The required modification to the coefficient $\beta_{\ell,\ell-1}$ will be understood in the following.} factor of $u'$. The dots represent monomials with more derivative factors.

We computed $\alpha_{\ell\ell}$ very simply in \eqref{eq:alphall_coeff}, via the recursion relation \eqref{eq:Rell_recursion}. One other coefficient which is straightforward to compute is the one of the highest single-derivative term, $\alpha_{\ell 1}$. Plugging this term into the recursion relation \eqref{eq:Rell_recursion} translates to\footnote{Again, this requires fixing one constant, which is immediate given the first few Gel'fand--Dikii polynomials.} \cite{gz90b, gz91}
\be
\label{eq:alphal1_coeff}
\alpha_{\ell+1,1} = \frac{1}{4} \alpha_{\ell 1} \qquad \Rightarrow \qquad \alpha_{\ell 1} = - 4^{-\ell}.
\ee
\noindent
The remaining coefficients are a bit trickier to compute \cite{gz91}. To do so, let us first reconsider the one-parameter transseries \textit{ansatz} for the order-$k$ multicritical model, \eqref{eq:trans-z-u}. Plugging in this \textit{ansatz} into the generic Gel'fand--Dikii potential \eqref{eq:Rl-alphall-alphali-betalj}, and expanding to \textit{first} order in the transmonomial combination 
\be
\label{eq:first-order-trans}
\sigma \exp \left( - A\, z^{\frac{2\ell+1}{2\ell}} \right)
\ee
\noindent
and to \textit{leading} order at large $z$, yields\footnote{Recall we are using the normalization where both $u_0^{(0)} = 1$ and $u_0^{(1)} = 1$.}
\be
R_{\ell} \left[ u \right] = \cdots + \sigma\, \rme^{- A\, z^{\frac{2\ell+1}{2\ell}}}\, z\, \left\{ \sum_{i=1}^{\ell} \alpha_{\ell i} \left( \frac{2\ell+1}{2\ell} \right)^{2\ell-2i} A^{2\ell-2i} \right\} + \cdots.
\ee
\noindent
It is appropriate to rewrite the instanton action slightly in order to fit this result,
\be
\label{eq:inst-act-ell-rho}
A = \frac{2\ell}{2\ell+1}\, \rho.
\ee
\noindent
Then, \textit{at this precise order}, the string equation \eqref{eq:multicritical_string_eq} may be simply written as a polynomial equation\footnote{This final polynomial expression, eventually determining the order-$k$ multicritical-theory instanton-actions, also explains the choice of initial normalization for the generic Gel'fand--Dikii KdV potentials in \eqref{eq:Rl-alphall}.}
\be
\label{eq:poly-inst-act-string-eq}
\CP_{k} (\rho) \equiv \sum_{i=1}^{k} \alpha_{k i}\, \rho^{2k-2i} = 0.
\ee
\noindent
This very same polynomial equation was obtained in \cite{gz91} via the WKB method. Herein, both the $\alpha_{k 1}$ and $\alpha_{k k}$ coefficients are already known. However, in order to determine the instanton actions (\textit{i.e.}, via \eqref{eq:inst-act-ell-rho}, the roots of this degree-$\left(2k-2\right)$ polynomial in $\rho$) one still needs to know the remaining coefficients---as we initially set out to find. The strategy in \cite{gz91}, which we briefly follow next, is that it turns out to be easier to first determine the polynomial $\CP_{k} (\rho)$ in closed form, rather than to first determine its $\alpha_{k i}$ coefficients. Let us see how.

Changing variables as $\rho^2 = \frac{4}{\sigma}$ yields
\be
\CP_{\ell} (\rho) = - \frac{1}{4\sigma^{\ell-1}}\, \NCP_{\ell} (\sigma),
\ee
\noindent
where $\NCP_{\ell} (\sigma)$ is a degree-$\left(\ell-1\right)$ polynomial in $\sigma$ (and  whose degree-zero coefficient equals one). Using the KdV-potentials property that \cite{gd75}
\be
\frac{\partial}{\partial u} R_{\ell} \left[ u \right] = - \left( \ell-\frac{1}{2} \right) R_{\ell-1} \left[ u \right],
\ee
\noindent
one can show \cite{gz91} that the $\NCP_{\ell} (\sigma)$ polynomials satisfy
\be
\NCP_{\ell}' (\sigma) = - \left( \ell-\frac{1}{2} \right) \NCP_{\ell-1} (\sigma).
\ee
\noindent
Solving this differential-recursion with the (aforementioned) ``initial condition'' $\NCP_{\ell} (0) = 1$ yields\footnote{This result corrects a small typo in \cite{gz91}.}
\be
\NCP_{\ell} (\sigma) = \frac{\Gamma \left( \ell+\frac{1}{2} \right)}{\Gamma \left( \ell \right) \Gamma \left( \frac{1}{2} \right)}\, \int_{0}^{1} \frac{\rmd s}{\sqrt{s}}\, \left( 1-s-\sigma \right)^{\ell-1}.
\ee
\noindent
Going back to our original polynomial for the instanton action \eqref{eq:poly-inst-act-string-eq}, we compute
\be
\CP_{\ell} (\rho) = \left(-1\right)^{\ell} \frac{\Gamma \left( \ell+\frac{1}{2} \right)}{2\, \Gamma \left( \ell \right) \Gamma \left( \frac{1}{2} \right)}\, {}_{2} F_{1} \left( 1-\ell, 1; \frac{3}{2} \,\left|\, \frac{\rho^2}{4} \right) \right.,
\ee
\noindent
which finally yields all\footnote{As a consistency check, indeed one has $\alpha_{\ell\ell} = \left( -1 \right)^{\ell} \frac{\Gamma \left( 2\ell \right)}{2^{2\ell}\, \Gamma \left( \ell \right)^2}$ and $\alpha_{\ell 1} = - \frac{1}{2^{2\ell}}$ as previously computed.} the $\alpha_{\ell i}$ coefficients:
\be
\label{eq:alphali_coeffs}
\alpha_{\ell i} = \left( -1 \right)^{i} \frac{\Gamma \left( 2\ell \right) \Gamma \left( \left( \ell-i \right) + 1 \right)}{2^{2\ell}\, \Gamma \left( \ell \right) \Gamma \left( i \right) \Gamma \left( 2 \left( \ell-i \right) + 2 \right)}.
\ee

Finally, let us address the $\beta_{\ell j}$ coefficients in \eqref{eq:Rl-alphall-alphali-betalj}. These were computed in \cite{gz91}, by requiring hermiticity at the level of the linearized string equation (which is an argument we will not dwell into). At the level of the Gel'fand--Dikii KdV potentials one obtains
\be
\label{eq:betalj_coeffs}
\beta_{\ell j} = 
\begin{cases}
\, \frac{1}{2} \left( 2\ell-2j \right) \left( j-1 \right) \alpha_{\ell j} & \text{ if }\, j = 2, \ldots, \ell-2, \\
\, \frac{1}{2} \left( \ell-2 \right) \alpha_{\ell,\ell-1} & \text{ if }\, j = \ell-1.
\end{cases}
\ee
\noindent
The only difference as compared to the linearized case \cite{gz91} is an extra factor of $\frac{1}{2}$ in the $\beta_{\ell,\ell-1}$ coefficient. This is associated to the presence of two factors of $u'$ in the third monomial of \eqref{eq:Rl-alphall-alphali-betalj}, which result in an extra factor of two when linearizing the string equation.

One may now repeat the same exercise that led to the (leading) nonperturbative content in the string equation, \eqref{eq:poly-inst-act-string-eq}. At such leading order one just recovers \eqref{eq:poly-inst-act-string-eq}, but where \textit{all} coefficients are now known. At first order in the transmonomial combination \eqref{eq:first-order-trans} and at \textit{next-to-leading order} at large $z$, one now finds
\be
R_{\ell} \left[ u \right] = \cdots + \sigma\, \rme^{- A\, z^{\frac{2\ell+1}{2\ell}}}\, z^{-\frac{1}{2\ell} - \frac{2\ell+1}{2\ell} \beta} \left( 2\beta-1 \right) \,\big\{ \,\cdots\, \big\}\, + \cdots.
\ee
\noindent
Herein, the dots within the curly brackets stand for an (odd) polynomial of degree-$\left(2\ell-3\right)$ in $\rho$, whose coefficients are irrelevant for our argument. As a result, at this next-to-leading order the string equation \eqref{eq:multicritical_string_eq} just sets $\beta = \frac{1}{2}$ (very much expected given \eqref{eq:A&beta-(2,3)} and \eqref{eq:A&beta-(2,5)}).

One may now return to the initial one-parameter transseries \textit{ansatz} for the order-$k$ multicritical theory, \eqref{eq:trans-z-u}, and rewrite it with the above newly-acquired information. It becomes
\be
\label{eq:trans-z-u-rho}
u \left( z, \sigma \right) = z^{\frac{1}{k}}\, \sum_{n=0}^{+\infty} \sigma^n \exp \left( - n\, \frac{2k}{2k+1}\, \rho\, z^{\frac{2k+1}{2k}} \right) z^{- \frac{2k+1}{4k} n}\, \sum_{g=0}^{+\infty} \frac{u_g^{(n)}}{z^{\frac{2k+1}{2k} g}},
\ee
\noindent
where $\rho$ is a root of the polynomial equation \eqref{eq:poly-inst-act-string-eq} with coefficients \eqref{eq:alphali_coeffs}. A transseries \textit{ansatz} for the free energy associated with this one-parameter specific-heat transseries would then be schematically of the form
\be
\label{eq:trans-F-one-parameter}
F \left( \kappa_{\text{s}}, \sigma \right) = \kappa_{\text{s}}^{-2}\, \sum_{n=0}^{+\infty} \sigma^n\, \rme^{- n\, \frac{2k}{2k+1}\,  \frac{\rho}{\kappa_{\text{s}}}}\, \kappa_{\text{s}}^{\frac{n}{2}}\, \sum_{g=0}^{+\infty} F_g^{(n)} \kappa_{\text{s}}^g.
\ee
\noindent
Note how this schematic structure is perfectly consistent with the original matrix model expectation \eqref{eq:MM-F(1)}. In particular, Stokes data---for which there are no closed-form expressions within the present string-equation realm---follows from \eqref{eq:MM-stokes-coeff-inf} or \eqref{eq:multicritical-stokes} once we make the bridge towards the spectral geometry formulation. This connection was also further discussed in \cite{msw07}.

So far we have not specified $\rho$. As we shall see now, doing so will have a dramatic influence on the final construction of the transseries solution \eqref{eq:trans-z-u-rho}. Consider the solution set of the polynomial equation \eqref{eq:poly-inst-act-string-eq} with real coefficients \eqref{eq:alphali_coeffs}. For example, for the $k=2$ or $(2,3)$ multicritical theory one finds
\be
\CP_{2} (\rho) = \frac{3}{8} - \frac{\rho^2}{16} = 0 \qquad \Rightarrow \qquad \rho = \pm \sqrt{6}.
\ee
\noindent
For the $k=3$ or $(2,5)$ multicritical theory one finds instead
\be
\CP_{3} (\rho) = - \frac{15}{32} + \frac{5 \rho^2}{32} - \frac{\rho^4}{64} = 0 \qquad \Rightarrow \qquad \rho = \pm \sqrt{5 \pm \rmi \sqrt{5}}.
\ee
\noindent
Via the relation between $\rho$ and the instanton action(s) $A$, \eqref{eq:inst-act-ell-rho}, these are the exact same results as obtained earlier via their respective string equations, \eqref{eq:k=2_string_eq} and \eqref{eq:k=3_string_eq}, in \eqref{eq:A&beta-(2,3)} and \eqref{eq:A&beta-(2,5)}. But now we can be completely general: for the order-$k$ multicritical theory one obtains a degree-$\left(2k-2\right)$ polynomial in $\rho$ with real coefficients, implying there will be $\left(2k-2\right)$ distinct roots, or, via \eqref{eq:inst-act-ell-rho}, that the order-$k$ multicritical model has $\left(2k-2\right)$ instanton actions. More precisely, with \eqref{eq:poly-inst-act-string-eq} being a degree-$\left(k-1\right)$ polynomial in $\rho^2$, one finds $\left(k-1\right)$ distinct instanton actions alongside their \textit{symmetric pairs}\footnote{This is in extension to the appearance of \textit{complex-conjugate pairs}, already seen via spectral geometry.}. This implies, in particular, that the Painlev\'e~I transseries should be a \textit{two}-parameter transseries \cite{gikm10, asv11}, and the Yang--Lee transseries should be a \textit{four}-parameter transseries. Furthermore, the order-$k$ multicritical should be a $\left(2k-2\right)$-parameter\footnote{Recall how the $\left(p,q\right) = \left(2,2k-1\right)$ minimal-model CFT has $\frac{1}{2} \left( p-1 \right) \left( q-1 \right) = k-1$ total primaries \cite{bpz84}. As such, the number of transseries parameters precisely matches twice the number of relevant operators in the theory.} (resonant---more on this shortly) transseries (such structures were discussed in, \textit{e.g.}, \cite{abs18}).

Let us organize the $\left(2k-2\right)$ instanton actions vectorially as
\bea
\label{eq:inst-vect-k=2}
k=2 \quad &\Rightarrow& \quad \boldsymbol{A} = \frac{4}{5} \left( \sqrt{6}, - \sqrt{6} \right), \\
\label{eq:inst-vect-k=3}
k=3 \quad &\Rightarrow& \quad \boldsymbol{A} = \frac{6}{7} \left( \sqrt{5 + \rmi \sqrt{5}}, - \sqrt{5 + \rmi \sqrt{5}}, \sqrt{5 - \rmi \sqrt{5}}, - \sqrt{5 - \rmi \sqrt{5}} \right), \\
&\vdots& \nonumber \\
\label{eq:inst-vect-k=k}
k \quad &\Rightarrow& \quad \boldsymbol{A} = \frac{2k}{2k+1} \left( \rho_1, - \rho_1, \cdots, \rho_{k-1}, - \rho_{k-1} \right).
\eea
\noindent
Then, following the general resurgent-transseries framework \cite{abs18}, the $\left(2k-2\right)$-parameter transseries describing the full, nonperturbative specific-heat, $u(z)$, solution to the order-$k$ multicritical-model string-equation \eqref{eq:multicritical_string_eq}, is given by
\be
\label{eq:general-k-u-trans}
u \left( z, \boldsymbol{\sigma} \right) = \sum_{\boldsymbol{n} \in \BN_{0}^{2k-2}} \boldsymbol{\sigma}^{\boldsymbol{n}} \exp \left( - \boldsymbol{n} \cdot \boldsymbol{A}\, z^{\frac{2k+1}{2k}} \right) \Phi_{\boldsymbol{n}} (z).
\ee
\noindent
Herein, $\boldsymbol{\sigma}^{\boldsymbol{n}} = \prod_{i=1}^{2k-2} \sigma_i^{n_i}$ are the $\left(2k-2\right)$ independent transseries parameters. They parametrize the $\left(2k-2\right)$ boundary conditions required to specify a unique solution to the nonlinear string-equation ODE. The $\Phi_{\boldsymbol{n}} (z)$ are asymptotic series associated with the $\boldsymbol{n}$ (generalized) multi-instanton sector. Note that in the present multi-parameter transseries case, the analogue of the $\beta$-dependent prefactor in \eqref{eq:trans-kappa-u} will no longer be of the simple form $\kappa_{\text{s}}^{n\beta}$ \cite{asv11}, and it is therefore convenient to absorb it into these asymptotic series (which will hence no longer necessarily start with constant term one). Schematically,
\be
\Phi_{\boldsymbol{n}} (z) \simeq \sum_{g=0}^{+\infty} \frac{u_g^{(\boldsymbol{n})}}{z^{\frac{2k+1}{2k} g}},
\ee
\noindent
but---as just mentioned---we are being cavalier on what concerns the many possible starting orders (see \cite{asv11} for further details, within the Painlev\'e~I context).

Now, the consequence that \eqref{eq:poly-inst-act-string-eq} is a polynomial in $\rho^2$, \textit{i.e.}, that instanton actions arise in symmetric pairs, is \textit{resonance} \cite{gikm10, asv11, abs18, bssv22}. What this means is that the instanton actions \eqref{eq:inst-vect-k=2}, \eqref{eq:inst-vect-k=3}, or \eqref{eq:inst-vect-k=k} are \textit{rationally dependent}:
\be
\left. \exists \,\, \boldsymbol{\mathfrak{n}} \neq \boldsymbol{0} \in \BZ^{2k-2} \,\, \right| \,\, \boldsymbol{\mathfrak{n}} \cdot \boldsymbol{A} = 0.
\ee
\noindent
In particular, the projection map $\mathfrak{P}$ which projects the $\BZ^{2k-2}$ transseries grid into the complex Borel plane $\BC$ specifying the location of Borel singularities,
\begin{align}
\label{eq:map-Zk-to-C}
\mathfrak{P} : \BZ^{2k-2} & \to \BC \nonumber \\
\boldsymbol{\ell} & \mapsto \boldsymbol{A} \cdot \boldsymbol{\ell},
\end{align}
\noindent
is no longer one-to-one, \textit{i.e.}, $\ker \mathfrak{P} \neq \boldsymbol{0}$ (see \cite{abs18} for further details). For example, if $k=2$ then $\boldsymbol{\mathfrak{n}} = \left( 1,1 \right)$, and if $k=3$ then $\boldsymbol{\mathfrak{n}}_1 = \left( 1,1,0,0 \right)$ and $\boldsymbol{\mathfrak{n}}_2 = \left( 0,0,1,1 \right)$; with their integer multiples generating the respective kernels. This has consequences both at the level of Borel singularities \cite{abs18}, and at the transseries level. Whereas multi-parameter transseries (resonant or not) such as \eqref{eq:general-k-u-trans} are usually written in standard ``rectangular framing'', with their sectors organized into an integer lattice, resonance naturally suggest a reorganization of this ``rectangular'' lattice along its ``diagonal'' resonant-kernel directions. In other words, resonance may be made explicit at transseries level with a straightforward rewriting into ``diagonal framing'' \cite{bssv22}. For example, if $k=2$ then \eqref{eq:general-k-u-trans} in ``diagonal framing'' is simply \cite{asv11, bssv22}
\bea
u \left( z, \sigma_1, \sigma_2 \right) &=& \sum_{n=0}^{+\infty} \sum_{m=0}^{+\infty} \sigma_1^{n} \sigma_2^{m}\, \rme^{- \left( n-m \right) A\, z^{\frac{5}{4}}}\, \Phi_{(n,m)} (z) = \\
&=& \Upphi^{(0)} (z,\uptau) + \sum_{k=1}^{+\infty} \sigma_1^{k}\, \rme^{- k A\, z^{\frac{5}{4}}}\, \Upphi^{(k)}_{+} (z,\uptau) + \sum_{k=1}^{+\infty} \sigma_2^{k}\, \rme^{+ k A\, z^{\frac{5}{4}}}\, \Upphi^{(k)}_{-} (z,\uptau). \nonumber
\eea
\noindent
Herein the instanton number is now explicit along the kernel direction, $\uptau \equiv \sigma_1 \sigma_2$, and the $\uptau$-``modulus''-dependent resonant asymptotic series are simply given by:
\bea
\Upphi^{(0)} (z,\uptau) &:=& \sum_{\ell=0}^{+\infty} \left( \sigma_1 \sigma_2 \right)^{\ell} \Phi_{(\ell,\ell)} (z), \\
\Upphi^{(k)}_{+} (z,\uptau) &:=& \sum_{\ell=0}^{+\infty} \left( \sigma_1 \sigma_2 \right)^{\ell} \Phi_{(\ell+k,\ell)} (z), \\
\Upphi^{(k)}_{-} (z,\uptau) &:=& \sum_{\ell=0}^{+\infty} \left( \sigma_1 \sigma_2 \right)^{\ell} \Phi_{(\ell,\ell+k)} (z).
\eea
\noindent
If, instead, $k=3$, the transseries in ``diagonal framing'' is now
\bea
u \left( z, \sigma_1, \sigma_2, \sigma_3, \sigma_4 \right) &=& \Upphi^{(0,0)} (z,\uptau_{\mathsf{I}}, \uptau_{\mathsf{II}}) + \\
&&
+ \sum_{k_1=0}^{+\infty}\!\!{}'\, \sigma_1^{k_1}\, \rme^{- k_1 A_1\, z^{\frac{7}{6}}} \sum_{k_2=0}^{+\infty}\!\!{}'\, \sigma_3^{k_2}\, \rme^{- k_2 A_2\, z^{\frac{7}{6}}}\, \Upphi^{(k_1,k_2)}_{++} (z,\uptau_{\mathsf{I}}, \uptau_{\mathsf{II}}) + \nonumber \\
&&
+ \sum_{k_1=0}^{+\infty}\!\!{}'\, \sigma_2^{k_1}\, \rme^{+ k_1 A_1\, z^{\frac{7}{6}}} \sum_{k_2=0}^{+\infty}\!\!{}'\, \sigma_4^{k_2}\, \rme^{+ k_2 A_2\, z^{\frac{7}{6}}}\, \Upphi^{(k_1,k_2)}_{--} (z,\uptau_{\mathsf{I}}, \uptau_{\mathsf{II}}) + \nonumber \\
&&
+ \sum_{k_1=0}^{+\infty}\!\!{}'\, \sigma_1^{k_1}\, \rme^{- k_1 A_1\, z^{\frac{7}{6}}} \sum_{k_2=0}^{+\infty}\!\!{}'\, \sigma_4^{k_2}\, \rme^{+ k_2 A_2\, z^{\frac{7}{6}}}\, \Upphi^{(k_1,k_2)}_{+-} (z,\uptau_{\mathsf{I}}, \uptau_{\mathsf{II}}) + \nonumber \\
&&
+ \sum_{k_1=0}^{+\infty}\!\!{}'\, \sigma_2^{k_1}\, \rme^{+ k_1 A_1\, z^{\frac{7}{6}}} \sum_{k_2=0}^{+\infty}\!\!{}'\, \sigma_3^{k_2}\, \rme^{- k_2 A_2\, z^{\frac{7}{6}}}\, \Upphi^{(k_1,k_2)}_{-+} (z,\uptau_{\mathsf{I}}, \uptau_{\mathsf{II}}), \nonumber
\eea
\noindent
where the primes in the double-sums indicate we must exclude the $(k_1,k_2)=(0,0)$ contribution, where the dependence is now on two ``moduli'' $\uptau_{\mathsf{I}} \equiv \sigma_1 \sigma_2$ and $\uptau_{\mathsf{II}} \equiv \sigma_3 \sigma_4$, and where the resonant asymptotic series are:
\bea
\Upphi^{(0,0)} (z,\uptau_{\mathsf{I}}, \uptau_{\mathsf{II}}) &:=& \sum_{\ell_1=0}^{+\infty} \sum_{\ell_2=0}^{+\infty} \left( \sigma_1 \sigma_2 \right)^{\ell_1} \left( \sigma_3 \sigma_4 \right)^{\ell_2} \Phi_{(\ell_1,\ell_1,\ell_2,\ell_2)} (z), \\
\Upphi^{(k_1,k_2)}_{++} (z,\uptau_{\mathsf{I}}, \uptau_{\mathsf{II}}) &:=& \sum_{\ell_1=0}^{+\infty} \sum_{\ell_2=0}^{+\infty} \left( \sigma_1 \sigma_2 \right)^{\ell_1} \left( \sigma_3 \sigma_4 \right)^{\ell_2} \Phi_{(\ell_1+k_1,\ell_1,\ell_2+k_2,\ell_2)} (z), \\
\Upphi^{(k_1,k_2)}_{--} (z,\uptau_{\mathsf{I}}, \uptau_{\mathsf{II}}) &:=& \sum_{\ell_1=0}^{+\infty} \sum_{\ell_2=0}^{+\infty} \left( \sigma_1 \sigma_2 \right)^{\ell_1} \left( \sigma_3 \sigma_4 \right)^{\ell_2} \Phi_{(\ell_1,\ell_1+k_1,\ell_2,\ell_2+k_2)} (z), \\
\Upphi^{(k_1,k_2)}_{+-} (z,\uptau_{\mathsf{I}}, \uptau_{\mathsf{II}}) &:=& \sum_{\ell_1=0}^{+\infty} \sum_{\ell_2=0}^{+\infty} \left( \sigma_1 \sigma_2 \right)^{\ell_1} \left( \sigma_3 \sigma_4 \right)^{\ell_2} \Phi_{(\ell_1+k_1,\ell_1,\ell_2,\ell_2+k_2)} (z), \\
\Upphi^{(k_1,k_2)}_{-+} (z,\uptau_{\mathsf{I}}, \uptau_{\mathsf{II}}) &:=& \sum_{\ell_1=0}^{+\infty} \sum_{\ell_2=0}^{+\infty} \left( \sigma_1 \sigma_2 \right)^{\ell_1} \left( \sigma_3 \sigma_4 \right)^{\ell_2} \Phi_{(\ell_1,\ell_1+k_1,\ell_2+k_2,\ell_2)} (z).
\eea
\noindent
The general pattern at arbitrary $k$, even if cumbersome, should be straightforward to write down. Returning to the resonant Borel singularities mentioned above, their disentangling is intricate even for the Painlev\'e~I case \cite{asv11, bssv22} and it should be rather complicated in the general order-$k$ multicritical case. In any case, for the purposes of the present paper, we shall simply make use of the projection map \eqref{eq:map-Zk-to-C} in order to solely focus on the location of na\"\i ve multiple-instanton Borel singularities (even in the general case). Let us see how.

\begin{figure}[t!]
\centering
     \begin{subfigure}[h]{0.4\textwidth}
         \centering
         \includegraphics[width=\textwidth]{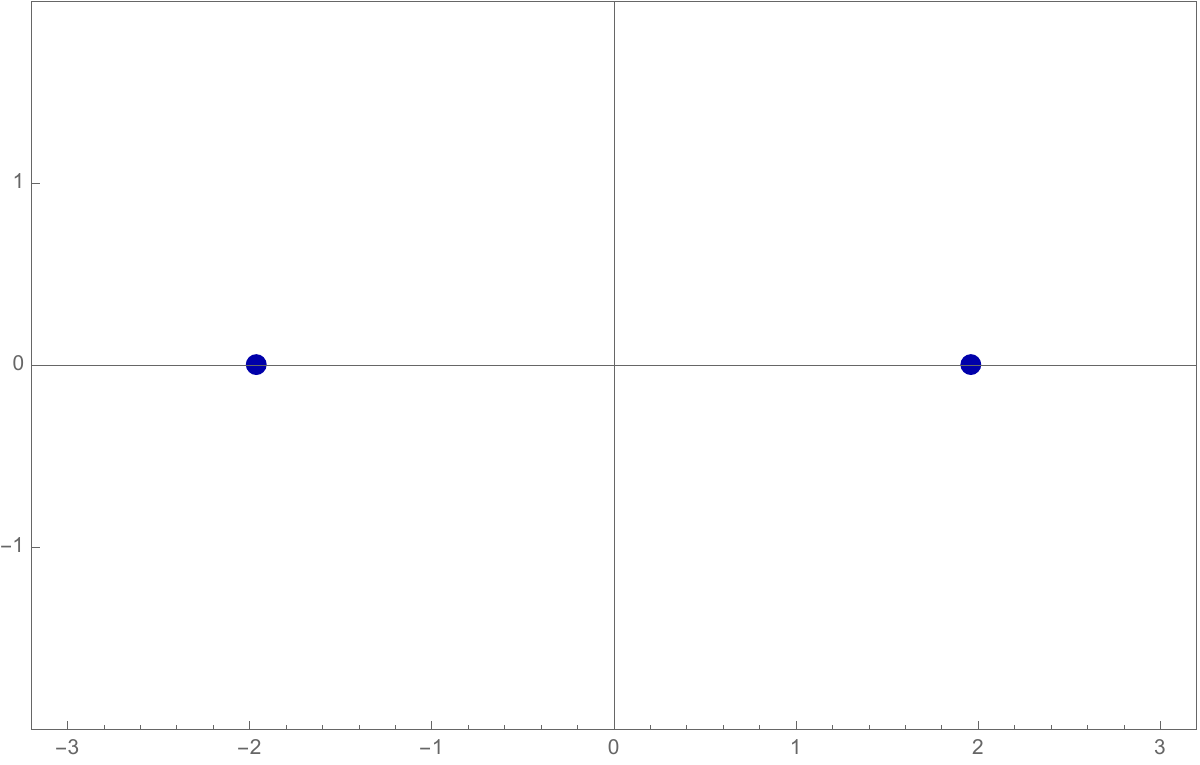}
     \end{subfigure}
\hspace{2mm}
     \begin{subfigure}[h]{0.57\textwidth}
         \centering
         \includegraphics[width=\textwidth]{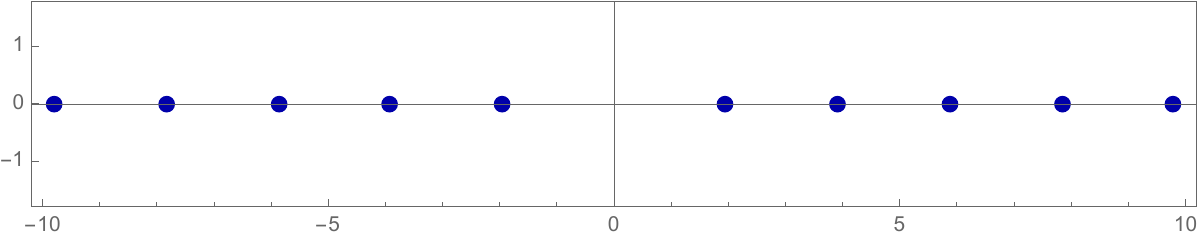}\\
         \vspace{3mm}
         \includegraphics[width=\textwidth]{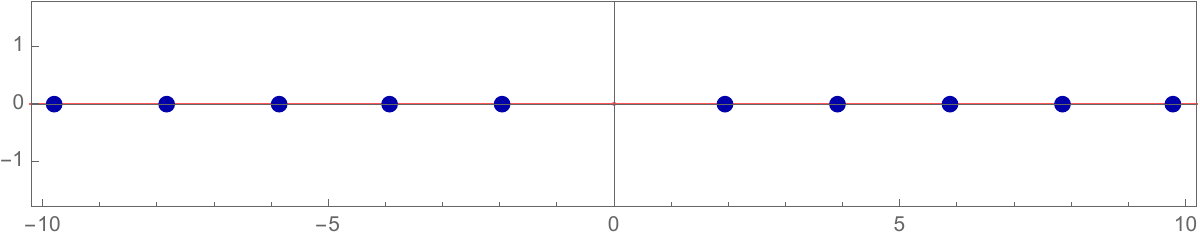}
     \end{subfigure}
\caption{These images represent the complex Borel plane for the $k=2$ or $(2,3)$ multicritical theory. On the left plot, the dots stand for the two (symmetric) instanton actions, \eqref{eq:A&beta-(2,3)} or \eqref{eq:inst-vect-k=2}. On the right-top plot, dots now stand for the multi-instanton Borel singularities descendent from the full transseries \eqref{eq:general-k-u-trans} via the projection map \eqref{eq:map-Zk-to-C} (see \cite{abs18} for details). The right-bottom plot is the same as the top one, but with the corresponding Stokes lines highlighted in red. The positive real axis is a Stokes line for this (even $k$) problem.}
\label{fig:k=2-inst-acts-&-stokes-lines}
\end{figure}

\begin{figure}[t!]
\centering
     \begin{subfigure}[h]{0.4\textwidth}
         \centering
         \includegraphics[width=\textwidth]{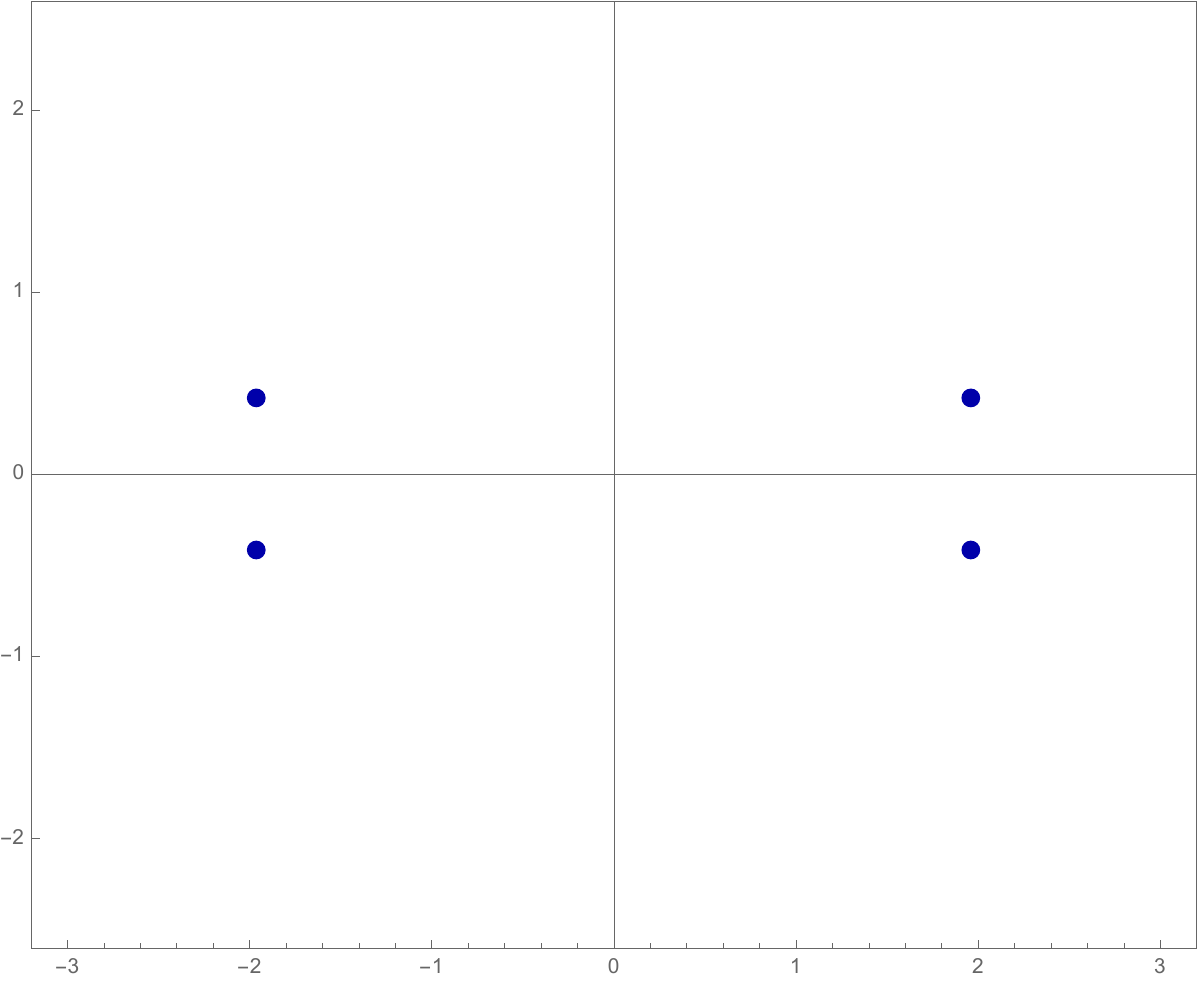}
     \end{subfigure}
\hspace{2mm}
     \begin{subfigure}[h]{0.57\textwidth}
         \centering
         \includegraphics[width=\textwidth]{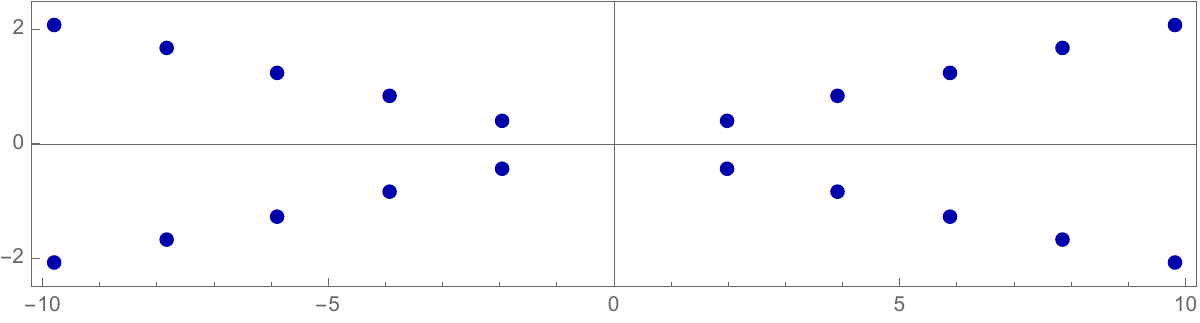}\\
         \vspace{3mm}
         \includegraphics[width=\textwidth]{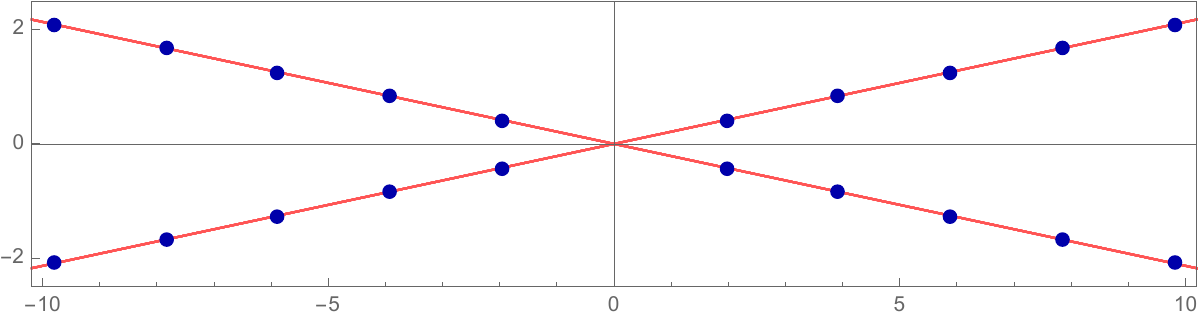}
     \end{subfigure}
\caption{These images represent the complex Borel plane for the $k=3$ or $(2,5)$ multicritical theory. On the left plot, the dots stand for the four (symmetric and complex conjugate) instanton actions, \eqref{eq:A&beta-(2,5)} or \eqref{eq:inst-vect-k=3}. On the right-top plot, dots now stand for the multi-instanton Borel singularities descendent from the full transseries \eqref{eq:general-k-u-trans} via the projection map \eqref{eq:map-Zk-to-C} (see \cite{abs18} for details). The right-bottom plot is the same as the top one, but with the corresponding Stokes lines highlighted in red. At this level of purely multiplicative Borel singularities, the positive real axis is apparently not a Stokes line for this (odd $k$) problem (but see the main text).}
\label{fig:k=3-inst-acts-&-stokes-lines}
\end{figure}

\begin{figure}[t!]
\centering
     \begin{subfigure}[h]{0.4\textwidth}
         \centering
         \includegraphics[width=\textwidth]{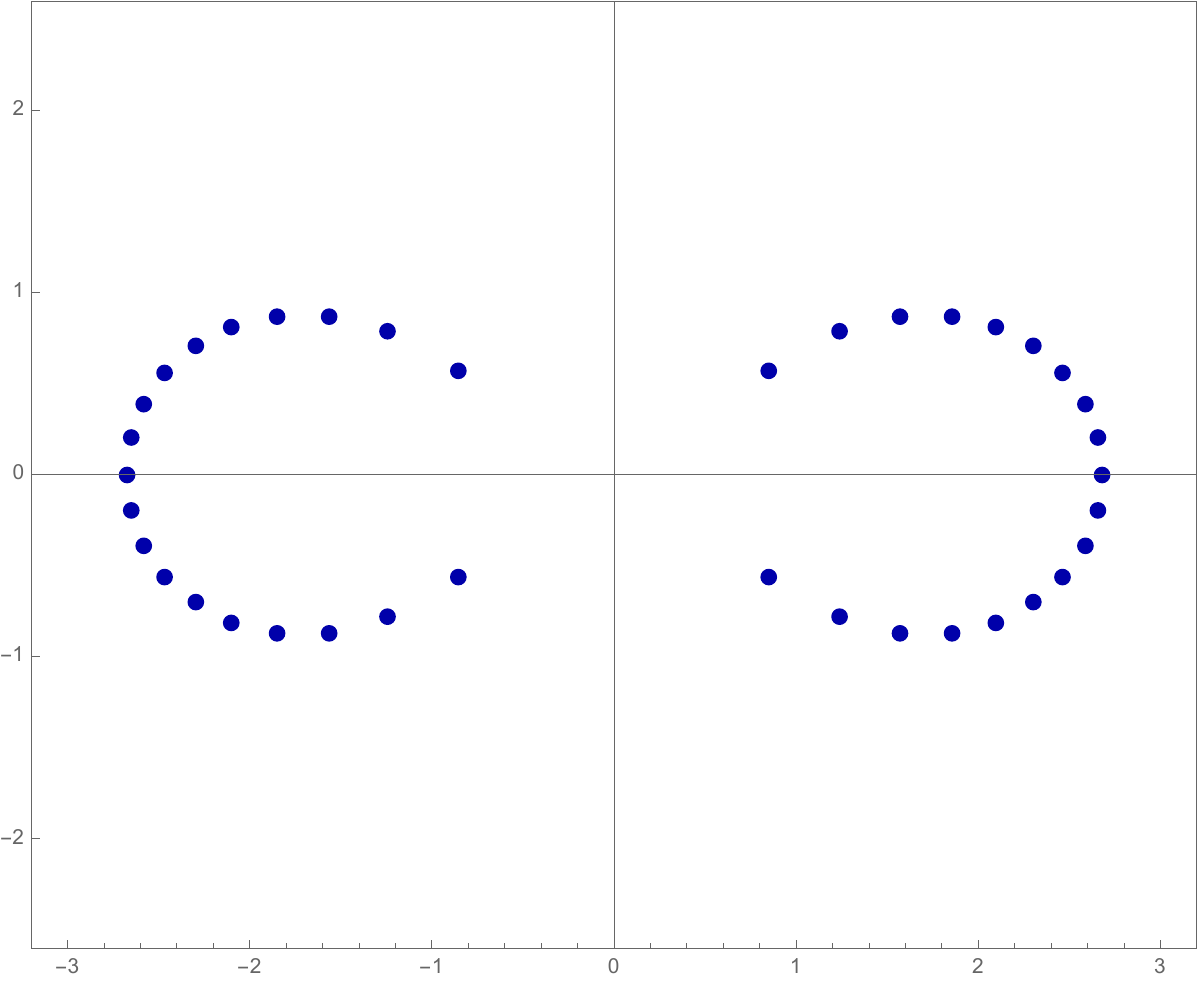}
     \end{subfigure}
\hspace{2mm}
     \begin{subfigure}[h]{0.57\textwidth}
         \centering
         \includegraphics[width=\textwidth]{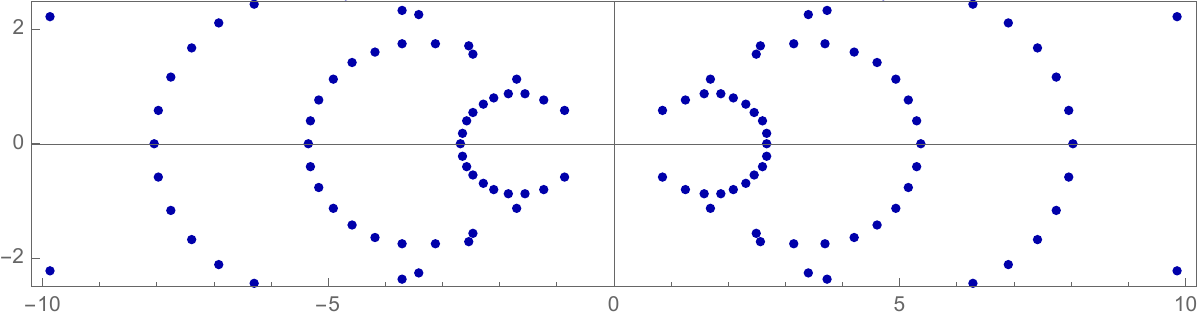}\\
         \vspace{3mm}
         \includegraphics[width=\textwidth]{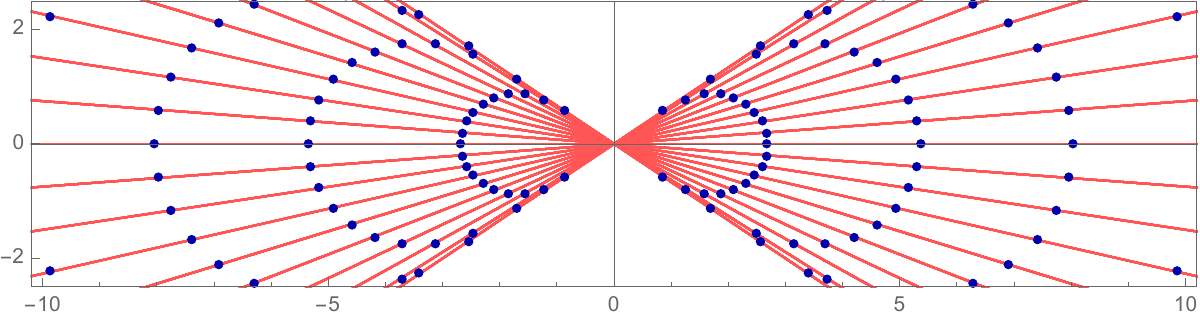}
     \end{subfigure}
\caption{These images represent the complex Borel plane for the $k=20$ multicritical theory. On the left plot, the dots stand for the $2 \times 20 - 2 = 38$ (symmetric and complex conjugate) instanton actions, \eqref{eq:inst-vect-k=k}. On the right-top plot, dots now stand for the multi-instanton Borel singularities descendent from the full transseries \eqref{eq:general-k-u-trans} via the projection map \eqref{eq:map-Zk-to-C} (see \cite{abs18} for details). The right-bottom plot is the same as the top one, but with the corresponding Stokes lines highlighted in red. The positive real axis is a Stokes line for this (even $k$) problem.}
\label{fig:k=20-inst-acts-&-stokes-lines}
\end{figure}

\begin{figure}[t!]
\centering
     \begin{subfigure}[h]{0.4\textwidth}
         \centering
         \includegraphics[width=\textwidth]{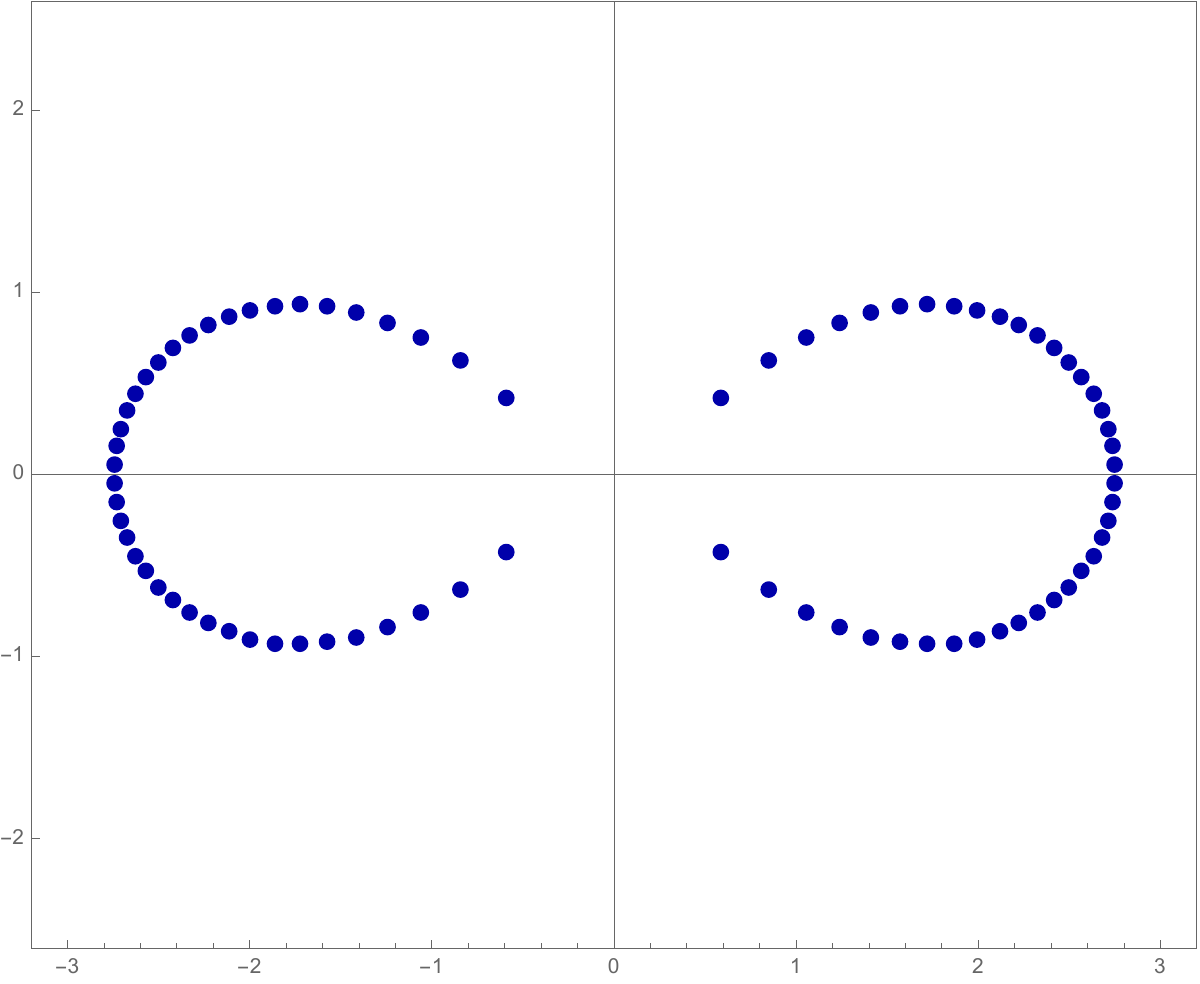}
     \end{subfigure}
\hspace{2mm}
     \begin{subfigure}[h]{0.57\textwidth}
         \centering
         \includegraphics[width=\textwidth]{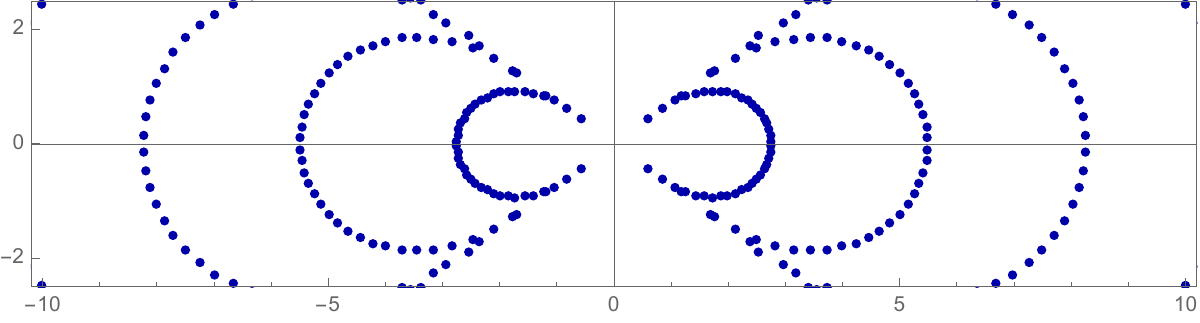}\\
         \vspace{3mm}
         \includegraphics[width=\textwidth]{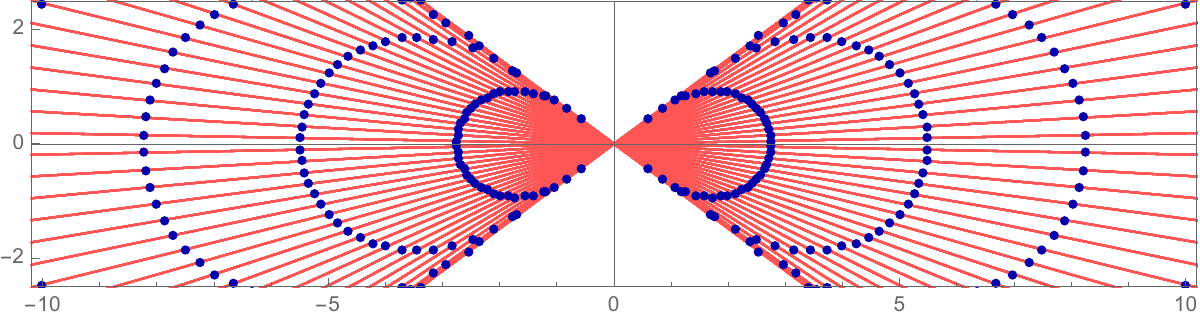}
     \end{subfigure}
\caption{These images represent the complex Borel plane for the $k=41$ multicritical theory. On the left plot, the dots stand for the $2 \times 41 - 2 = 80$ (symmetric and complex conjugate) instanton actions, \eqref{eq:inst-vect-k=k}. On the right-top plot, dots now stand for the multi-instanton Borel singularities descendent from the full transseries \eqref{eq:general-k-u-trans} via the projection map \eqref{eq:map-Zk-to-C} (see \cite{abs18} for details). The right-bottom plot is the same as the top one, but with the corresponding Stokes lines highlighted in red. At this level of purely multiplicative Borel singularities, the positive real axis is apparently not a Stokes line for this (odd $k$) problem (but see the main text).}
\label{fig:k=41-inst-acts-&-stokes-lines}
\end{figure}

In figure~\ref{fig:k=2-inst-acts-&-stokes-lines} we discuss the multi-instanton Borel singularities associated to the $k=2$ or $(2,3)$ multicritical theory, and in figure~\ref{fig:k=3-inst-acts-&-stokes-lines} we do the same for the $k=3$ or $(2,5)$ theory. These Borel singularities organize themselves along rays on the complex plane, known as Stokes lines, which we also plot (see, \textit{e.g.}, \cite{abs18}). One clear difference between the even $k=2$ and odd $k=3$ cases, is that in the former the positive/negative real axes are Stokes lines, whereas in the latter apparently they are not (with obvious consequences on what concerns Borel summability along $\BR^+$). This can be also seen at large values\footnote{In fact, it can be seen for generic values of $k$. It was shown in \cite{gz90b, gz91} that the polynomial \eqref{eq:poly-inst-act-string-eq} always has a real positive solution for $\rho^2$ when $k$ is even, whereas there are no real solutions for $\rho^2$ when $k$ is odd.} of $k$. In figure~\ref{fig:k=20-inst-acts-&-stokes-lines} we discuss the multi-instanton Borel singularities associated to the (even) $k=20$ multicritical theory, and in figure~\ref{fig:k=41-inst-acts-&-stokes-lines} the same for the (odd) $k=41$ one. One caveat in these figures is that we are just plotting integer multiples of the instanton actions (na\"\i ve ``primary sheet'' Borel singularities). There are also (``secondary sheet'') Borel singularities associated to their generic linear combinations \cite{s07,s14}. In fact, once this is taken into account, and because instanton actions arise in complex-conjugate pairs, it turns out that the positive/negative real axes are \textit{always} Stokes lines\footnote{Albeit to be fully rigorous, one still needs to compute Stokes data associated to these would-be singularities.}.

%%%%%%%%%%%%%%%%%%%%%%%%%%%%%%%%%%%%%%%%%%%%%%%%%%%%%%%%%%%%%%%%%
%%%%%%%%%%%%%%%%%%%%%%%%%%%%%%%%%%%%%%%%%%%%%%%%%%%%%%%%%%%%%%%%%

%%%%%%%%%%%%%%%%%%%%%%%%%%%%%%%%%%%%%%%%%%%%%%%%%%%%%%%%%%%%%%%%%
%%%%%%%%%%%%%%%%%%%%%%%%%%%%%%%%%%%%%%%%%%%%%%%%%%%%%%%%%%%%%%%%%

%%%%%%%%%%%%%%%%%%%%%%%%%%%%%%%%%%%%%%%%%%%%%%%%%%%%%%%%%%%%%%%%%
%%%%%%%%%%%%%%%%%%%%%%%%%%%%%%%%%%%%%%%%%%%%%%%%%%%%%%%%%%%%%%%%%

%%%%%%%%%%%%%%%%%%%%%%%%%%%%%%%%%%%%%%%%%%%%%%%%%%%%%%%%%%%%%%%%%
%%%%%%%%%%%%%%%%%%%%%%%%%%%%%%%%%%%%%%%%%%%%%%%%%%%%%%%%%%%%%%%%%

\begin{figure}[t!]
\centering
     \begin{subfigure}[h]{0.45\textwidth}
         \centering
         \includegraphics[width=\textwidth]{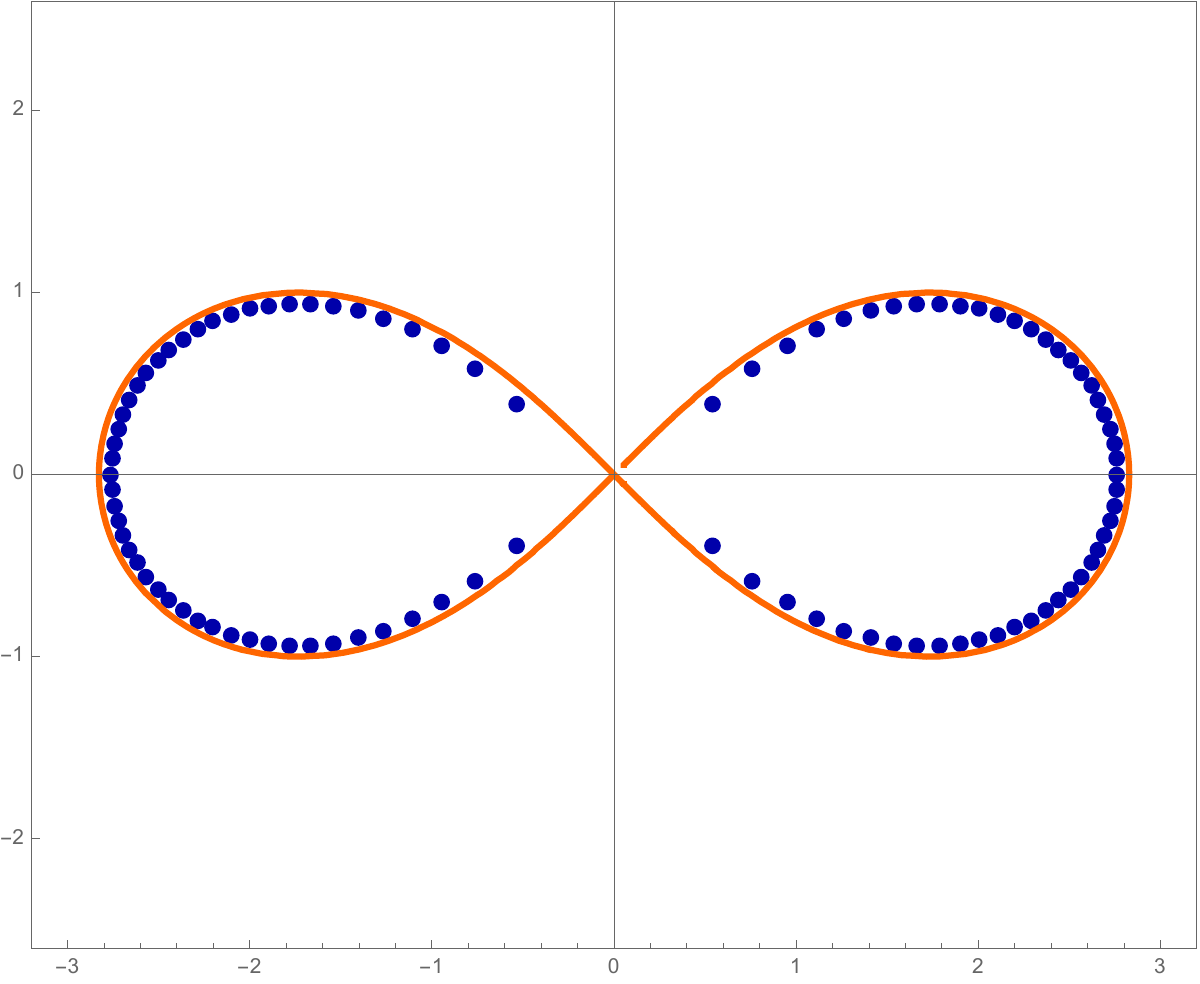}
     \end{subfigure}
\hspace{6mm}
     \begin{subfigure}[h]{0.45\textwidth}
         \centering
         \includegraphics[width=\textwidth]{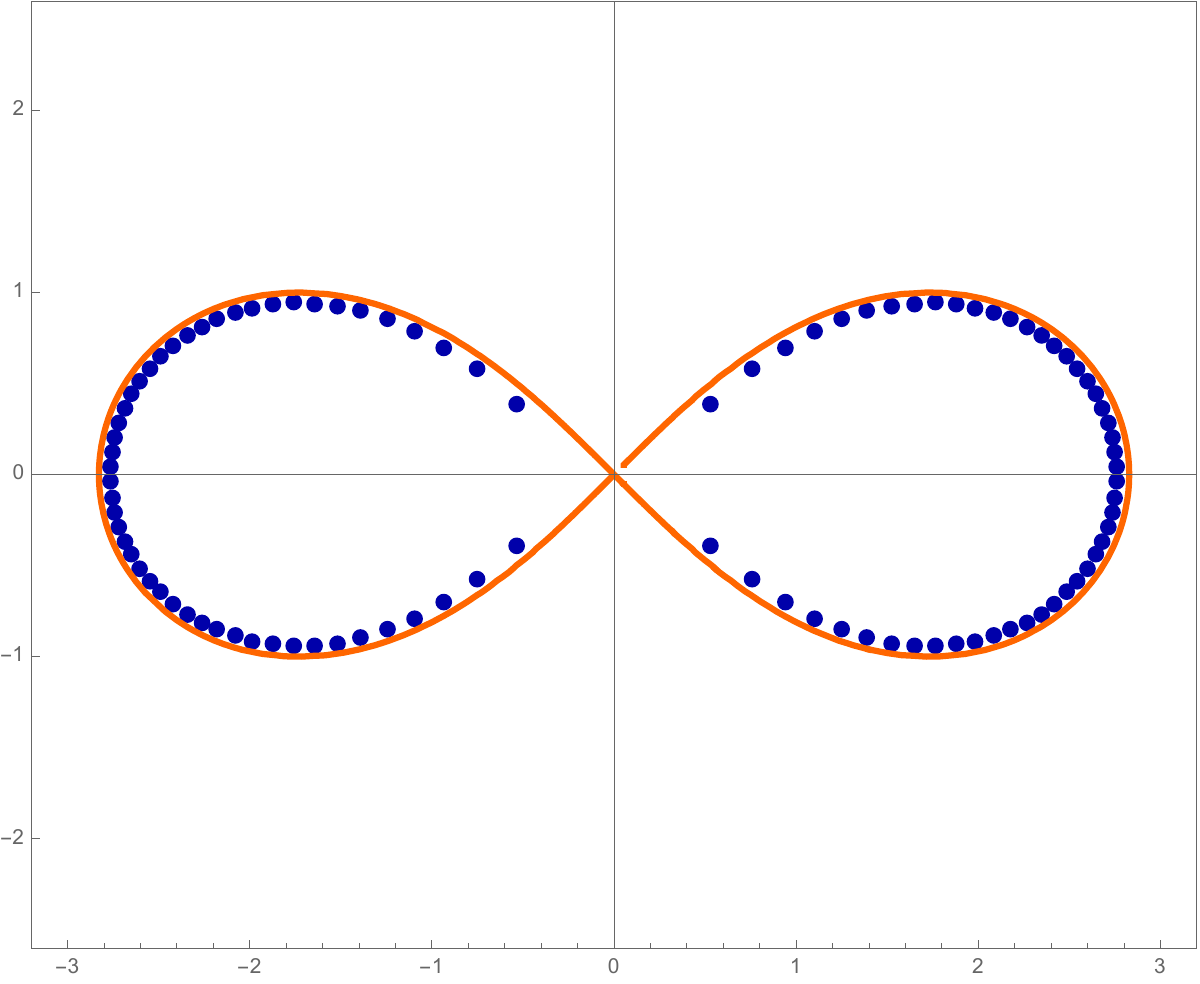}
     \end{subfigure}
\caption{Instanton actions (on the complex Borel plane) for the $k=50$ and the $k=51$ multicritical theories (left plot, right plot, respectively). Their corresponding non-trivial saddles were plotted back in figure~\ref{fig:multicrit_saddles}. As already discussed, even $k$ leads to a real instanton action (and its symmetric pair), whereas odd $k$ does not. As $k$ grows the distribution of these instanton actions asymptote to the Bernoulli-lemniscate discussed in the text, also plotted in both figures.}
\label{fig:multi-inst-action-lemniscate}
\end{figure}

Closed-form expressions at generic $k$ are however out of reach: as already discussed back in subsection~\ref{subsec:multicritical} there are no generic expressions for the roots of hypergeometric polynomials, hence no generic formulae for multicritical instanton actions. But, just as discussed in subsection~\ref{subsec:multicritical}, something may be said about their $k\to+\infty$ limit: they asymptote to the $\abs{z^2-4} = 4$ Bernoulli-lemniscate on the $z$ complex plane\footnote{In standard cartesian coordinates this is $\left( x^2+y^2 \right)^2 = 8 \left( x^2-y^2 \right)$, making the lemniscate equation clear.}, from its inside \cite{zsw12}. This is illustrated in figure~\ref{fig:multi-inst-action-lemniscate}.

%%%%%%%%%%%%%%%%%%%%%%%%%%%%%%%%%%%%%%%%%%%%%%%%%%%%%%%%%%%%%%%%%
%%%%%%%%%%%%%%%%%%%%%%%%%%%%%%%%%%%%%%%%%%%%%%%%%%%%%%%%%%%%%%%%%

Let us end the nonperturbative discussion of the string equations with two comments concerning the resurgent-transseries analysis of generic, order-$k$ multicritical models. The first comment we have already briefly mentioned: if we properly explore the multi-sheeted structure of Borel singularities, either even or odd $k$ the positive real axis is always a Stokes line. This implies generic multicritical models, with all instantons turned on, are never Borel resummable along the positive real axis. Of course given the number of Stokes lines (and consequently, Stokes transitions) in the general $k$ theory (look back at figures~\ref{fig:k=20-inst-acts-&-stokes-lines} and~\ref{fig:k=41-inst-acts-&-stokes-lines}), Borel summability across the complex plane will be equally intricate in either even or odd cases.

The second comment concerns resurgent large-order relations (which we will address in detail in section~\ref{sec:large_order}). Focusing solely on the perturbative sector \eqref{eq:u_pert_exp-2}, the na\"\i ve \textit{leading} large-order growth of the specific-heat perturbative coefficients should be of the form \cite{gz91, asv11, abs18}
\be
\label{eq:multi-large-order-naive}
u^{(0)}_{g \gg 1} \simeq \frac{S_{1}}{2\pi\rmi}\, \frac{\Gamma \left( 2g-\beta \right)}{A^{2g-\beta}}\, u^{(1)}_{0} + \cdots.
\ee
\noindent
Herein, $u^{(0)}_{g}$ are the perturbative coefficients in the asymptotic expansion \eqref{eq:u_pert_exp-2}, $S_{1}$ is the Stokes coefficient\footnote{Let us mention again that, being fully precise, we should consider the appropriate Borel residue or Stokes vector \cite{abs18}, but for the purposes of this work we shall stick with the simplified terminology of ``Stokes coefficients''.}, $A$ is the instanton action, $\beta = \frac{1}{2}$ the characteristic exponent we have already computed, and $u^{(1)}_{0}$ the leading coefficient of the one-instanton sector in \eqref{eq:trans-z-u} (which we have normalized to one in \eqref{eq:u(1)(2,3)} and \eqref{eq:u(1)(2,5)}, following standard transseries conventions). Now the first question one may have is: which instanton action?, out of the $\left(2k-2\right)$ possibilities in \eqref{eq:inst-act-ell-rho} and \eqref{eq:poly-inst-act-string-eq}, \textit{i.e.}, in \eqref{eq:inst-vect-k=k}. The correct answer is \textit{all}, at least if we are to have a \textit{complete} resurgent large-order relation \cite{abs18}, in which case the above equation needs appropriate modification. But our point is that even if we solely focus on the \textit{leading} contribution to the large-order growth, the above na\"\i ve large-order relation also requires modification. In this case, the leading contribution is associated to the Borel singularities closest\footnote{The actual distance to the origin of the leading Borel singularity(ies) changes with $k$. For example, in the $k=2$ Painlev\'e~I case the distance is $d \approx 1.95959\ldots$, whereas in the $k=3$ Yang--Lee case this distance is $d \approx 2.00601\ldots$. Then, as $k$ grows the distance shrinks: for $k=10$ it is $d \approx 1.38842\ldots$, for $k=50$ it is $d \approx 0.661924\ldots$, for $k=100$ it is $d \approx 0.471571\ldots$, for $k=500$ it is $d \approx 0.212145\dots$, and for $k=1000$ it is $d \approx 0.150119\ldots$.} to the origin. In the Painlev\'e~I case in figure~\ref{fig:k=2-inst-acts-&-stokes-lines} that is two of them (studied in detail in \cite{gikm10, asv11}), but in the generic case, either figures~\ref{fig:k=3-inst-acts-&-stokes-lines}, \ref{fig:k=20-inst-acts-&-stokes-lines}, or~\ref{fig:k=41-inst-acts-&-stokes-lines}, that means \textit{four}\footnote{Englobing both symmetric and complex-conjugate pairs, a generic feature of multicritical Borel singularities.} of them (a situation which is highly reminiscent of what happened in the cases of topological strings on the resolved conifold and Chern--Simons gauge theory on lens spaces in \cite{ps09, ars14}). The leading Stokes data for these contributions should be computable from \eqref{eq:MM-stokes-coeff-inf} following \cite{msw07}, and this we have done earlier in \eqref{eq:S1F1SG(2,3)}, \eqref{eq:S1F1SG(2,5)}, and generically \eqref{eq:multicritical-stokes} (we study these from large-order in section~\ref{sec:large_order}). Furthermore, complete large-order relations describing the growth of the perturbative sector up to \textit{multi}-instanton (exponentially suppressed) contributions should be obtainable, as well as resurgent large-order relations for the (resonant) multi-instanton sectors of the generic order-$k$ theory. But we would expect all these to be extremely intricate, and as such out of the scope of the present paper. The same considerations hold for generic (complete) Stokes data (see \cite{asv11, abs18, bssv22}). In particular, it might well be that the techniques recently developed in \cite{bssv22} to compute the complete Stokes data of the Painlev\'e~I case might be extendable to the full Stokes data of generic order-$k$ multicritical cases.

%%%%%%%%%%%%%%%%%%%%%%%%%%%%%%%%%%%%%%%%%%%%%%%%%%%%%%%%%%%%%%%%%
\subsection{String Equations and General KdV Times}\label{subsec:general-KdV}
%%%%%%%%%%%%%%%%%%%%%%%%%%%%%%%%%%%%%%%%%%%%%%%%%%%%%%%%%%%%%%%%%

As just discussed in the previous subsection, there is a string equation describing the (specific heat of the) exact $k$th \textit{multicritical} model, given by \eqref{eq:multicritical_string_eq}. But, \textit{off-criticality}, one may also consider renormalization-group flows in-between these different multicritical theories \cite{gm90a, ds90, bdss90, dss90}. Such interpolating flows are described by the general, ``massive'', string equation\footnote{Akin to what we commented for the multicritical string equation in \eqref{eq:k+1/2_string_eq_norm}, also herein one sometimes finds an alternative normalization in the literature,
\be
\sum_{k=1}^{+\infty} \left( k+\frac{1}{2} \right) t_k R_k \left[ u \right] = z.
\ee
\noindent
As in \eqref{eq:S_action_multi}, it has the advantage that is also quite simple to write down an action for this string equation \cite{gz90b},
\be
\CS = \int \rmd z \left( \sum_{k=1}^{+\infty} t_{k} R_{k+1} + t_{0} R_{1} \right) = \sum_{k=0}^{+\infty} t_{k} \int \rmd z\, R_{k+1}
\ee
\noindent
(herein we have set $t_{0} \equiv -4z$ for a more compact final expression).}
\be
\label{eq:massive_string_eq}
\sum_{k=1}^{+\infty} t_k R_k \left[ u \right] = z.
\ee
\noindent
Herein the $\{ t_k \}$ are the KdV times (this is basically a limit of the Novikov equation \eqref{eq:Novikov} \cite{gd75}). Clearly, the specific-heat solution to the above ``massive'' string equation now gains dependence on all these KdV times\footnote{Hence this equation also describes 2d topological gravity with \textit{arbitrary} KdV times \cite{dw90, w91, k92, iz92, dw18}. However our focus in this subsection is solely on the flow in-between distinct multicritical theories. We shall then discuss 2d topological gravity with arbitrary KdV times later in section~\ref{sec:large_order}.} (and which might be computed via the higher KdV flows \eqref{eq:KdV_flow}).

For example, the flow from the $k=3$ model, described by the Yang--Lee equation \eqref{eq:k=3_string_eq}, to the $k=2$ model, described by the Painlev\'e~I equation \eqref{eq:k=2_string_eq}, may be described by the massive string-equation \eqref{eq:massive_string_eq} with only non-zero $\left\{ t_2, t_3 \right\}$ KdV times. This is a one-parameter flow (one could, for instance, adequately normalize $t_3 = - \frac{32}{5}$ via \eqref{eq:multicritical_string_eq} and parametrize the flow solely with $t_2$)---but this seemingly simplest problem already leads to a rather intricate nonlinear ODE (see below). An alternative approach would be to follow \cite{dss90} and rewrite this flow as a KdV differential flow using \eqref{eq:KdV_flow} (\textit{e.g.}, describing the $t_2$-dependence of the specific heat).

In order to explicitly write down string equations for these ``massive'' models (or, later on, for minimal strings), one first needs to address the genus expansion of their corresponding solutions in the present context where there is explicit dependence on several variables. To illustrate this point, start by going back to the Painlev\'e~I equation \eqref{eq:k=2_string_eq} and its perturbative solution \eqref{eq:pert(2,3)-u}. The resulting perturbative free-energy \eqref{eq:pert(2,3)-F} matches the spectral-geometry genus expansion, \eqref{eq:multicrit_free_en-2}-\eqref{eq:multicrit_free_en-3}-\eqref{eq:multicrit_free_en-4} in \eqref{eq:Fds}, as we identified the string-equation variable $z$ with the double-scaled string-coupling $\kappa_{\text{s}}$ via \eqref{eq:kappa-to-z}. This interchangeable role between $z$ and $\kappa_{\text{s}}$ is \textit{a property of multicritical models}, and it ultimately arises from the very simple relation between $z$ and the classical (planar) specific heat $u_0$ via the classical (leading) string equation \eqref{eq:planar_multi_string-eq},
\be
\label{eq:u0=z1/k-easy}
u_0^k = z \qquad \Rightarrow \qquad u_0 = z^\frac{1}{k} \equiv \kappa_{\text{s}}^{-\frac{2}{2k+1}};
\ee
\noindent
where we recall \eqref{eq:planar_u_&_F} to make the bridge to the string genus-expansion. As more KdV times get turned on, the classical string-equation becomes more and more intricate. The simplest way to handle this as we move forward is to just \textit{undo} the identification of $z$ and $\kappa_{\text{s}}$ in \eqref{eq:kappa-to-z}. The string coupling is then explicitly introduced in the string equation from scratch, which is done via the following rescaling\footnote{This essentially amounts to introducing a factor of $g_{\text{s}}$ with every derivative appearing in the string equation.} of all KdV times and $z$-variable in \eqref{eq:massive_string_eq}
\be
\label{eq:t&x_KdV_rescale}
t_k \rightarrow \frac{t_k}{g_{\text{s}}}, \qquad z \rightarrow \frac{z}{g_{\text{s}}}.
\ee
\noindent
Note that we are now denoting the ``massive'' (and eventually, minimal) double-scaled string-coupling by $g_{\text{s}}$. This should not be confused with the same notation in the off-critical matrix model in \eqref{eq:ZN}; it still is double-scaled in the sense of $\kappa_{\text{s}}$ in \eqref{eq:double-scaled-string-coupling}---we are just no longer using the $\kappa_{\text{s}}$-notation as it was always identified with the variable $z$ in  the discussion up to now. In the Painlev\'e~I example we are illustrating this point with, one obtains in the new notation
\be
u^2 - \frac{1}{3}\, g_{\text{s}}^2\, u'' = z,
\ee
\noindent
leading to the perturbative expansions
\be
u_{(2,3)} \simeq z^{\frac{1}{2}} - \frac{1}{24 z^{2}}\, g_{\text{s}}^{2} - \frac{49}{1152 z^{\frac{9}{2}}}\, g_{\text{s}}^{4} - \frac{1225}{6912 z^{7}}\, g_{\text{s}}^{6} - \frac{4412401}{2654208 z^{\frac{19}{2}}}\, g_{\text{s}}^{8} - \frac{73560025}{2654208 z^{12}}\, g_{\text{s}}^{10} - \cdots
\ee
\noindent
and
\be
F_{\text{ds}} \simeq - \frac{4 z^{\frac{5}{2}}}{30}\, \frac{1}{g_{\text{s}}^{2}} - \frac{1}{48} \log z + \frac{7}{2880 z^{\frac{5}{2}}}\, g_{\text{s}}^{2} + \frac{245}{82944 z^{5}}\, g_{\text{s}}^{4} + \frac{259553}{19906560 z^{\frac{15}{2}}}\, g_{\text{s}}^{6} + \frac{1337455}{10616832 z^{10}}\, g_{\text{s}}^{8} + \cdots.
\ee
\noindent
These are of course completely equivalent to \eqref{eq:pert(2,3)-u} and \eqref{eq:pert(2,3)-F}, now in the $z$ and $g_{\text{s}}$ notation. At the level of the Gel'fand--Dikii KdV potentials, the recursion relation \eqref{eq:Rell_recursion} is now\footnote{We keep using the exact same notation, in spite of the extra factor in the string coupling $g_{\text{s}}$.} simply
\be
R_{\ell+1}' = \frac{1}{4}\, g_{\text{s}}^2\, R_{\ell}''' - u\, R_{\ell}' - \frac{1}{2} u'\, R_{\ell},
\ee
\noindent
and a few first of these are, as expected,
\bea
\label{eq:GD-R2-gs}
R_2 &=& \frac{1}{16} \left( 3 u^2 - g_{\text{s}}^2\, u'' \right), \\
\label{eq:GD-R3-gs}
R_3 &=& - \frac{1}{64} \left( 10 u^3 - 10 g_{\text{s}}^2\, u u'' - 5 g_{\text{s}}^2 \left( u' \right)^2 + g_{\text{s}}^4\, u'''' \right).
\eea

Let us go back to our generic ``massive'' string equation \eqref{eq:massive_string_eq}, with all KdV times turned on. The classical (genus zero) string equation is\footnote{For now we have absorbed the $\frac{1}{k}\, \alpha_{kk}$ factors from the KdV potentials \eqref{eq:Rl-alphall} into the KdV times (see below).}
\be
\label{eq:classical-massive-string-eq}
\sum_{k=1}^{+\infty} t_k u_0^k = z,
\ee
\noindent
and the aforementioned simple (multicritical) correspondence between $z$ and $\kappa_{\text{s}}$ is lost---which just reflects the fact that the simple solution for $u_0 (z)$ in \eqref{eq:u0=z1/k-easy} is lost\footnote{For example, if we have non-vanishing KdV times only up to order $K$, then the solutions to the classical string-equation are the $K$ roots of the resulting degree-$K$ polynomial, with dependence on the KdV times, $u_0 \equiv u \left( z, t_{1}, \ldots, t_{K} \right)$. In terms of the maximal real root $u_0$ \cite{bz08}, the genus-zero free energy becomes (more on this below)
\be
F_0 \left( u_0, \left\{ t_k \right\} \right) = - \sum_{i=1}^{+\infty} \sum_{j=1}^{+\infty} \frac{i j}{\left( i+1 \right) \left( i+j+1 \right)}\, t_i t_j\, u_0^{i+j+1}.
\ee
}. This problem propagates to all higher genera: using the perturbative \textit{ansatz} (replacing \eqref{eq:u_pert_exp}-\eqref{eq:u_pert_exp-2})
\be
\label{eq:u_pert_exp-gs}
u(z) \simeq \sum_{g=0}^{+\infty} u_g (z)\, g_{\text{s}}^{2g}
\ee
\noindent
in \eqref{eq:massive_string_eq}, the coefficients $u_g \equiv u_g (z)$ all have, in principle, a \textit{complicated} algebraic dependence on $z$. But this is ultimately due to \eqref{eq:classical-massive-string-eq}---which also means that this problem is easily solvable by working with $u_0$ from now on, rather than $z$ \cite{biz80, iz92}. In this variable, all recursively-constructed coefficients $u_g \equiv u_g (u_0)$ become \textit{rational functions} of $u_0$, and the perturbative-level ``multicritical simplicity'' is recovered. In fact, from a pragmatic, purely perturbative-asymptotic standpoint, one need not worry in explicitly solving \eqref{eq:classical-massive-string-eq}; an implicit solution\footnote{All one needs to worry about are $z$-derivatives of $u_0 (z)$, which follow from the implicit definition alone \eqref{eq:classical-massive-string-eq}.} will do fine.

The one final point to address is to ensure that the construction outlined above will match the equivalent perturbative (and nonperturbative) constructions arising from spectral geometry. Having just introduced explicit $g_{\text{s}}$-dependence in the string equation, one still has to introduce explicit $z$-dependence on the spectral geometry side. The spectral curve with explicit dependence on all KdV times, and including $z$ via $u_0$, takes the form \cite{e16}
\bea
\label{eq:full_spect_curve-x}
x &=& 2 \upzeta^2 - u_0, \\
\label{eq:full_spect_curve-y}
y &=& \sum_{k=1}^{+\infty} t_k\, u_0^{\frac{2k-1}{2}}\, Q_k \left(\frac{\upzeta}{\sqrt{u_0}}\right).
\eea
\noindent
Herein the $Q_k (\upzeta)$ are the $(2,2k-1)$ multicritical spectral-curves appearing in \eqref{eq:multicrit_spect_curve_hyp-y},
\be
\label{eq:multicrit_spect_curve_hyp2B}
Q_k (\upzeta) = 2\sqrt{2}\, k\, {}_2F_1 \left( 1-k, 1; \left. \frac{3}{2}\, \right| 2\upzeta^2 \right) \upzeta,
\ee
\noindent
and $u_0$ satisfies the classical string equation \eqref{eq:classical-massive-string-eq}. It was shown in \cite{e16} that applying the topological recursion to this spectral curve yields the \textit{exact same} results as those obtained from the ``massive'' string equation, as long as the KdV times further include an adequate normalization: they must include an additional factor of $\frac{k}{\alpha_{kk}}$, in such a way that the coefficient of the $u^k$ term in the $k$th Gel'fand--Dikii KdV potential is $1 \times t_k$ (recall the normalization \eqref{eq:Rl-alphall}, and think of multicritical string-equations, \textit{e.g.}, \eqref{eq:k=2_string_eq} and \eqref{eq:k=3_string_eq}, as the ``massive'' string-equation building-blocks). This ensures ``KdV compatibility'' of spectral curve and string equation.

%%%%%%%%%%%%%%%%%%%%%%%%%%%%%%%%%%%%%%%%%%%%%%%%%%%%%%%%%%%%%%%%%
\subsubsection*{Perturbative Content via String Equations}
%%%%%%%%%%%%%%%%%%%%%%%%%%%%%%%%%%%%%%%%%%%%%%%%%%%%%%%%%%%%%%%%%

Let us finally go back to our initial example dealing with the (one parameter) flow $k=3 \rightarrow k=2$, which we now know will be described by the ``massive'' string equation \cite{dss90}
\be
\label{eq:flow-k=3->k=2}
t_2 \left( u^2 - \frac{1}{3} g_{\text{s}}^2\, u'' \right) + t_3 \left( u^3 - g_{\text{s}}^2\, u u'' - \frac{1}{2} g_{\text{s}}^2 \left( u' \right)^2 + \frac{1}{10} g_{\text{s}}^4\, u'''' \right) = z.
\ee
\noindent
This was solved numerically in \cite{dss90}, via the KdV differential flow equation \eqref{eq:KdV_flow}. Herein, we may directly tackle \eqref{eq:flow-k=3->k=2} both at perturbative and nonperturbative levels. The classical equation is
\be
\label{eq:classical_eq:flow-k=3->k=2}
t_2\, u_0^2 + t_3\, u_0^3 = z,
\ee
\noindent
defining $u_0$ as a function of $t_2$, $t_3$, and $z$, and which will be the variable to use in the following. For the corresponding genus-zero free energy one uses \eqref{eq:u-to-F_ds}-\eqref{eq:u-to-F_ds-2} in the $z$ and $g_{\text{s}}$ variables, \textit{i.e.},
\be
\label{eq:u-to-F_ds-gs}
g_{\text{s}}^2 F_{\text{ds}}'' (z) = -\frac{1}{2} u (z).
\ee
\noindent
Note that there is no need to solve \eqref{eq:classical_eq:flow-k=3->k=2} explicitly; just change integration variables $\rmd z \to \rmd u_0$ with the jacobian following straightforwardly from the implicit equation \eqref{eq:classical_eq:flow-k=3->k=2}. It follows
\be
F_0 (u_0,t_2,t_3) = - \frac{u_0^5}{840} \left( 112\, t_2^2 + 245\, t_2 t_3 u_0 + 135\, t_3^2 u_0^2 \right).
\ee
\noindent
When $t_2=0$ we find the genus-zero free energy of the $k=3$ ot $(2,5)$ multicritical theory described by the Yang--Lee equation \eqref{eq:k=3_string_eq}, which was earlier computed in \eqref{eq:pert(2,5)-F}. When $t_3=0$, we find instead the genus-zero free energy of the $k=2$ ot $(2,3)$ multicritical theory described by the Painlev\'e~I equation \eqref{eq:k=2_string_eq}, computed in \eqref{eq:pert(2,3)-F}. This is obvious from \eqref{eq:flow-k=3->k=2} and \eqref{eq:classical_eq:flow-k=3->k=2}.

The full perturbative content of the ``massive'' string equation \eqref{eq:flow-k=3->k=2} also follows rather automatically. One just needs to follow the procedure outlined in subsection~\ref{subsec:multi_string_eq}: start with the above perturbative \textit{ansatz} \eqref{eq:u_pert_exp-gs}, turning the ``massive'' string equation into a recursion relation for its coefficients, and solve iteratively. One finds for the specific heat
\bea
u_{3\to2} &\simeq& u_0 - \frac{1}{6 u_0^4 \left( 2 t_2 + 3 t_3 u_0 \right)^4} \Big\{ 4 t_2^2 + 18 t_2 t_3 u_0 + 27 t_3^2 u_0^2 \Big\}\, g_{\text{s}}^{2} - \\
&&
\hspace{-25pt}
- \frac{t_2 + 3 t_3 u_0}{36 u_0^9 \left( 2 t_2 + 3 t_3 u_0 \right)^9} \Big\{ 784 t_2^4 + 5184 t_2^3 t_3 u_0 + 14580 t_2^2 t_3^2 u_0^2 + 20412 t_2 t_3^3 u_0^3 + 15309 t_3^4 u_0^4 \Big\}\, g_{\text{s}}^{4} - \nonumber \\
&&
\hspace{-25pt}
- \frac{1}{216 u_0^{14} \left( 2 t_2 + 3 t_3 u_0 \right)^{14}} \Big\{ 627200 t_2^8 + 9251776 t_2^7 t_3 u_0 + 60518112 t_2^6 t_3^2 u_0^2 + \nonumber \\
&&
\hspace{-25pt}
+ 230119056 t_2^5 t_3^3 u_0^3 + 559807848 t_2^4 t_3^4 u_0^4 + 902697372 t_2^3 t_3^5 u_0^5 + 965608614 t_2^2 t_3^6 u_0^6 + \nonumber \\
&&
\hspace{-25pt}
+ 661191336 t_2 t_3^7 u_0^7 + 247946751 t_3^8 u_0^8 \Big\}\, g_{\text{s}}^{6} - \cdots. \nonumber
\eea
\noindent
Note how a na\"\i ve iterative solution yields coefficients which also depend on $z$-derivatives of previous coefficients---but these may always be explicitly computed via the implicit classical equation \eqref{eq:classical_eq:flow-k=3->k=2}, hence ending up with the above rational solution in $u_0$. For the free energy, we obtain
\bea
F_{3\to2} &\simeq& - \frac{u_0^5}{840} \Big\{ 112 t_2^2 + 245 t_2 t_3 u_0 + 135 t_3^2 u_0^2 \Big\}\, \frac{1}{g_{\text{s}}^2} - \frac{1}{24} \log u_0 \left( 2 t_2 + 3 t_3 u_0 \right) + \\
&&
+ \frac{t_2 + 3 t_3 u_0}{720 u_0^5 \left( 2 t_2 + 3 t_3 u_0 \right)^5} \Big\{ 56 t_2^2 + 162 t_2 t_3 u_0 + 243 t_3^2 u_0^2 \Big\}\, g_{\text{s}}^2 + \nonumber \\
&&
+ \frac{1}{18144 u_0^{10} \left( 2 t_2 + 3 t_3 u_0 \right)^{10}} \Big\{ 54880 t_2^6 + 598752 t_2^5 t_3 u_0 + 2776032 t_2^4 t_3^2 u_0^2 + \nonumber \\
&&
+ 7074216 t_2^3 t_3^3 u_0^3 + 10707552 t_2^2 t_3^4 u_0^4 + 9723402 t_2 t_3^5 u_0^5 + 4861701 t_3^6 u_0^6 \Big\}\, g_{\text{s}}^4 + \cdots \nonumber
\eea
\noindent
(with the double-integration always computed as described above for the genus-zero case). One may easily check that as $(t_2,t_3)$ flow from $(0,1) \to (1,0)$, this perturbative free-energy flows from the $(2,5)$ perturbative free-energy \eqref{eq:pert(2,5)-F} to the $(2,3)$ perturbative free-energy \eqref{eq:pert(2,3)-F}.

%%%%%%%%%%%%%%%%%%%%%%%%%%%%%%%%%%%%%%%%%%%%%%%%%%%%%%%%%%%%%%%%%
\subsubsection*{Nonperturbative Content via String Equations}
%%%%%%%%%%%%%%%%%%%%%%%%%%%%%%%%%%%%%%%%%%%%%%%%%%%%%%%%%%%%%%%%%

\begin{figure}[t!]
\centering
         \includegraphics[width=1\textwidth]{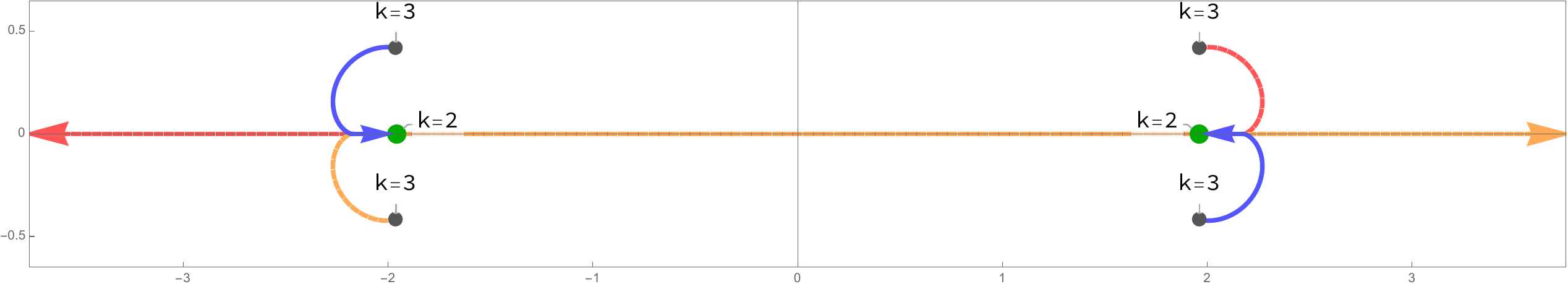}
\caption{We plot the KdV-flow trajectories of all instanton actions \eqref{eq:A-3-to-2} on the complex Borel plane, as the KdV times $(t_2,t_3)$ flow from $(0,1) \to (1,0)$. This is done by  evaluation of \eqref{eq:A-3-to-2}, so as to have $A_{3\to2} \left(\pm,\pm\right)$ depend on a $(0,1) \to (1,0)$ linear trajectory alone (we set $u_0=1$ in the plot). Both the Yang--Lee actions \eqref{eq:A&beta-(2,5)} (in dark gray) and the Painlev\'e~I actions \eqref{eq:A&beta-(2,3)} (in green) are also highlighted in the figure. It is clear how the ``massive'' $\left(\pm,-\right)$ nonperturbative sectors have the expected $k=3 \to k=2$ flow (the blue trajectories), whereas the $\left(\pm,+\right)$ nonperturbative sectors decouple to infinity (the divergent red/orange trajectories).}
\label{fig:t2-flow_borel-plane}
\end{figure}

Proceeding towards the nonperturbative content of the ``massive'' string equation \eqref{eq:flow-k=3->k=2} is rather more intricate. The procedure is still the same as that outlined in subsection~\ref{subsec:multi_string_eq}: one accesses multi-instanton sectors by solving the string equation with a nonperturbative transseries \textit{ansatz}, based upon the weight of instanton effects \eqref{eq:D-inst-exp}. This is implemented with the one-parameter transseries \textit{ansatz} \eqref{eq:trans-kappa-u} now adapted to the $z$ and $g_{\text{s}}$ variables,
\be
\label{eq:trans-massive-gs-u}
u \left( g_{\text{s}}, \sigma \right) = \sum_{n=0}^{+\infty} \sigma^n\, \rme^{- \frac{n A(z)}{g_{\text{s}}}}\, \sum_{g=0}^{+\infty} u_g^{(n)}(z)\, g_{\text{s}}^g.
\ee
\noindent
Plugging this one-parameter transseries into the ``massive'' string equation \eqref{eq:flow-k=3->k=2}, however, now yields much more intricate differential and algebraic relations. One finds \textit{four} instanton actions
\bea
\label{eq:A-3-to-2}
A_{3\to2} &=& \pm \int \rmd u_0\, u_0 \left( 2 t_2 + 3 t_3 u_0 \right) \sqrt{\frac{5 t_2 + 15 t_3 u_0 \pm \sqrt{5} \sqrt{5 t_2^2 - 6 t_2 t_3 u_0 - 9 t_3^2 u_0^2}}{3 t_3}} = \\
&=& \pm \frac{2}{189 t_3^2} \left( 81 t_3^3 u_0^3 + 81 t_2 t_3^2 u_0^2 + 3 t_2^2 t_3 u_0 - 5 t_2^3 \mp t_2^2 \sqrt{5} \sqrt{5 t_2^2 - 6 t_2 t_3 u_0 - 9 t_3^2 u_0^2} \right) \times \nonumber \\
&&
\times \sqrt{\frac{5 t_2 + 15 t_3 u_0 \pm \sqrt{5} \sqrt{5 t_2^2 - 6 t_2 t_3 u_0 - 9 t_3^2 u_0^2}}{3 t_3}} \nonumber
\eea
\noindent
(the overall $\pm$ signs are independent from the square-root $\pm$ signs). This is a non-trivial function, but because we are ultimately interested in the KdV flow let us first explicitly evaluate it near its endpoints. Near $t_2 \sim 0$ one finds (we are relabeling the above actions as $A_{3\to2} \left(\pm,\pm\right)$ for clarity)
\bea
A_{3\to2} \left(\pm,+\right) &=& \pm \frac{6}{7} \sqrt{5+\rmi\sqrt{5}}\, \left( t_3\, u_0^{7/2} + \frac{7}{6}\, t_2\, u_0^{5/2} + \frac{7}{216} \left( 5-2\rmi\sqrt{5} \right) \frac{t_2^2}{t_3}\, u_0^{3/2} + \cdots \right), \\
A_{3\to2} \left(\pm,-\right) &=& \pm \frac{6}{7} \sqrt{5-\rmi\sqrt{5}}\, \left( t_3\, u_0^{7/2} + \frac{7}{6}\, t_2\, u_0^{5/2} + \frac{7}{216} \left( 5+2\rmi\sqrt{5} \right) \frac{t_2^2}{t_3}\, u_0^{3/2} + \cdots
\right),
\eea
\noindent
On the other hand, near $t_3 \sim 0$ one finds
\bea
A_{3\to2} \left(\pm,+\right) &=& \pm \sqrt{\frac{10}{3}}\, \left( - \frac{20}{189}\, \frac{t_2^{7/2}}{t_3^{5/2}} + \frac{t_2^{3/2}}{t_3^{1/2}}\, u_0^2 + \frac{7}{5}\, t_2^{1/2}\, t_3^{1/2}\, u_0^3 + \frac{9}{40}\, \frac{t_3^{3/2}}{t_2^{1/2}}\, u_0^4 + \cdots \right), \\
A_{3\to2} \left(\pm,-\right) &=& \pm \frac{4}{5} \sqrt{6}\, \left( t_2\, u_0^{5/2} + \frac{33}{28}\, t_3\, u_0^{7/2} + \frac{27}{160}\, \frac{t_3^2}{t_2}\, u_0^{9/2} + \cdots \right).
\eea
\noindent
As expected, when $t_2=0$ we find the nonperturbative instanton-action structure of the $k=3$  multicritical theory, which was earlier computed in \eqref{eq:A&beta-(2,5)}. But when $t_3=0$ we have a surprise. Whereas the $\left(\pm,-\right)$ instanton sectors do flow to the $k=2$ multicritical theory instanton-actions in \eqref{eq:A&beta-(2,3)}, there are two other sectors $\left(\pm,+\right)$ which do not have a $k=2$ match in this limit. These sectors are, however, \textit{singular} in the $t_3 \to 0$ limit, hence disappearing on the Borel plane and decoupling from the theory at this endpoint of the flow. The trajectories of all $\left(\pm,\pm\right)$ instanton-actions on the complex Borel plane under the KdV flow are shown in figure~\ref{fig:t2-flow_borel-plane}.

%%%%%%%%%%%%%%%%%%%%%%%%%%%%%%%%%%%%%%%%%%%%%%%%%%%%%%%%%%%%%%%%%
%%%%%%%%%%%%%%%%%%%%%%%%%%%%%%%%%%%%%%%%%%%%%%%%%%%%%%%%%%%%%%%%%

In order to better understand the fate of all these nonperturbative sectors at the end of the flow, let us turn to the spectral-geometry analysis---in particular to the computation of Stokes data associated to these sectors, which allow us to go beyond the nonperturbative information encoded in the instanton actions we just addressed. Indeed, if we consider the spectral curve associated to the $k=3\rightarrow k=2$ flow,
\bea
\label{eq:full_spect_curve-t2t3-x}
x &=& 2 \upzeta^2 - u_0, \\
\label{eq:full_spect_curve-t2t3-y}
y &=& \frac{4}{3}\sqrt{2}\, t_2\, \upzeta \left( 3u_0 - 4\upzeta^{2} \right)-\frac{2}{5}\sqrt{2}\, t_3\, \upzeta \left( 15u_0^{2} - 40u_0\upzeta^{2} + 32\upzeta^{4} \right),
\eea
\noindent
which is constructed via \eqref{eq:full_spect_curve-x}-\eqref{eq:full_spect_curve-y}, it is then laborious but straightforward to use \eqref{eq:MM-stokes-coeff-inf} in order to compute the Stokes coefficients associated to the $\left(+,+\right)$ and $\left(+,-\right)$ sectors (recall how spectral geometry computes ``half'' the nonperturbative data). Again, we obtain non-trivial functions, but of which we are only interested in displaying their behavior near $t_2 \sim 0$,
\bea
S_1 \cdot F^{(1)}_{1} \left(+,+\right) &=& - \frac{\left(25+5\rmi \sqrt{5}\right)^{1/4}}{2\sqrt{2\pi}\left(\sqrt{5}-5\rmi\right)^{3/2}}\frac{1}{t_3^{1/2}u_0^{7/2}}+\cdots, \\
S_1 \cdot F^{(1)}_{1} \left(+,-\right) &=& - \frac{\left(25-5\rmi \sqrt{5}\right)^{1/4}}{2\sqrt{2\pi}\left(\sqrt{5}+5\rmi\right)^{3/2}}\frac{1}{t_3^{1/2}u_0^{7/2}}+\cdots,
\eea
\noindent
and near $t_3 \sim 0$,
\bea
S_1 \cdot F^{(1)}_{1} \left(+,+\right) &=& \frac{3\left(\frac{3}{2}\right)^{3/4}}{20\cdot 5^{1/4}\sqrt{\pi}}\frac{t_3^{5/4}}{t_2^{7/4}}+\cdots, \\
S_1 \cdot F^{(1)}_{1} \left(+,-\right) &=& \frac{\rmi}{4 \cdot 6^{3/4} \sqrt{\pi}}\frac{1}{t_2^{1/2}u_0^{5/4}}+\cdots.
\eea
\noindent
In parallel with what occurred for the instanton actions, when $t_2=0$ we obtain the expected Stokes coefficients of the $k=3$ multicritical model, in agreement with, say,  \eqref{eq:S1F1SG(2,5)}. Instead, when $t_3 \to 0$ only the $\left(+,-\right)$ sector flows to the $k=2$ multicritical-theory Stokes coefficient which was computed in \eqref{eq:S1F1SG(2,3)}. The $\left(+,+\right)$ Stokes coefficient flows to zero, in perfect accordance with the pattern which had already emerged above in the context of the instanton actions.

It would also be interesting to compute the specific-heat and free-energy one-instanton sectors for this problem, in parallel with what was done in subsection~\ref{subsec:multi_string_eq}. Note how also herein the instanton actions arise in symmetric pairs, \eqref{eq:A-3-to-2}, implying that the problem is \textit{resonant}, and that the corresponding transseries is actually a \textit{four}-parameter \textit{resonant}-transseries. It would also be rather interesting to understand what exactly is the ultimate fate of all nonperturbative sectors in this transseries, as it flows from $k=3$ to $k=2$ multicritical models (in particular, the nonperturbative sectors which do not flow to $k=2$). However, this analysis leads to rather intricate calculations, and is certainly behind the mainstream and scope of this work.

%%%%%%%%%%%%%%%%%%%%%%%%%%%%%%%%%%%%%%%%%%%%%%%%%%%%%%%%%%%%%%%%%
\subsubsection*{Convenient Change of Basis}
%%%%%%%%%%%%%%%%%%%%%%%%%%%%%%%%%%%%%%%%%%%%%%%%%%%%%%%%%%%%%%%%%

Let us make one final comment on spectral geometry with general KdV times. What we have done so far was to write the generic spectral curve, \eqref{eq:full_spect_curve-x}-\eqref{eq:full_spect_curve-y}, on a spectral-curve ``basis'' whose basis-elements are multicritical spectral-curves \eqref{eq:multicrit_spect_curve_hyp2B}, and where the KdV times are the ``coefficients'' of this basis expansion---let us call it the ``multicritical basis''. As we build up to subsection~\ref{subsec:JT_string_eq}, it turns out that another (eventually more convenient for what follows) basis to expand a generic spectral curve will be required (and which was already partially hinted at in \eqref{eq:multicrit_spect_curv_re-x}-\eqref{eq:multicrit_spect_curv_re-y}). This is given by the monomials \cite{t:mt20}
\be
\label{eq:multicrit_spect_curve-monomial}
\widetilde{Q}_n (\upzeta) := (-1)^{n+1}\, 2^{3n-\frac{3}{2}}\, \frac{\Gamma \left( n \right) \Gamma \left( n+1 \right)}{\Gamma \left( 2n \right)}\, \upzeta^{2n-1}.
\ee
\noindent
These new basis elements---let us call it the ``monomial basis''---may of course be linearly expanded in terms of the old basis; the change of basis being:
\be
\widetilde{Q}_n (\upzeta) = \left(-1\right)^{n}\, \sum_{m=1}^{n} \binom{n}{m} \left(-1\right)^{m} Q_m (\upzeta)
\ee
\noindent
(this is basically an inversion of the hypergeometric function back to monomials \cite{olbc10}). In the following, we will apply this change of basis to the spectral curves of minimal strings---and rescaled versions thereof---which are relevant for the study of the JT-gravity string-equation. In all such cases, we will only be interested in the $z=0$ limit of all computed quantities, which corresponds to $u_0=1$ (this is associated to the choice of conformal background; see below). 

A generic spectral curve may be written in the new monomial-basis as
\be
y (\upzeta) = \sum_{n=1}^{+\infty} t_n\, Q_n (\upzeta) = \sum_{m=1}^{+\infty} \widetilde{Q}_m (\upzeta)\, \sum_{i=0}^{+\infty} \binom{i+m}{m}\, t_{i+m} \equiv \sum_{m=1}^{+\infty} \widetilde{t}_m\, \widetilde{Q}_m(\upzeta),
\ee
\noindent
where we introduced new KdV times $\widetilde{t}_p (k)$ as the coordinates in the new basis, satisfying\footnote{Recall that $t_p (k) = 0$ when $p>k$, hence we explicitly truncate the infinite sums.}:
\be
\label{eq:change-basis_KdV-times}
\widetilde{t}_p (k) = \sum_{n=0}^{k-p} \binom{n+p}{p}\, t_{n+p} (k), \qquad t_p (k) = \sum_{n=0}^{k-p} \left(-1\right)^{n} \binom{n+p}{p}\, \widetilde{t}_{n+p} (k).
\ee
\noindent
It turns out that, at the level of the string equation, we can take this change of basis into account in a simple way. This can be verified by considering the genus-zero classical string-equation \eqref{eq:classical-massive-string-eq} with the above change of KdV coordinates,
\be
z = \sum_{k=1}^{+\infty} t_k u_0^k = \sum_{k=1}^{+\infty} \widetilde{t}_k \sum_{i=1}^k \left(-1\right)^{k-i}\binom{k}{i}\, u_0^i = \sum_{k=1}^{+\infty} \widetilde{t}_k \left\lbrace \left( u_0-1 \right)^k - \left(-1\right)^{k} \right\rbrace.
\ee
\noindent
In the end this takes the simple (inital) form
\be
\label{eq:mon_base_string_eq}
\sum_{k=1}^{+\infty} \widetilde{t}_k \widetilde{u}_0^k = \widetilde{z},
\ee
\noindent
if we define
\be
\label{eq:UandZchange}
\widetilde{u}_0 := u_0-1 \qquad \text{and} \qquad \widetilde{z}:= z + \sum_{j=1}^{+\infty} \left(-1\right)^{j} \widetilde{t}_j.
\ee
\noindent
This means that the ``new'' KdV times $\widetilde{t}_k$ may be associated to a string equation which is related in a simple way to the ``old'' one. Working with this new monomial basis is particularly convenient when, as we shall do in the following subsections to establish contact with JT gravity, one has to deal with rescalings of $x$ at the level of the spectral curve. Since the $\widetilde{Q}_k$ are just monomials in $\upzeta$, the new KdV times $\widetilde{t}_k$ rescale in a straightforward way upon rescaling $x$, while the transformation rule for the $t_k$ is much more involved.

%%%%%%%%%%%%%%%%%%%%%%%%%%%%%%%%%%%%%%%%%%%%%%%%%%%%%%%%%%%%%%%%%
\subsection{String Equations for Minimal String Theories}\label{subsec:minimal_string_eq}
%%%%%%%%%%%%%%%%%%%%%%%%%%%%%%%%%%%%%%%%%%%%%%%%%%%%%%%%%%%%%%%%%

Recall how back in section~\ref{sec:spec_curv} we moved from multicritical models in subsection~\ref{subsec:multicritical} to minimal strings in subsection~\ref{subsec:minimal}, by essentially changing from the \textit{multicritical} closed-string background to the \textit{conformal} closed-string background \cite{mss91, ss03}. At the level of spectral geometry, this was implemented via perturbation of the multicritical spectral-curves \eqref{eq:multi-to-minimal} by a very specific choice of KdV times \eqref{eq:min-str-kdv-times}. Having addressed multicritical string-equations in subsection~\ref{subsec:multi_string_eq}, and then how they get rearranged by specific choices of KdV times in the previous subsection, we may now finally arrive at string equations\footnote{In other words, string equations whose solutions directly match minimal-string-theoretic CFT calculations.} for minimal strings in the conformal background.

The string equation for the general $(2,2k-1)$ minimal string is obtained from the ``massive'' string equation \eqref{eq:massive_string_eq} by fixing all the KdV times via spectral geometry \eqref{eq:multi-to-minimal} (as was described in the previous subsection) to the conformal background \eqref{eq:min-str-kdv-times}. This string equation is then
\be
\label{eq:minimal-string_string_eq}
\sum_{p=0}^{\left[\frac{k-1}{2}\right]} t_{k-2p}(k) \cdot \frac{k-2p}{\alpha_{k-2p,k-2p}} \cdot R_{k-2p} \left[ u \right] = z
\ee
\noindent
(recall we needed to include the additional factor of $\frac{k}{\alpha_{kk}}$ so that the $u^k$ multiplying-factor in the $k$th KdV potential is the identity, and we have made this explicit in the above formula). As in the multicritical case \eqref{eq:multicritical_string_eq}, the minimal-string string-equation only really depends on $k$, hence is in this sense universal. For example, for the $k=2$ or $(2,3)$ minimal string theory (with conformal KdV time given by $t_{2} (2) = - \frac{3}{4\sqrt{2}}$) it follows
\be
\label{eq:P1-minimal-conventions}
t_{2}(2)\, \frac{2}{\alpha_{22}} R_{2} \left[ u \right] = - \frac{3}{4\sqrt{2}} \left( u^2 - \frac{1}{3} g_{\text{s}}^2\, u'' \right) = z.
\ee
\noindent
This example is basically the same as in the $(2,3)$ multicritical case: if one rescales $u \mapsto t_2^{-2/5} u$, $z \mapsto t_2^{1/5} z$ this minimal-string string-equation precisely becomes the multicritical Painlev\'e~I equation \eqref{eq:k=2_string_eq}. The last step in reaching the conformal background further requires fixing $z$; in this case to $-t_0 (2)$, \textit{i.e.}, $z = - \frac{3}{4\sqrt{2}}$, or, equivalently, $u_0=1$ \cite{ss03}. Note that one finds $t_{2} (2) + t_0 (2) = 0$, a generic property of conformal backgrounds. The story starts to be more interesting as we move up in $k$. For the $k=3$ or $(2,5)$ minimal string theory one finds instead (with conformal KdV times now given by $t_{3} (3) = \frac{5}{4\sqrt{2}}$ and $t_{1} (3) = - \frac{5}{4\sqrt{2}}$)
\be
\label{eq:(2,5)-minimal-string-eq}
\frac{5}{4\sqrt{2}} \left( - u + u^3 - g_{\text{s}}^2\, u u'' - \frac{1}{2} g_{\text{s}}^2 \left( u' \right)^2 + \frac{1}{10} g_{\text{s}}^4\, u'''' \right) = z. 
\ee
\noindent
This already departs from the $(2,5)$ multicritical case, where we only found the third Gel'fand--Dikii KdV potential leading up to the Yang--Lee equation \eqref{eq:k=3_string_eq}. For this case of odd $k$ one already finds $t_{3} (3) + t_{1} (3) = 0$, implying (\textit{e.g.}, from the classical string-equation) that the last step in reaching the conformal background now requires fixing $z$ to $0$ or, equivalently, $u_0=1$ \cite{ss03}. Of course as we go higher in $k$ the minimal-string string-equations differ more and more from their multicritical counterparts. Generically, for even $k$ the last step in reaching the conformal background entails setting $z = - t_0(k)$ or $u_0=1$; and, for odd $k$, the final setting being that of $z=0$ or $u_0=1$ (\textit{i.e.}, the sum of all KdV times, $t_0$ included in the even case, always vanishing).

%%%%%%%%%%%%%%%%%%%%%%%%%%%%%%%%%%%%%%%%%%%%%%%%%%%%%%%%%%%%%%%%%
\subsubsection*{Perturbative Content via String Equations}
%%%%%%%%%%%%%%%%%%%%%%%%%%%%%%%%%%%%%%%%%%%%%%%%%%%%%%%%%%%%%%%%%

Let us directly address\footnote{A numerical analysis of this $(2,5)$ minimal-string model was carried out in \cite{mmss04}.} \eqref{eq:(2,5)-minimal-string-eq} both at perturbative and nonperturbative levels. Starting off with the classical string-equation, we find
\be
\frac{5}{4\sqrt{2}} \left( u_0^3 - u_0 \right) = z
\ee
\noindent
implicitly defining $u_0$ as a function of $z$. As already discussed, this is the variable we want to use for computations---albeit one should keep in mind that, as just explained above, in the final result we still need to tune-in to the conformal background by setting $z=0$ corresponding to $u_0=1$. For the corresponding genus-zero free energy, just proceed as in the previous ``massive'' string-equation example. It follows 
\be
F_0 (u_0) = - \frac{5 u_0^3}{5376} \left( 135\, u_0^4 - 189\, u_0^2 + 70 \right),
\ee
\noindent
in which case the conformal-background free energy, corresponding to the $(2,5)$ minimal string theory, is 
\be
\lim_{u_0 \to 1} F_0 (u_0) \equiv F_0 = - \frac{5}{336}.
\ee

The full perturbative content of the minimal string theory \eqref{eq:(2,5)-minimal-string-eq} also follows rather swiftly. The procedure is always the same and we need not repeat it. One  first finds for the specific heat:
\bea
u_{(2,5)} (g_{\text{s}}) &\simeq& u_0 - \frac{16}{25 \left( 3 u_0^2 - 1 \right)^4} \Big\{ 9 u_0^2 + 1 \Big\}\, g_{\text{s}}^2 - \frac{768 u_0}{625 \left( 3 u_0^2 - 1 \right)^9} \Big\{ 1701 u_0^4 + 1458 u_0^2 + 109 \Big\}\, g_{\text{s}}^4 - \nonumber \\
&&
\hspace{-20pt}
- \frac{4096}{15625 \left( 3 u_0^2 - 1 \right)^{14}} \Big\{ 9183213 u_0^8 + 21579858 u_0^6 + 7278336 u_0^4 + 471678 u_0^2 + 3043 \Big\}\, g_{\text{s}}^6 - \nonumber \\
&&
\hspace{-20pt}
- \frac{196608 u_0}{1953125 \left( 3 u_0^2 - 1 \right)^{19}} \Big\{ 48337157106 u_0^{10} + 263278861029 u_0^8 + 207027170352 u_0^6 + \nonumber \\
&&
\hspace{-20pt}
+ 44174336742 u_0^4 + 2658798870 u_0^2 + 32146285 \Big\}\, g_{\text{s}}^8 - \nonumber \\
&&
\hspace{-20pt}
- \frac{3145728}{48828125 \left( 3 u_0^2 - 1 \right)^{24}} \Big\{ 161199334416267 u_0^{14} + 2657228606609220 u_0^{12} + \nonumber \\
&&
\hspace{-20pt}
+ 3957905812088907 u_0^{10} + 1763707595790708 u_0^8 + 284674916842929 u_0^6 + \nonumber \\
&&
\hspace{-20pt}
+ 16373612574900 u_0^4 + 271126625145 u_0^2 + 567523220 \Big\}\, g_{\text{s}}^{10} - \cdots,
\eea
\noindent
which in the $u_0 \to 1$ conformal background becomes
\be
u_{(2,5)} (g_{\text{s}}) \simeq 1 - \frac{2}{5}\, g_{\text{s}}^2 - \frac{4902}{625}\, g_{\text{s}}^4 - \frac{9629032}{15625}\, g_{\text{s}}^6 - \frac{212065676394}{1953125}\, g_{\text{s}}^8 - \frac{1657755294838368}{48828125}\, g_{\text{s}}^{10} - \cdots.
\ee
\noindent
For the free energy one first\footnote{Of course at this stage such explicit $u_0$-dependence is no longer required---it was only needed to perform the double-integration \eqref{eq:u-to-F_ds-gs} that brought us here; but, at the end of the day, we are only interested in the $u_0 \to 1$ conformal-background limit. We leave it here solely for convenience of the reader in this example.} obtains:
\bea
F_{\text{ds}} (g_{\text{s}}) &\simeq& - \frac{5 u_0^3}{5376} \Big\{ 135 u_0^4 - 189 u_0^2 + 70 \Big\}\, \frac{1}{g_{\text{s}}^2} - \frac{1}{24} \log \left( 3 u_0^2 - 1 \right) + \\
&&
+ \frac{2 u_0}{125 \left( 3 u_0^2 - 1 \right)^5} \Big\{ 81 u_0^2 + 29 \Big\}\, g_{\text{s}}^2 + \frac{32}{13125 \left(3 u_0^2 - 1 \right)^{10}} \Big\{ 180063 u_0^6 + 289413 u_0^4 + \nonumber \\
&&
+ 46413 u_0^2 + 583 \Big\}\, g_{\text{s}}^4 + \frac{256 u_0}{78125 \left( 3 u_0^2 - 1 \right)^{15}} \Big\{ 127876077 u_0^8 + 555372612 u_0^6 + \nonumber \\
&&
+ 292380030 u_0^4 + 33418980 u_0^2 + 671165 \Big\}\, g_{\text{s}}^6 + \nonumber \\
&&
+ \frac{4096}{1953125 \left( 3 u_0^2 - 1 \right)^{20}} \Big\{ 247618025217 u_0^{12} + 3515395531068 u_0^{10} + 3997100920869 u_0^8 + \nonumber \\
&&
+ 1222856804304 u_0^6 + 115597983195 u_0^4 + 2888356020 u_0^2 + 8613055 \Big\}\, g_{\text{s}}^8 + \cdots,
\eea
\noindent
which in the conformal-background limit finally becomes
\be
F_{\text{ds}} (g_{\text{s}}) \simeq - \frac{5}{336}\, \frac{1}{g_{\text{s}}^2} - \frac{1}{12} \log 2 + \frac{11}{200}\, g_{\text{s}}^2 + \frac{64559}{52500}\, g_{\text{s}}^4 + \frac{63107429}{625000}\, g_{\text{s}}^6 + \frac{71105204951}{3906250}\, g_{\text{s}}^8 + \cdots.
\ee
\noindent
This free energy for the $(2,5)$ minimal string theory is in perfect agreement with the topological recursion results in subsection~\ref{subsec:minimal}, \eqref{eq:minimal_free_en-2}-\eqref{eq:minimal_free_en-3}-\eqref{eq:minimal_free_en-4}, once we set $k=3$.

%%%%%%%%%%%%%%%%%%%%%%%%%%%%%%%%%%%%%%%%%%%%%%%%%%%%%%%%%%%%%%%%%
\subsubsection*{Nonperturbative Content via String Equations}
%%%%%%%%%%%%%%%%%%%%%%%%%%%%%%%%%%%%%%%%%%%%%%%%%%%%%%%%%%%%%%%%%

Moving-on towards the nonperturbative content of our minimal string theory, lies somewhat in-between the complexity of what was done for multicritical models in subsection~\ref{subsec:multi_string_eq} and what was (only partially) done for ``massive'' models in subsection~\ref{subsec:general-KdV}. In any case, the procedure should be well familiar to the reader by now. Let us start-off with the one-parameter transseries \textit{ansatz}
\be
\label{eq:trans-ms-gs-u}
u \left( g_{\text{s}}, \sigma \right) = \sum_{n=0}^{+\infty} \sigma^n\, \rme^{- \frac{n A(z)}{g_{\text{s}}}}\, \sum_{g=0}^{+\infty} u_g^{(n)}(z)\, g_{\text{s}}^g.
\ee
\noindent
Plugging this transseries \textit{ansatz} into the $(2,5)$ minimal-string string-equation \eqref{eq:(2,5)-minimal-string-eq} immediately yields (at leading nonperturbative order) the ODE
\be
\label{eq:(2,5)-minimal-A-ODE}
A'(z)^{4} - 10 u_0\, A'(z)^{2} - 10 \left( 1-3u_0^2 \right) = 0,
\ee
\noindent
from which one obtains the instanton action(s)
\bea
\label{eq:(2,5)-minimal-A-solution}
A_{(2,5)} &=& \pm \frac{5}{4\sqrt{2}} \int \rmd u_0 \left( 1 - 3 u_0^2 \right) \sqrt{5 u_0 \pm \sqrt{5} \sqrt{2 - u_0^2}} = \\
&=& \pm \frac{5}{42\sqrt{2}} \left( 8 u_0 - 9 u_0^3 \pm \sqrt{5} \sqrt{2 - u_0^2} \right) \sqrt{5 u_0 \pm \sqrt{5} \sqrt{2 - u_0^2}} \nonumber
\eea
\noindent
(the overall $\pm$ signs are independent from the square-root $\pm$ signs). These are \textit{four} independent solutions, organized in symmetric pairs---the by-now familiar hallmark of resonance \cite{gikm10, asv11, abs18, bssv22}, which is also making its appearance in the present minimal-string context. In the $u_0 \to 1$ conformal background limit these become
\bea
A_{(2,5)} \left(\pm,+\right) &=& \pm \frac{5}{21}\, \sqrt{\frac{1}{2} \left( 5 - \sqrt{5} \right)}, \\
A_{(2,5)} \left(\pm,-\right) &=& \mp \frac{5}{21}\, \sqrt{\frac{1}{2} \left( 5 + \sqrt{5} \right)}.
\eea
\noindent
It is immediate to check that the (positive) $A_{(2,5)} \left(-,-\right)$ and the (negative) $A_{(2,5)} \left(-,+\right)$ match the minimal-string instanton-actions previously obtained via spectral geometry in subsection~\ref{subsec:minimal}, \eqref{eq:A-(2,5)_ms_SG} (respectively). What we now find is, in addition, their resonant \textit{symmetric} pairs. Further note that the relation between $A_{(2,5)}$ and $u_0$ is not rational, in which case $u_0$ is no longer a good working variable going forward into the nonperturbative realm. A more convenient variable to work with turns out to be
\be
U = \frac{\sqrt{5}}{u_0} \sqrt{2-u_0^2},
\ee
\noindent
in terms of which the above instanton actions become
\be
A_{(2,5)} = \pm \frac{5}{21} \left( \frac{5}{2} \right)^{3/4} \frac{\left( 5 \pm U \right)^{5/2} \left( - 2 \pm U \right)}{\left( 5+U^2 \right)^{7/4}}.
\ee

Focusing on the $A_{(2,5)} \left(-,-\right)$ one-instanton sector (corresponding to the ``plus-plus'' minimal-string instanton action), we can proceed to compute $u^{(1)}_0 (z)$. At first order in the transmonomial combination $\sigma\, \rme^{-\frac{A(z)}{g_{\text{s}}}}$ and at next-to-leading order in the string coupling $g_{\text{s}}$ we find a linear, first-order, homogeneous ODE; which is solvable via elementary integration as
\be
\label{eq:u(1)0-ODE-soltn}
u^{(1)}_0 = \frac{\left( 5+U^2 \right)^{3/8}}{U^{1/2} \left( 5-U \right)^{1/4}}.
\ee
\noindent
At next-to-next-to-leading order in the string coupling $g_{\text{s}}$ one finds again a linear and first-order, albeit now \textit{non-homogeneous}\footnote{Whose corresponding \textit{homogeneous} part coincides with the linear ODE yielding \eqref{eq:u(1)0-ODE-soltn}.} ODE; a pattern which repeats to all higher orders. Upon straightforward integration of these increasingly-complicated ODEs one finds\footnote{There is a strong resemblance with the (analogue) quartic matrix-model calculations in \cite{m08, asv11}.}, up to four loops\footnote{Up to this order, a pattern emerges which could likely be general and of help in automating the procedure going into higher orders. All computed coefficients are of the form
\be
u^{(1)}_g = \frac{\left( 5+U^2 \right)^{\frac{1}{8} \left( 14g+3 \right)}}{U^{\frac{1}{2} \left( 6g+1 \right)} \left( 5-U \right)^{\frac{1}{4} \left( 10g+1 \right)} \left( 5+U \right)^{2g + \frac{g-1}{2} \left( g \mod 2 \right)}}\, \NCU^{(1)}_g (U),
\ee
\noindent
where $\NCU^{(1)}_g (U)$ is a polynomial of degree $4g + \frac{g-1}{2} \left( g \mod 2 \right)$ in $U$.},
\bea
u^{(1)}_1 &=& \frac{1}{12 \cdot 2^{1/4} \cdot 5^{3/4}}\, \frac{\left( 5+U^2 \right)^{17/8}}{U^{7/2} \left( 5-U \right)^{11/4} \left( 5+U \right)^{2}} \times \\
&&
\times \left( - 11 U^4 -160 U^3 + 165 U^2 + 550 U - 2500 \right), \nonumber \\
u^{(1)}_2 &=& \frac{1}{1440 \sqrt{10}}\, \frac{\left( 5+U^2 \right)^{31/8}}{U^{13/2} \left( 5-U \right)^{21/4} \left( 5+U \right)^{4}} \times \\
&&
\times \left( 1921 U^8 + 30592 U^7 + 109090 U^6 - 213220 U^5 + 181625 U^4 + 5049500 U^3 - \right. \nonumber \\
&&
\left. - 12582500 U^2 - 29750000 U + 96250000 \right), \nonumber \\ 
u^{(1)}_3 &=& \frac{1}{259200 \cdot 2^{3/4} \cdot 5^{1/4}}\, \frac{\left( 5+U^2 \right)^{45/8}}{U^{19/2} \left( 5-U \right)^{31/4} \left( 5+U \right)^{7}} \times \\
&&
\times \left( - 1715795 U^{13} - 25533439 U^{12} - 545347785 U^{11} - 1141514155 U^{10} - 7592723675 U^9 - \right. \nonumber \\
&&
- 3333014325 U^8 - 17704132875 U^7 + 82903980375 U^6 + 352228016250 U^5 + \nonumber \\
&&
+ 1524167975000 U^4 - 3098239375000 U^3 + 14151637500000 U^2 + 10948437500000 U - \nonumber \\
&&
\left. 
- 53178125000000 \right). \nonumber
\eea
\noindent
In the $u_0 \to 1$ conformal-background limit, the specific-heat one-instanton sector associated with the $A_{(2,5)} \left(-,-\right)$ instanton-action is then
\bea
u^{(1)}_{(2,5)} (g_{\text{s}}) &\simeq& \left( 3+\sqrt{5} \right)^{1/8} - \frac{1}{24 \cdot 2^{7/8} \cdot 5^{3/4}} \left( 39+5\sqrt{5} \right) \left( 2+\sqrt{5} \right)^{1/4} g_{\text{s}} + \\
&&
+ \frac{1}{57600} \left( -6009+10543\sqrt{5} \right) \left( 123+55\sqrt{5} \right)^{1/8} g_{\text{s}}^2 - \nonumber \\
&&
- \frac{1}{2073600 \cdot 2^{7/8} \cdot 5^{3/4}} \left( 16210465 + 4426549 \sqrt{5} \right) \left( 2+\sqrt{5} \right)^{1/4} g_{\text{s}}^4 + \cdots. \nonumber
\eea

Moving-on towards the one-instanton nonperturbative free-energies, a novelty arises. First, reconsider the generic form of this one-instanton sector of the free energy we are looking for, \textit{e.g.}, \eqref{eq:MM-F(1)}, but now with explicit $g_{\text{s}}$ and $z$ dependences accounted for:
\bea
F^{(1)} (g_{\text{s}}, z) &\simeq& \rme^{-\frac{A(z)}{g_{\text{s}}}}\, \sum_{g=0}^{+\infty} F_g^{(1)} (z)\, g_{\text{s}}^g = \\
&=& \rme^{-\frac{A(z)}{g_{\text{s}}}}\, \left\{ F_0^{(1)}(z) + g_{\text{s}} F_1^{(1)}(z) + g_{\text{s}}^2 F_2^{(1)}(z) + g_{\text{s}}^3 F_3^{(1)}(z) + \cdots \right\}. \nonumber
\eea
\noindent
Now, in the present double-scaling limit, the relation between specific-heat and free-energy \eqref{eq:u-to-F_ds-gs} is differential in $z$, \textit{not} in $g_{\text{s}}$ (unlike for instance in subsection~\ref{subsec:multi_string_eq} where $z$ and $g_{\text{s}}$ were identified). To see what this entails in an example, write it out explicitly for the one-instanton sector of the specific heat,
\bea
\label{(2,5)u-1inst}
u^{(1)} (g_{\text{s}},z) &\simeq& \rme^{-\frac{A(z)}{g_{\text{s}}}} \left\{ u_0^{(1)}(z) + g_{\text{s}} u_1^{(1)}(z) + \cdots \right\} = \\
&&
\hspace{-60pt}
= -2\,\rme^{-\frac{A(z)}{g_{\text{s}}}} \left\{ A'(z)^2\, F_0^{(1)}(z) + g_{\text{s}} \left( A'(z)^2\, F_1^{(1)}(z) - 2A'(z)\, \partial_z F_0^{(1)}(z) - A''(z)\, F_0^{(1)} (z) \right) + \cdots\right\}. \nonumber
\eea
\noindent
Comparing first and second lines of \eqref{(2,5)u-1inst} then yields, iteratively,
\bea
F_0^{(1)}(z) &=& -\frac{1}{2}\frac{u_0^{(1)}(z)}{A'(z)^2}, \\
F_1^{(1)}(z) &=& -\frac{1}{2}\frac{1}{A'(z)^2} \left( u_1^{(1)}(z) + 2\frac{\partial_z u_0^{(1)}(z)}{A'(z)} - 3 \frac{A''(z)\, u_0^{(1)}(z)}{A'(z)^{2}} \right),
\eea
\noindent
and so on; where, in practice, the relation between specific-heat and free-energy has now became \textit{algebraic}. More precisely, and in contrast with most of the corresponding calculations in subsection~\ref{subsec:multi_string_eq}, herein no actual integration needs to be performed\footnote{Again, there is a strong resemblance with the (analogue) quartic matrix-model calculations in \cite{m08, asv11}. The present calculations are however simpler due to the differential relation between $u (z)$ and $F_{\text{ds}}(z)$ in \eqref{eq:u-to-F_ds-gs}; whereas in the matrix model \cite{m08, asv11} the corresponding relation is of finite-difference form.} in order to obtain the nonperturbative free-energies out of the nonperturbative specific-heat. Repeated iteration of this procedure, as applied to the specific case in hand, yields the first few free-energies of the $A_{(2,5)}(-,-)$ one-instanton sector\footnote{Again, these functions are no longer really required; but only their conformal background $u_0 \to 1$ limit.}. They are:
\bea
F_0^{(1)} &=& \frac{1}{10^{1/2}}\, \frac{\left( 5+U^2 \right)^{7/8}}{U^{1/2} \left( 5-U \right)^{5/4}}, \\
F_1^{(1)} &=& \frac{1}{60 \cdot 2^{3/4} \cdot 5^{1/4}}\, \frac{\left( 5+U^{2} \right)^{21/8}}{U^{7/2} \left( 5-U \right)^{15/4} \left( 5+U \right)^{2}} \times \\
&&
\times \left( 107 U^4 + 784 U^3 + 315 U^2 - 1750 U + 2500 \right), \nonumber \\
F_2^{(1)} &=& \frac{1}{72000}\, \frac{\left( 5+U^2 \right)^{35/8}}{U^{13/2} \left( 5-U \right)^{25/4} \left( 5+U \right)^{4}} \times \\
&&
\times \left( 241349 U^8 + 3560480 U^7 + 17047850 U^6 + 8207500 U^5 - 30195875 U^4 + \right. \nonumber \\
&&
\left. + 48107500 U^3 + 42087500 U^2 - 358750000 U + 481250000 \right), \nonumber \\
F_3^{(1)} &=& \frac{1}{518400 \cdot 2^{1/4} \cdot 5^{3/4}}\, \frac{\left( 5+U^{2} \right)^{49/8}}{U^{19/2} \left( 5-U \right)^{35/4} \left( 5+U \right)^{7}} \times \\
&&
\times \left( - 48012659 U^{13} - 1345391215 U^{12} - 15039543225 U^{11} - 81174878875 U^{10}  - \right. \nonumber \\
&&
- 202539920075 U^9 - 58770757125 U^8 + 180000721125 U^7 - 309460469625 U^6 + \nonumber \\
&&
+ 398016866250 U^5 + 1487988025000 U^4 - 8229904375000 U^3 + 5142637500000 U^2 + \nonumber \\
&&
\left. + 33470937500000 U - 53178125000000 \right), \nonumber
\eea
\noindent
which in the conformal-background limit $U \to \sqrt{5}$ become, in the (full) free-energy one-instanton contribution,
\bea
F^{(1)}_{\text{ds}} (g_{\text{s}}) &\simeq & \frac{2^{3/8} \cdot 5^{1/8}}{\left( 5-\sqrt{5} \right)^{5/4}}\, g_{\text{s}}^{1/2}\, \rme^{\frac{1}{g_{\text{s}}}\, \frac{5}{21} \sqrt{\frac{1}{2} \left( 5+\sqrt{5} \right)}} \left( 1 - \frac{1}{120} \sqrt{\frac{107791 + 49201 \sqrt{5}}{2 \sqrt{5}}}\, g_{\text{s}} + \right. \\
&&
+ \left. \frac{7 \left(196475 + 69697 \sqrt{5}\right)}{288000}\, g_{\text{s}}^2 - \frac{\sqrt{\frac{1}{2} \left( 5 + \sqrt{5} \right)} \left( 340372753 + 149564293 \sqrt{5} \right)}{20736000}\, g_{\text{s}}^3 + \cdots \right). \nonumber
\eea
\noindent
Matching against what was earlier obtained in subsection~\ref{subsec:minimal}, these free-energy two and three-loop coefficients around the one-instanton sector $A_{(2,5)} \left(-,-\right)$ precisely match the spectral-geometry results \eqref{eq:F12F11SG_ms_(2,5)}-\eqref{eq:F13F11SG_ms_(2,5)}. Recall, again, that string equations do not directly determine Stokes data, but only from large-order asymptotics---and we shall return to this point later.

Let us make one comment. In the transseries \textit{ansatz} \eqref{eq:trans-ms-gs-u} (or, for that matter, \eqref{eq:trans-massive-gs-u}) there was no need to include a characteristic-exponent $\beta$ as in \eqref{eq:trans-kappa-u}. This could na\"\i vely seem to be in contradiction with the generic matrix-model structure of the one-instanton sector, \eqref{eq:MM-F(1)}. The reason why there is no conflict is because herein the $g_{\text{s}}$-square-root term actually arises from a (instanton saddle-point) gaussian integration (basically, from ``inside'' the square-root in \eqref{eq:MM-stokes-coeff}) \cite{msw07}. This factor hence arises within the Stokes coefficient (see as well \cite{m08, asv11}).

Having earlier shown how generic multicritical models are resonant (recall for instance \eqref{eq:inst-act-ell-rho}-\eqref{eq:poly-inst-act-string-eq} and the discussion that ensued), and having now shown that the $(2,5)$ minimal string theory is also resonant, the next obvious question is whether one can show that \textit{generic} minimal string theories are \textit{all} resonant. The procedure to prove this is quite similar to what we did back in subsection~\ref{subsec:multi_string_eq}. Of course obtaining full nonperturbative results at generic $k$ will be rather intricate; not only due to the lack of an explicit closed-form expression for the order-$k$ minimal-string string-equation \eqref{eq:minimal-string_string_eq}---very much like in the multicritical case---but now with the added annoyance of having to work with explicit $u_0$-dependence in order to eventually reach the conformal-background. We can still make some progress precisely along the lines of the exact same question for multicritical models. Back there we learned that \textit{part} of the \textit{generic} Gel'fand--Dikii KdV potentials could be explicitly determined: \eqref{eq:Rl-alphall-alphali-betalj} with coefficients \eqref{eq:alphali_coeffs} and \eqref{eq:betalj_coeffs}. Translated to the present analysis, this implies that (part of) the \textit{generic} minimal-string string-equation may be written as
\bea
&&
\hspace{-15pt}
\label{eq:leading+next-to-leading_MS_string_eq}
\sum_{p=0}^{\left[\frac{k-1}{2}\right]} t_{k-2p}(k) \left( u^{k-2p} + \frac{k-2p}{\alpha_{k-2p,k-2p}} \sum_{i=1}^{k-2p-1} \alpha_{k-2p, i}\, g_{\text{s}}^{2k-4p-2i}\, u^{i-1}\, u^{(2k-4p-2i)} + \right. \\
&&
\hspace{+65pt}
\left.
+ \frac{k-2p}{\alpha_{k-2p,k-2p}} \sum_{j=2}^{k-2p-1} \beta_{k-2p, j}\, g_{\text{s}}^{2k-4p-2j}\, u^{j-2}\, u'\, u^{(2k-4p-2j-1)} + \cdots \right) = z. \nonumber
\eea
\noindent
Plugging the minimal-string one-paramter transseries \textit{ansatz} \eqref{eq:trans-ms-gs-u} into the above minimal-string string-equation, and expanding to first order in the transmonomial $\sigma \exp \left( - \frac{A}{g_{\text{s}}} \right)$ and to leading (zeroth) order in the string coupling, yields:
\be
\label{eq:poly-ODE-inst-act-string-eq}
\cdots + \sigma\, \rme^{-\frac{A}{g_{\text{s}}}}\, u_0^{(1)} \left\{\, \sum_{n=0}^{k-1} \left( \sum_{\ell=k-2\left[\frac{k-1}{2}\right]}^{k} \ell\, t_{\ell} (k)\, \frac{\alpha_{\ell,\ell-n}}{\alpha_{\ell\ell}}\, u_0^{\ell-n-1} \right) A'(z)^{2n} \right\} + \cdots = z.
\ee
\noindent
At this precise order the generic minimal-string string-equation gets rewritten as a first-order nonlinear ODE for $A(z)$; explicitly given by a degree-$(2k-2)$ polynomial in $A'(z)$ with real coefficients. This polynomial ODE generalizes \eqref{eq:poly-inst-act-string-eq} from the multicritical to the minimal string case. For example, for the $k=2$ or $(2,3)$ minimal string theory one finds the very familiar
\be
- \frac{3 u_0}{2\sqrt{2}} + \frac{\sqrt{2}}{9 u_0^2}\, A'(u_0)^2 = 0 \qquad \Rightarrow \qquad A(u_0) = \pm \frac{3\sqrt{3}}{5}\, u_0^{5/2},
\ee
\noindent
where we changed variables $z \to u_0$ via the classical string-equation obtained from \eqref{eq:P1-minimal-conventions}. For the $k=3$ or $(2,5)$ minimal string theory one finds instead
\be
\frac{5}{4\sqrt{2}} - \frac{15 u_0^2}{4\sqrt{2}} + \frac{5 u_0}{4\sqrt{2}}\, A'(z)^2 - \frac{1}{8\sqrt{2}}\, A'(z)^4 = 0.
\ee
\noindent
This is just \eqref{eq:(2,5)-minimal-A-ODE}. To get a better grasp on the role of the $u_0 \to 1$ conformal-background limit, consider the solutions to this equation, \eqref{eq:(2,5)-minimal-A-solution}. The trajectories of the resulting four $(\pm,\pm)$ instanton-actions on the complex Borel plane, under the ``flow'' dictated by the conformal-background limit $u_0 \to 1$, are shown in figure~\ref{fig:(2,5)-conf-back-limit}. The generic trend should now be clear. The leading nonperturbative string equation \eqref{eq:poly-ODE-inst-act-string-eq} is a polynomial of degree-$(k-1)$ in $A'(u_0)^2$, yielding $(k-1)$ distinct, $u_0$-dependent, instanton actions---which ``flow'' to their respective conformal-background spectral-geometry results \eqref{eq:min-string-action-n,k} under $u_0 \to 1$---, \textit{alongside} their \textit{symmetric} pairs. This implies, quite similarly to its multicritical counterpart, that the order-$k$ minimal-string should be described by a $(2k-2)$-parameter \textit{resonant} transseries.

%%%%%%%%%%%%%%%%%%%%%%%%%%%%%%%%%%%%%%%%%%%%%%%%%%%%%%%%%%%%%%%%%
\begin{figure}[t!]
\centering
         \includegraphics[width=0.975\textwidth]{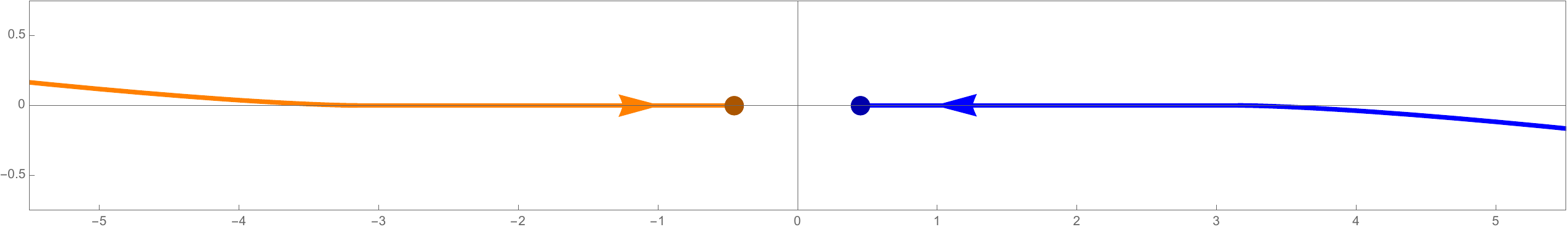}\\
         \vspace{3mm}
         \includegraphics[width=0.975\textwidth]{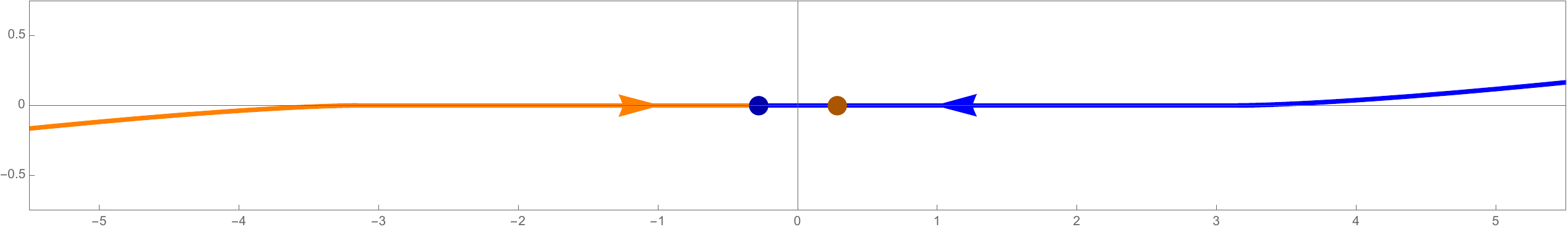}
\caption{Plot the $u_0 \to 1$ ``conformal-background flow'' of all $(2,5)$ instanton actions \eqref{eq:(2,5)-minimal-A-solution} on the complex Borel plane (we split the plot in two figures simply to avoid cumbersomeness of overlapping trajectories). The blue trajectories end up at the (positive) $A_{(2,5)} \left(-,-\right)$ (top) and (negative) $A_{(2,5)} \left(-,+\right)$ (bottom) instanton actions computed by spectral geometry \eqref{eq:A-(2,5)_ms_SG}. The orange trajectories end up at their respective symmetric pairs, as dictated by the (even-powered) leading nonperturbative string equation \eqref{eq:poly-ODE-inst-act-string-eq}. The resulting transseries is resonant.}
\label{fig:(2,5)-conf-back-limit}
\end{figure}
%%%%%%%%%%%%%%%%%%%%%%%%%%%%%%%%%%%%%%%%%%%%%%%%%%%%%%%%%%%%%%%%%

Organizing the (conformal background) $(2k-2)$ instanton actions vectorially (akin to what we did in the multicritical case), and following the general resurgent transseries framework \cite{abs18}, the $(2k-2)$-parameter transseries describing the full, nonperturbative specific-heat, $u(g_{\text{s}})$, solution to the order-$k$ minimal-string string-equation \eqref{eq:minimal-string_string_eq}, is hence given by\footnote{We have omitted any $z$- or $u_0$-dependence as we are assuming the conformal-background limit has been taken.}
\be
u \left( g_{\text{s}}, \boldsymbol{\sigma} \right) = \sum_{\boldsymbol{n} \in \BN_{0}^{2k-2}} \boldsymbol{\sigma}^{\boldsymbol{n}} \exp \left( - \frac{\boldsymbol{n} \cdot \boldsymbol{A}}{g_{\text{s}}} \right)\, \sum_{g=0}^{+\infty} u_g^{(\boldsymbol{n})}\, g_{\text{s}}^g.
\ee
\noindent
The notation is the same as in subsection~\ref{subsec:multi_string_eq} so we need not repeat it. The same holds for the discussion concerning resonance. The one point which is worth mentioning concerns the structure of Borel singularities and corresponding Stokes lines. All Borel singularities are now upon the real axis\footnote{This is a feature of the conformal background, which gets lost as we move to general backgrounds \cite{mmss04} (\textit{e.g.}, as in the multicritical case we addressed earlier; albeit Borel singularities still appear in both symmetric and complex-conjugate pairs amounting to a nonperturbative \textit{real} correction to the perturbative expansion).}, which means there are only two Stokes lines: the $\BR^+$ and the $\BR^-$ rays. Whereas this is a simpler picture than the corresponding one in subsection~\ref{subsec:multi_string_eq}, the Stokes automorphisms associated to these Stokes lines will however be considerably more intricate than their corresponding multicritical counterparts (see, \textit{e.g.}, \cite{abs18} for a discussion). Finally, the large-order comment we made at the end of subsection~\ref{subsec:multi_string_eq} also holds now (with the appropriate modifications on the location of Borel singularities), and we shall return to this point in the next section.

The one last thing we wish to address concerns the next-to-leading nonperturbative string equation; or, in other words, the role of the ``$\beta$-term'' in \eqref{eq:leading+next-to-leading_MS_string_eq}. Repeating the exercise that led to \eqref{eq:poly-ODE-inst-act-string-eq}, \textit{i.e.}, expanding to first order in the transmonomial $\sigma \exp \left( - \frac{A}{g_{\text{s}}} \right)$ but now to next-to-leading (first) order in the string coupling, yields:
\bea
\cdots &+& \sigma\, \rme^{-\frac{A}{g_{\text{s}}}}\, g_{\text{s}}\, \Bigg\{\, - \sum_{n=0}^{k-1} \sum_{\ell=k-2\left[\frac{k-1}{2}\right]}^{k} \ell\, n\, t_{\ell} (k)\, \frac{\alpha_{\ell,\ell-n}}{\alpha_{\ell\ell}} \bigg( 2 u_0^{\ell-n-1}\, A'(z)^{2n-1} \left( u_0^{(1)} (z) \right)' + \\
&+& \Big\{ \left( 2n-1 \right) u_0\, A''(z) + \left( \ell-n-1 \right) u_0'\, A'(z) \Big\}\, u_0^{\ell-n-2}\, A'(z)^{2n-2}\, u_0^{(1)} (z) \bigg) \Bigg\} + \cdots = z.
\nonumber
\eea
\noindent
Note how, at this order, both $u_1^{(1)}(z)$ and $u^{(1)}_0 (z)$ unknowns initially appear; but the terms proportional to $u_1^{(1)}(z)$ match the equation obtained at previous order \eqref{eq:poly-ODE-inst-act-string-eq}; hence vanish---and one is left with an equation for $u^{(1)}_0 (z)$ alone. This pattern also occurs at higher orders, and is another feature of resonance \cite{asv11}. As such, at this next-to-leading (nonperturbative) order the string equation \eqref{eq:leading+next-to-leading_MS_string_eq} becomes a (homogeneous) first-order linear ODE for the leading one-instanton coefficient, $u^{(1)}_0 (z)$. Its general solution is given by
\be
u^{(1)}_0 (z) \propto \left( \sum_{n=0}^{k-1} \sum_{\ell=k-2\left[\frac{k-1}{2}\right]}^{k} \ell\, n\, t_{\ell} (k)\, \frac{\alpha_{\ell,\ell-n}}{\alpha_{\ell\ell}}\, u_0^{\ell-n-1}\, A'(z)^{2n-1} \right)^{-\frac{1}{2}},
\ee
\noindent
where we still have to specify the solution for the instanton action\footnote{Note that this expression entails a sum over \textit{odd} powers of $A'(z)$, whereas in \eqref{eq:poly-ODE-inst-act-string-eq} the sum was over \textit{even} powers---hence leading to the symmetric (resonant) solutions $\sim \pm A(z)$. This expression then explicitly shows how the sign of the instanton action also matters in the construction of its associated instanton sector.} from \eqref{eq:poly-ODE-inst-act-string-eq} (and a choice of appropriate ``working variable'' as in the example \eqref{eq:u(1)0-ODE-soltn}, rather than the $z$-variable used above). For example, for the $k=2$ or $(2,3)$ minimal string theory one finds, \textit{e.g.},
\be
u^{(1)}_0 (z) \propto \frac{1}{\sqrt{A'(z)}} \propto \frac{1}{\left( 6 u_0 \right)^{1/4}}.
\ee
\noindent
For the $k=3$ or $(2,5)$ minimal string theory one finds instead, \textit{e.g.},
\be
u^{(1)}_0 (z) \propto \frac{1}{\sqrt{A'(z)^3 - 5 u_0 A'(z)}} \propto \frac{1}{\left( \sqrt{5} \sqrt{2-u_0^2} \right)^{1/2} \left( 5 u_0 - \sqrt{5} \sqrt{2-u_0^2} \right)^{1/4}}.
\ee
\noindent
This is just \eqref{eq:u(1)0-ODE-soltn}, herein written in the original $z$-variable via $u_0$. Going to higher orders would then be similar to the rest of the $(2,5)$ example which we worked out earlier.

%%%%%%%%%%%%%%%%%%%%%%%%%%%%%%%%%%%%%%%%%%%%%%%%%%%%%%%%%%%%%%%%%
\subsection{On the String Equation for JT Gravity}\label{subsec:JT_string_eq}
%%%%%%%%%%%%%%%%%%%%%%%%%%%%%%%%%%%%%%%%%%%%%%%%%%%%%%%%%%%%%%%%%

We can finally address the connection between the minimal string and JT gravity \cite{sss19, ss19, os19, j19, t:mt20, gv21}, now explicitly at the level of the string equations. Recall from subsection~\ref{subsec:deformations} how we obtained the spectral curve of JT gravity out of an appropriate limit of the $(2,2k-1)$ ``rescaled'' minimal string, via \eqref{eq:min-string-to-JT-spec-curv}. We now wish to implement this precise same limit at the level of the minimal-string string-equation. As already pointed-out in subsection~\ref{subsec:general-KdV}, due to the aforementioned rescaling of the spectral curve it turns out that it is much more convenient to work in the spectral-geometry ``monomial basis'' \eqref{eq:multicrit_spect_curve-monomial}. This means that we must first ``change coordinates'' from the standard conformal-background KdV-times \eqref{eq:min-str-kdv-times}---which we herein recall for convenience of the reader,
\be
t_p (k) = \left(-1\right)^{k} \frac{2^{p-\frac{5}{2}} \left(2k-1\right) \sqrt{\pi}}{\Gamma \left(p+1\right) \Gamma \left(\frac{3-k-p}{2}\right) \Gamma \left(\frac{2+k-p}{2}\right)}
\ee
\noindent
---to the monomial-basis conformal-background KdV-times computed via \eqref{eq:change-basis_KdV-times}. These are:
\be
\label{eq:widetildetpk_monomial-KdV}
\widetilde{t}_p (k) = \left(-1\right)^{k+1} \frac{\left(2k-1\right) \Gamma \left(k+p-1\right)}{2^{p+\frac{1}{2}}\, \Gamma \left(p+1\right) \Gamma \left(p\right) \Gamma \left(k-p+1\right)}.
\ee
\noindent
The rescaling in \eqref{eq:min-string-to-JT-spec-curv} is now simple to implement on this ``monomial KdV basis'', by appropriately rescaling them according to \eqref{eq:multicrit_spect_curve-monomial}. The resulting ``rescaled KdV times'' are simply\footnote{Note that in this basis, string equations for minimal strings will no longer display an even/odd split of relevant KdV times; such as that observed in, \textit{e.g.}, \eqref{eq:P1-minimal-conventions} (even-labelled times) versus \eqref{eq:(2,5)-minimal-string-eq} (odd-labelled times).} \cite{t:mt20}
\be
\widetilde{\mathsf{t}}_p (k) = \frac{(-1)^{k-1}}{2\pi} \left(\frac{2\sqrt{2}\pi}{2k-1}\right)^{2p-1} \widetilde{t}_p (k) = \frac{2^{2p-3}\pi^{2p-2}}{\left( 2k-1\right)^{2p-2}}\, \frac{\Gamma \left(k+p-1\right)}{\Gamma \left(p+1\right) \Gamma \left(p\right) \Gamma \left(k-p+1\right)}.
\ee
\noindent
The punch-line of these rewritings is that, now, these final KdV times do have a well-defined (finite) $k \to +\infty$ limit. One obtains (see as well \cite{os19, j19, t:mt20})
\be
\label{eq:k=infty_KdV_times}
\widetilde{\mathsf{t}}_p \equiv \widetilde{\mathsf{t}}_p (\infty) = \lim_{k \to +\infty} \widetilde{\mathsf{t}}_p (k) = \frac{\pi^{2p-2}}{2\, \Gamma \left(p+1\right) \Gamma \left(p\right)}.
\ee
\noindent
One immediate consistency-check is just to recompute the JT spectral curve \eqref{eq:min-string-to-JT-spec-curv}, using the spectral-geometry monomial basis \eqref{eq:multicrit_spect_curve-monomial} and the above KdV times. It follows, as expected,
\be
y (x) = \sum_{n=1}^{+\infty} \,\widetilde{\mathsf{t}}_n\, \widetilde{Q}_n \left( \sqrt{\frac{x+1}{2}} \right) = \frac{1}{2\pi} \sin \left( 2\pi \sqrt{x+1} \right).
\ee

The final step is now elementary: the JT-gravity string-equation follows from the $k \to +\infty$ limit of the minimal-string string-equation \eqref{eq:minimal-string_string_eq} with KdV times \eqref{eq:k=infty_KdV_times}. This yields\footnote{Now also easily understood as 2d topological gravity on the background of fixed KdV times $\left\{ \widetilde{\mathsf{t}}_p \right\}$ \cite{os19}.}
\be
\label{eq:JT-string-equation-implicit}
\sum_{p=1}^{+\infty} \,\widetilde{\mathsf{t}}_p \cdot \frac{p}{\alpha_{pp}} \cdot R_{p} \left[ u \right] = z.
\ee
\noindent
In order to be fully explicit, we still have to use what we know concerning generic Gel'fand--Dikii KdV potentials, \textit{i.e.}, \eqref{eq:Rl-alphall-alphali-betalj} with coefficients \eqref{eq:alphali_coeffs} and \eqref{eq:betalj_coeffs}; which have in fact already been written for the minimal string in \eqref{eq:leading+next-to-leading_MS_string_eq}. This implies that (part of) the exact JT-gravity string-equation is
\be
\label{eq:JT-string-equation-explicit}
\sum_{p=1}^{+\infty} \,\widetilde{\mathsf{t}}_p \left( u^{p} + \frac{p}{\alpha_{pp}} \sum_{i=1}^{p-1} \alpha_{p i}\, g_{\text{s}}^{2p-2i}\, u^{i-1}\, u^{(2p-2i)} + \frac{p}{\alpha_{pp}} \sum_{j=2}^{p-1} \beta_{p j}\, g_{\text{s}}^{2p-2j}\, u^{j-2}\, u'\, u^{(2p-2j-1)} + \cdots 
\right) = z,
\ee
\noindent
including contributions at all genera, both perturbative and nonperturbative alike. The different sums, in the different components of this equation, may be explicitly evaluated. The first is the genus-zero no-derivatives monomial, which immediately sums up\footnote{This sum may also be evaluated just up to order $k$, yielding the genus-zero string equation for the minimal strings discussed in the previous subsection \cite{t:mt20}, as
\be
\sum_{p=1}^{k} \,\widetilde{\mathsf{t}}_p (k)\, u^{p} = \frac{1}{2} u\, {}_2F_1 \left( 1-k, k; 2 \left| - \frac{4\pi^2u}{\left( 2k-1 \right)^2} \right. \right).
\ee
} to the known result \cite{z07, z08, os19, j19, t:mt20}
\be
\label{eq:JT-string-eq-no-derivative-term}
\sum_{p=1}^{+\infty} \,\widetilde{\mathsf{t}}_p\, u^{p} = \frac{1}{2\pi} \sqrt{u}\, I_1 \left( 2\pi \sqrt{u} \right).
\ee
\noindent
Herein $I_{\nu}(z)$ is the modified Bessel function of the first kind (\textit{e.g.}, \cite{olbc10}). Note how this result is somewhat reminiscent of the leading contribution to the circular Wilson loop in $\CN=4$ supersymmetric Yang--Mills theory (SYM) \cite{esz00}. This is also a rather concise result, but the following terms will not be as compact-looking. Because of the double-sum, there are two ways to look at the next, single-derivative ``$\alpha_{pi}$-term'' monomials. One may think of this contribution just as written in the expression above---\textit{i.e.}, as an expansion in the string coupling, or a derivative expansion---in which case one evaluates the ``$i$-sum'' first to find
\bea
&&
\hspace{-20pt}
\label{eq:JT-string-eq-single-derivative-term}
\sum_{p=1}^{+\infty} \,\widetilde{\mathsf{t}}_p\, \frac{p}{\alpha_{pp}} \sum_{i=1}^{p-1} \alpha_{p i}\, g_{\text{s}}^{2p-2i}\, u^{i-1}\, u^{(2p-2i)} = \sum_{k=1}^{+\infty} \left(-1\right)^k \pi^k\, \frac{\left(k+1\right)!}{\left(2k+2\right)!}\, \frac{I_{k} \left( 2\pi \sqrt{u} \right)}{u^{k/2}}\, g_{\text{s}}^{2k}\, u^{(2k)} = \\
&&
\hspace{20pt}
= - \frac{\pi}{12}\, \frac{I_{1} \left( 2\pi \sqrt{u} \right)}{u^{1/2}}\, g_{\text{s}}^2\, u'' + \frac{\pi^2}{120}\, \frac{I_{2} \left( 2\pi \sqrt{u} \right)}{u}\, g_{\text{s}}^4\, u'''' - \frac{\pi^3}{1680}\, \frac{I_{3} \left( 2\pi \sqrt{u} \right)}{u^{3/2}}\, g_{\text{s}}^{6}\, u^{(6)} + \cdots. \nonumber
\eea
\noindent
Interestingly enough, the Bessel structure which was already featured in the first, no-derivatives monomial, \eqref{eq:JT-string-eq-no-derivative-term}, is also very clearly appearing herein as well. Further, this expression is now very reminiscent of the all-orders $\CN=4$ SYM circular Wilson loop (with $1/N^2$ expansion simply computed via the Gaussian matrix model) \cite{dg00}. Borrowing from that particular line of research \cite{o06}, one may use the Bessel generating-function
\be
\label{eq:bessel-generating-fct}
\rme^{\pi \left( u\, \omega + \frac{1}{\omega} \right)} = \sum_{m \in \BZ} \frac{I_m \left( 2\pi \sqrt{u} \right)}{\left( \sqrt{u} \right)^m}\, \frac{1}{\omega^m}
\ee
\noindent
to rewrite \eqref{eq:JT-string-eq-single-derivative-term} as\footnote{Using this trick, one may also rewrite the genus-zero contribution \eqref{eq:JT-string-eq-no-derivative-term} as
\be
\frac{1}{2\pi} \sqrt{u}\, I_1 \left( 2\pi \sqrt{u} \right) = \oint \frac{\rmd\omega}{2\pi\rmi}\, \rme^{\pi \left( u\, \omega + \frac{1}{\omega} \right)} \frac{u}{2\pi}.
\ee
}
\be
\sum_{p=1}^{+\infty} \,\widetilde{\mathsf{t}}_p\, \frac{p}{\alpha_{pp}} \sum_{i=1}^{p-1} \alpha_{p i}\, g_{\text{s}}^{2p-2i}\, u^{i-1}\, u^{(2p-2i)} = \oint \frac{\rmd\omega}{2\pi\rmi}\, \rme^{\pi \left( u\, \omega + \frac{1}{\omega} \right)} \sum_{k=1}^{+\infty} \left(-1\right)^k \pi^k\, \frac{\left(k+1\right)!}{\left(2k+2\right)!}\, \omega^{k-1}\, g_{\text{s}}^{2k}\, u^{(2k)}.
\ee
\noindent
Note how the sum in the integrand may be exactly evaluated, albeit producing a rather nonlocal (and non-integrable) result:
\be
\sum_{k=1}^{+\infty} \left(-1\right)^k \pi^k\, \frac{\left(k+1\right)!}{\left(2k+2\right)!}\, \omega^{k-1}\, g_{\text{s}}^{2k}\, u^{(2k)} = - \frac{u}{2\omega} + \frac{D\!F \left( \frac{1}{2} \sqrt{\omega\pi}  g_{\text{s}} \frac{\rmd}{\rmd z} \right)}{\omega^{3/2} \sqrt{\pi} g_{\text{s}} \frac{\rmd}{\rmd z}}\, u.
\ee
\noindent
Herein $D\!F(z)$ is the Dawson integral we have already seen in subsection~\ref{subsec:deformations}. The alternative way to look at the single-derivative ``$\alpha_{pi}$-term'' monomials in \eqref{eq:JT-string-equation-explicit} would be to evaluate the ``$p$-sum'' first. In that case we find\footnote{The eventual appearance of non-local trigonometric terms was already anticipated in \cite{os19}. Herein we obtain their non-local generating function via hypergeometric functions (alongside all corresponding coefficients).}
\bea
&&
\hspace{-20pt}
\label{eq:JT-string-eq-single-derivative-term-2}
\sum_{p=1}^{+\infty} \,\widetilde{\mathsf{t}}_p\, \frac{p}{\alpha_{pp}} \sum_{i=1}^{p-1} \alpha_{p i}\, g_{\text{s}}^{2p-2i}\, u^{i-1}\, u^{(2p-2i)} = - \sum_{k=1}^{+\infty} u^{k-1}\, \frac{\pi^{2k}}{12 \Gamma \left( k \right) \Gamma \left( k+2 \right)} \times \\
&&
\hspace{20pt}
\times \left\{ {}_1F_2 \left( 2; \left. \frac{5}{2}, 2+k \,\right| -\frac{\pi^2 g_{\text{s}}^2}{4}\, \frac{\rmd^2}{\rmd z^2} \right) + k\, {}_1F_2 \left( 1; \left. \frac{5}{2}, 2+k \,\right| -\frac{\pi^2 g_{\text{s}}^2}{4}\, \frac{\rmd^2}{\rmd z^2} \right) \right\} g_{\text{s}}^2 \frac{\rmd^2 u}{\rmd z^2} = \nonumber \\
&&
\hspace{-20pt}
= \frac{1}{2} \left( \frac{\sin \left( \pi g_{\text{s}} \frac{\rmd}{\rmd z} \right)}{\pi g_{\text{s}} \frac{\rmd}{\rmd z}} - 1 \right) u - u \left( 
\frac{\cos \left( \pi g_{\text{s}} \frac{\rmd}{\rmd z} \right) - 1}{g_{\text{s}}^2 \frac{\rmd^2}{\rmd z^2}} + \frac{\pi^2}{2} \right) u - \cdots. \nonumber
\eea
\noindent
This expansion is a bit less transparent than the previous Bessel-expansion one, albeit again making explicit the nonlocal nature of the resulting string equation. Moving-on to the single-derivative-and-single-$u'$ ``$\beta_{pj}$-term'' monomial, there are of course the same two ways to look at it given the double-sum. In the string-coupling expansion we find
\bea
&&
\hspace{-20pt}
\label{eq:JT-string-eq-single-derivative-single-u'-term}
\sum_{p=1}^{+\infty} \,\widetilde{\mathsf{t}}_p\, \frac{p}{\alpha_{pp}} \sum_{j=2}^{p-1} \beta_{p j}\, g_{\text{s}}^{2p-2j}\, u^{j-2}\, u'\, u^{(2p-2j-1)} = \left(-1\right) \pi^{2}\, \frac{2!}{4!}\, \frac{I_{2} \left( 2\pi \sqrt{u} \right)}{u}\, g_{\text{s}}^{2}\, \frac{1}{2}\, u'\, u' + \\
&&
\hspace{20pt}
+ \sum_{k=2}^{+\infty} \left(-1\right)^k \pi^{k+1}\, \frac{k \left(k+1\right)!}{\left(2k+2\right)!}\, \frac{I_{k+1} \left( 2\pi \sqrt{u} \right)}{u^{(k+1)/2}}\, g_{\text{s}}^{2k}\, u'\, u^{(2k-1)} = \nonumber \\
&&
\hspace{-20pt}
= - \frac{\pi^2}{24}\, \frac{I_{2} \left( 2\pi \sqrt{u} \right)}{u}\, g_{\text{s}}^2 \left( u' \right)^2 + \frac{\pi^3}{60}\, \frac{I_{3} \left( 2\pi \sqrt{u} \right)}{u^{3/2}}\, g_{\text{s}}^4\, u' u''' - \frac{\pi^4}{560}\, \frac{I_{4} \left( 2\pi \sqrt{u} \right)}{u^2}\, g_{\text{s}}^{6}\, u' u^{(5)} + \cdots. \nonumber
\eea
\noindent
Clearly, the Bessel pattern keeps repeating itself---which is probably indicating this is a generic feature of the full string equation. The resemblance to the $\CN=4$ SYM circular Wilson loop will also remain clean once we add both terms into the string equation (see below). In spite of having only two ``data points'', one likely hypothesis is that the following (higher derivatives factors/monomials) terms in the string equation will involve sums over the Bessel combinations
\be
\sim \frac{I_{k+2} \left( 2\pi \sqrt{u} \right)}{\left( \sqrt{u} \right)^{k+2}}, \qquad \sim \frac{I_{k+3} \left( 2\pi \sqrt{u} \right)}{\left( \sqrt{u} \right)^{k+3}}, \qquad \text{and so on}.
\ee
\noindent
Using the Bessel generating-function \eqref{eq:bessel-generating-fct} one may rewrite \eqref{eq:JT-string-eq-single-derivative-single-u'-term} as
\bea
&&
\sum_{p=1}^{+\infty} \,\widetilde{\mathsf{t}}_p\, \frac{p}{\alpha_{pp}} \sum_{j=2}^{p-1} \beta_{p j}\, g_{\text{s}}^{2p-2j}\, u^{j-2}\, u'\, u^{(2p-2j-1)} = \\
&&
= \oint \frac{\rmd\omega}{2\pi\rmi}\, \rme^{\pi \left( u\, \omega + \frac{1}{\omega} \right)} \left( \left(-1\right) \pi^{2}\, \frac{2!}{4!}\, \omega\, g_{\text{s}}^{2}\, \frac{1}{2}\, u'\, u' + \sum_{k=2}^{+\infty} \left(-1\right)^k \pi^{k+1}\, \frac{k \left(k+1\right)!}{\left(2k+2\right)!}\, \omega^k\, g_{\text{s}}^{2k}\, u'\, u^{(2k-1)} \right). \nonumber
\eea
\noindent
Note how the sum in this integrand may also be exactly evaluated, with the result again involving the Dawson integral, but we do not find the final expression particularly illuminating and omit it. If one were to alternatively do the ``$p$-sum'' first instead, one would find
\bea
&&
\hspace{-20pt}
\label{eq:JT-string-eq-single-derivative-single-u'-term-2}
\sum_{p=1}^{+\infty} \,\widetilde{\mathsf{t}}_p\, \frac{p}{\alpha_{pp}} \sum_{j=2}^{p-1} \beta_{p j}\, g_{\text{s}}^{2p-2j}\, u^{j-2}\, u'\, u^{(2p-2j-1)} = - \frac{\pi^2}{24}\, \frac{I_{2} \left( 2\pi \sqrt{u} \right)}{u}\, g_{\text{s}}^2\, u'\, \frac{\rmd u}{\rmd z} -
\\
&&
\hspace{20pt}
- \sum_{k=2}^{+\infty} u^{k-2}\, u'\, \frac{\pi^{2k} g_{\text{s}}^2}{12 \Gamma \left( k-1 \right) \Gamma \left( k+2 \right)} \left\{ {}_2F_3 \left( 2, 2; 1, \left. \frac{5}{2}, 2+k \,\right| -\frac{\pi^2 g_{\text{s}}^2}{4}\, \frac{\rmd^2}{\rmd z^2} \right) + \right. \nonumber \\
&&
\hspace{-20pt}
\left.
+ k\, {}_1F_2 \left( 2; \left. \frac{5}{2}, 2+k \,\right| -\frac{\pi^2 g_{\text{s}}^2}{4}\, \frac{\rmd^2}{\rmd z^2} \right) - \left( k+1 \right) \right\} \frac{\rmd u}{\rmd z} = \cdots + u' \left( \frac{\cos \left( \pi g_{\text{s}} \frac{\rmd}{\rmd z} \right) - 1}{g_{\text{s}}^2 \frac{\rmd^3}{\rmd z^3}} \right) u + \cdots, \nonumber
\eea
\noindent
again making explicit the nonlocal nature of the resulting string equation.

Assembling together the different terms computed above, we finally obtain a glimpse of the general structure which the exact JT-gravity string-equation will be built of. In the (explicit) Bessel representation, we find:
\bea
&&
\frac{1}{2\pi} \sqrt{u}\, I_1 \left( 2\pi \sqrt{u} \right) - \frac{\pi}{12} \left( \frac{I_{1} \left( 2\pi \sqrt{u} \right)}{u^{1/2}}\, u'' + \frac{\pi}{2} \frac{I_{2} \left( 2\pi \sqrt{u} \right)}{u} \left( u' \right)^2 \right) g_{\text{s}}^2 + \frac{\pi^2}{120} \left( \frac{I_{2} \left( 2\pi \sqrt{u} \right)}{u}\, u'''' + \right. \nonumber \\
&&
\left.
+ 2\pi\, \frac{I_{3} \left( 2\pi \sqrt{u} \right)}{u^{3/2}}\, u' u''' + \cdots \right) g_{\text{s}}^4 - \frac{\pi^3}{1680} \left( \frac{I_{3} \left( 2\pi \sqrt{u} \right)}{u^{3/2}}\, u^{(6)} + 3\pi\, \frac{I_{4} \left( 2\pi \sqrt{u} \right)}{u^2}\, u' u^{(5)} + \cdots \right) g_{\text{s}}^{6} + \cdots + \nonumber \\
&&
\label{eq:JT-gravity-string-equation-Bessel}
+ \left(-1\right)^k \frac{\pi^k \left(k+1\right)!}{\left(2k+2\right)!} \left( \frac{I_{k} \left( 2\pi \sqrt{u} \right)}{u^{k/2}}\, u^{(2k)} + k\pi\, \frac{I_{k+1} \left( 2\pi \sqrt{u} \right)}{u^{(k+1)/2}}\, u' u^{(2k-1)} + \cdots \right) g_{\text{s}}^{2k} + \cdots = z.
\eea
\noindent
As mentioned earlier, the resemblance to the Wilson loop calculation in \cite{dg00} is now rather clean. With hindsight this is probably somewhat expected: the relevant correlation functions of JT-gravity involve the bulk thermal partition-functions $Z(\beta)$, which have the matrix-model representation \eqref{eq:macro-loop-MM} \cite{sss19}. These matrix-model macroscopic loops \cite{bdss90, sss19, os19} are precisely the Wilson loop insertions in the (Gaussian) matrix model of \cite{dg00}. It would be very interesting to make this reasoning more precise, and fully unravel the would-be matrix-model structure behind the complete JT-gravity string-equation. On the other hand, unfortunately, obtaining such a complete, closed-form expression is yet out of reach. It is however clear from above that the JT-gravity string-equation may be written in a string-coupling derivative-expansion, where, at order $g_{\text{s}}^{2k}$, one very likely finds the derivative ``building blocks''
\be
\propto \frac{I_{k+n} \left( 2\pi \sqrt{u} \right)}{\left( \sqrt{u} \right)^{k+n}}\, \underbrace{u^{(\delta_1)} u^{(\delta_2)} \cdots u^{(\delta_s)}}_{\sum_{i=1}^{s} \delta_i = 2k,\, \delta_i \neq 0},
\ee
\noindent
but where the relation between $n$ and $s$ is not set by our analysis (for each $k$ there is an upper bound on $n$, likely dictated by the number of different building-blocks at each order; and there is an upper bound on $s$, when all $\delta_i=1$ in which case $s=2k$). Further, precisely pinpointing the coefficients in front of these building-blocks will also have to be left for future work. Nonetheless, the number of such building-blocks at fixed order $g_{\text{s}}^{2k}$ is basically dictated by the number of integer partitions, $p(2k)$---\textit{e.g.}, for the first three terms illustrating \eqref{eq:JT-gravity-string-equation-Bessel}, one has $p(2)=2$, $p(4)=5$, and $p(6)=11$. Explicitly, at order $g_{\text{s}}^4$ the full structure very likely is:
\bea
&&
\hspace{-40pt}
\frac{I_{2} \left( 2\pi \sqrt{u} \right)}{u}\, u'''' + 2\pi\, \frac{I_{3} \left( 2\pi \sqrt{u} \right)}{u^{3/2}}\, u' u''' + \epsilon_1\, \frac{I_{2+n_1} \left( 2\pi \sqrt{u} \right)}{u^{\left(2+n_1\right)/2}}\, u'' u'' + \\
&&
\hspace{80pt}
+\, \epsilon_2\, \frac{I_{2+n_2} \left( 2\pi \sqrt{u} \right)}{u^{\left(2+n_2\right)/2}}\, u'' u' u' + \epsilon_3\, \frac{I_{2+n_3} \left( 2\pi \sqrt{u} \right)}{u^{\left(2+n_3\right)/2}}\, u' u' u' u'. \nonumber
\eea

As a final remark, note that the above expressions may be equivalently written making use of the Bessel generating-function \eqref{eq:bessel-generating-fct}. Focusing \textit{solely on the integrand, without exponential kernel} of such expression (and which should be rather obvious from the discussion above), one now finds $\omega$-polynomials at each order $g_{\text{s}}^{2k}$. They are:
\bea
g_{\text{s}}^{0} \qquad &\to& \qquad \frac{1}{2\pi}\, u, \\
g_{\text{s}}^{2} \qquad &\to& \qquad - \frac{\pi}{12} \left( u'' + \frac{\pi}{2}\, \omega \left( u' \right)^2 \right), \\
g_{\text{s}}^{4} \qquad &\to& \qquad \frac{\pi^2}{120} \left( \omega\, u'''' + 2\pi\, \omega^2\, u' u''' + \cdots \right), \\
g_{\text{s}}^{6} \qquad &\to& \qquad - \frac{\pi^3}{1680} \left( \omega^2\, u^{(6)} + 3\pi\, \omega^3\, u' u^{(5)} + \cdots \right), \\
&\vdots& \nonumber \\
g_{\text{s}}^{2k} \qquad &\to& \qquad \left(-1\right)^k \frac{\pi^k \left(k+1\right)!}{\left(2k+2\right)!} \left( \omega^{k-1}\, u^{(2k)} + k\pi\, \omega^k\, u' u^{(2k-1)} + \cdots \right).
\eea
\noindent
There is not much more to say about these expressions but to repeat \textit{verbatim} what we already said concerning their (equivalent) explicit Bessel representation. The only one extra comment we can make is that these polynomials will likely be analogues of the change of basis construction, from Buchholz-polynomials\footnote{The analogues of the Buchholz-polynomial-basis $\left\{ \mathfrak{p}_n (g_{\text{s}}) \right\}$ from \cite{o06}, in the present context, would be:
\bea
\mathfrak{p}_1 (g_{\text{s}}) &=& - \frac{\pi}{12} u'' g_{\text{s}}^2 + \frac{1}{2\pi} u, \\
\mathfrak{p}_2 (g_{\text{s}}) &=& \frac{\pi^2}{120} u'''' g_{\text{s}}^4 - \frac{\pi^2}{24} \left( u' \right)^2 g_{\text{s}}^2, \\
\mathfrak{p}_3 (g_{\text{s}}) &=& - \frac{\pi^3}{1680} u^{(6)} g_{\text{s}}^6 + \frac{\pi^3}{60} u' u''' g_{\text{s}}^4 + \cdots.
\eea
} to monomials, appearing in the $\CN=4$ SYM Wilson loop context \cite{o06}. One possibility (also raised in \cite{os19} and with additional support in \cite{eggls23}) is that all this non-local structure we find may be rewritten as a \textit{finite-difference equation}; but exactly what such final string-equation structure may be remains to be unravelled. 

%%%%%%%%%%%%%%%%%%%%%%%%%%%%%%%%%%%%%%%%%%%%%%%%%%%%%%%%%%%%%%%%%
\subsubsection*{Perturbative Content via the String Equation}
%%%%%%%%%%%%%%%%%%%%%%%%%%%%%%%%%%%%%%%%%%%%%%%%%%%%%%%%%%%%%%%%%

Let us begin by addressing the JT-gravity string-equation at the perturbative level. Start with the classical string-equation, with well-known expression \cite{z07, z08, os19, j19, t:mt20}
\be
\frac{1}{2\pi} \sqrt{u_0}\, I_1 \left( 2\pi \sqrt{u_0} \right) = z.
\ee
\noindent
This implicitly defines $u_0$ as a function of $z$ and is the good variable for perturbative calculations. The corresponding genus-zero free energy then follows in the usual fashion\footnote{Note that because of \eqref{eq:UandZchange}, the relation between free energy and specific heat in the monomial basis becomes $F''=-\frac{1}{2}(u+1)$. Moreover, in order to obtain the free energy from the specific heat, one has to integrate twice from $-1$, instead of from $0$ as in the rest of this section.},
\bea
F_0 (u_0) &=& \frac{1}{48\pi^4} \Bigg\{ \pi^2 u_0^2 I_0 \left( 2\pi \sqrt{u_0} \right)^2-\pi^2 I_0 \left( 2\pi \rmi \right)^2 -2\pi u_0^{3/2} I_0 \left( 2\pi \sqrt{u_0} \right) I_1 \left( 2\pi \sqrt{u_0} \right)-\\
&& -2\pi \rmi\, I_0 \left( 2\pi \rmi \right) I_1 \left( 2\pi \rmi \right)+u_0\left(1-\pi^2\left(3+4 u_0\right)\right)I_1 \left( 2\pi \sqrt{u_0} \right)^2+\left(1+\pi^2\right)I_1 \left( 2\pi \rmi \right)^2+\nonumber\\
&& +6\pi\sqrt{u_0}\, I_1 \left( 2\pi \sqrt{u_0} \right)\Big[u_0\, I_2 \left( 2\pi \sqrt{u_0} \right)+I_2 \left( 2\pi \rmi\right)\Big]\Bigg\},\nonumber
\eea
\noindent
where the final step entails reaching the JT-gravity conformal-background\footnote{In the present basis for KdV times, the conformal-background limit now sets $u_0 = 0 = z$.} $u_0 \to 0$ as:
\be
F_0 \equiv \lim_{u_0 \to 0} F_0 (u_0) = \frac{1}{48\pi^4}\left(-\pi^2 I_0 \left( 2\pi \rmi \right)^2 -2\pi \rmi\, I_0 \left( 2\pi \rmi \right) I_1 \left( 2\pi \rmi \right)+\left(1+\pi^2\right)I_1 \left( 2\pi \rmi \right)^2\right).
\ee
\noindent
Note that the associated Weil--Petersson volume to this JT-gravity free energy is essentially undefined (see, \textit{e.g.}, \cite{sss19}). It would therefore be interesting to understand the geometric interpretation of the above quantity. Reproducing the full perturbative content of JT gravity is, however, out of reach in the present context, due to the lack of a complete (exact) JT-gravity string-equation.

%%%%%%%%%%%%%%%%%%%%%%%%%%%%%%%%%%%%%%%%%%%%%%%%%%%%%%%%%%%%%%%%%
\subsubsection*{Nonperturbative Content via the String Equation}
%%%%%%%%%%%%%%%%%%%%%%%%%%%%%%%%%%%%%%%%%%%%%%%%%%%%%%%%%%%%%%%%%

The last topic we wish to address concerns the nonperturbative content of JT gravity, herein solely out of its string equation (a proper study will appear in \cite{eggls23}). Not having the full string equation we cannot expect to go extremely far, but the fact that we know the ``$\alpha_{pi}$'' and ``$\beta_{pj}$''-terms in the JT-gravity string-equation \eqref{eq:JT-string-equation-explicit}, alongside our minimal-string (nonperturbative) experience from subsection~\ref{subsec:minimal_string_eq}, seem to indicate that at the very least we should be able to compute instanton action(s) together with the (corresponding) leading one-instanton coefficient(s).

Start-off with the one-parameter transseries \textit{ansatz} 
\be
\label{eq:trans-JT-gs-u}
u \left( g_{\text{s}}, \sigma \right) = \sum_{n=0}^{+\infty} \sigma^n\, \rme^{- \frac{n A(z)}{g_{\text{s}}}}\, \sum_{g=0}^{+\infty} u_g^{(n)}(z)\, g_{\text{s}}^g
\ee
\noindent
(this is precisely the same transseries \textit{ansatz} as \eqref{eq:trans-ms-gs-u}, the one we used earlier for the minimal string). Plugging it into the JT-gravity string-equation \eqref{eq:JT-string-equation-explicit} and expanding to first order in the transmonomial $\sigma \exp \left( - \frac{A}{g_{\text{s}}} \right)$ and to leading (zeroth) order in the string coupling, yields\footnote{Note how the $k=0$ term in the sum arises from the classical component of the JT-gravity string-equation.}:
\be
\label{eq:ODE-JT-inst-act-string-eq}
\cdots + \sigma\, \rme^{-\frac{A}{g_{\text{s}}}}\, u_0^{(1)} \left\{\, \sum_{k=0}^{+\infty} \left( \left(-1\right)^k \pi^k\, \frac{\left(k+1\right)!}{\left(2k+2\right)!}\, \frac{I_{k} \left( 2\pi \sqrt{u_0} \right)}{u_0^{k/2}} \right) A'(z)^{2k} \right\} + \cdots = z.
\ee
\noindent
One may compare this result with the minimal-string analogue, \eqref{eq:poly-ODE-inst-act-string-eq}. Also herein we see how, at this precise order, the JT-gravity string-equation is getting rewritten as a first-order nonlinear ODE for $A(z)$---albeit now as an infinite sum in $A'(z)$ rather than as a polynomial. This also makes it much harder to solve, now. Using the Bessel generating-function \eqref{eq:bessel-generating-fct}, we may rewrite the instanton-action equation coming out of \eqref{eq:ODE-JT-inst-act-string-eq} as
\be
\label{eq:ODE-JT-inst-act-string-eq-integral-rep}
\oint \frac{\rmd\omega}{2\pi\rmi}\, \rme^{\pi \left( u_0\, \omega + \frac{1}{\omega} \right)}\, \frac{D\!F \left( \frac{1}{2} \sqrt{\omega\pi}\, A'(z) \right)}{\omega^{3/2} \sqrt{\pi}\, A'(z)} = 0.
\ee
\noindent
Recall that $D\!F(z)$ is the Dawson integral we encountered twice before (whose zeroes have no closed-form expression). Hence, unfortunately, this seems to be as far as we can currently go in evaluating the instanton action(s) of JT gravity directly from its string equation. This is not a significant roadblock as they may still, obviously, be obtained from spectral geometry \cite{eggls23}---and follow from \eqref{eq:spectral-geo-JT-grav-Ank} discussed earlier. The one extra piece of information we now obtain from the JT-gravity string-equation is that, via either \eqref{eq:ODE-JT-inst-act-string-eq} or \eqref{eq:ODE-JT-inst-act-string-eq-integral-rep}, the resulting JT-gravity transseries will be an infinite-parameter \textit{resonant} transseries. This is easy to check from \eqref{eq:ODE-JT-inst-act-string-eq} which solely involves even powers of $A'(z)$, or, from \eqref{eq:ODE-JT-inst-act-string-eq-integral-rep} as $\frac{D\!F(z)}{z}$ is an even function.

Proceeding with the nonperturbative transseries analysis to next-to-leading order, \textit{i.e.}, expanding in \eqref{eq:JT-string-equation-explicit} to first order in the transmonomial $\sigma \exp \left( - \frac{A}{g_{\text{s}}} \right)$ and now further to first order in the string coupling, we find:
\bea
\cdots &+& \sigma\, \rme^{-\frac{A}{g_{\text{s}}}}\, g_{\text{s}}\, \Bigg\{\, - \sum_{k=0}^{+\infty} \sum_{i=1}^{+\infty} \,\widetilde{\mathsf{t}}_{k+i}\, \frac{k \left( k+i \right)}{\alpha_{k+i,k+i}}\, \alpha_{k+i,i} \bigg( 2 u_0^{i-1}\, A'(z)^{2k-1} \left( u_0^{(1)} (z) \right)' +  \\
&+& \Big\{ \left( 2k-1 \right) u_0\, A''(z) + \left( i-1 \right) u_0'\, A'(z) \Big\}\, u_0^{i-2}\, A'(z)^{2k-2}\, u_0^{(1)} (z) \bigg) \Bigg\} + \cdots = z.
\nonumber
\eea
\noindent
Again, this result is very comparable to the minimal-string analogue. Also herein, at this order, both $u_1^{(1)}(z)$ and $u_0^{(1)} (z)$ unknowns initially appear, but the term proportional to $u_1^{(1)}(z)$ matches the equation obtained at previous order \eqref{eq:ODE-JT-inst-act-string-eq} and hence vanishes---leaving us with an equation for $u_0^{(1)} (z)$ alone. This (generic) pattern is another hallmark of resonance \cite{asv11}. In this way, at this (nonperturbative) order, the JT-gravity string-equation becomes a (homogeneous) first-order linear ODE for the leading one-instanton coefficient $u_0^{(1)} (z)$. Its general solution is given by:
\be
u_0^{(1)} (z) \propto \left( \sum_{k=0}^{+\infty} \left(-1\right)^k \pi^k\, \frac{k \left(k+1\right)!}{\left(2k+2\right)!}\, \frac{I_{k} \left( 2\pi \sqrt{u_0} \right)}{u_0^{k/2}} A'(z)^{2k-1} \right)^{-\frac{1}{2}}
\ee
\noindent
(this could be further rewritten by explicit evaluation of the sum, in the spirit of \eqref{eq:ODE-JT-inst-act-string-eq-integral-rep} and also involving the Dawson integral, but we do not find the final expression particularly illuminating in this context and omit it). Going to higher orders is, for the moment, out of reach---at least until we find the exact JT-gravity string-equation.

%%%%%%%%%%%%%%%%%%%%%%%%%%%%%%%%%%%%%%%%%%%%%%%%%%%%%%%%%%%%%%%%%
%%%%%%%%%%%%%%%%%%%%%%%%%%%%%%%%%%%%%%%%%%%%%%%%%%%%%%%%%%%%%%%%%
\section{Resurgent Asymptotics: Perturbative versus Nonperturbative}\label{sec:large_order}
%%%%%%%%%%%%%%%%%%%%%%%%%%%%%%%%%%%%%%%%%%%%%%%%%%%%%%%%%%%%%%%%%
%%%%%%%%%%%%%%%%%%%%%%%%%%%%%%%%%%%%%%%%%%%%%%%%%%%%%%%%%%%%%%%%%

At this stage we have computed quite some nonperturbative data, in a plethora of multicritical and minimal string models (instanton actions, Stokes data, perturbative expansions around instanton saddles). One interesting technical point in resurgence is that all these data are \textit{testable}, via large-order resurgent asymptotics (see \cite{abs18} for thorough discussions). In this section this is what we shall do: validate our nonperturbative results from a resurgent large-order analysis of the growth of perturbation theory in these models. But, in order to achieve this, one must first \textit{generate} perturbative data. As we have discussed---specially in appendix~\ref{app:top_rec}---one way to generate perturbative data is via the topological recursion; but this is a computationally slow process (hence hard to reach ``true large-order''). A computationally faster alternative is to use directly the string equations, \textit{e.g.}, as thoroughly addressed for the Painlev\'e~I and~II cases in \cite{msw07, m08, msw08, gikm10, asv11, sv13, bssv22}. But as we now move towards generic string equations, and as discussed at length in the previous section, these equations are written down (in either multicritical or minimal string setting), \textit{for each fixed order} $k$, out of the corresponding Gel'fand--Dikii KdV potential---themselves computed recursively; without a closed-form expression at arbitrary order $k$. So, we seem to have a trade-off. The topological recursion may be run at \textit{generic} order $k$, but it is computationally \textit{slow}. String equations are computationally \textit{fast}, but one may only run a \textit{fixed} order $k$ at a time. Might there be some middle ground to proceed?

One adequate algorithmic procedure was already introduced quite early-on, in \cite{gm90b}. It makes use of the nonlinear ODE for the resolvent \eqref{eq:nonlinear_ODE_resolv}---hence in some sense setting up a string-equation-like calculation at \textit{generic} order $k$. This algorithm is however quite attached to the multicritical problem. One other very efficient  algorithm was introduced by Zograf in \cite{z07, z08} (see \cite{os19} as well), which exploits various properties of the free energy $F\left(\left\{t_k\right\}\right)$ of general 2d topological gravity \cite{dw90, w91, k92, iz92, dw18}---\textit{i.e.}, the 2d gravity model whose KdV times are all turned-on and left unspecified. While originally devoted to the computation of Weil--Petersson volumes---which are obtained via the same choice of KdV times which yields JT gravity---it can be easily generalized to any \textit{arbitrary} choice of KdV times as we shall show in the following. This allows us to address both multicritical and minimal-string models with the exact same procedure (and even eventually address any 2d topological gravity\footnote{Likely even alongside their multi-boundary correlation functions, similarly to what was done in \cite{os20a}.} in future work).

Once a good number of perturbative data is generated through the algorithm, there comes the question of using it to numerically test the nonperturbative data we have computed analytically in the previous sections. We already briefly mentioned resurgent large-order relations in subsection~\ref{subsec:multi_string_eq}, and we gave a na\"\i ve expression for the leading large-order growth of the perturbative coefficients for the specific heat, in \eqref{eq:multi-large-order-naive}. As we now make such resurgent large-order relations precise, the models we are considering result in slightly more complicated large-order formulae than the na\"\i ve \eqref{eq:multi-large-order-naive}, simply by the fact that there are multiple instanton actions with the same absolute value and, therefore, multiple singularities on the complex Borel plane which are equally distant from the origin (hence with equal weight). As already illustrated in subsection~\ref{subsec:multi_string_eq}, $(2,2k-1)$ multicritical models with $k\geq 3$ feature four symmetric and complex conjugate instanton actions, all of which equally contributing to the leading large-order growth of the perturbative coefficients of the free energy, which then take the form\footnote{We start herein following the notation in \cite{msw07}.} (see, \textit{e.g.}, \cite{msw07, ps09, ars14} for other examples of similar asymptotics)
\be
\label{eq:multi-large-order-F}
F^{(0)}_{g \gg 1} \simeq \frac{\Gamma \left( 2g-\beta_F \right)}{\abs{A}^{2g-\beta_F}}\, \mu^{(1)}\, \cos \left( \left(2g-\beta_F\right) \theta_A + \theta_1 \right) \left.\Big\{ 1+\mathcal{O} \left(g^{-1}\right) \right.\Big\}.
\ee
\noindent
Herein, $A = \abs{A}\, \rme^{\rmi\theta_A}$ is the instanton action in the first quadrant of the complex plane, and $\beta_F$ the characteristic exponent\footnote{Precisely, $\beta_F$ includes both our previous $\beta$ and one-instanton free-energy perturbative starting-order \cite{msw07, asv11}.} of the free energy. Also, following the notation of \cite{msw07}, we have combined the Stokes coefficient $S_1$ and the one-loop contribution to the one-instanton sector $F_1^{(1)}$ into
\be
4\,\frac{S_1\cdot F_1^{(1)}}{2\pi\rmi} \equiv \mu^{(1)}\,\rme^{\rmi\theta_1}.
\ee
\noindent
The factor of $4$ is associated to the four instanton actions which contribute with same weight to the leading large-order growth. Things are slightly less complicated in the case of $(2,2k-1)$ minimal strings (and the $(2,3)$ model which is common to multicritical and minimal models). There, as we have shown, all instanton actions are real and come in pairs of opposite sign. Hence, there are only two closest singularities to the origin of the complex Borel plane, located on the real line (hence these models are in the class of examples in \cite{gikm10, asv11, sv13, bssv22}). The leading contribution to the large-order growth of perturbative coefficients now takes the simpler form
\be
\label{eq:minimal-large-order-F}
F^{(0)}_{g \gg 1} \simeq \frac{ S_1}{\rmi\pi}\, \frac{\Gamma \left( 2g-\beta_F \right)}{A^{2g-\beta_F}}\, F_1^{(1)} \left.\Big\{ 1+\mathcal{O} \left(g^{-1}\right) \right.\Big\}, 
\ee
\noindent
where the factor of $2$ with respect to the na\"\i ve expression \eqref{eq:multi-large-order-naive} comes from having to take into account two instanton actions. In this simpler setting, it is easy to construct sequences which converge to the nonperturbative data we would like to test, at large $g$. For example, the instanton action $A$ can be tested by considering \cite{msw07}
\be
\label{eq:Qg}
Q_g=\frac{F^{(0)}_{g+1}}{4g^{2}F^{(0)}_{g}}=\frac{1}{A^2}+\mathcal{O}\left(g^{-1}\right).
\ee
\noindent
Similarly, one can test the characteristic exponent $\beta_F$ and the first two contributions to the one-instanton sector (Stokes coefficient included) using, respectively, \cite{msw07}
\bea
\label{eq:Bg}
B_g &=& \frac{1}{2}-g\left(A^2\frac{F^{(0)}_{g+1}}{4g^{2}F^{(0)}_{g}}-1\right) = \beta_F + \mathcal{O} \left(g^{-1}\right), \\
\label{eq:M1g}
M_{1,g} &=& \frac{A^{2g-\beta_F}F^{(0)}_{g}}{\Gamma(2g-\beta_F)} = \frac{S_1\cdot F_1^{(1)}}{\rmi\pi} + \mathcal{O}\left(g^{-1}\right), \\
\label{eq:M2g}
M_{2,g} &=& \frac{2g}{A}\left(\frac{\rmi\pi\, A^{2g-\beta_F}\, F_g^{(0)}}{S_1\cdot F_1^{(1)}\, \Gamma(2g-\beta_F)}-1\right) = \frac{F_2^{(1)}}{F_1^{(1)}} + \mathcal{O}\left(g^{-1}\right).
\eea
\noindent
On top of this, the speed of convergence of the above sequences can be dramatically improved through the use of Richardson transforms (see \cite{msw07, ps09, ars14, abs18} for an introduction and applications in very similar contexts). On the other hand, unfortunately, in the case of $k>2$ multicritical models, the oscillating nature of the large-order relation \eqref{eq:multi-large-order-F} does not allow for the construction of the above converging sequences \eqref{eq:Qg}-\eqref{eq:Bg}-\eqref{eq:M1g}-\eqref{eq:M2g}, nor for the use of Richardson transforms. Nevertheless, some tests of nonperturbative data can still be carried out. For example, the phase of the instanton action $\theta_A$ can be obtained through the following converging sequence
\bea
\Theta_g &=& \frac{\abs{A}^{2}}{\left(2g-\beta_F+1\right) \left(2g-\beta_F\right)}\, \frac{F^{(0)}_{g+1}}{F^{(0)}_g} + \frac{\left(2g-\beta_F-1\right) \left(2g-\beta_F-2\right)}{\abs{A}^2}\, \frac{F^{(0)}_{g-1}}{F^{(0)}_g} = \nonumber \\
\label{eq:Thetag}
&=& 2 \cos \left(2\theta_A\right) + \mathcal{O}\left(g^{-1}\right). 
\eea

%%%%%%%%%%%%%%%%%%%%%%%%%%%%%%%%%%%%%%%%%%%%%%%%%%%%%%%%%%%%%%%%%
\subsection{Multicritical Large-Order: Instanton Actions and Stokes Data}\label{subsec:multi_large_order}
%%%%%%%%%%%%%%%%%%%%%%%%%%%%%%%%%%%%%%%%%%%%%%%%%%%%%%%%%%%%%%%%%

Let us now see how the Zograf algorithm may be adapted to the computation of perturbative free energies for $(2,2k-1)$ multicritical models. We will start by reviewing the algorithm and then show how it may be enlarged in the sense of being applicable to any, arbitrary choice of KdV times. Running the risk of notational cluttering, but in order to be as consistent as possible with the existing literature, in this section we shall also introduce a new set of KdV times $\left\{ \tau_k \right\}$ simply related to the ``monomial basis'' KdV times $\left\{ \widetilde{t}_k \right\}$ we introduced back in subsection~\ref{subsec:general-KdV},
\be
\label{eq:zograf_times}
\tau_k = \left(-1\right)^k k!\, 2^{k-1}\, \widetilde{t}_k, \qquad k \geq 2,
\ee
\noindent
and
\be
\tau_0=\frac{1}{2} \widetilde{t}_0, \qquad \tau_1 = 1-\widetilde{t}_1.
\ee
\noindent
The perturbative expansion for the free energy of general 2d topological gravity is then
\be
F \left( \{\tau_k\} \right) \simeq \sum_{g=0}^{+\infty} F_g \left( \{\tau_k\} \right) g_{\text{s}}^{2g-2}.
\ee

A crucial step on which the algorithm heavily relies is the fact that we can replace these KdV times with a new set of variables $I_k$ defined as \cite{iz92}
\be
\label{eq:I_k-def}
I_k \left(\mathfrak{u}_0,\{\tau_k\}\right) := \sum_{n=0}^{+\infty} \tau_{n+k}\, \frac{\mathfrak{u}_0^n}{n!},
\ee
\noindent
with $\mathfrak{u}_0$ being related to the ``monomial basis'' genus-zero specific-heat through $-2\mathfrak{u}_0=\widetilde{u}_0$. It immediately follows that combining \eqref{eq:zograf_times} and \eqref{eq:I_k-def} with the genus-zero string-equation \eqref{eq:mon_base_string_eq} yields
\be
I_0=\widetilde{u}_0
\ee
\noindent
(where we also set $\widetilde{z}=-\widetilde{t}_0$). In these variables, it turns out that the perturbative coefficients of the free energy with $g\geq 2$ become polynomials in $\left( 1-I_1 \right)^{-1}$ and all the other $I_k$ with $k\geq 2$ \cite{iz92} (see also \cite{os19}). A second fundamental property of the free energy of 2d topological gravity, that the algorithm is based upon, is the fact that its specific heat $u = g_{\text{s}}^2\, \frac{\partial^2F}{\partial \tau_0^2}$ satisfies the KdV equation \cite{gd75} (recall \eqref{eq:KdV})
\be
\label{eq:KdV-equation}
\frac{\partial u}{\partial \tau_1} = \frac{\partial}{\partial \tau_0} \left(\frac{1}{2} u^2 + \frac{1}{12}\, g_{\text{s}}^2\, \frac{\partial^2 u}{\partial\tau_0^2}\right).
\ee
\noindent
From now on, we shall refer to derivatives with respect to $\tau_0$ and $\tau_1$ as $\partial_0$ and $\partial_1$, respectively. Next, in order to use these properties to recursively (and efficiently!) generate the perturbative free energies of some specified 2d topological gravity model, say corresponding to some choice of KdV times $\tau^\star_{k\geq 2}$, introduce the pair of variables $(y,t)$ defined as:
\be
y := \mathfrak{u}_0 \left( \left\{ \tau_0, \tau_1, \tau^\star_{k\geq 2} \right\} \right)
\ee
\noindent
and
\be
\label{eq:zog-t-def}
t := 1 - I_1 \left( y, \left\{ \tau_0, \tau_1, \tau^\star_{k\geq 2} \right\} \right).
\ee
\noindent
It is also convenient to introduce the function $f_2$, in parallel with \eqref{eq:I_k-def} and defined as
\be
\label{eq:f_2-def}
f_2 (y) := \sum_{n=0}^{+\infty} \tau_{n+2}\, \frac{y^n}{n!}.
\ee

%%%%%%%%%%%%%%%%%%%%%%%%%%%%%%%%%%%%%%%%%%%%%%%%%%%%%%%%%%%%%%%%%
\begin{figure}[t!]
\centering
     \begin{subfigure}[h]{0.49\textwidth}
         \centering
         \includegraphics[width=\textwidth]{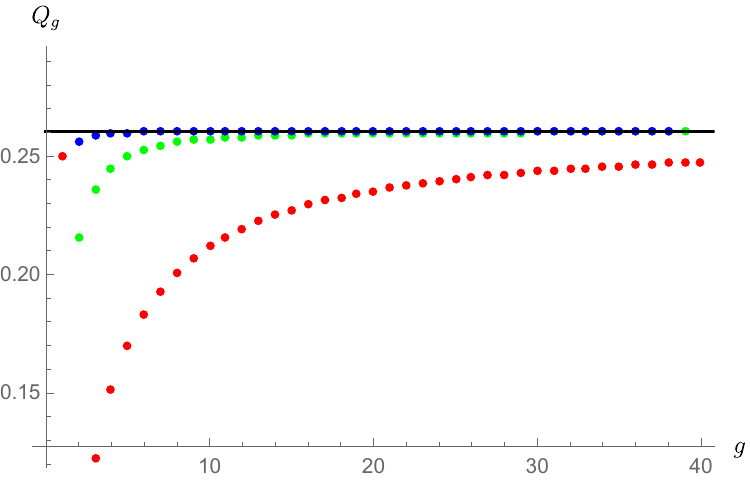}
     \end{subfigure}
\hspace{0mm}
     \begin{subfigure}[h]{0.49\textwidth}
         \centering
         \includegraphics[width=\textwidth]{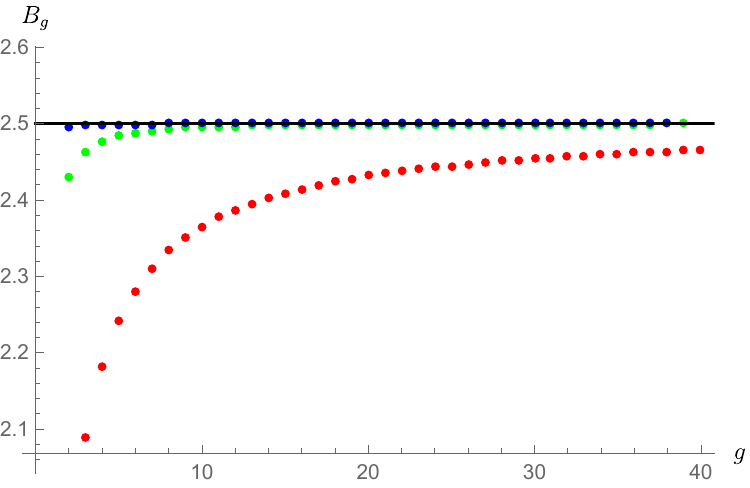}
     \end{subfigure}
\caption{These images plot the sequences $Q_g$ in \eqref{eq:Qg} (left) and $B_g$ in \eqref{eq:Bg} (right) for the $(2,3)$ multicritical model. In red we plot the original sequences, whereas green and blue plot the first and second Richardson transforms, respectively. The black lines correspond to the predicted, analytical values. In both cases, the relative error at highest precision is of the order $\sim 10^{-6}$.}
\label{fig:pure-gravity-1}
\end{figure}
%%%%%%%%%%%%%%%%%%%%%%%%%%%%%%%%%%%%%%%%%%%%%%%%%%%%%%%%%%%%%%%%%

%%%%%%%%%%%%%%%%%%%%%%%%%%%%%%%%%%%%%%%%%%%%%%%%%%%%%%%%%%%%%%%%%
\begin{figure}[t!]
\centering
     \begin{subfigure}[h]{0.49\textwidth}
         \centering
         \includegraphics[width=\textwidth]{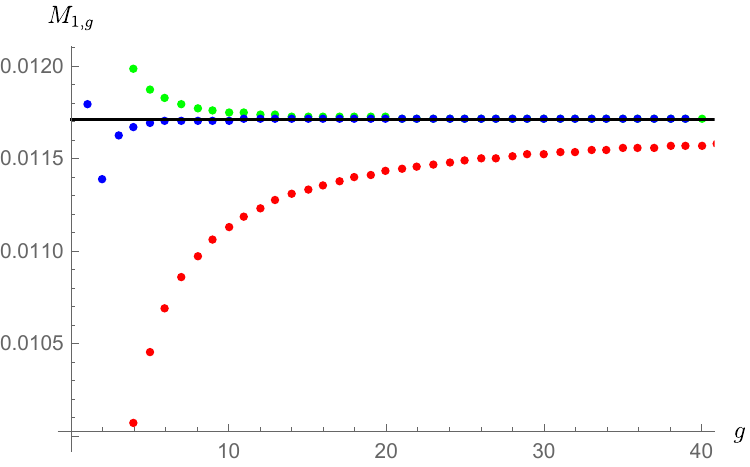}
     \end{subfigure}
\hspace{0mm}
     \begin{subfigure}[h]{0.49\textwidth}
         \centering
         \includegraphics[width=\textwidth]{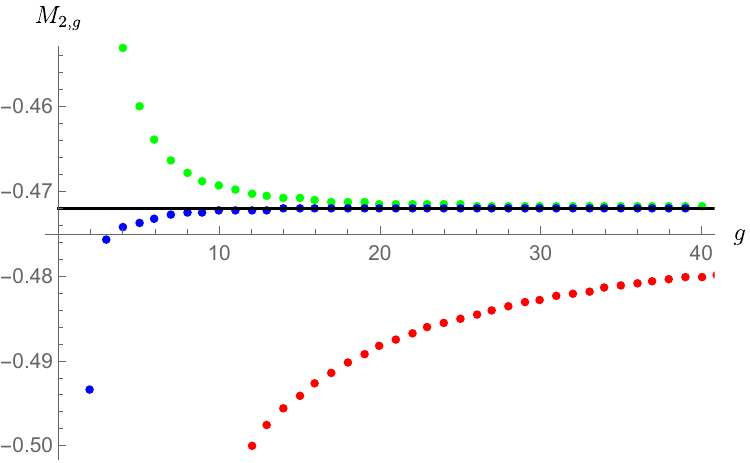}
     \end{subfigure}
\caption{These images plot the sequences $M_{1,g}$ in \eqref{eq:M1g} (left) and $M_{2,g}$ in \eqref{eq:M2g} (right) for the $(2,3)$ multicritical model. Red plots the original sequences, whereas green and blue plot the first and second Richardson transforms, respectively. The black lines correspond to the predicted, analytical values. In both cases, the relative error at highest precision is of the order $\sim 10^{-6}$.}
\label{fig:pure-gravity-2}
\end{figure}
%%%%%%%%%%%%%%%%%%%%%%%%%%%%%%%%%%%%%%%%%%%%%%%%%%%%%%%%%%%%%%%%%

In the above set-up, we are now essentially working in a configuration on which all KdV times---except $\tau_0$ and $\tau_1$---are set to the specific values corresponding to the model of interest, and on which we replaced the two remaining ``free'' KdV times, $\tau_0$ and $\tau_1$, by the new variables $y$ and $t$. The partial derivatives with respect to $\tau_0$ and $\tau_1$ appearing in \eqref{eq:KdV-equation} are easily written in terms of the new variables \cite{z07,z08},
\be
\label{eq:zograf-derivatives}
\partial_0 = \frac{1}{t} \left(\frac{\partial}{\partial y} - f_2(y)\, \frac{\partial}{\partial t}\right), \qquad \partial_1 = - \frac{\partial}{\partial t} + y\, \partial_0.
\ee
\noindent
For convenience and consistency with the literature, let us further denote by $\uppsi(y,t)$ the free energy in which we have fixed all KdV times (as above) except $\tau_0$ and $\tau_1$, and in which we explicitly emphasize the dependence in the new variables $y$ and $t$. Clearly, it also admits the usual perturbative expansion
\be
\uppsi(y,t) \simeq \sum_{g=0}^{+\infty} \uppsi_g (y,t)\, g_{\text{s}}^{2g-2},
\ee
\noindent
as does its specific heat\footnote{On the other hand, because it is always clear from context, we keep the specific-heat notation fixed throughout.} $u = g_{\text{s}}^2\, \partial_0^2 \uppsi(y,t)$ \cite{z07,z08}
\be
u (y,t) \simeq y + \sum_{g=1}^{+\infty} \partial_0^2\uppsi_g (y,t)\, g_{\text{s}}^{2g-2}.
\ee
\noindent
This expansion can now finally be plugged into the KdV equation \eqref{eq:KdV-equation} which, after a couple of straightforward manipulations, produces a recursion for the $\uppsi_g$,
\be
\label{eq:psi-recursion}
- \partial_t \partial_0 \uppsi_g = \frac{1}{2} \sum_{j=1}^{g-1} \partial_0^2 \uppsi_j\, \partial_0^2 \uppsi_{g-j} + \frac{1}{12} \partial_0^4 \uppsi_{g-1}, \qquad g\geq 2.
\ee
\noindent
The recursion itself involves only derivatives of the $\uppsi_g (y,t)$, hence can be easily integrated to directly yield these coefficients (by using the aforementioned properties of the $F_{g} \left( \{ \tau_k \} \right)$). Indeed, it turns out that as a result of \eqref{eq:zog-t-def} and the fact that the $F_{g} \left( \{\tau_k\} \right)$ are polynomials in $\left( 1-I_1 \right)^{-1}$, the $n$-th derivative of $\uppsi_g(y,t)$ with respect to $\tau_0$ may be written as a Laurent polynomial in $t$ of the form \cite{z07,z08}
\be
\label{eq:laurent}
\partial_0^n \uppsi_g (y,t) = \sum_{k=2g+n-1}^{5g+2n-5} \frac{f_{g,n,k}(y)}{t^{k}}.
\ee
\noindent
Using the initial condition
\be
\label{eq:zograf-initial}
\uppsi_1 (y,t) = -\frac{1}{24} \log t \quad \Rightarrow \quad \partial_0 \uppsi_1 (y,t) = \frac{1}{24}\, \frac{f_2(y)}{t^{2}},
\ee
\noindent
one can then recursively determine all the coefficients $f_{g,0,k}(y)$ by simply plugging the Laurent polynomial \eqref{eq:laurent} into the recursion \eqref{eq:psi-recursion}. The final step of the algorithm is setting the two remaining ``free'' KdV times $\tau_0$ and $\tau_1$ to the values corresponding to the model of interest---hence obtaining the perturbative free-energy coefficients we set out to compute. In the strict context of the algorithm, it is always convenient to set them both to zero; corresponding to the choice $(y,t)=(0,1)$. At the level of the ``monomial basis'' KdV times, this corresponds to
\be
\widetilde{t}_0=0, \qquad \widetilde{t}_1=1.
\ee
\noindent
Note that while setting $\widetilde{t}_0=0$ is required in all models we are interested in, the same cannot be said for the choice $\widetilde{t}_1=1$. However, this caveat may be easily solved by considering the equivalent model in which we perform the rescalings
\be
\label{eq:zograf-rescaling}
\widetilde{t}_k \rightarrow \frac{\widetilde{t}_k}{\widetilde{t}_1}, \qquad  g_{\text{s}} \rightarrow \widetilde{t}_1\, g_{\text{s}}.
\ee

%%%%%%%%%%%%%%%%%%%%%%%%%%%%%%%%%%%%%%%%%%%%%%%%%%%%%%%%%%%%%%%%%
\begin{figure}[t!]
\centering
     \begin{subfigure}[h]{0.49\textwidth}
         \centering
         \includegraphics[width=\textwidth]{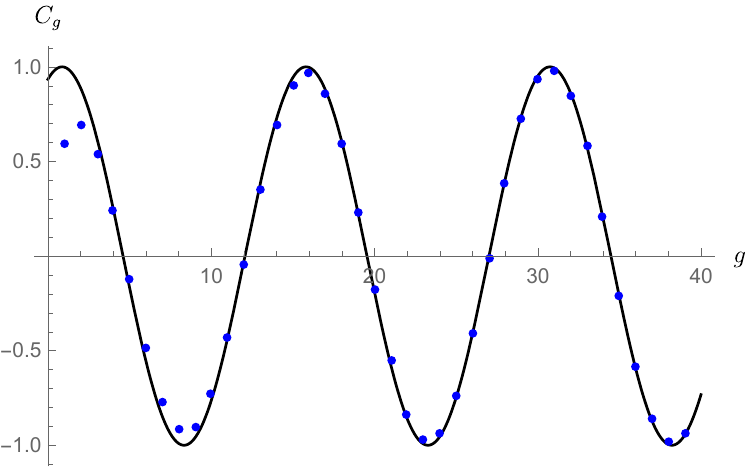}
     \end{subfigure}
\hspace{0mm}
     \begin{subfigure}[h]{0.49\textwidth}
         \centering
         \includegraphics[width=\textwidth]{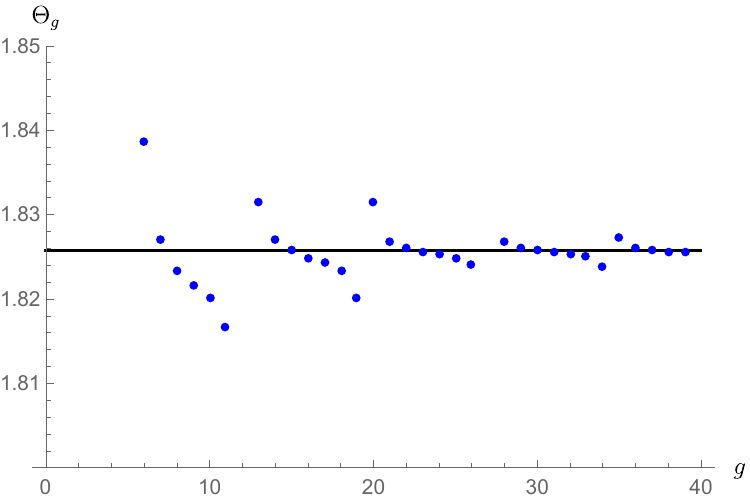}
     \end{subfigure}
\caption{These images plot the sequences $C_{g}$ in \eqref{eq:Cg} (left) and $\Theta_{g}$ in \eqref{eq:Thetag} (right) for the $(2,5)$ multicritical model. On the left plot, the black line corresponds to the function $\cos \left( \left(2g-\beta_F\right) \theta_A + \theta_1 \right)$, whereas on the right plot it corresponds to the predicted value $2 \cos \left( 2\theta_1 \right)$. Both plots show excellent agreement at large order, with the relative error at highest precision on the left being of order $\sim 10^{-2}$, and that on the right being of the order $\sim 10^{-4}$.
}
\label{fig:(2,5)nonpert}
\end{figure}
%%%%%%%%%%%%%%%%%%%%%%%%%%%%%%%%%%%%%%%%%%%%%%%%%%%%%%%%%%%%%%%%%

%%%%%%%%%%%%%%%%%%%%%%%%%%%%%%%%%%%%%%%%%%%%%%%%%%%%%%%%%%%%%%%%%
\begin{figure}[t!]
\centering
     \begin{subfigure}[h]{0.49\textwidth}
         \centering
         \includegraphics[width=\textwidth]{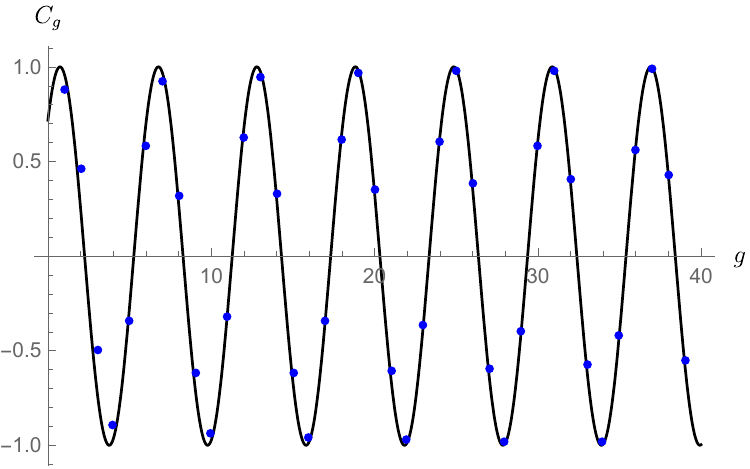}
     \end{subfigure}
\hspace{0mm}
     \begin{subfigure}[h]{0.49\textwidth}
         \centering
         \includegraphics[width=\textwidth]{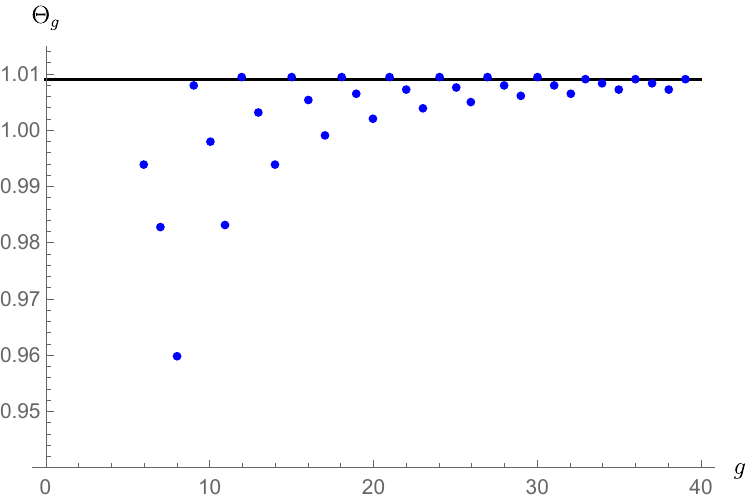}
     \end{subfigure}
\caption{These images plot the sequences $C_{g}$ in \eqref{eq:Cg} (left) and $\Theta_{g}$ in \eqref{eq:Thetag} (right) for the $(2,19)$ multicritical model. On the left plot, the black line corresponds to the function $\cos \left( \left(2g-\beta_F\right) \theta_A + \theta_1 \right)$, whereas on the right plot it corresponds to the predicted value $2 \cos \left( 2\theta_1 \right)$. Both plots show excellent agreement at large order, with the relative error at highest precision on the left being of order $\sim 10^{-2}$, and that on the right being of the order $\sim 10^{-4}$.
}
\label{fig:(2,19)nonpert}
\end{figure}
%%%%%%%%%%%%%%%%%%%%%%%%%%%%%%%%%%%%%%%%%%%%%%%%%%%%%%%%%%%%%%%%%

%%%%%%%%%%%%%%%%%%%%%%%%%%%%%%%%%%%%%%%%%%%%%%%%%%%%%%%%%%%%%%%%%
\begin{figure}[t!]
\centering
     \begin{subfigure}[h]{0.49\textwidth}
         \centering
         \includegraphics[width=\textwidth]{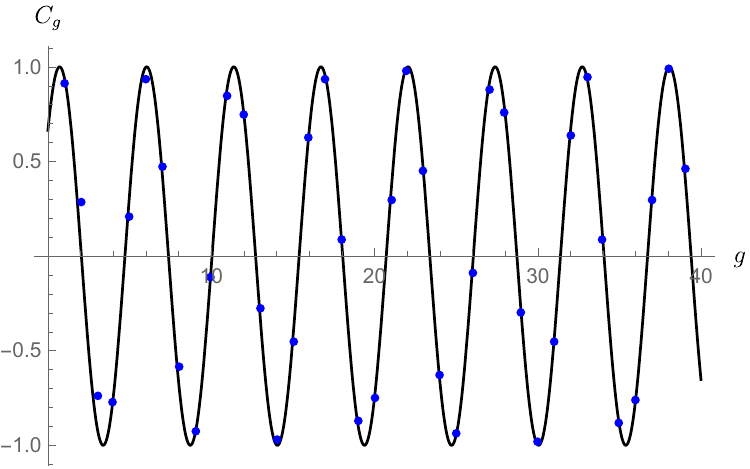}
     \end{subfigure}
\hspace{0mm}
     \begin{subfigure}[h]{0.49\textwidth}
         \centering
         \includegraphics[width=\textwidth]{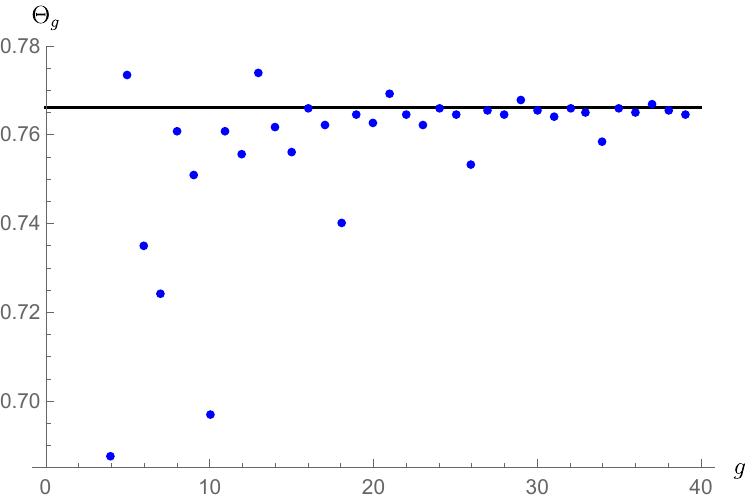}
     \end{subfigure}
\caption{These images plot the sequences $C_{g}$ in \eqref{eq:Cg} (left) and $\Theta_{g}$ in \eqref{eq:Thetag} (right) for the $(2,39)$ multicritical model. On the left plot, the black line corresponds to the function $\cos \left( \left(2g-\beta_F\right) \theta_A + \theta_1 \right)$, whereas on the right plot it corresponds to the predicted value $2 \cos \left( 2\theta_1 \right)$. Both plots show excellent agreement at large order, with the relative error at highest precision on the left being of order $\sim 10^{-2}$, and that on the right being of the order $\sim 10^{-3}$.
}
\label{fig:(2,39)nonpert}
\end{figure}
%%%%%%%%%%%%%%%%%%%%%%%%%%%%%%%%%%%%%%%%%%%%%%%%%%%%%%%%%%%%%%%%%

Let us illustrate how the algorithm works by explicitly computing $F_2$ in the general multicritical model. The ``monomial basis'' KdV times for the $(2,2k-1)$ multicritical model are given by
\be
\label{eq:multicrit-times}
\widetilde{t}_p (k) = 2^{-p}\, \sqrt{\pi}\, \binom{k}{p}\, \frac{(2p-1)!!}{\Gamma\left(p+\frac{1}{2}\right)}.
\ee
\noindent
Using definitions \eqref{eq:zograf_times} and \eqref{eq:f_2-def}, and performing the necessary rescaling \eqref{eq:zograf-rescaling}, we find
\be
f_2(y) = \sum_{p=2}^{k} \left(-1\right)^{p} p \left(p-1\right) \sqrt{\pi}\, \binom{k}{p}\, \frac{(2p-1)!!}{2k\,\Gamma\left(p+\frac{1}{2}\right)}\, y^{p-2},
\ee
\noindent
which is the only model-dependent input required by the algorithm. For $g=2$, the recursion \eqref{eq:psi-recursion} takes the simple form
\be
\label{eq:g=2-recursion}
- \partial_t \partial_0 \uppsi_2 = \frac{1}{2} \partial_0^2 \uppsi_1\, \partial_0^2 \uppsi_1 + \frac{1}{12} \partial_0^4 \uppsi_1.
\ee
\noindent
Applying the partial derivative with respect to $\tau_0$, as written in \eqref{eq:zograf-derivatives}, to the initial condition \eqref{eq:zograf-initial}, we obtain the following expressions for the quantities appearing on the right-hand-side of \eqref{eq:g=2-recursion}: 
\bea
\partial_0^2 \uppsi_1(y,t) &=& \frac{1}{12} f_2(y)^2\, t^{-4}, \\
\partial_0^4 \uppsi_1(y,t) &=& 2 f_2(y)^4\, t^{-8} + \frac{59}{24} f_2(y)^2\, f_2'(y)\, t^{-7} + \nonumber \\
&&
+ \frac{1}{24} \left( 7 f_2'(y)^2 + 11 f_2(y)\, f_2''(y) \right) t^{-6} + \frac{1}{24} f_2'''(y)\, t^{-5}.
\eea
\noindent
The integration with respect to $t$ is trivial and yields the following coefficients for the Laurent polynomial \eqref{eq:laurent} associated to $\partial_0 \uppsi_2(y,t)$:
\bea
f_{2,1,7}(y) &=& \frac{7}{288} f_2(y)^4,\\
f_{2,1,6}(y) &=& \frac{5}{144} f_2(y)^2\, f_2'(y),\\
f_{2,1,5}(y) &=& \frac{29}{5760} f_2'(y)^2 + \frac{11}{1440} f_2(y)\, f_2''(y),\\
f_{2,1,4}(y) &=& \frac{1}{1152} f_2'''(y).
\eea
\noindent
One finally just needs to use the above results in order to obtain the $f_{2,0,j}(y)$ ($j=3,4,5$) coefficients of $\uppsi_2(y,t)$; hence to finally determine the genus-two free energy. This is done by making use of the following equation \cite{z07, z08}
\be
\partial_0 \uppsi_2(y,t) = \frac{1}{t} \left( \frac{\partial}{\partial y} - f_2(y)\, \frac{\partial}{\partial t} \right) \sum_{j=3}^5 f_{2,0,j}(y)\, t^{-j} = \sum_{j=4}^7 f_{2,1,j}(y)\, t^{-j},
\ee
\noindent
where we made use of the Laurent expansions \eqref{eq:laurent} for both $\uppsi_2(y,t)$ and $\partial_0 \uppsi_2(y,t)$. Directly comparing corresponding powers of $t$ yields
\bea
f_{2,0,5}(y) &=& \frac{7}{1440} f_2(y)^3,\\
f_{2,0,4}(y) &=& \frac{29}{5760} f_2(y)\, f_2'(y),\\
f_{2,0,3}(y) &=& \frac{1}{1152} f_2''(y).
\eea
\noindent
Setting $t=1$ and $y=0$ we finally obtain the genus-two free energy:
\be
\label{eq:g=2-zograf}
F_2 = \frac{1}{\left( 2\widetilde{t}_1\right)^2} \times \frac{56\, \widetilde{t}_2^3 - 87\,\widetilde{t}_1\, \widetilde{t}_2\, \widetilde{t}_3 + 30\, \widetilde{t}_1^2\, \widetilde{t}_4}{180\, \widetilde{t}_1^3},
\ee
\noindent
where we have used the definition of $f_2(y)$, \eqref{eq:f_2-def}, and recast everything in terms of the ``monomial basis'' KdV times using \eqref{eq:zograf_times}. Note that the prefactor in \eqref{eq:g=2-zograf} comes from a different choice of conventions in our definitions of free energy and of $g_{\text{s}}$ with respect to \cite{z07,z08}, and from the rescaling \eqref{eq:zograf-rescaling}. Finishing to specify the model we are interested in (\textit{i.e.}, the $(2,2k-1)$ multicritical model with KdV times fixed as \eqref{eq:multicrit-times}) yields at last
\be
F_2 (k) = \frac{1}{1440 k} \left(2k+3\right) \left(k-1\right).
\ee
\noindent
This is in perfect agreement with the result from topological recursion, \eqref{eq:multicrit_free_en-2}.

We ran our algorithm up to\footnote{Let us note that the algorithm is efficient enough to be easily run past $g=40$ on any laptop. For the purposes of this present paper, our results are already extremely sharp at genus $40$; hence enough.} genus $g=40$, for different values of multicritical order $k$ going from 1 to 20; in order to test the many nonperturbative data we computed in the previous sections. In the $k=2$ case, the simpler form \eqref{eq:minimal-large-order-F} of the large-order resurgent asymptotics allows for the construction of the sequences \eqref{eq:Qg}-\eqref{eq:Bg}-\eqref{eq:M1g}-\eqref{eq:M2g}, and therefore for high-precision tests of the instanton action $A$, of the characteristic exponent $\beta_F$, and of the first two contributions to the one-instanton sector (Stokes data included), all of which are plotted in figures~\ref{fig:pure-gravity-1} and~\ref{fig:pure-gravity-2} (see as well \cite{msw07, msw08, gikm10, asv11, bssv22}). For $k>2$ oscillations kick-in. We then tested the phase $\theta_A$ of the instanton action via \eqref{eq:Thetag}, and plotted it for $k=3$, $k=10$, and $k=20$, in figures~\ref{fig:(2,5)nonpert},~\ref{fig:(2,19)nonpert}, and~\ref{fig:(2,39)nonpert}; alongside the sequence
\be
\label{eq:Cg}
C_g = F_{g}^{(0)}\frac{\abs{A}^{2g-\beta_F}}{\mu^{(1)}\Gamma(2g-\beta_F)} = \cos \left( \left(2g-\beta_F\right) \theta_A + \theta_1 \right) \left.\Big\{ 1+\mathcal{O} \left(g^{-1}\right) \right.\Big\},
\ee
\noindent
which works as a simultaneous test of instanton action, characteristic exponent, and first loop around the one-instanton. All numerics support our analytics to extreme high precision.

%%%%%%%%%%%%%%%%%%%%%%%%%%%%%%%%%%%%%%%%%%%%%%%%%%%%%%%%%%%%%%%%%
\subsection{Minimal String Large-Order: Instanton Actions and Stokes Data}\label{subsec:minimal_large_order}
%%%%%%%%%%%%%%%%%%%%%%%%%%%%%%%%%%%%%%%%%%%%%%%%%%%%%%%%%%%%%%%%%

%%%%%%%%%%%%%%%%%%%%%%%%%%%%%%%%%%%%%%%%%%%%%%%%%%%%%%%%%%%%%%%%%
\begin{figure}[t!]
\centering
     \begin{subfigure}[h]{0.49\textwidth}
         \centering
         \includegraphics[width=\textwidth]{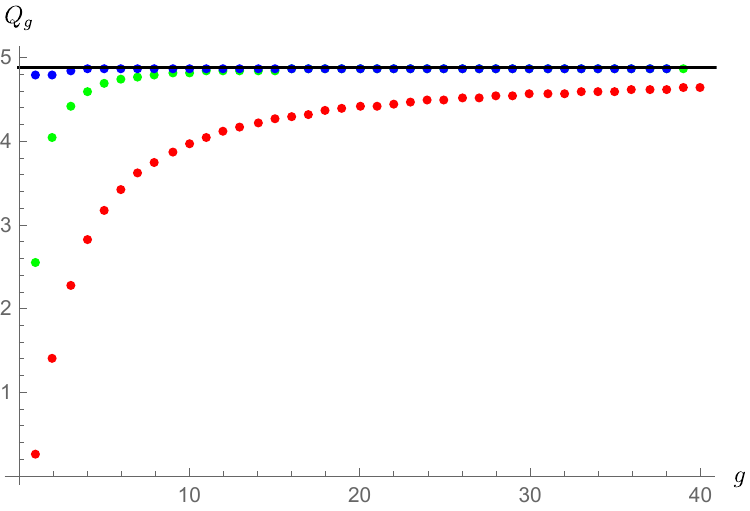}
     \end{subfigure}
\hspace{0mm}
     \begin{subfigure}[h]{0.49\textwidth}
         \centering
         \includegraphics[width=\textwidth]{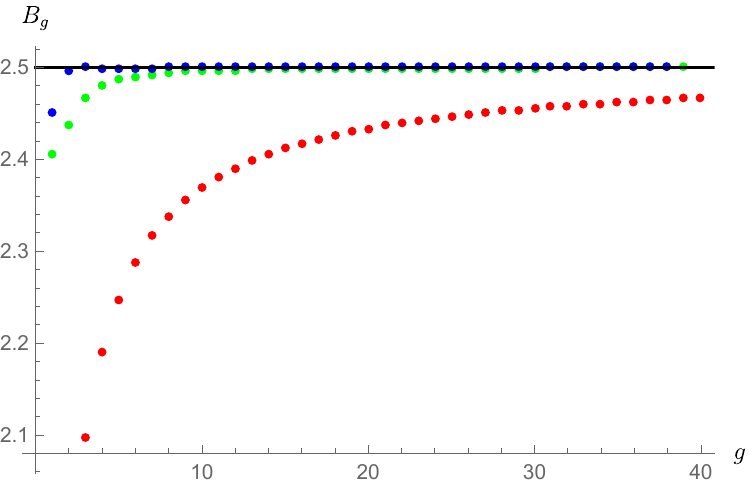}
     \end{subfigure}
\caption{These images plot the sequences $Q_g$ in \eqref{eq:Qg} (left) and $B_g$ in \eqref{eq:Bg} (right) for the $(2,5)$ minimal string. In red we plot the original sequences, whereas green and blue plot first and second Richardson transforms, respectively. The black lines correspond to the predicted, analytical values. In both cases the relative error at highest precision is of the order $\sim 10^{-6}$.}
\label{fig:(2,5)MS-1}
\end{figure}
%%%%%%%%%%%%%%%%%%%%%%%%%%%%%%%%%%%%%%%%%%%%%%%%%%%%%%%%%%%%%%%%%

%%%%%%%%%%%%%%%%%%%%%%%%%%%%%%%%%%%%%%%%%%%%%%%%%%%%%%%%%%%%%%%%%
\begin{figure}[t!]
\centering
     \begin{subfigure}[h]{0.49\textwidth}
         \centering
         \includegraphics[width=\textwidth]{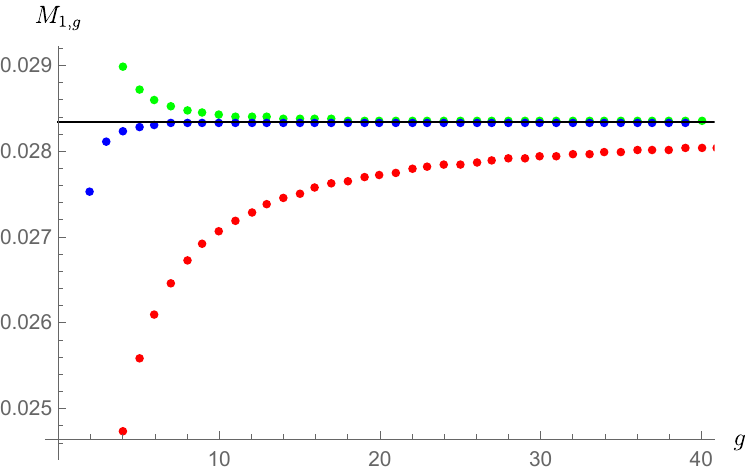}
     \end{subfigure}
\hspace{0mm}
     \begin{subfigure}[h]{0.49\textwidth}
         \centering
         \includegraphics[width=\textwidth]{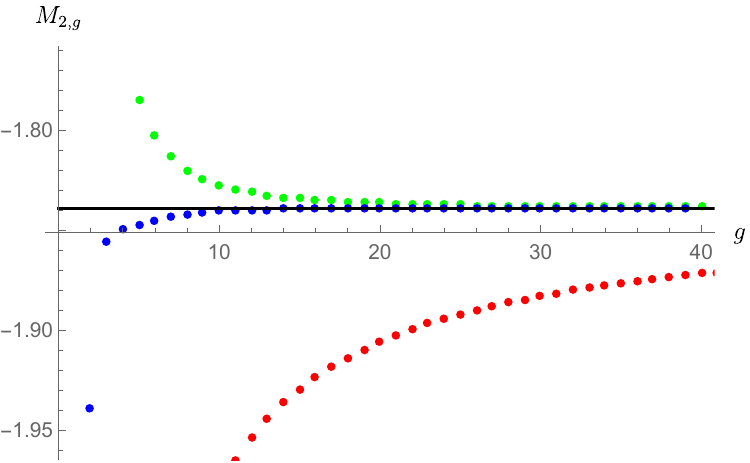}
     \end{subfigure}
\caption{These images plot the sequences $M_{1,g}$ in \eqref{eq:M1g} (left) and $M_{2,g}$ in \eqref{eq:M2g} (right) for the $(2,5)$ minimal string. Red plots the original sequences, whereas green and blue plot first and second Richardson transforms, respectively. The black lines correspond to the predicted, analytical values. Last-genus relative errors are of orders $\sim 10^{-6}$ (left) and $\sim 10^{-5}$ (right).}
\label{fig:(2,5)MS-2}
\end{figure}
%%%%%%%%%%%%%%%%%%%%%%%%%%%%%%%%%%%%%%%%%%%%%%%%%%%%%%%%%%%%%%%%%

%%%%%%%%%%%%%%%%%%%%%%%%%%%%%%%%%%%%%%%%%%%%%%%%%%%%%%%%%%%%%%%%%
\begin{figure}[t!]
\centering
     \begin{subfigure}[h]{0.49\textwidth}
         \centering
         \includegraphics[width=\textwidth]{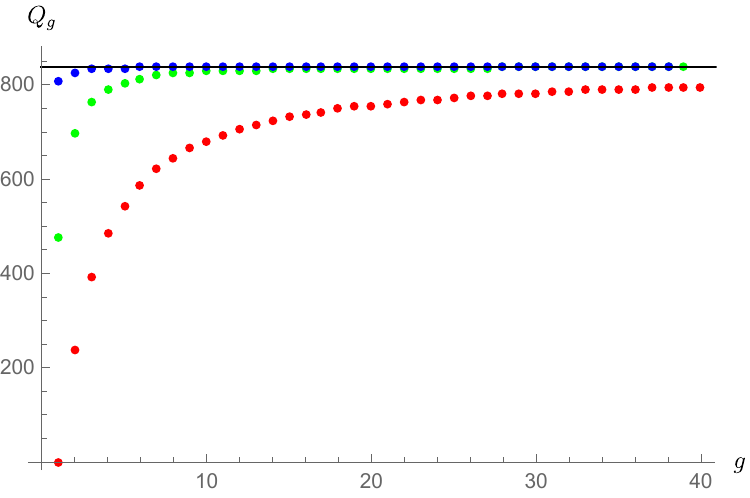}
     \end{subfigure}
\hspace{0mm}
     \begin{subfigure}[h]{0.49\textwidth}
         \centering
         \includegraphics[width=\textwidth]{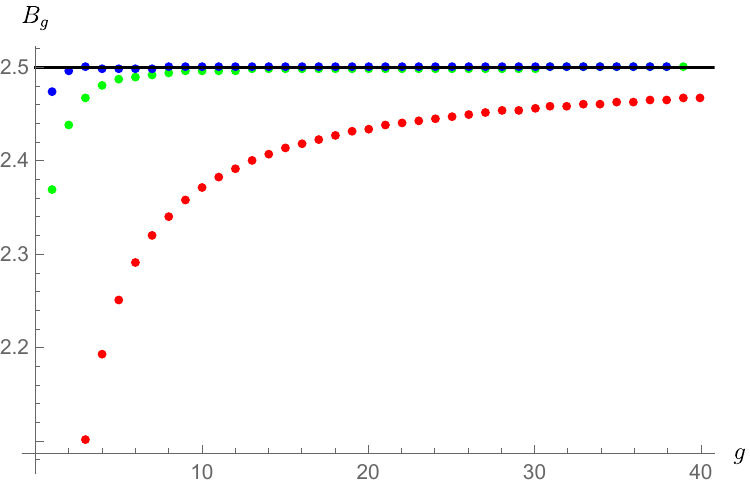}
     \end{subfigure}
\caption{These images plot the sequences $Q_g$ in \eqref{eq:Qg} (left) and $B_g$ in \eqref{eq:Bg} (right) for the $(2,19)$ minimal string. In red we plot the original sequences, whereas green and blue plot first and second Richardson transforms, respectively. The black lines correspond to the predicted, analytical values. In both cases the relative error at highest precision is of the order $\sim 10^{-6}$.}
\label{fig:(2,19)MS-1}
\end{figure}
%%%%%%%%%%%%%%%%%%%%%%%%%%%%%%%%%%%%%%%%%%%%%%%%%%%%%%%%%%%%%%%%%

%%%%%%%%%%%%%%%%%%%%%%%%%%%%%%%%%%%%%%%%%%%%%%%%%%%%%%%%%%%%%%%%%
\begin{figure}[t!]
\centering
     \begin{subfigure}[h]{0.49\textwidth}
         \centering
         \includegraphics[width=\textwidth]{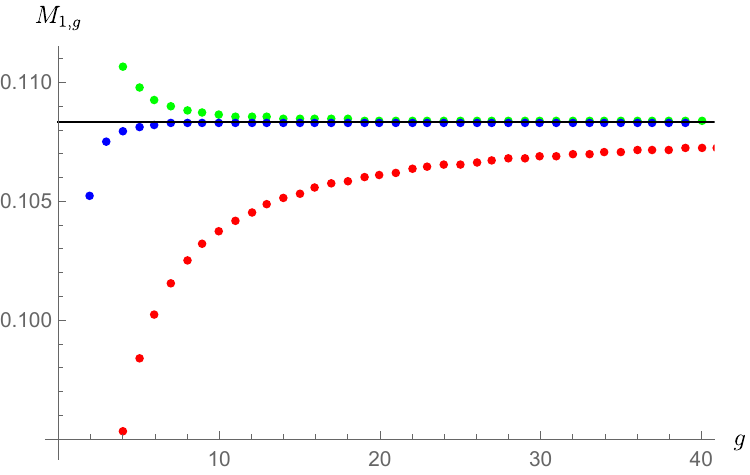}
     \end{subfigure}
\hspace{0mm}
     \begin{subfigure}[h]{0.49\textwidth}
         \centering
         \includegraphics[width=\textwidth]{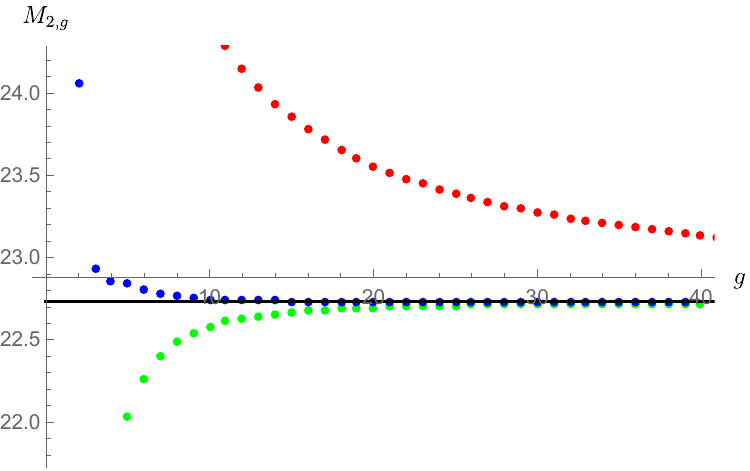}
     \end{subfigure}
\caption{These images plot the sequences $M_{1,g}$ in \eqref{eq:M1g} (left) and $M_{2,g}$ in \eqref{eq:M2g} (right) for the $(2,19)$ minimal string. Red plots the original sequences, whereas green and blue plot first and second Richardson transforms, respectively. The black lines correspond to the predicted, analytical values. Last-genus relative errors are of orders $\sim 10^{-6}$ (left) and $\sim 10^{-5}$ (right).}
\label{fig:(2,19)MS-2}
\end{figure}
%%%%%%%%%%%%%%%%%%%%%%%%%%%%%%%%%%%%%%%%%%%%%%%%%%%%%%%%%%%%%%%%%

%%%%%%%%%%%%%%%%%%%%%%%%%%%%%%%%%%%%%%%%%%%%%%%%%%%%%%%%%%%%%%%%%
\begin{figure}[t!]
\centering
     \begin{subfigure}[h]{0.49\textwidth}
         \centering
         \includegraphics[width=\textwidth]{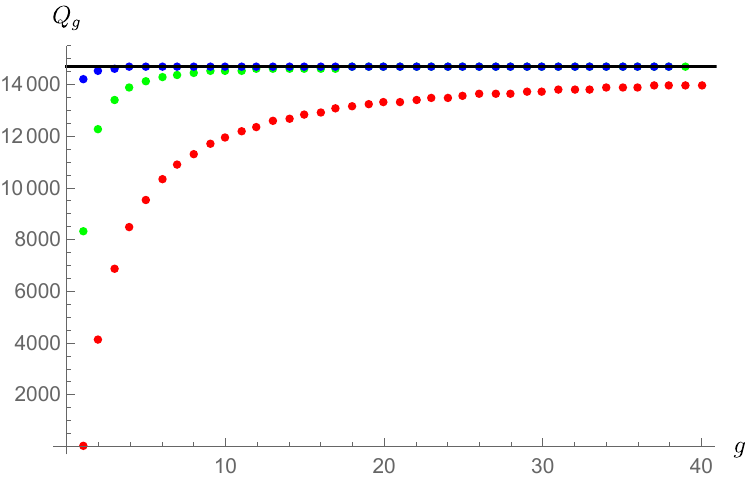}
     \end{subfigure}
\hspace{0mm}
     \begin{subfigure}[h]{0.49\textwidth}
         \centering
         \includegraphics[width=\textwidth]{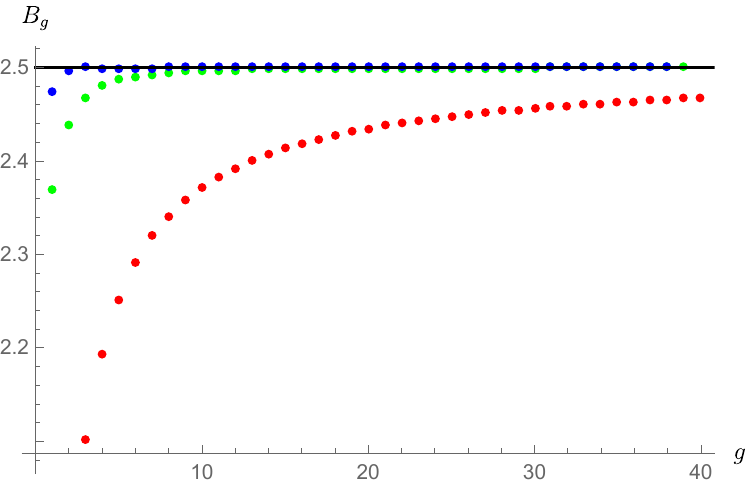}
     \end{subfigure}
\caption{These images plot the sequences $Q_g$ in \eqref{eq:Qg} (left) and $B_g$ in \eqref{eq:Bg} (right) for the $(2,39)$ minimal string. In red we plot the original sequences, whereas green and blue plot first and second Richardson transforms, respectively. The black lines correspond to the predicted, analytical values. In both cases the relative error at highest precision is of the order $\sim 10^{-6}$.}
\label{fig:(2,39)MS-1}
\end{figure}
%%%%%%%%%%%%%%%%%%%%%%%%%%%%%%%%%%%%%%%%%%%%%%%%%%%%%%%%%%%%%%%%%

%%%%%%%%%%%%%%%%%%%%%%%%%%%%%%%%%%%%%%%%%%%%%%%%%%%%%%%%%%%%%%%%%
\begin{figure}[t!]
\centering
     \begin{subfigure}[h]{0.49\textwidth}
         \centering
         \includegraphics[width=\textwidth]{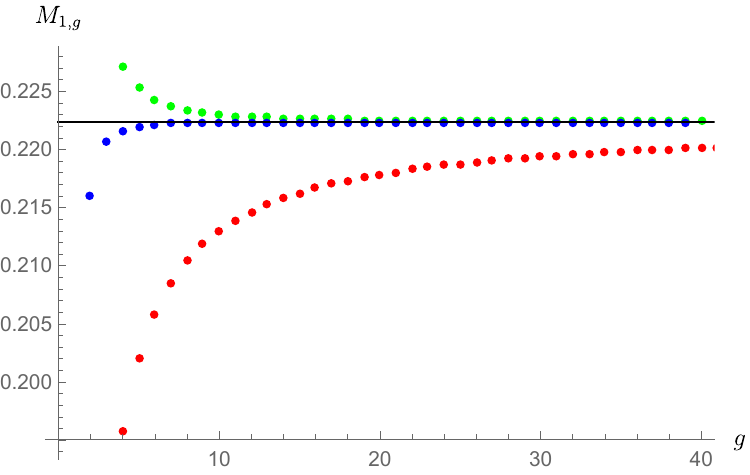}
     \end{subfigure}
\hspace{0mm}
     \begin{subfigure}[h]{0.49\textwidth}
         \centering
         \includegraphics[width=\textwidth]{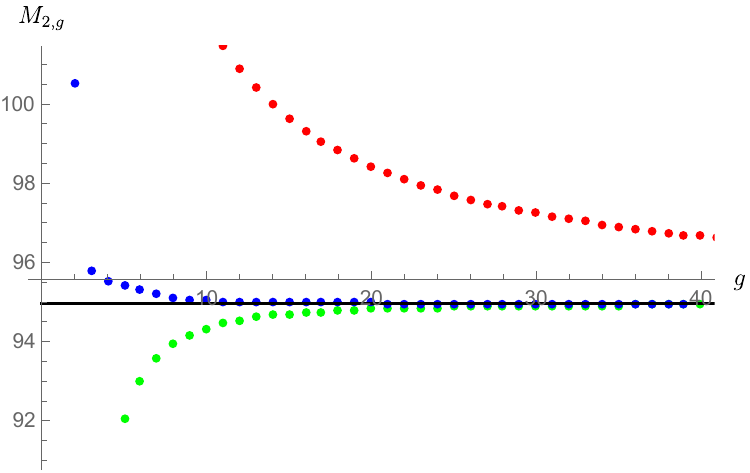}
     \end{subfigure}
\caption{These images plot the sequences $M_{1,g}$ in \eqref{eq:M1g} (left) and $M_{2,g}$ in \eqref{eq:M2g} (right) for the $(2,39)$ minimal string. Red plots the original sequences, whereas green and blue plot first and second Richardson transforms, respectively. The black lines correspond to the predicted, analytical values. Last-genus relative errors are of orders $\sim 10^{-6}$ (left) and $\sim 10^{-5}$ (right).}
\label{fig:(2,39)MS-2}
\end{figure}
%%%%%%%%%%%%%%%%%%%%%%%%%%%%%%%%%%%%%%%%%%%%%%%%%%%%%%%%%%%%%%%%%

In light of the previous subsection, running the algorithm to compute perturbative data of $(2,2k-1)$ minimal strings is now absolutely straightforward. The ``monomial basis'' KdV times for minimal strings were already introduced in subsection~\ref{subsec:JT_string_eq}, in \eqref{eq:widetildetpk_monomial-KdV}, and we recall them herein
\be
\widetilde{t}_p (k) = \left(-1\right)^{k+1} \frac{\left(2k-1\right) \Gamma \left(k+p-1\right)}{2^{p+\frac{1}{2}}\, \Gamma \left(p+1\right) \Gamma \left(p\right) \Gamma \left(k-p+1\right)}.
\ee
\noindent
The resulting $f_2(y)$ function then takes the form
\be
f_2(y) = \sum_{p=2}^{k} \left(-1\right)^{p} \frac{\Gamma \left(k+p-1\right)}{\Gamma \left(p-1\right) \Gamma \left(p\right) \Gamma \left(k-p+1\right)}\, y^{p-2}.
\ee
\noindent
Plugging it into the general formula \eqref{eq:g=2-zograf}, and using the expression for $\widetilde{t}_1(k)$ of $(2k-1)$ minimal strings, \textit{i.e.},
\be
\widetilde{t}_1(k) = \frac{2k-1}{\sqrt{2}},
\ee
\noindent
we immediately obtain the genus-two perturbative free-energy
\be
F_2 (k) = \frac{1}{8640}\, \frac{k \left(k-1\right)}{\left(2k-1\right)^2} \left(30 - 67 k + 110 k^2 - 86 k^3 + 43 k^4\right).
\ee
\noindent
This is in perfect agreement with the result from topological recursion, \eqref{eq:minimal_free_en-2}.

%%%%%%%%%%%%%%%%%%%%%%%%%%%%%%%%%%%%%%%%%%%%%%%%%%%%%%%%%%%%%%%%%
\begin{figure}[t!]
\centering
     \begin{subfigure}[h]{0.49\textwidth}
         \centering
         \includegraphics[width=\textwidth]{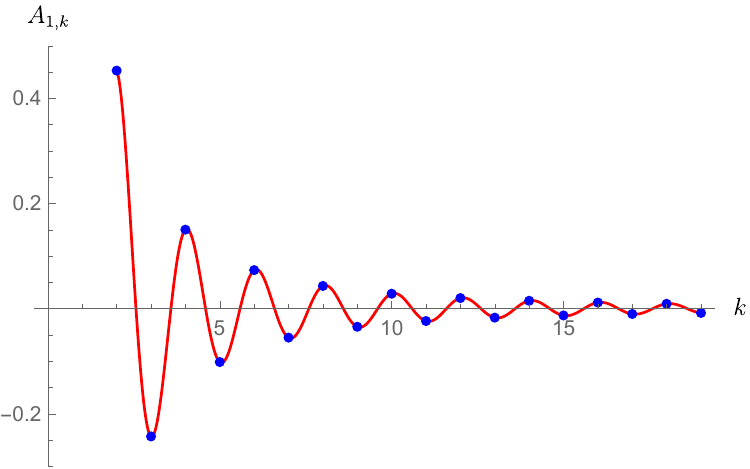}
     \end{subfigure}
\hspace{0mm}
     \begin{subfigure}[h]{0.49\textwidth}
         \centering
         \includegraphics[width=\textwidth]{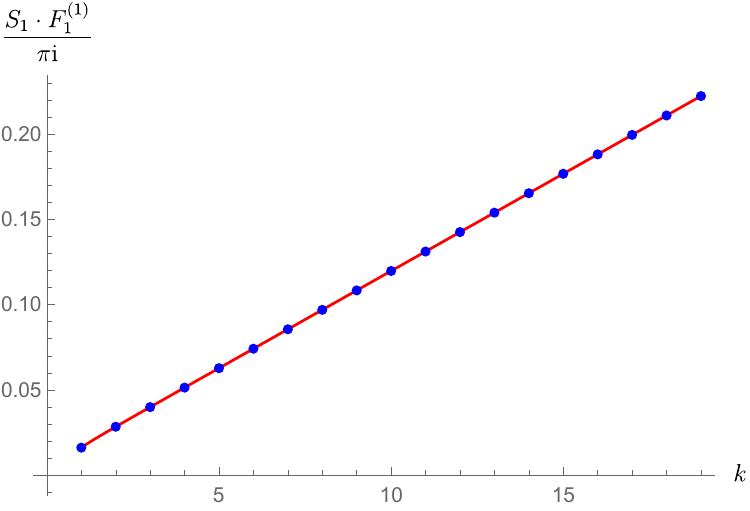}
     \end{subfigure}
\caption{For varying values of $k$ from $2$ to $20$, we plot in blue the last value (\textit{i.e.}, highest genus) of the second Richardson transform of the sequences $Q_g$ in \eqref{eq:Qg} (left plot) and $M_{1,g}$ in \eqref{eq:M1g} (right plot). In red we then plot the functions predicting the minimal-string instanton actions in \eqref{eq:min-string-action-n,k} (left) and Stokes coefficients in \eqref{eq:min-string-stokes-n,k} (right). The agreement is extremely clear.}
\label{fig:k-func-plots}
\end{figure}
%%%%%%%%%%%%%%%%%%%%%%%%%%%%%%%%%%%%%%%%%%%%%%%%%%%%%%%%%%%%%%%%%

As we did in the case of minimal models, we ran the algorithm up to $g=40$ for values of $k$ running from $2$ to $20$ in order to perform numerical tests of the many nonperturbative data we computed in sections~\ref{sec:spec_curv} and \ref{sec:string_eqs}. As we already mentioned, the case of $(2,2k-1)$ minimal strings is slightly simpler than that of multicritcal models with $k>2$: for all values of $k$ we can construct the sequences \eqref{eq:Qg}-\eqref{eq:Bg}-\eqref{eq:M1g}-\eqref{eq:M2g} and use Richardson transforms in order to improve on their convergence. We plotted our results for $k=3$ in figures~\ref{fig:(2,5)MS-1} and~\ref{fig:(2,5)MS-2}, for $k=10$ in figures~\ref{fig:(2,19)MS-1} and~\ref{fig:(2,19)MS-2}, and for $k=20$ in figures~\ref{fig:(2,39)MS-1} and~\ref{fig:(2,39)MS-2}. All of them show excellent agreement with the nonperturbative data we obtained analytically in the previous sections. Finally, since in the case of $(2,2k-1)$ minimal strings we also have at our disposal formulae \eqref{eq:min-string-action-n,k} and \eqref{eq:min-string-stokes-n,k}, which give us the instanton action and the one-loop contribution (with Stokes coefficient included) explicitly as functions of $k$, we plotted them both in figure~\ref{fig:k-func-plots} together with the corresponding numerical data at varying $k$. Again, all numerics support our analytic results to extreme high precision.

%%%%%%%%%%%%%%%%%%%%%%%%%%%%%%%%%%%%%%%%%%%%%%%%%%%%%%%%%%%%%%%%%
%%%%%%%%%%%%%%%%%%%%%%%%%%%%%%%%%%%%%%%%%%%%%%%%%%%%%%%%%%%%%%%%%
\section{Black Hole Geometries and Their Perturbations}\label{sec:black_holes}
%%%%%%%%%%%%%%%%%%%%%%%%%%%%%%%%%%%%%%%%%%%%%%%%%%%%%%%%%%%%%%%%%
%%%%%%%%%%%%%%%%%%%%%%%%%%%%%%%%%%%%%%%%%%%%%%%%%%%%%%%%%%%%%%%%%

There are many reasons for the recent great deal of interest in JT gravity. Among all these, perhaps most are, in some sense, strictly two dimensional---for instance its seminal role in the uncovering and understanding of ensemble holography; and its contender-status to an exactly-solvable theory of quantum gravity. But there are also \textit{higher-dimensional} reasons to be interested in JT gravity, as it describes the throat-dynamics of near-extremal (low temperature) black holes in dimension $d>2$, \textit{e.g.}, see \cite{ao93, fnn00, nsstv18, mtv18, mstv19a, gmt19, it20, hitz20}. In this way, the same now holds true for minimal string theories: they are no longer only interesting as ``very solvable'' 2d string theories, but, when regarded as deformations of JT gravity, they gain new light as further describing deformations of the aforementioned JT-like higher-dimensional\footnote{Which means we will take the standpoint of avoiding arguments of the type \cite{kms04} in this section.} black holes; see \cite{w20a, t:mt20, h:mt20, w20b, tuw20}. This is what we shall turn to next. Note that, with the exception of some comments in subsection~\ref{subsec:deformations}, up to now we have been mainly interested in building our way up from minimal strings towards JT gravity. In this section we shall be interested in the exact opposite direction, \textit{i.e.}, solely regarding minimal strings as \textit{deformations} of JT gravity.  

The (Euclidean) JT-action describing 2d dilaton-gravity is given by \cite{t83, j84}
\be
\label{eq:JT-action}
\CS_{\text{JT}} = - S_0\, \chi (M) - \frac{1}{2} \int_M \rmd^2 x\, \sqrt{g}\, \phi \left( R+2 \right) - \int_{\partial M} \rmd s\ \sqrt{h}\, \phi \left( K-1 \right).
\ee
\noindent
Herein $M$ is the 2d bulk manifold with boundary $\partial M$, $2\pi\chi (M) = \frac{1}{2} \int_M \rmd^2 x\, \sqrt{g}\, R + \int_{\partial M} \rmd s\, \sqrt{h}\, K$ its (topological\footnote{Hence why \textit{pure} 2d gravity is trivial.}) Euler characteristic, and $\phi$ is the dilaton field---which may be regarded as a Lagrange multiplier enforcing bulk hyperbolicity. The non-trivial dynamics then takes place on the one-dimensional fluctuating boundary, as described by the Schwarzian action (see, \textit{e.g.}, \cite{t20, f20} for reviews and references). There is a rather natural deformation of standard JT gravity, with no additional fields but where the linear dilaton is replaced by a non-trivial potential $W(\phi)$ \cite{w20a, h:mt20, w20b}. This is given by
\be
\label{eq:deformed-JT-action}
\CS_{\text{JT-}W} = \cdots - \frac{1}{2} \int_M \rmd^2 x\, \sqrt{g} \left( \phi R + W(\phi) \right) - \cdots.
\ee
\noindent
Herein we are only displaying the required bulk dilaton modification to the standard JT-action \eqref{eq:JT-action}. Writing $W(\phi) = 2\phi + U(\phi)$ to make direct contact with the JT-action \eqref{eq:JT-action}, the spacetime $U(\phi)$ deformation translates\footnote{The present formula assumes $U(0)=0$; see \cite{w20b} for details otherwise.} to a very precise matrix model deformation\footnote{At least in the one-cut phase where the deformation is taken to be small \cite{w20b}. Not much is yet known about possible multi-cut phases of these matrix models.}, and vice-versa \cite{w20b},
\be
\rho_{0,U} (\lambda) = \rme^{S_0} \left\{ \frac{1}{4\pi^2} \sinh \left( 2\pi \sqrt{\lambda} \right) + \frac{1}{8\pi\sqrt{\lambda}} \left( \rme^{2\pi\sqrt{\lambda}}\, U (\sqrt{\lambda}) + \rme^{-2\pi\sqrt{\lambda}}\, U (-\sqrt{\lambda}) \right) \right\}.
\ee
\noindent
It is precisely in the ``vice-versa'' direction which we are interested in, so as to write down the spacetime actions associated to the minimal-string-theoretic models we have studied earlier and that, in particular, we wrote as deformations of the JT spectral-curve back in subsection~\ref{subsec:deformations}. In this scenario, it turns out that the dilaton potential associated to the spacetime action of the $(2,2k-1)$ minimal string model takes the form \cite{t:mt20, tuw20} (see as well \cite{koy17} for an earlier treatment of the same action in an unrelated context)
\be
\label{eq:minimal-deformed-JT-action}
\CS_{\text{JT-}k} = \cdots - \frac{1}{2} \int_M \rmd^2 x\, \sqrt{g} \left( \phi R + \frac{2k-1}{2\pi} \sinh \frac{4\pi\phi}{2k-1} \right) - \cdots.
\ee
\noindent
It is a straightforward exercise to check that---as expected---this (bulk dilaton) action reduces to the (bulk dilaton) JT action \eqref{eq:JT-action} in the $k \to +\infty$ limit.

%%%%%%%%%%%%%%%%%%%%%%%%%%%%%%%%%%%%%%%%%%%%%%%%%%%%%%%%%%%%%%%%%
\subsection{Linear Perturbations of JT Gravity}\label{subsec:JT_perturb}
%%%%%%%%%%%%%%%%%%%%%%%%%%%%%%%%%%%%%%%%%%%%%%%%%%%%%%%%%%%%%%%%%

Linear perturbations of spacetime geometries have a long and rich history, but one which is too large to review herein---we refer the reader to, \textit{e.g.}, \cite{ns04} for a brief overview and list of references. For stable geometries, the ``ring down'' return to equilibrium is characterized by damped, single frequency oscillations; the so-called \textit{quasinormal modes}---again we refer the reader to \cite{ns04} for overview and references. Let us just mention in passing that the general $d$-dimensional stability problem was only settled somewhat recently \cite{ik03a, ik03b, ik03c} (see \cite{ik11} for a review), with their associated (asymptotic) quasinormal frequencies addressed in \cite{cns04, ns04}; with their corresponding gravitational greybody-factors computed in \cite{hns07}; and with the extension to (higher-dimensional) string-theoretic dilaton-gravity worked out in \cite{ms06}. Of course in the present 2d dilaton-gravity context, the problem is much simpler \cite{kkm04}---and has in fact already been partially addressed in the JT gravity context \cite{bsb20}, which we shall build-upon in the following.

To set the stage, let us remain in the $d$-dimensional (non-dilatonic) setting for a paragraph. Neglecting the precise form of the spacetime action (we shall focus on 2d deformed JT-gravity \eqref{eq:deformed-JT-action}-\eqref{eq:minimal-deformed-JT-action} below), consider a spherically-symmetric spacetime metric solution
\be
\mathsf{g} = - f(r)\, \rmd t \otimes \rmd t + \frac{\rmd r \otimes \rmd r}{f(r)} + r^2\, \rmd\Omega_{d-2}^2
\ee
\noindent
(also ignoring the precise form of $f(r)$; explicit formulae will follow). In this background geometry, the wave-equation for a massless, uncharged, scalar field $\upvarphi$, is simply
\be
\frac{1}{\sqrt{-g}} \partial_\mu \left( \sqrt{-g}\, g^{\mu\nu} \partial_{\nu} \upvarphi \right) = 0
\ee
\noindent
(with $g = \det g_{\mu\nu}$). Decomposing $\upvarphi$ in $d$-dimensional spherical harmonics and then further isolating the radial component $\psi_\omega (r)$ via a plane-wave expansion in $\rme^{\rmi\omega t}$, the above wave-equation gets recast as a Schr\"odinger-like equation \cite{cdl02}:
\be
\label{eq:d-dim-tortoise-Schrodinger}
- \frac{\rmd^2\psi_\omega}{\rmd x^2} (x) + V(x)\, \psi_\omega (x) = \omega^2 \psi_{\omega} (x).
\ee
\noindent
In here $x$ is the \textit{tortoise coordinate}, which---just like the potential $V(x)$---is determined from the function $f(r)$ in the background spacetime metric,
\be
\rmd x = \frac{\rmd r}{f(r)}.
\ee
\noindent
This (very useful) coordinate has the features of keeping asymptotic infinity (or the cosmological horizon if in a de~Sitter geometry) at $x=+\infty$, and sending a black-hole (outer) event-horizon to $x=-\infty$ (finding $f(x) > 0$, $\forall x \in \BR$). We will be explicit on the potential $V(x)$ below. The \textit{discrete} component of the spectrum of \eqref{eq:d-dim-tortoise-Schrodinger} is obtained when imposing outgoing boundary conditions both at asymptotic infinity and black-hole horizon, \textit{i.e.},
\bea
\lim_{x \to -\infty} \psi_\omega (x) &\sim& \rme^{+\rmi\omega x}, \\
\lim_{x \to +\infty} \psi_\omega (x) &\sim& \rme^{-\rmi\omega x}.
\eea
\noindent
The role of asymptotic infinity in the case of AdS is of course different, as it gets replaced by the AdS boundary---effectively acting as an infinite-well (see \cite{ns04} and below). As such, one natural requirement becomes that of vanishing boundary conditions at the wall of the AdS ``box''. This discrete spectrum $\omega_n \in \BC$ is the quasinormal spectrum (again, see, \textit{e.g.}, \cite{ns04} for further details).

The 2d dilaton-gravity version of this black-hole problem was addressed in \cite{kkm04} (and more recently applied specifically to JT gravity in \cite{bsb20}) and is very simple. The linear-dilaton black-hole Schwarzschild-like solution to bulk JT gravity \eqref{eq:JT-action} is \cite{lk94, ao93}
\bea
\label{eq:JT-BH-g}
\mathsf{g} &=& - f(r)\, \rmd t \otimes \rmd t + \frac{\rmd r \otimes \rmd r}{f(r)} \equiv - \left( r^2-2M \right) \rmd t \otimes \rmd t + \frac{\rmd r \otimes \rmd r}{r^2-2M}, \\
\label{eq:JT-BH-phi}
\phi &=& r,
\eea
\noindent
with $M$ its mass and $R_{\text{h}} = \sqrt{2M}$ its horizon radius. The tortoise coordinate follows,
\be
\label{eq:JT-BH-tortoise}
x = \int \frac{\rmd r}{f(r)} = - \frac{1}{R_{\text{h}}}\, \text{arccoth} \left( \frac{r}{R_{\text{h}}} \right).
\ee
\noindent
The wave-equation for a massless, uncharged, scalar field now further includes a dilaton-coupling\footnote{This coupling is best understood from a higher-dimensional perspective \cite{kkm04}.} $\CF(\phi)$ as \cite{kkm04}
\be
\frac{1}{\sqrt{-g}\, \CF(\phi)} \partial_\mu \left( \sqrt{-g}\, \CF(\phi)\, g^{\mu\nu} \partial_{\nu} \upvarphi \right) = 0.
\ee
\noindent
With the natural choice $\CF (\phi) = \phi$, one may isolate the equation for the radial component of $\upvarphi$, $\psi_\omega (r)$, as the Schr\"odinger-like \eqref{eq:d-dim-tortoise-Schrodinger} with potential given by \cite{kkm04, bsb20} (where $r$ is implicitly a function of $x$)
\bea
\label{eq:JT-potential}
V (r) &=& \frac{f}{2\CF} \left\{ f \CF'' + f' \CF' - \frac{f}{2\CF} \left( \CF' \right)^2 \right\} = \\
&=& \frac{3}{4} r^2 - \frac{1}{2} R_{\text{h}}^2 - \frac{R_{\text{h}}^4}{4 r^2} = \\
\label{eq:JT-potential-3}
&=& \frac{3 R_{\text{h}}^2}{4 \sinh^2 \left( R_{\text{h}}\, x \right)} + \frac{R_{\text{h}}^2}{4 \cosh^2 \left( R_{\text{h}}\, x \right)}.
\eea
\noindent
In the second equality we specified for the black-hole solution \eqref{eq:JT-BH-g}-\eqref{eq:JT-BH-phi}, and in the third equality we replaced for the black-hole tortoise coordinate \eqref{eq:JT-BH-tortoise}. It was shown in \cite{bsb20} that the Schr\"odinger-like equation \eqref{eq:d-dim-tortoise-Schrodinger} with above potential \eqref{eq:JT-potential-3} may be rewritten in hypergeometric form, which then essentially yields, upon adequate Schwarzschild-AdS boundary conditions, the (purely imaginary) quasinormal spectrum of this geometry:
\be
\label{eq:JT-BH-QNM}
\omega_n = - 2 \rmi R_{\text{h}}\, n, \qquad n \in \BN^{\times}.
\ee

Some of the most relevant quantities of interest in semi-classical black-hole physics include their Hawking temperature and entropy. These were addressed for general 2d dilaton-gravity theories in \cite{gkl94} (see also \cite{w20a}), where it was shown that both quantities depend only upon the value of the dilaton field at the black-hole horizon. For the black-hole solution \eqref{eq:JT-BH-g}-\eqref{eq:JT-BH-phi}, they naturally turn out to depend only on the horizon radius $R_{\text{h}}$, as
\bea
\label{eq:BH-temp}
T_{\text{BH}} &=& \frac{1}{4\pi}\, W (R_{\text{h}}), \\
\label{eq:BH-entr}
S_{\text{BH}} &=& 2\pi R_{\text{h}}.
\eea 
\noindent
In the case of JT gravity, these become
\bea
T_{\text{JT-BH}} &=& \frac{1}{2\pi} \sqrt{2M},\\
S_{\text{JT-BH}} &=& 2\pi \sqrt{2M}.
\eea 
\noindent
Finally, it was shown in \cite{w20a} that these types of black hole solutions exist if and only if the dilaton potential satisfies
\be
\label{eq:exist-cond}
\int_{\phi_{\text{h}}}^{\phi} \rmd\phi'\, W(\phi') > 0, \qquad \forall_{\phi > \phi_{\text{h}}}.
\ee
\noindent
Further, out of such solutions, the only thermodynamically stable ones are those which satisfy \cite{w20a}
\be
\label{eq:therm-cond}
W'(\phi_{\text{h}}) > 0.
\ee
\noindent
Both these conditions are clearly satisfied in the present case of JT gravity.

On top of the above black hole discussion, it is also interesting to consider the purely AdS$_2$ solution to bulk JT gravity---which is basically still of the form \eqref{eq:JT-BH-g} but with different function $f(r)$. One now has:
\bea
\label{eq:JT-AdS-g}
\mathsf{g} &=& - \left( 1 + \abs{\Lambda}\, r^2 \right) \rmd t \otimes \rmd t + \frac{\rmd r \otimes \rmd r}{1 + \abs{\Lambda}\, r^2}, \\
\label{eq:JT-AdS-phi}
\phi &=& r,
\eea
\noindent
herein parametrized by the negative cosmological-constant\footnote{Which we had normalized $\Lambda = -2$ in the JT action \eqref{eq:JT-action}, but which we will keep explicit in the following.} $\Lambda$. This solution immediately  yields the tortoise coordinate
\be
x = \frac{1}{\sqrt{\abs{\Lambda}}}\, \arctan \left( \sqrt{\abs{\Lambda}}\, r \right).
\ee
\noindent
This further implies that, in terms of this coordinate, space is restricted to the finite interval $(0,\frac{\pi}{2\sqrt{\abs{\Lambda}}})$. The potential associated to the corresponding Schr\"odinger-like equation \eqref{eq:d-dim-tortoise-Schrodinger} now takes the form
\bea
\label{eq:JT-potential-AdS}
V (r) &=& \frac{3}{4} \Lambda^2\, r^2 + \frac{1}{2} \abs{\Lambda} - \frac{1}{4r^{2}} = \\
\label{eq:JT-potential-AdS-2}
&=& \frac{3\abs{\Lambda}}{4 \cos^2 \left( \sqrt{\abs{\Lambda}}\, x \right)} - \frac{\abs{\Lambda}}{4 \sin^2 \left( \sqrt{\abs{\Lambda}}\, x \right)}.
\eea

\begin{figure}[t!]
\centering
         \includegraphics[width=0.65\textwidth]{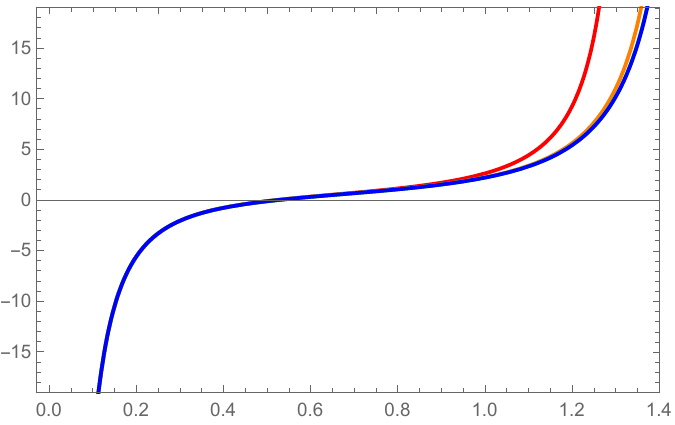}
\caption{Plots of the Schr\"odinger potential $V(x)$ in \eqref{eq:d-dim-tortoise-Schrodinger} for the AdS$_{2}$ solution of the $(2,2k-1)$ minimal string. With cosmological constant fixed as $\Lambda=-1$, we plot the JT gravity potential \eqref{eq:JT-potential-AdS-2} in blue, and then the minimal string potentials \eqref{eq:MS-potential-AdS} for $k=2$ (red), $k=6$ (orange), and $k=26$ (green). This last one is unseen to the naked-eye, basically overlapped by the JT result.}
\label{fig:ads_pot}
\end{figure}

The comparison of this potential with its higher-dimensional cousins is interesting. Higher-dimensional metric-perturbations ($d\geq 4$) typically split into three types: tensor, vector, and scalar-type perturbations \cite{ik03a}; and for pure AdS backgrounds this results in infinite-well-like potentials (with rounded edges). In special dimensions, however, this generic result may come with some tweaks---and in dimension $d=5$, scalar-type perturbations actually lead to an ``asymmetric'' infinite-well, which is not bounded from below at the AdS boundary \cite{ns04}. Interestingly enough, the pure AdS$_{2}$ JT-gravity potential \eqref{eq:JT-potential-AdS-2} is precisely of this type, this time around not bounded from below at the AdS center---this is illustrated in figure~\ref{fig:ads_pot}. But, moreover, purely-AdS $d=5$ scalar-type perturbations (with vanishing boundary conditions at the AdS wall) yield a \textit{continuous normal mode} spectrum \cite{ns04}. Interestingly enough the \textit{same} also happens herein. Consider the Schr\"odinger-like equation \eqref{eq:d-dim-tortoise-Schrodinger} with the above potential \eqref{eq:JT-potential-AdS-2}. Introducing the coordinate
\be
z = \sin^{2} \left( \sqrt{\abs{\Lambda}}\, x \right)
\ee
\noindent
this Schr\"odinger equation gets reduced to a hypergeometric form; immediately yielding the general solution
\bea
\psi_{\omega} (z) &=& \left(1-z\right)^{a-\frac{1}{4}}\, z^{\frac{1}{4}-b} \left\{ C_1 \times {}_{2}F_1 \left(\left.a-b-\frac{\omega}{2\sqrt{\left|\Lambda\right|}},a-b+\frac{\omega}{2\sqrt{\left|\Lambda\right|}};1-2b\,\right|z\right) + \right. \nonumber \\
&&
\left. + C_2 \times z^{2b}\, {}_{2}F_1 \left(\left.a+b-\frac{\omega}{2\sqrt{\left|\Lambda\right|}},a+b+\frac{\omega}{2\sqrt{\left|\Lambda\right|}};1+2b\,\right|z\right) \right\},
\eea
\noindent
with
\be
a = \frac{1}{4} \left(2-\rmi\sqrt{2}\right), \qquad b = \frac{\sqrt{2}}{4},
\ee
\noindent
and where $C_1$ and $C_2$ are integration constants. In order to have a solution which is regular at the origin, $z=0$, we have to set $C_1=0$. The second condition, namely that the solution needs to vanish at the boundary, $z=1$, is then automatically satisfied. This is in complete parallel with what happened in \cite{ns04}. Therefore, there is no restriction on the spectrum of linear perturbations, much like in the case of scalar-type perturbations in $d=5$ \cite{ns04}, and the \textit{normal}-mode spectrum is continuous,
\be
\label{eq:JT-AdS2-QNM}
\omega \in \BR.
\ee

What we wish to address next are the possible minimal-string deformations---as in \eqref{eq:minimal-deformed-JT-action}---to these above results. We start with the pure AdS case, and  follow-up to the black hole problem.

%%%%%%%%%%%%%%%%%%%%%%%%%%%%%%%%%%%%%%%%%%%%%%%%%%%%%%%%%%%%%%%%%
%%%%%%%%%%%%%%%%%%%%%%%%%%%%%%%%%%%%%%%%%%%%%%%%%%%%%%%%%%%%%%%%%

%%%%%%%%%%%%%%%%%%%%%%%%%%%%%%%%%%%%%%%%%%%%%%%%%%%%%%%%%%%%%%%%%
\subsection{Perturbations of JT Gravity and Minimal AdS Geometries}\label{subsec:AdS_perturb}
%%%%%%%%%%%%%%%%%%%%%%%%%%%%%%%%%%%%%%%%%%%%%%%%%%%%%%%%%%%%%%%%%

Let us first address the $(2,2k-1)$ minimal-string deformation of the above AdS$_2$ solution to bulk JT gravity, \eqref{eq:JT-AdS-g}. For generic potential, $W(\phi)$, the solution is still of the form \eqref{eq:JT-BH-g}-\eqref{eq:JT-BH-phi} only that the function $f(r)$ appearing in the metric becomes \cite{lk94, gkl94, kkm04, w20a}
\be
f(r) = J(\phi) + C, \qquad \text{with} \qquad J(\phi) = \int \rmd \phi\, W(\phi). 
\ee
\noindent
The integration constant $C$ parametrizes the solution---depending on its sign it leads to either an AdS$_2$-like ($C>0$) or Schwarzschild-like ($C<0$) solution. For the $(2,2k-1)$ minimal string AdS$_2$-like solution, and after a rescaling of $r$, the ``metric function'' takes the form
\be
\label{eq:MS-AdS-metric}
f(r) = 1 - \left( 2k-1 \right)^{2} + \left( 2k-1 \right)^{2} \cosh \frac{\sqrt{2\abs{\Lambda}}\, r}{2k-1}.
\ee
\noindent
One can straightforwardly check that this leads to the undeformed result \eqref{eq:JT-AdS-g} as $k$ grows to infinity. Its associated tortoise coordinate is
\be
\label{eq:MS-AdS-tortoise}
x = \frac{\sqrt{2} \left( 2k-1 \right)}{\sqrt{\abs{\Lambda}}\, \sqrt{2 \left( 2k-1 \right)^{2} - 1}}\, \arctan \left( \sqrt{2 \left( 2k-1 \right)^2 - 1}\, \tanh \frac{\sqrt{\abs{\Lambda}}\, r}{\sqrt{2} \left( 2k-1 \right)} \right).
\ee
\noindent
Finally, using \eqref{eq:MS-AdS-metric} in \eqref{eq:JT-potential}, the potential for the $(2,2k-1)$ minimal-string AdS$_2$ Schr\"odinger-like equation \eqref{eq:d-dim-tortoise-Schrodinger} follows as
\bea
\label{eq:MS-potential-AdS}
V (r) &=& \frac{1}{4r^2}\, \left( 1 - \left( 2k-1 \right)^{2} + \left( 2k-1 \right)^{2}  \cosh \frac{\sqrt{2\abs{\Lambda}}\, r}{2k-1} \right) \times \\
&&
\times
\left( - 1 + \left( 2k-1 \right)^{2} - \left( 2k-1 \right)^{2}  \cosh \frac{\sqrt{2\abs{\Lambda}}\, r}{2k-1} + 2 \sqrt{2\abs{\Lambda}}\, r \left( 2k-1 \right) \sinh \frac{\sqrt{2\abs{\Lambda}}\, r}{2k-1} \right) = \nonumber \\
\label{eq:MS-potential-AdS-XX}
&=& \frac{3\abs{\Lambda}}{4 \cos^2 \left( \sqrt{\abs{\Lambda}}\, x \right)} - \frac{\abs{\Lambda}}{4 \sin^2 \left( \sqrt{\abs{\Lambda}}\, x \right)} + \\
&&
+ \frac{\abs{\Lambda}}{96 k^2} \left( \frac{3}{\sin^2 \left( \sqrt{\abs{\Lambda}}\, x \right)}\left( 1 - \sqrt{\abs{\Lambda}}\, x\, \cot \left( \sqrt{\abs{\Lambda}}\, x \right) \right) - \right. \nonumber \\
&&
\left.
- \frac{1}{2 \cos^4 \left( \sqrt{\abs{\Lambda}}\, x \right)} \left( 2 + 9 \sqrt{\abs{\Lambda}}\, x\, \sin \left( 2 \sqrt{\abs{\Lambda}}\, x \right) - 34 \sin^2 \left( \sqrt{\abs{\Lambda}}\, x \right) \right) \right) + \cdots. \nonumber
\eea
\noindent
This is a clear deformation of the JT AdS potential \eqref{eq:JT-potential-AdS-2}, specially as written in the second equality, \eqref{eq:MS-potential-AdS-XX}, and is very much in the spirit of the large multicritical-order expansions we considered back in subsection~\ref{subsec:deformations}; \textit{e.g.}, recall the expansion of the minimal-string spectral curve in \eqref{eq:rescal-spect-curv}. As such deformations, and without great surprise, all these $(2,2k-1)$ minimal-string (exact, \textit{i.e.}, as in \eqref{eq:MS-potential-AdS}) potentials have rather similar properties as the corresponding JT-gravity potential---and as also illustrated in figure~\ref{fig:ads_pot}. In this sense, it is also reasonable to expect that all these minimal strings will have real-continuous corresponding normal spectra, akin to the JT normal spectrum \eqref{eq:JT-AdS2-QNM}. It would be interesting to make this statement precise, and specifically identify the would-be role of the ``$k$-deformation'' in such real spectra.

%%%%%%%%%%%%%%%%%%%%%%%%%%%%%%%%%%%%%%%%%%%%%%%%%%%%%%%%%%%%%%%%%
\subsection{Perturbations of JT Gravity and Minimal Black Hole Geometries}\label{subsec:black-hole_perturb}
%%%%%%%%%%%%%%%%%%%%%%%%%%%%%%%%%%%%%%%%%%%%%%%%%%%%%%%%%%%%%%%%%

\begin{figure}[t!]
\centering
         \includegraphics[width=0.65\textwidth]{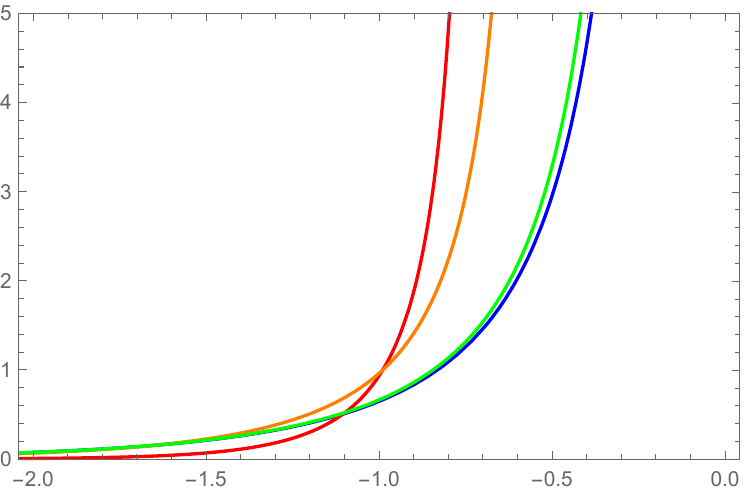}
\caption{Plots of the Schr\"odinger potential $V(x)$ in \eqref{eq:d-dim-tortoise-Schrodinger} for the black holes of the $(2,2k-1)$ minimal string. With black-hole mass fixed as $M=\frac{1}{2}$, we plot the JT gravity potential \eqref{eq:JT-potential-3} in blue, and  minimal string potentials \eqref{eq:MS-potential} for $k=2$ (red), $k=6$ (orange), and $k=26$ (green).}
\label{fig:schw_pot}
\end{figure}

As a last point, we arrive at the black-hole Schwarzschild-like solution of the minimal string or deformed JT-action \eqref{eq:minimal-deformed-JT-action}. As always, the solution is still of the form \eqref{eq:JT-BH-g}-\eqref{eq:JT-BH-phi}, but where now the function $f(r)$ appearing in the metric is \cite{lk94, gkl94, kkm04, w20a}
\be
\label{eq:MS-f}
f(r) = \frac{(2k-1)^{2}}{8\pi^2} \left( \cosh \frac{4\pi r}{2k-1} - \cosh \frac{4\pi R_{\text{h}}}{2k-1} \right)
\ee
\noindent
(a clear deformation of \eqref{eq:JT-BH-g} at large $k$). Herein, the black-hole horizon radius is now related to the mass\footnote{Note that the $k\to+\infty$ limit is taken at fixed black-hole mass, hence the black-hole radius varies with $k$.} as
\be
R_{\text{h}} = \frac{2k-1}{4\pi}\, \arccosh \left( 1 + \frac{16 \pi^2 M}{\left(2k-1\right)^{2}} \right),
\ee
\noindent
becoming the standard $R_{\text{h}} = \sqrt{2M}$ at infinite $k$. In the following, we shall  use $M$ to parametrize the black-hole solution, as this choice yields more compact expressions. The tortoise coordinate associated to this solution is
\be
\label{eq:MS-BH-tortoise}
x = - \frac{2k-1}{\sqrt{2M}\sqrt{\left( 2k-1 \right)^{2} + 8\pi^2M}}\, \text{arctanh} \left( \frac{2 \sqrt{2M} \pi}{\sqrt{\left( 2k-1 \right)^{2} + 8\pi^2M}}\, \coth \frac{2\pi r}{2k-1} \right).
\ee
\noindent
Using the standard procedure one may then compute the Schr\"odinger potential \eqref{eq:d-dim-tortoise-Schrodinger} for the minimal-string black-hole, in both $r$ and $x$ coordinates, as
\bea
\label{eq:MS-potential}
V (r) &=& \frac{1}{256 \pi^4 r^2}\, \left( 16 \pi^2 M + \left( 2k-1 \right)^{2} - \left( 2k-1 \right)^{2} \cosh \frac{4 \pi r}{2k-1} \right) \times \\
&&
\times
\left( - 16 \pi^2 M - \left( 2k-1 \right)^{2} + \left( 2k-1 \right)^{2} \cosh \frac{4 \pi r}{2k-1} - 8 \pi r \left( 2k-1 \right) \sinh \frac{4 \pi r}{2k-1} \right) = \nonumber \\
&=& \frac{3M}{2 \sinh^2 \left( \sqrt{2M}\, x \right)} + \frac{M}{2 \cosh^2 \left( \sqrt{2M}\, x \right)} + \\
&&
+ \frac{\pi^2 M^{2}}{3 k^2} \left( \frac{3}{\cosh^2 \left( \sqrt{2M}\, x \right)} \left( 1 - \sqrt{2M}\, x\, \tanh \left( \sqrt{2M}\, x \right) \right) - \right. \nonumber \\
&&
\left.
- \frac{1}{2 \sinh^4 \left( \sqrt{2M}\, x \right)} \left( 2 + 9 \sqrt{2M}\, x\, \sinh \left( 2 \sqrt{2M}\, x \right) - 34 \cosh^2 \left( \sqrt{2M}\, x \right) \right) \right) + \cdots. \nonumber
\eea
\noindent
Akin to \eqref{eq:MS-potential-AdS-XX}, this is a clear deformation of the JT black-hole potential \eqref{eq:JT-potential-3}, and again very much in the spirit of the JT deformations of subsection~\ref{subsec:deformations} (all in all, not even so different from the previous subsection). As expected, all the exact minimal-string potentials are hence very similar to the corresponding JT-gravity black hole potential---and this is illustrated in figure~\ref{fig:schw_pot}. The minimal-string quasinormal spectra should now amount to a rather clean ``$k$-deformation'' of the spectrum in \eqref{eq:JT-BH-QNM}, and it would be interesting to compute it.

%%%%%%%%%%%%%%%%%%%%%%%%%%%%%%%%%%%%%%%%%%%%%%%%%%%%%%%%%%%%%%%%%
%%%%%%%%%%%%%%%%%%%%%%%%%%%%%%%%%%%%%%%%%%%%%%%%%%%%%%%%%%%%%%%%%

It is also worth pointing out that both conditions for existence \eqref{eq:exist-cond} and thermodynamical stability \eqref{eq:therm-cond} of our minimal-string black-hole solutions are satisfied. Moreover, one can easily compute their temperature and entropy using \eqref{eq:BH-temp} and \eqref{eq:BH-entr}. They are straightforward:
\bea
T_{\text{JT-}k} &=& \frac{2k-1}{8\pi^2}\, \sinh \frac{4\pi R_{\text{h}}}{2k-1},\\
S_{\text{JT-}k} &=& 2\pi R_{\text{h}}.
\eea 
\noindent
The existence of black hole solutions of this form is particularly tantalizing when we ponder on the $k=2$ case, corresponding to $c=0$ 2d quantum gravity which, as we have seen in section~\ref{sec:string_eqs}, is fully described at quantum-level by the Painlev\'e~I equation. From the matrix model perspective this is in fact the simplest model among those we considered so far, and therefore the most amenable to the study of its (resurgent) nonperturbative content through the techniques outlined in most of this work. This gives us strong indications that we may be able to fully describe the (mind-boggling) quantum black-hole hiding inside Painlev\'e~I in the near future.

%%%%%%%%%%%%%%%%%%%%%%%%%%%%%%%%%%%%%%%%%%%%%%%%%%%%%%%%%%%%%%%%%
\acknowledgments
We would like to thank
Salvatore~Baldino,
Bertrand~Eynard,
Elba~Garcia-Failde,
Danilo~Lewa\'{n}ski,
Adrien~Ooms,
Nicolas~Orantin,
Maximilian~Schwick,
Noam~Tamarin,
Roberto~Vega,
for useful discussions, comments and/or correspondence. RS would further like to thank CERN-TH for extended hospitality, where parts of this work were conducted. This research was supported in part by FCT-Portugal grants CAMGSD~UIDB/04459/2020 and PTDC/MAT-OUT/28784/2017. This paper is partly a result of the ERC-SyG project, Recursive and Exact New Quantum Theory (ReNewQuantum) funded by the European Research Council (ERC) under the European Union's Horizon 2020 research and innovation programme, under grant agreement 810573.
%%%%%%%%%%%%%%%%%%%%%%%%%%%%%%%%%%%%%%%%%%%%%%%%%%%%%%%%%%%%%%%%%

\newpage

%%%%%%%%%%%%%%%%%%%%%%%%%%%%%%%%%%%%%%%%%%%%%%%%%%%%%%%%%%%%%%%%%
%%%%%%%%%%%%%%%%%%%%%%%%%%%%%%%%%%%%%%%%%%%%%%%%%%%%%%%%%%%%%%%%%
\appendix
%%%%%%%%%%%%%%%%%%%%%%%%%%%%%%%%%%%%%%%%%%%%%%%%%%%%%%%%%%%%%%%%%
%%%%%%%%%%%%%%%%%%%%%%%%%%%%%%%%%%%%%%%%%%%%%%%%%%%%%%%%%%%%%%%%%

%%%%%%%%%%%%%%%%%%%%%%%%%%%%%%%%%%%%%%%%%%%%%%%%%%%%%%%%%%%%%%%%%
%%%%%%%%%%%%%%%%%%%%%%%%%%%%%%%%%%%%%%%%%%%%%%%%%%%%%%%%%%%%%%%%%
\section{Topological Recursion: Basics and Data}\label{app:top_rec}
%%%%%%%%%%%%%%%%%%%%%%%%%%%%%%%%%%%%%%%%%%%%%%%%%%%%%%%%%%%%%%%%%
%%%%%%%%%%%%%%%%%%%%%%%%%%%%%%%%%%%%%%%%%%%%%%%%%%%%%%%%%%%%%%%%%

This appendix includes a brief overview of the topological-recursion \cite{eo07a, eo08, eo09}, and presents the complete set of perturbative and nonperturbative data which we have computed for this paper (starting with the different spectral curves discussed in the main body of the text).

Consider a compact Riemann surface $\Sigma$ and two meromorphic functions $x,y : \BC \to \Sigma$, such that
\be
\CE \left ( x(\upzeta), y(\upzeta) \right) = 0, \qquad \forall \upzeta \in \BC.
\ee
\noindent
This algebraic relation defines an algebraic curve $\CE$, which we shall refer to as the spectral curve. For the purpose of this paper we can restrict ourselves to genus-zero spectral curves with only one branch-point (\textit{i.e.}, a zero of $\rmd x(\upzeta)$), located at $\upzeta=0$. In all cases we consider, we also have that this branch point is simple, which means that in its vicinity $y(\upzeta)$ behaves like a square-root
\be
y(\upzeta) \sim \sqrt{x(\upzeta)-x(0)}.
\ee
\noindent
To the spectral curve $\CE$ one can associate a family of multi-differential-forms, labelled by a pair of positive integers $g$ and $h$, which we denote by $\omega_{g,h}$. These differentials $\omega_{g,h}$ are, among other interesting properties, symplectic invariants of the spectral curve $\CE$ in the sense that they are left unchanged under $\text{PSL}(2,\BC)$ transformations of the curve;
\be
x \mapsto \frac{ax+b}{cx+d}, \qquad y \mapsto \frac{(cx+d)^{2}}{ad-bc}\, y, \qquad a,b,c,d \in \BC.
\ee

In this context, the topological recursion is a construction which allows for the computation of these symplectic invariants iteratively, as a recursion in their Euler characteristic $\chi=2-2g-h$. In fact, once we specify the two invariants with highest Euler characteristic,
\be
\label{eq:top_rec_app_init}
- 2\, \omega_{0,1} \left(\upzeta_1\right) := y \left(\upzeta_1\right) \mathrm{d} x \left(\upzeta_1\right), \qquad \omega_{0,2} \left(\upzeta_1,\upzeta_2\right) := \frac{\mathrm{d}\upzeta_1\, \mathrm{d}\upzeta_2}{(\upzeta_1-\upzeta_2)^2},
\ee
\noindent
then the multi-differentials $\omega_{g,h}$ are obtained in terms of all the other invariants with higher Euler characteristic, through the (topological recursion) recursive relation:
\bea
\omega_{g,h} \left(\upzeta_1,J\right) &=& \underset{\upzeta \to 0}{\text{Res}}\, \Bigg\{ \frac{\frac{1}{2} \int_{-\upzeta}^{\upzeta} \omega_{0,2} \left(\upzeta_{1},\bullet\right)}{\omega_{0,1} \left(\upzeta\right) - \omega_{0,1} \left(-\upzeta\right)}\, \Big( \omega_{g-1,h+1} \left(\upzeta,-\upzeta,J\right) + \nonumber \\
&&
\hspace{75pt}
+ \sum_{\substack{m+m'=g \\ I\sqcup I'=J}}' \omega_{m,|I|+1} \left(\upzeta,I\right)\, \omega_{m',|I'|+1} \left(-\upzeta,I'\right) \Big) \Bigg\}.
\label{eq:top_rec_app}
\eea
\noindent
Herein $J=\left\{ \upzeta_2, \ldots ,\upzeta_h \right\}$, and the prime in the summation indicates we should not include either one of the $\left(I=J;m=g\right)$ and $\left(I'=J; m'=g\right)$ contributions. One special simplification occurs as, for all spectral curves addressed in this paper, we always have
\be
y \left(\upzeta\right) = - y \left(-\upzeta\right) \qquad \text{and} \qquad \mathrm{d}x \left(\upzeta\right) = 4 \upzeta\, \mathrm{d}\upzeta.
\ee
\noindent
When further combined with \eqref{eq:top_rec_app_init} this implies that we may rewrite \eqref{eq:top_rec_app} in a simpler fashion,
\bea
\omega_{g,h} \left(\upzeta_1,J\right) &=& \underset{\upzeta \to 0}{\text{Res}}\, \Bigg\{ \frac{1}{4 y(\upzeta)}\, \frac{1}{\upzeta_1^2-\upzeta^{2}}\, \Big( \omega_{g-1,h+1} \left(\upzeta,-\upzeta,J\right) + \nonumber \\
&&
\hspace{75pt}
+ \sum_{\substack{m+m'=g \\ I\sqcup I'=J}}' \omega_{m,|I|+1} \left(\upzeta,I\right)\, \omega_{m',|I'|+1} \left(-\upzeta,I'\right) \Big) \Bigg\}.
\label{eq:top_rec_app2}
\eea

In the case of spectral curves associated to matrix models, which is basically the only focus in this paper, the $\omega_{g,h}$ symplectic invariants are related to the multi-resolvent correlation-functions of the matrix model defined in equations \eqref{eq:init_cond}-\eqref{eq:hatted-def} from \eqref{eq:Wh-genus}. One simply has
\be
\omega_{g,h} \left(\upzeta_{1}, \ldots, \upzeta_{h}\right) = \widehat{W}_{g,h} \left(\upzeta_{1}, \ldots, \upzeta_{h}\right) \rmd \upzeta_{1} \cdots \rmd \upzeta_{h},
\ee
\noindent
thus leading to the recursion \eqref{eq:top-rec-hat} for the $\widehat{W}_{g,h}$ used in the main text. The genus-$g$ free energies are also symplectic invariants of the spectral curve, and for $g>1$ they are obtained through the equation
\be
\label{eq:free-en-app}
\left( 2-2g \right) F_g  = \underset{\upzeta \to 0}{\text{Res}}\, \Phi \left(\upzeta\right)\, \omega_{g,1} \left(\upzeta\right),
\ee
\noindent
with $\rmd \Phi = \omega_{0,1}$---and this is clearly equivalent to \eqref{eq:free-en-hat}. Symplectic invariants obtained via the topological recursion may also be used to compute nonperturbative data associated to the instanton sectors of the matrix model, through a procedure which essentially consists in a generalization of the techniques in \cite{msw07, msw08}, and which will be further discussed elsewhere \cite{eggls23}.

%%%%%%%%%%%%%%%%%%%%%%%%%%%%%%%%%%%%%%%%%%%%%%%%%%%%%%%%%%%%%%%%%
\subsection{Topological-Recursion Perturbative Data}\label{subapp:pert}
%%%%%%%%%%%%%%%%%%%%%%%%%%%%%%%%%%%%%%%%%%%%%%%%%%%%%%%%%%%%%%%%%

Let us list all \textit{perturbative} higher-genus free-energies we have computed via the topological recursion. For the order-$k$ multicritical model introduced in subsection~\ref{subsec:multicritical}, we have computed the recursion up to genus $g=6$. The results are:
\bea
F_2 (k) &=& \frac{1}{1440 k} \left(2k+3\right) \left(k-1\right), \\
F_3 (k) &=& \frac{1}{362880 k^3} \left(2k+3\right) \left(k-1\right) \left(31 + 189 k + 196 k^2 + 4 k^3\right), \\
F_4 (k) &=& \frac{1}{43545600 k^5} \left(2k+3\right) \left(k-1\right) \times \\
&&
\times \left(2312 + 26769 k + 102342 k^2 + 151529 k^3 + 75564 k^4 - 2340 k^5 - 3376 k^6\right), \nonumber \\
F_5 (k) &=& \frac{1}{2874009600 k^7} \left(2k+3\right) \left(k-1\right) \times \\
&&
\times \left(239652 + 4008927 k + 25909952 k^2 + 82722337 k^3 + 137638194 k^4 + 112876346 k^5 + \right. \nonumber \\
&&
\left.
+ 32210864 k^6 - 7317408 k^7 - 3912160 k^8 - 177504 k^9 \right), \nonumber \\
F_6 (k) &=& \frac{1}{94152554496000 k^{9}} \left(2k+3\right) \left(k-1\right) \times \\
&&
\times \left(23512816512 + 509338440552 k + 4581584171447 k^2 + 22354103808946 k^3 + \right. \nonumber \\
&&
+ 64710382632956 k^4 + 113735064708698 k^5 +117755729609013 k^6 + \nonumber \\
&&
+ 62675952710588 k^7 + 6764765217432 k^8 - 8243508564928 k^9 - 2970481706032 k^{10} - \nonumber \\
&&
\left.
- 135640896832 k^{11} + 25626075648 k^{12} \right). \nonumber
\eea
\noindent
There are some clear patterns worth mentioning. All computed free energies are of the form
\be
F_g (k) = \frac{1}{k^{2g-3}} \left(2k+3\right) \left(k-1\right) \NCF_{g} (k),
\ee
\noindent
where $\NCF_{g} (k)$ is a polynomial in $k$ of degree $3g-6$.

Moving-on to the case of minimal strings coupled with $\left(2,2k-1\right)$ minimal matter, introduced in subsection~\ref{subsec:minimal}, we have also computed the recursion up to genus $g=6$. The results are:
\bea
F_2 (k) &=& \frac{1}{8640}\, \frac{k \left(k-1\right)}{\left(2k-1\right)^2} \left(30 - 67 k + 110 k^2 - 86 k^3 + 43 k^4\right), \\
F_3 (k) &=& \frac{1}{58060800}\, \frac{k \left(k-1\right)}{\left(2k-1\right)^4} \times \\
&&
\times \left(336000 - 1432000 k + 4251576 k^2 - 8784420 k^3+ 14206510 k^4 - 17769995 k^5 + \right. \nonumber \\
&&
\left.
+ 17500403 k^6 - 13101230 k^7 + 7247840 k^8 - 2648355 k^9 + 529671 k^{10}\right), \nonumber \\
F_4 (k) &=& \frac{1}{31603654656000} \frac{k \left(k-1\right)}{\left(2k-1\right)^6} \times \\
&&
\times \left(1280240640000 - 7604222976000 k + 29932974316800 k^2 - 86544914688000 k^3 + \right. \nonumber \\
&&
+ 202262973821712 k^4 - 391913713370816 k^5+643422480514408 k^6 - \nonumber \\
&&
- 900500072437616 k^7 + 1077995918402833 k^8 - 1099742332191848 k^9 + \nonumber \\
&&
+ 949539278079604 k^{10} - 684464640926248 k^{11} + 403772326359038 k^{12} - \nonumber \\
&&
- 188563656370936 k^{13} + 66122182536188 k^{14} - 15673806936136 k^{15} + \nonumber \\
&&
\left.
+ 1959225867017 k^{16}\right), \nonumber \\
F_5 (k) &=& \frac{1}{1001203779502080000} \frac{k \left(k-1\right)}{\left(2k-1\right)^8} \times \\
&&
\times \left(669207387340800000 - 5006018897510400000 k + 23882868863497728000 k^2 - \right. \nonumber \\
&&
- 84847230712253260800 k^3 + 245922696032336375040 k^4 - \nonumber \\
&&
- 603153147810115018944 k^5 + 1282516798936116296064 k^6 - \nonumber \\
&&
- 2395244227449449183920 k^7 + 3962246055509530839680 k^8 - \nonumber \\
&&
- 5830604743581800268404 k^9 + 7647275872601625516424 k^{10} - \nonumber \\
&&
- 8936723028122980217445 k^{11} + 9285597429340289197635 k^{12} - \nonumber \\
&&
- 8543514960714629065219 k^{13} + 6918600235434376200929 k^{14} - \nonumber \\
&&
- 4888821461355640633250 k^{15} + 2978785141156427480830 k^{16} - \nonumber \\
&&
- 1539329673189975264474 k^{17} + 658956538365811558314 k^{18} - \nonumber \\
&&
- 225544974194444314585 k^{19} + 58286999041187136815 k^{20} - \nonumber \\
&&
\left.
- 10209286177640772959 k^{21} + 928116925240070269 k^{22} \right), \nonumber \\
F_6 (k) &=& \frac{1}{721587587962739097600000} \frac{k \left(k-1\right)}{\left(2k-1\right)^{10}} \times \\
&&
\times \left(14630104828862595072000000 - 130982959897686805708800000 k + \right. \nonumber \\
&&
+ 725939422483920351436800000 k^2 - 3001709823215219451740160000 k^3 +  \nonumber \\
&&
+ 10145893884982322119923916800 k^4 - 29238203083225004250201169920 k^5 + \nonumber \\
&&
+ 73753573253353378752199507200 k^6 - 165467812512056644303593964800 k^7 + \nonumber \\
&&
+ 333711917931728631993990292224 k^8 - 609207968242216758059638490816 k^9 + \nonumber \\
&&
+ 1011276691521847523941209320224 k^{10} - 1530649889147633960278176498576 k^{11} + \nonumber \\
&&
+ 2115398059110212015747774582688 k^{12} - 2670177150035303900870800059396 k^{13} + \nonumber \\
&&
+ 3076249941453241595395568062089 k^{14} - 3229530261640508671445952867486 k^{15} + \nonumber \\
&&
+ 3081746691432956015861562234783 k^{16} - 2663579255295998554126813837656 k^{17} + \nonumber \\
&&
+ 2075511277148901128743897092009 k^{18} - 1449310360952428264234383118746 k^{19} + \nonumber \\
&&
+ 899940195927478251857413895463 k^{20} - 491963655044990471631626378436 k^{21} + \nonumber \\
&&
+ 233672385560860553169848637219 k^{22} - 94740542523985724845968068946 k^{23} + \nonumber \\ 
&&
+ 31984509124152690237334987653 k^{24} - 8666652528585983584027735776 k^{25} + \nonumber \\
&&
+ 1776350651917788131073941259 k^{26} - 246596682843508754186897446 k^{27} + \nonumber \\
&&
\left.
+ 17614048774536339584778389 k^{28} \right). \nonumber
\eea
\noindent
Also here there are some clear patterns worth mentioning. All computed free energies are of the form
\be
F_g (k) = \frac{k \left(k-1\right)}{\left(2k-1\right)^{2g-2}}\, \NCF_{g} (k),
\ee
\noindent
where $\NCF_{g} (k)$ is a polynomial in $k$ now of degree $6g-8$.

The next case concerns JT gravity, which was introduced in subsection~\ref{subsec:deformations} (whose spectral curve and associated resurgent structure will be studied in closely-related work \cite{eggls23}). The situation is now simpler as compared to the previous examples, as the free energies are no-longer polynomials (in $k$) but just numbers. As such, herein we have computed the recursion up to genus $g=7$. The results are:
\bea
F_2 &=& \frac{43\, \pi^{6}}{2160}, \\
F_3 &=& \frac{176557\, \pi^{12}}{1209600}, \\
F_4 &=& \frac{1959225867017\, \pi^{18}}{493807104000}, \\
F_5 &=& \frac{84374265930915479\, \pi^{24}}{355541114880000}, \\
F_6 &=& \frac{2516292682076619940682627\, \pi^{30}}{100667911267123200000}, \\
F_7 &=& \frac{57836500609415964441264863965730519\, \pi^{36}}{14128121232007335641088000000}.
\eea
\noindent
The only pattern worth mentioning is that all computed free energies are of the form $F_g \propto \pi^{6g-6}$.

Our final case pertains to the spectral curve we dubbed the ``Dawson curve'' in subsection~\ref{subsec:deformations} (and which is further studied in \cite{gos21}). Again, the free energies are just numbers and this is also a simpler situation as compared to multicritical models or minimal strings. We have computed the recursion up to genus $g=6$; and the results are:
\bea
F_2 &=& \frac{\alpha^{2}}{720}, \\
F_3 &=& \frac{\alpha^{4}}{45360}, \\
F_4 &=& -\frac{211 \alpha^{6}}{1360800}, \\
F_5 &=& -\frac{1849 \alpha^{8}}{14968800}, \\
F_6 &=& \frac{16683643 \alpha^{10}}{30648618000}.
\eea

%%%%%%%%%%%%%%%%%%%%%%%%%%%%%%%%%%%%%%%%%%%%%%%%%%%%%%%%%%%%%%%%%
\subsection{Topological-Recursion Nonperturbative Data}\label{subapp:nonpert}
%%%%%%%%%%%%%%%%%%%%%%%%%%%%%%%%%%%%%%%%%%%%%%%%%%%%%%%%%%%%%%%%%

Let us now list all \textit{nonperturbative} higher-genus free energies we have computed via the nonperturbative topological recursion (but we refer to \cite{eggls23} for details on these computations). These data also include instanton actions and Stokes data, as already explained in subsection~\ref{subsec:multicritical}. There are, however, two points that we need to mention. The first; one needs to start by specifying an instanton sector, and this  requires computing the respective instanton actions. Whereas in the minimal string case this may be done generically in closed-form, the same is not true for the multicritical models. This means one has to proceed step-by-step in enumerating the different instanton actions for each specific order-$k$ model. The second; the nonperturbative topological recursion starts off from a specific instanton sector, which is to say it lists coefficients around a specific, selected nonperturbative sector---and not for \textit{all} nonperturbative sectors at the same time (hence we will only look at some selected sectors, in each case).

For the case of pure 2d quantum gravity, or the $(2,3)$ multicritical theory first introduced in subsection~\ref{subsec:multicritical}, we have computed the recursion up to four loops (around the $A = \frac{4}{5}\sqrt{6}$ one-instanton sector). The results are:
\bea
A &=& \frac{4}{5}\sqrt{6}, \\
S_1 \cdot F^{(1)}_{1} &=& \frac{\rmi}{4 \cdot 6^{\frac{3}{4}} \sqrt{\pi}}, \\
\frac{F^{(1)}_{2}}{F^{(1)}_{1}} &=& -\frac{37}{32 \sqrt{6}}, \\
\frac{F^{(1)}_{3}}{F^{(1)}_{1}} &=& \frac{6433}{12288}, \\
\frac{F^{(1)}_{4}}{F^{(1)}_{1}} &=& -\frac{12741169}{5898240 \sqrt{6}}.
\eea

The one other interesting multicritical model we have addressed at length in the main body of the paper is the case of Yang--Lee edge singularity, or the $(2,5)$ multicritical theory, also first introduced in subsection~\ref{subsec:multicritical}. For this case, we have computed the recursion up to four loops (around both $A_{\pm}$ one-instanton sectors listed below). The results are:
\bea
A_{\pm} &=& \frac{6}{7}\sqrt{5\pm \rmi \sqrt{5}}, \\
S_1 \cdot F^{(1)}_{1} &=& - \frac{\left( 25 \pm 5\rmi \sqrt{5} \right)^{\frac{1}{4}}} {2\sqrt{2\pi} \left( \sqrt{5} \mp 5\rmi \right)^{\frac{3}{2}}}, \\
\frac{F^{(1)}_{2}}{F^{(1)}_{1}} &=& \frac{63 \mp 8\rmi \sqrt{5}}{48 \sqrt{5 \pm \rmi\sqrt{5}}}, \\
\frac{F^{(1)}_{3}}{F^{(1)}_{1}} &=& \frac{13375 \mp 10963 \rmi \sqrt{5}}{46080}, \\
\frac{F^{(1)}_{4}}{F^{(1)}_{1}} &=& - \frac{3125 \left( 5454855674 \sqrt{5} \pm 4381239845 \rmi \right)}{432 \sqrt{5 \pm \rmi\sqrt{5}} \left( \sqrt{5} \pm \rmi \right)^9 \left( \sqrt{5} \mp  5 \rmi \right)^{10}}.
\eea

Finally---and also for comparison with the previous example---we consider the $(2,5)$ minimal string theory example, first introduced in subsection~\ref{subsec:minimal}. For this case, we have computed the recursion up to four loops (around the $A_{(1,3)}$ one-instanton sector listed below). The results are:
\bea
A_{(1,3)} &=&\frac{5}{21} \sqrt{\frac{1}{2} \left( 5 + \sqrt{5} \right)}, \\
S_1 \cdot F^{(1)}_{1} &=& \frac{1}{2^{\frac{1}{4}} \left( 5 - \sqrt{5} \right)^{\frac{3}{4}} \sqrt{5 \left( \sqrt{5} - 1 \right) \pi}}, \\
\frac{F^{(1)}_{2}}{F^{(1)}_{1}} &=& - \frac{1}{120}\, \sqrt{\frac{107791 + 49201 \sqrt{5}}{2 \sqrt{5}}}, \\
\frac{F^{(1)}_{3}}{F^{(1)}_{1}} &=& \frac{7 \left( 196475 + 69697 \sqrt{5} \right)}{288000}, \\
\frac{F^{(1)}_{4}}{F^{(1)}_{1}} &=& - \frac{\sqrt{\frac{1}{2} \left( 5 + \sqrt{5} \right)} \left( 340372753 + 149564293 \sqrt{5} \right)}{20736000}.
\eea

%%%%%%%%%%%%%%%%%%%%%%%%%%%%%%%%%%%%%%%%%%%%%%%%%%%%%%%%%%%%%%%%%
%%%%%%%%%%%%%%%%%%%%%%%%%%%%%%%%%%%%%%%%%%%%%%%%%%%%%%%%%%%%%%%%%
\section{Resolvent of the Harmonic Oscillator}\label{app:harm_resolv}
%%%%%%%%%%%%%%%%%%%%%%%%%%%%%%%%%%%%%%%%%%%%%%%%%%%%%%%%%%%%%%%%%
%%%%%%%%%%%%%%%%%%%%%%%%%%%%%%%%%%%%%%%%%%%%%%%%%%%%%%%%%%%%%%%%%

For completeness of the presentation in the main text, let us herein briefly study the asymptotic properties and the Gel'fand--Dikii expansion of the (diagonal component of the) integral kernel of the resolvent \eqref{eq:resolvents}, in the very simple and explicit example of the harmonic oscillator. Within the conventions stated in the main text, and with frequency $\omega=2$, the harmonic potential is $u(z)=z^2$ and the hamiltonian operator is
\be
\label{eq:hamiltonian-x^2}
\mathsf{H} = - \frac{\rmd^2}{\rmd z^2} + z^2,
\ee
\noindent
with spectrum $\lambda_n = -2n-1$. The integral kernel of this operator is
\be
\label{eq:harmonic-resolvent}
R_\lambda (x,y) = \bra{x} \frac{1}{- \frac{\rmd^2}{\rmd z^2} + \left( z^2 + \lambda \right) \1} \ket{y}.
\ee

Standard textbooks usually follow a slightly different route and rather first discuss the time-evolution operator $\textsf{U} \left( t_{\text{f}}, t_{\text{i}} \right) \ket{\psi(t_{\text{i}})} = \ket{\psi(t_{\text{f}})}$ alongside its kernel (or propagator)
\be
\label{eq:qm-propagator}
\left. K \left( x, t_{x} \,\right| y, t_{y} \right) = \bra{x} \textsf{U} \left( t_{x}, t_{y} \right) \ket{y}.
\ee
\noindent
Any quantum mechanics textbook will compute the propagator for the harmonic oscillator. In our conventions this is (with $\tau \equiv \Delta t = t_{x} - t_{y}$)
\be
\left. K \left( x, t_{x} \,\right| y, t_{y} \right) = \frac{1}{\sqrt{2\pi\rmi \sin \left( 2 \tau \right)}} \exp \left\{ \frac{\rmi}{2 \sin \left( 2 \tau \right)} \left( \left( x^2 + y^2 \right) \cos \left( 2 \tau \right) - 2 x y \right) \right\}.
\ee
\noindent
The diagonal component of this kernel is simply
\be
\label{eq:harmonic-propagator}
K_{\tau} (z) \equiv \left. K \left( z, t_{x} \,\right| z, t_{y} \right) = \frac{1}{\sqrt{2\pi\rmi \sin 2\tau}} \exp \left\{ - \rmi z^2 \tan \tau \right\},
\ee
\noindent
with straightforward power-series expansion about $\tau \sim 0$ given by
\bea
K_{\tau} (z) &=& \frac{1}{\sqrt{\rmi\pi}} \left( \frac{1}{2}\, \tau^{-\frac{1}{2}} - \frac{\rmi}{2} z^2\, \tau^{\frac{1}{2}} - \frac{1}{12} \left( 3 z^4 - 2 \right) \tau^{\frac{3}{2}} + \frac{\rmi}{12} z^2 \left( z^4 - 4 \right) \tau^{\frac{5}{2}} + \right. \nonumber \\
&&
\label{eq:harmonic-propagator-expansion}
\left.
+ \frac{1}{240} \left( 5 z^8 - 60 z^4 + 12 \right) \tau^{\frac{7}{2}} - \frac{\rmi}{720} z^2 \left( 3 z^8 - 80 z^4 + 124\right) \tau^{\frac{9}{2}} + \cdots \right).
\eea
\noindent
This expression already encodes the Gel'fand--Dikii ``magic''. In fact, with $u(z) = z^2$, the first few Gel'fand--Dikii KdV potentials, \eqref{eq:GD-R0}-\eqref{eq:GD-R1}-\eqref{eq:GD-R2}-\eqref{eq:GD-R3} and so on, become:
\bea
R_0 &=& \frac{1}{2}, \\
R_1 &=& - \frac{1}{4} z^2, \\
R_2 &=& \frac{1}{16} \left( 3 z^4 - 2 \right), \\
R_3 &=& - \frac{5}{32} z^2 \left( z^4 - 4 \right), \\
R_4 &=& \frac{7}{256} \left( 5 z^8 - 60 z^4 + 12 \right), \\
R_5 &=& - \frac{21}{512} z^2 \left( 3 z^8 - 80 z^4 + 124 \right).
\eea
\noindent
It is clear that these polynomials exactly match \eqref{eq:harmonic-propagator-expansion}, at each order, up to an overall coefficient.

The final step (to verify the missing overall coefficient) is now elementary. The (harmonic) resolvent \eqref{eq:harmonic-resolvent} may be written in terms of the propagator \eqref{eq:qm-propagator} via a simple Fourier transform, from time to energy representations---hence turning the small-time expansion \eqref{eq:harmonic-propagator-expansion} into a large-energy expansion. Focusing solely on the kernel-diagonals, one has
\be
\label{eq:K-to-R-FT}
R_\lambda (z) = \rmi \int_{0}^{+\infty} \rmd\tau\, \rme^{-\rmi \lambda \tau}\, K_{\tau} (z).
\ee
\noindent
Using the explicit harmonic-oscillator formulae \eqref{eq:harmonic-propagator}-\eqref{eq:harmonic-propagator-expansion} it immediately follows
\be
R_\lambda (z) = \frac{R_{0}}{\lambda^{\frac{1}{2}}} + \frac{R_{1}}{\lambda^{\frac{3}{2}}} + \frac{R_{2}}{\lambda^{\frac{5}{2}}} + \frac{R_{3}}{\lambda^{\frac{7}{2}}}  + \frac{R_{4}}{\lambda^{\frac{9}{2}}} + \cdots,
\ee
\noindent
which is indeed the Gel'fand--Dikii asymptotic expansion \eqref{eq:resolv_asymp} \cite{gd75}, as we wanted to show. To verify that this is in fact an \textit{asymptotic} expansion (after all, \eqref{eq:harmonic-propagator-expansion} is a standard Taylor expansion) all one needs to do is to check the growth of the generic terms in the expansion, via Fourier transform \eqref{eq:K-to-R-FT} of \eqref{eq:harmonic-propagator-expansion}. The generic term in \eqref{eq:harmonic-propagator-expansion} is a polynomial in $z$, times a numerical coefficient---which does not grow factorially-fast as it arises from a Taylor expansion---, times $\tau^{\frac{2k-1}{2}}$ with $k \in \BN_0$. The Fourier transform of this monomial is
\be
\rmi \int_{0}^{+\infty} \rmd \tau\, \rme^{-\rmi \lambda \tau}\, \tau^{\frac{2k-1}{2}} = \rmi^{\frac{1}{2}-k}\, \frac{\Gamma \left( k+\frac{1}{2} \right)}{\lambda^{k+\frac{1}{2}}},
\ee
\noindent
which gives rise to factorial growth of the generic coefficient; hence the asymptotic expansion.

%%%%%%%%%%%%%%%%%%%%%%%%%%%%%%%%%%%%%%%%%%%%%%%%%%%%%%%%%%%%%%%%%
%%%%%%%%%%%%%%%%%%%%%%%%%%%%%%%%%%%%%%%%%%%%%%%%%%%%%%%%%%%%%%%%%
\newpage
%%%%%%%%%%%%%%%%%%%%%%%%%%%%%%%%%%%%%%%%%%%%%%%%%%%%%%%%%%%%%%%%%
%%%%%%%%%%%%%%%%%%%%%%%%%%%%%%%%%%%%%%%%%%%%%%%%%%%%%%%%%%%%%%%%%

\bibliographystyle{plain}
%\bibliography{papers}

\end{document}